\documentclass{appolb} 

\usepackage{hyperref}
\hypersetup{
    colorlinks,
    linkcolor={red!50!black},
    citecolor={blue!50!black},
    urlcolor={blue!80!black}
}

\usepackage{fancyhdr}
\usepackage{doi}

\usepackage{rotating} 

\usepackage[T1]{fontenc} 
\usepackage{wrapfig} 
\usepackage{amsmath}
\usepackage{xcolor}
\usepackage[title]{appendix}
\usepackage{tikz}
\newcommand*\circled[1]{\tikz[baseline=(char.base)]{
            \node[shape=circle,draw,inner sep=2pt] (char) {#1};}}

\usepackage{chngcntr}
\counterwithin{figure}{section} 
\counterwithin{table}{section} 

\usepackage{multirow}
\usepackage{soul}
\usepackage{cancel}
\usepackage{ marvosym }
\begin{document}

\eqsec  

\title{Probing nucleon spin structure in deep-inelastic scattering, proton-proton collisions and Drell-Yan processes%
\thanks{
Presented at the 61. Cracow School of Theoretical Physics (Electron-Ion Collider Physics)
}
}
\author{Caroline Riedl
\address{University of Illinois at Urbana-Champaign (USA)}
} 
\maketitle
\begin{abstract}
A pedagogical summary of current and past experimental results of spin-dependent nucleon structure prior to the arrival of the Electron-Ion Collider is attempted. After an introduction, results from fixed-target experiments at SLAC, Fermilab, Jefferson Lab, CERN and DESY and collider experiments from RHIC are presented, starting with the longitudinal spin structure of the nucleon, followed by generalized parton distributions (GPDs), which map the proton in transverse position space. The final part discusses transverse proton or parton spin and transverse parton momenta (TMDs), and their their (spin-orbit) correlations, which are addressed by a multitude of recent experimental results. The GPDs and TMDs provide complementary pathways to mapping multi-dimensional nucleon structure.
\end{abstract}
\vspace{\baselineskip} 
  

\pagestyle{fancy}

\fancyhf{}                               

\fancyhead[LE]{{\leftmark}} 
\fancyhead[RO]{\slshape\nouppercase{\rightmark}} 
\fancyfoot[LE,RO]{\bfseries\thepage}           

\vspace{10pt}
\noindent\rule{\textwidth}{1pt}
\tableofcontents
\noindent\rule{\textwidth}{1pt}
\vspace{10pt}

\section{Introduction}

\subsection{The proton is not so ordinary}

Only about 4\% of the matter of the universe are nowadays considered to be composed of ``ordinary'' Standard-Model particles. While matter made of non-Standard-Model particles has attracted considerable attention in  recent  years,  the  proton  as  prominent  constituent  of  the  visible  universe  is  not  as  well known as one may naively expect. Some fundamental questions about the proton yet remain to be answered - what is the proton’s radius?  Does the proton decay?  Why is the proton as 3-quark compound so heavy, as compared to 2-quark compounds, the mesons?  While these questions are discussed elsewhere, this document attempts to summarize the current status of the following questions: \emph{where does the proton spin come from?} Secondly, \emph{do partons undergo orbital motion?}, and lastly, \emph{what is the multi-dimensional picture of the proton in transverse-momentum and position space?}

To that end, we will first take a few steps back in history and will review the methods by that the structure of matter has been revealed during the 20th century. Then we will look at contemporary efforts of mapping proton structure. 

\subsection{Proton spin puzzle} 

The spin structure of the proton and other hadrons is experimentally explored by analyzing the types of particles and their angular distributions produced in (1) lepton-nucleon deep inelastic scattering (DIS), (2) proton-proton collisions (pp), (3) the hadron-hadron Drell-Yan (DY) process, and (4) electron-positron annihilation (ee). The proton is a spin-1/2 fermion composed of three spin-1/2 valence quarks. The initial motivation for the aforementioned broad set of measurements was the finding in the late 1980's by the EMC collaboration at CERN \cite{emc1988} that the proton's spin of 1/2 (in units of the reduced Planck constant $\hbar$) is not the result of a simple spin-algebraic combination of the spin of its three spin-1/2 valence quarks \cite{Jaffe1995}. Expressing the contribution of the quark spins as $\Delta\Sigma$, that of the gluons by $\Delta G$, and the total angular momentum of partons as $\mathcal{L}$, one can write down the spin puzzle as: 
\begin{equation}
\frac{1}{2}=\frac{1}{2}\Delta\Sigma + \Delta G + L.
\label{eq:spinpuzle}
\end{equation}

While a plethora of additional information is available nowadays from other fixed-target experiments at DESY, CERN, Jefferson Lab, Fermilab and SLAC and from RHIC collider experiments, some missing pieces are needed for a full assembly of the proton spin puzzle. We will below discuss contributions to the proton spin from the spin of the quarks, from the spin of the gluons, and from orbital angular momentum of quarks and gluons. 

\subsection{A short guide to this paper}

The reader is expected to be familiar with the material in the appendix. If you are a beginner, it is recommended to review some or all sections of the appendix first. 
Throughout this paper, the convention $c=\hbar=1$ is used. When written out, momenta have units of GeV/$c$ and masses GeV/$c^2$.
\section{From elastic to deep-inelastic scattering}
\label{sec:dis}
In this chapter, we will start from the classical Rutherford formula and will step by step develop more generic expressions, thereby crossing the border from elastic scattering to deep-inelastic lepton-proton scattering \cite{rith2006}.  
\subsection{Elastic scattering}
\label{sec:elascat}
We are looking at elastic scattering: 
\begin{equation}
\text{A} \text{B} \rightarrow \text{A}^\prime \text{B}^\prime,
\label{eq:elscat}
\end{equation}
where B is a nuclear target at rest in the laboratory and A is a beam of particles, for example electrons. The scattering process is elastic if the invariant mass (App.~\ref{sec:relativistickine}) of the target with mass $M$ is the same before and after scattering, $M^2=W^2$, where $W$ is the 4-momentum of B$^\prime$. This means the nucleon or nucleus is not excited. Then there exists, at fixed beam energy, a unique correlation between the energy $E^\prime$ of the scattered beam particle A$^\prime$ and the scattering angle $\theta$ (Fig.~\ref{fig:rutherford} left):
\begin{equation}
E^\prime=\frac{E}{1+\frac{E}{M}(1-\cos\theta)}.
\label{eq:easlticond}
\end{equation}
In other words, $E^\prime$ and $\theta$ are not independent of each other. With $\nu=E-E^\prime$ (the energy transfer by the beam particle) and $Q^2=-q^2$ (the squared 4-momentum transfer by the beam particle, see appendix Tab.~\ref{tab:elscat}), the \emph{condition for elastic scattering} is
\begin{equation}
2M\nu-Q^2=0,
\label{eq:mainelastic}
\end{equation}
which is an alternative formulation of Eq.~\ref{eq:easlticond}. For more in-depth details about the kinematics in elastic scattering, see App.~\ref{sec:elastic}.

\subsection{Refining Rutherford's formula}
\begin{figure}
\begin{center}
\includegraphics[width=0.56\textwidth]{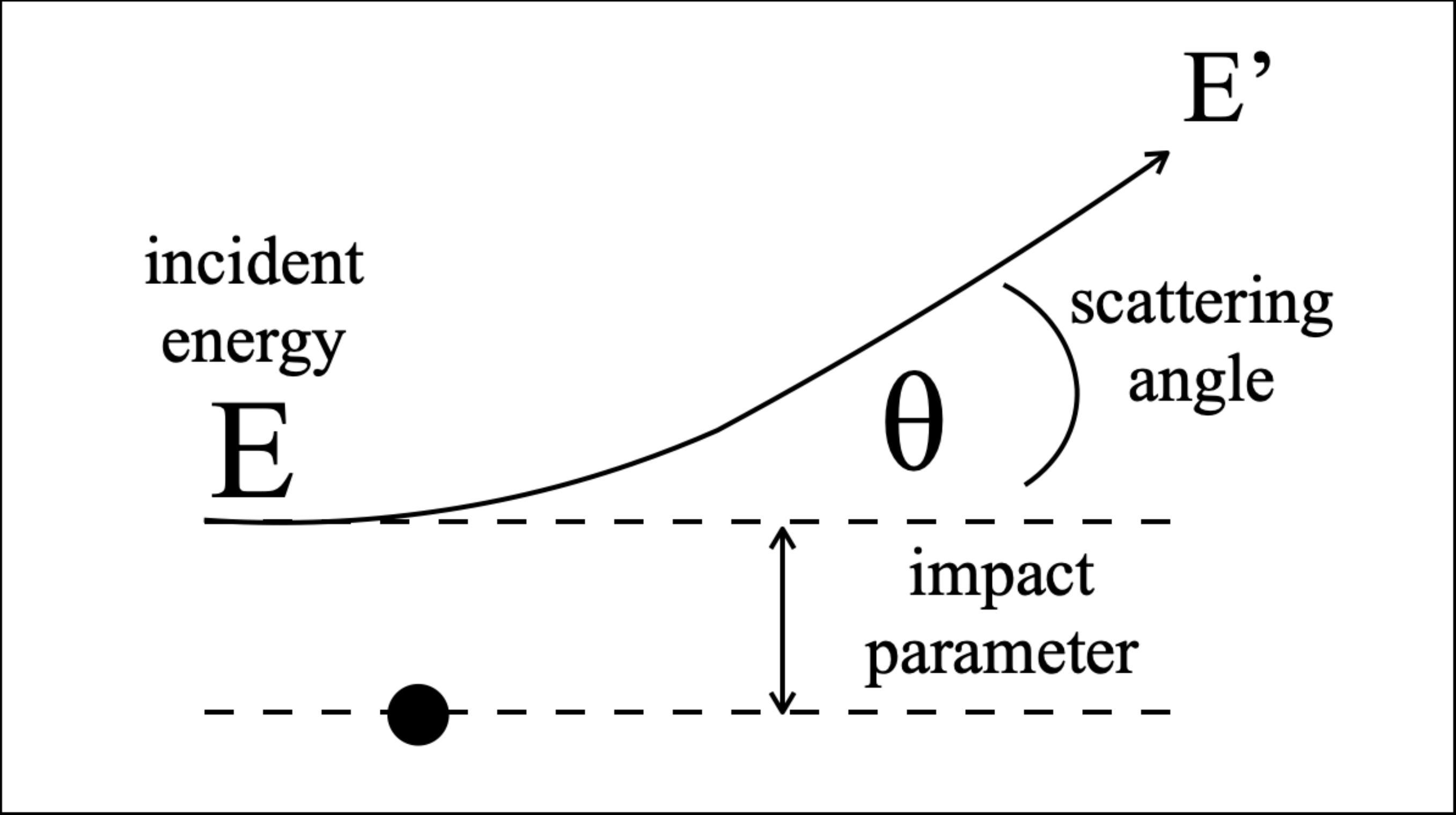}
\fbox{\includegraphics[width=0.40\textwidth]{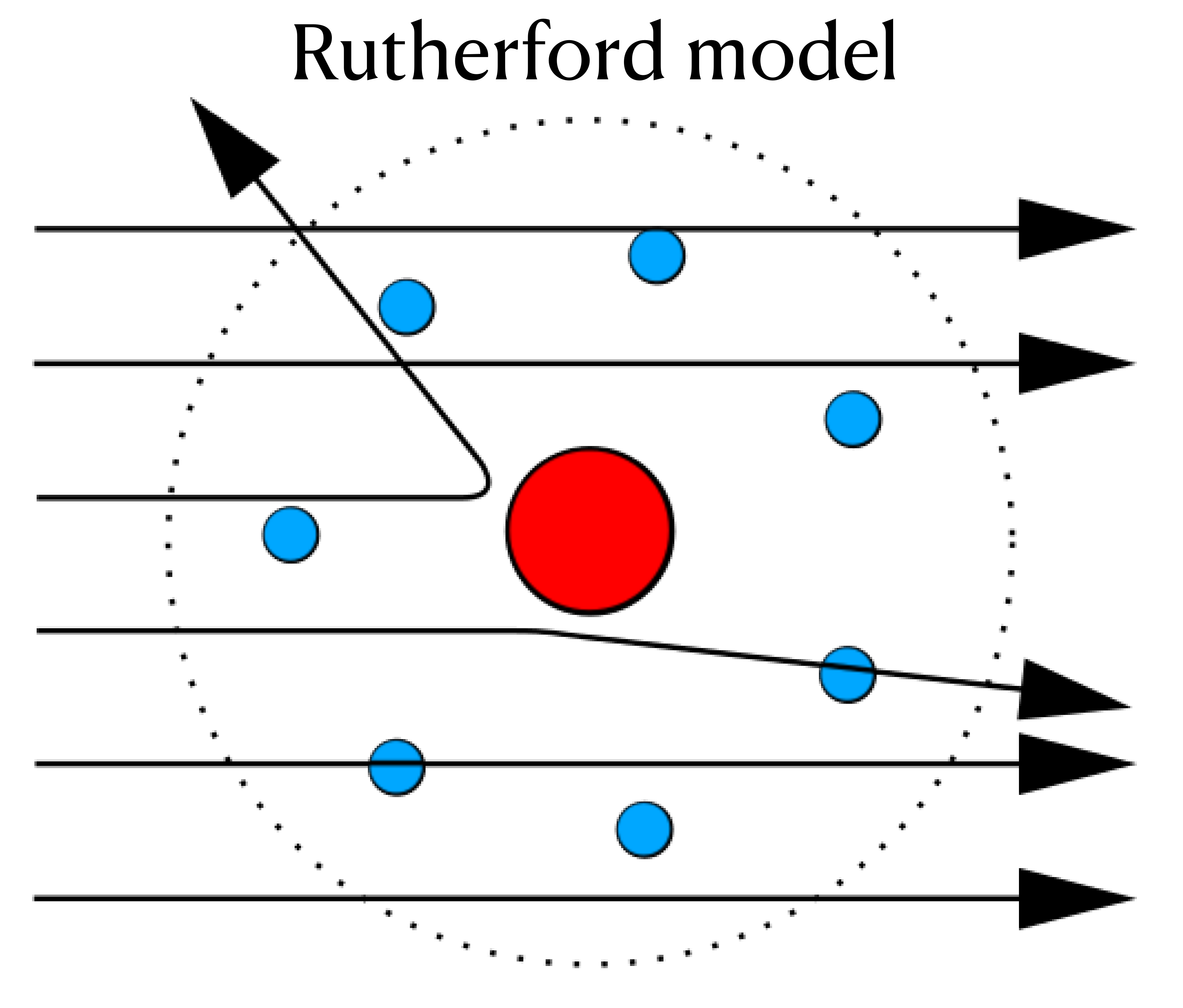}}
\end{center}
\caption[Rutherford scattering]{Left: incident (outgoing) energy $E$ ($E^\prime$) and polar scattering angle $\theta$. Right: depiction of the Rutherford atomic model with nucleus (red), shell electrons (blue), and incident and scattered $\alpha$-particles (black lines). }
\label{fig:rutherford}
\end{figure}

The Geiger–Marsden experiments (also called the Rutherford gold-foil experiments) from the first decade of the 20th century are a famous example of elastic scattering. They opened the stage for the formulation of the first modern atomic model (Fig.~\ref{fig:rutherford} right). Alpha ($\alpha$) particles from the radioactive decay of radium were sent on a thin gold foil, i.e., in Eq.~\ref{eq:elscat}, A~$\equiv\alpha$ and B~$\equiv$~Au. The Rutherford cross section (differential in the solid angle $\Omega$) is derived from the classical electric Coulomb potential and is based on purely geometric considerations: 
\begin{equation}
\left(\frac{\mathrm{d}\sigma}{\mathrm{d}\Omega}\right)_{\mathrm{Rutherford}}=\frac{Z^2\alpha_{\mathrm{em}}^2(\hbar c)^2}{4E^2}\cdot\frac{1}{\sin^4(\frac{\theta}{2})},
\label{eq:rutherford}
\end{equation}
with $\alpha_{\mathrm{em}}$ the electromagnetic fine structure constant (Eq.~\ref{eq:alphaem}) and $Z$ the electric charge number of the nucleus. 
Using alpha particles in the MeV range, it was possible for the first time to resolve structures ($\sim10^{-12}$~m) smaller than the atomic radius ($\sim10^{-10}$~m), but not quite yet the nucleus ($\sim10^{-15}$~m). 

The alpha particles in the Rutherford experiment did not quite have relativistic energies yet, and moreover they were spinless particles. For relativistic energies, the Rutherford cross section from Eq.~\ref{eq:rutherford} has to be modified by spin effects. The modification accounts for the conservation of helicity $h$ (Eq.~\ref{eq:appheli}), which contains the scalar product of spin vector and 3-momentum vector. This becomes relevant when we consider beam particles with spin~$\neq 0$ like electrons (or generically charged leptons), which we will do from now on. The Rutherford cross section modified for relativistic energies and taking into account helicity conservation is called Mott cross section \cite{rith2006}:
\begin{equation}
\left(\frac{\mathrm{d}\sigma}{\mathrm{d}\Omega}\right)_{\mathrm{Mott}}^*=\left(\frac{\mathrm{d}\sigma}{\mathrm{d}\Omega}\right)_{\mathrm{Rutherford}}\cdot\underbrace{\cos^2(\frac{\theta}{2})}_{\bigoplus\mathrm{spin}},
\label{eq:mott}
\end{equation}
where we have introduced an additional term that forbids scattering by 180$^\circ$. This assumes a spinless target, which due to total angular momentum conservation cannot absorb the spin transfer from the beam to the target. The flip of the electron's spin direction is required by helicity conservation. 

The improved Mott cross section accounts also for the recoil of the nucleus and its extension: 
\begin{equation}
\left(\frac{\mathrm{d}\sigma}{\mathrm{d}\Omega}\right)_{\mathrm{Mott}}=\left(\frac{\mathrm{d}\sigma}{\mathrm{d}\Omega}\right)_{\mathrm{Mott}}^*\cdot\underbrace{\frac{E^\prime}{E}}_{\bigoplus\mathrm{recoil}}\cdot\underbrace{\left|F(|\vec{q}\,|^2)\right|^2}_{\bigoplus\mathrm{extension}}.
\label{eq:mottimproved}
\end{equation}
The quantity $F(|\vec{q}\,|^2)$ is the form factor, which we will study next. 
\subsection{Form factors}
\label{sec:formfactors}

Form factors $F(|\vec{q}\,|^2)$ encode the information about the geometrical shape of the object that is scattered off. They depend on the squared 3-momentum transfer $|\vec{q}\,|^2$ (see appendix Tab.~\ref{tab:elscat}) and are in principle connected to radial charge distributions in position space $f(\vec{r}\,)$ via Fourier transforms \cite{rith2006}:
\begin{eqnarray}
F(|\vec{q}\,|^2)&\sim&\int\mathrm{d}^3\vec{r}\; e^{i\frac{\vec{q}\cdot\vec{r}}{\hbar}}f(\vec{r}\,),\label{eq:ftr}\\
\updownarrow&&\mathrm{Fourier\;\;transform}\nonumber\\
f(\vec{r}\,)&\sim&\int\mathrm{d}^3\vec{q}\; e^{i\frac{\vec{q}\cdot\vec{r}}{\hbar}}F(|\vec{q}\,|^2).\label{eq:ftq}
\end{eqnarray}
The integrals in Eqs.~\ref{eq:ftr} and \ref{eq:ftq} cover all possible values in $\mathrm{d}^3\vec{r}$ respectively $\mathrm{d}^3\vec{q}$ from 0 to infinity. Driven by the maximum available beam energy, the form factors can only be measured over a limited range in squared 3-momentum transfer. Therefore it is in reality not possible to analytically infer the radial charge distributions $f(\vec{r}\,)$ from the Fourier transform of the measured form factors $F(|\vec{q}\,|^2)$. Instead, parameterizations (model descriptions depending on a set of parameters) of $f(\vec{r}\,)$ are chosen that translate best into the measured $F(|\vec{q}\,|^2)$.

\begin{figure}
\begin{center}
\includegraphics[width=0.48\textwidth]{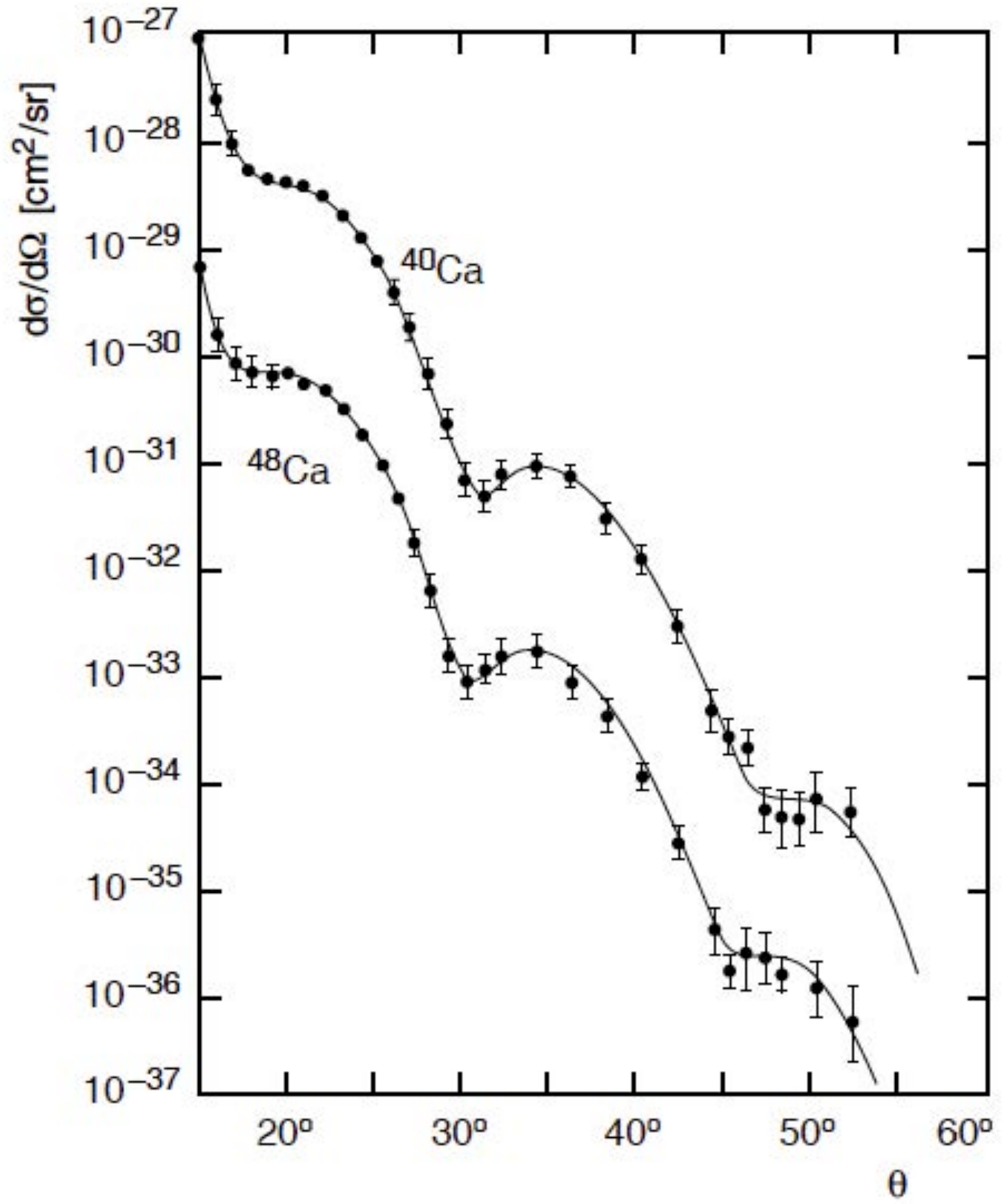}
\includegraphics[width=0.48\textwidth]{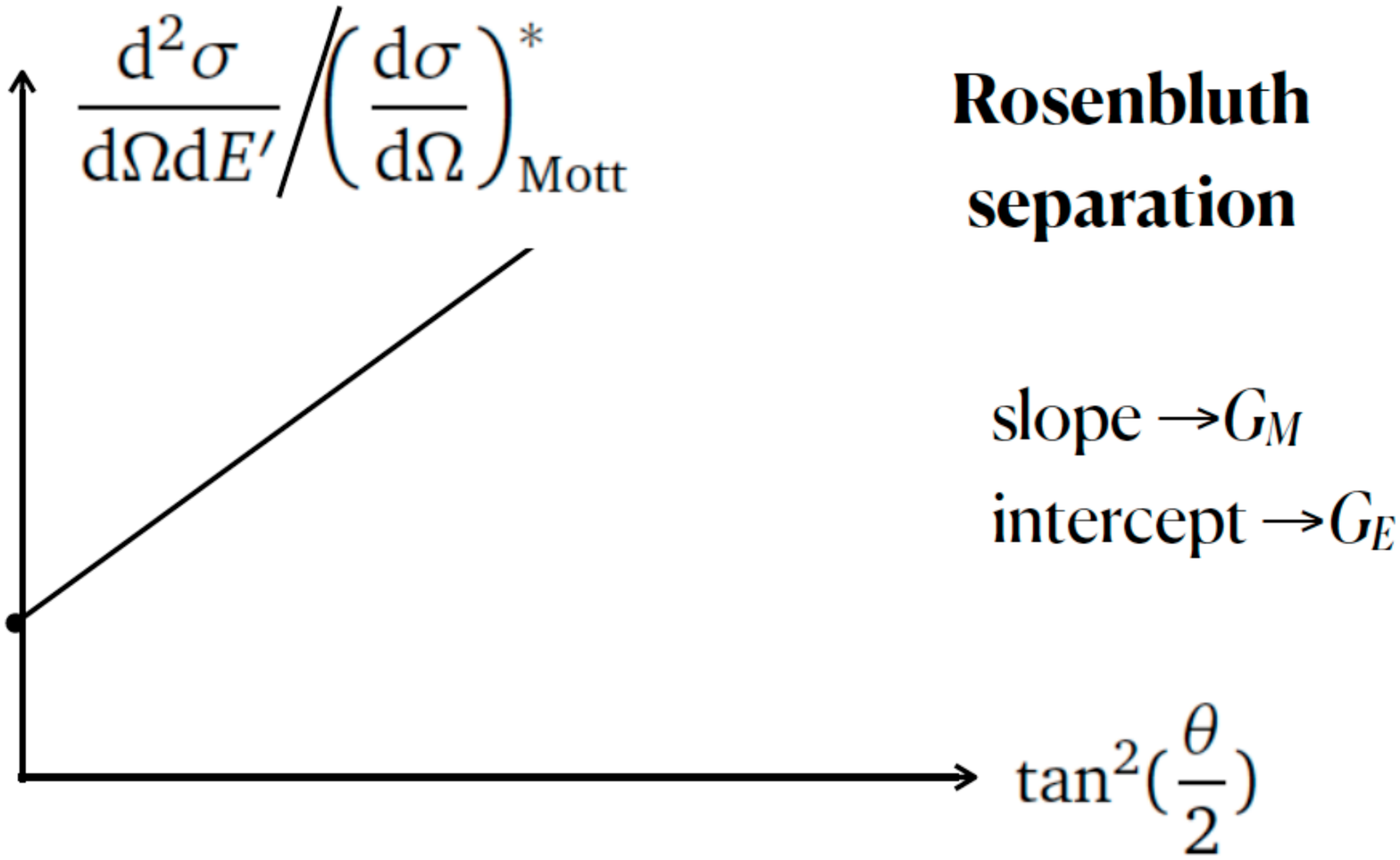}
\end{center}
\caption[Elastic cross section; Rosenbluth separation]{Left: cross section for elastic scattering of electrons off calcium isotopes at SLAC. Scattering angle and 3-momentum transfer are related as $\sin^4(\theta/2)\sim|\vec{q}\,|^4$. Figure taken from Ref.~\cite{SLAC1967}. Right: principle of Rosenbluth separation.}
\label{fig:slac}
\end{figure}

Figure~\ref{fig:slac} shows cross section measurements in elastic scattering on calcium isotopes, exhibiting the typical distinct diffraction pattern for sharply localized objects such as heavy nuclei. The oscillating shape of the form factor translates into a sphere with diffuse surface in position space - a calcium nucleus. From the location of the first minimum of the form factor, the nucleus' mean charge radius can be deferred. Since the minima in Fig.~\ref{fig:slac} are shifted to smaller values of $\theta$ (thus $|\vec{q}\,|$) it can be concluded that the $^{48}$Ca nucleus is larger than the $^{40}$Ca nucleus. 
Lighter nuclei like $^6$Li behave Gaussian in both momentum and position space, while the proton is a dipole in momentum space and an exponential in position space \cite{rith2006}. In general, extended charge distributions have steeply falling form factors, and vice versa slowly falling form factors for small objects. A constant distribution in momentum space translates into a point-like distribution in position space. This relation sounds at first not very exciting, but it will soon allow us to draw groundbreaking conclusions and it is therefore important to keep it in mind.   

\subsection{Scattering off a particle with spin: magnetic interaction}
\label{sec:magint}
So far, we have neglected the interaction of the lepton's electric current with the target's magnetic moment. If the target nucleon or nucleus has spin~$\neq 0$, there is magnetic interaction through the particle's magnetic moment $\vec{\mu}$:
\begin{equation}
\vec{\mu}=g\frac{e}{2m}\vec{s},
\end{equation}
with $e$ its electrical charge, $m$ its mass, $\vec{s}$ its spin vector and $g$ the g-factor. From the Dirac equation\footnote{The Dirac equation appears in relativistic quantum mechanics. It treats space and time linearly and on the same footing, unlike the Schr\"odinger equation does.}, one gets $g=2$ for a point-like particle, often referred to as ``Dirac particle''. Structures without charge have $g=0$. 
One defines the Bohr (nuclear) magneton as if the proton were a Dirac particle:
\begin{equation}
\mu_{\text N}=\frac{e}{2M}\hbar=3.1525\cdot 10^{14}\;\mathrm{MeV/T}.
\end{equation}
However, proton and neutron are not point-like and they have a substructure, resulting in their anomalous magnetic moments \footnote{What about electron and muon anomalous magnetic moments?  The Fermilab - BNL \emph{g-2 measurements} indicate a $g\neq 2$ at currently 4.2 standard deviations \cite{gm22021}. Deviations from 2 are however not an indication of lepton substructure, but of higher-order contributions from beyond-standard-model particles.}:
\begin{eqnarray}
\mu_{\text p}=2.79\mu_{\text N}&,&g_{\text p}=5.58\neq 2,\\
\mu_{\text n}=-1.91\mu_{\text N}&,&g_{\text n}=-3.82\neq 0.
\end{eqnarray}
Taking into account the magnetic moment, the Mott cross section from Eq.~\ref{eq:mott} has to be modified for a spin-1/2, point-like ($g=2$) targets as follows: 
\begin{equation}
\frac{\mathrm{d}^2\sigma}{\mathrm{d}\Omega\mathrm{d}E^\prime}=\left(\frac{\mathrm{d}\sigma}{\mathrm{d}\Omega}\right)_{\mathrm{Mott}}^*\cdot\left[\underbrace{1}_{\substack{\text{electric} \\ \text{effects}}}+2\tau\underbrace{\tan^2(\frac{\theta}{2})}_{\text{magnetic effects}}\right]\cdot\underbrace{\delta\left(\frac{Q^2}{2M}-\nu\right)}_{\mathrm{elastic\;\;condition}},
\end{equation}
where we have added the elastic condition Eq.~\ref{eq:mainelastic} acting as a delta-function and a term accounting for magnetic effects $\sim\sin^2(\theta/2)$, which divided by the $\cos^2(\theta/2)$ for helicity conservation from the Mott cross section (which we pulled in front of the brackets) gives $\tan^2(\theta/2)$. Also note that the cross section is now double differential in $\Omega$ and $E^\prime$. 

For a spin-1/2 target with extended structure, we again find the form factor from Eq.~\ref{eq:mottimproved}, which we call electric form factor $F_{\text E}$, since it was introduced in a scattering formula solely derived from the electric Coulomb force. In addition, since we are taking into account magnetic effects for spin-afflicted targets, we now have a magnetic form factor $F_{\text M}$:
\begin{equation}
\frac{\mathrm{d}^2\sigma}{\mathrm{d}\Omega\mathrm{d}E^\prime}=\left(\frac{\mathrm{d}\sigma}{\mathrm{d}\Omega}\right)_{\mathrm{Mott}}^*\cdot\left[\underbrace{F_{\text E}^2(|\vec{q}\,|^2)}_{\substack{\text{electric} \\ \text{form factor}}}+2\tau\tan^2(\frac{\theta}{2})\cdot \underbrace{F_{\text M}^2(|\vec{q}\,|^2)}_{\substack{\text{magnetic} \\ \text{form factor}}}\right]\cdot\delta\left(\frac{Q^2}{2M}-\nu\right),
\label{eq:fefm}
\end{equation}
with $\tau=Q^2/(4M^2)$.
\subsection{Electric and magnetic form factors of the nucleon (Rosenbluth)}

Equation~\ref{eq:fefm} is traditionally written in a form referred to as the Rosenbluth formula in elastic lepton-nucleon scattering with the electric and magnetic form factors of the nucleon, $G_{\text E}(Q^2)$ and $G_{\text M}(Q^2)$, respectively: 
\begin{eqnarray}
\frac{\mathrm{d}^2\sigma}{\mathrm{d}\Omega\mathrm{d}E^\prime}=\left(\frac{\mathrm{d}\sigma}{\mathrm{d}\Omega}\right)_{\mathrm{Mott}}^*\cdot\left[\frac{G_{\text E}^2(Q^2)+\tau G_{\text M}^2(Q^2)}{1+\tau}\right.&&\nonumber \\
\left.+2\tau\tan^2(\frac{\theta}{2})\cdot G_{\text M}^2(Q^2)\right]
&\cdot&\delta\left(\frac{Q^2}{2M}-\nu\right).
\label{eq:rosenbluth}
\end{eqnarray}
Because both 3-momentum and energy are exchanged in the recoil process, it is more useful to formulate the form factors in dependence of the Lorentz-invariant squared 4-momentum $Q^2\equiv-q^2=4EE^\prime\sin^2(\theta/2)$, instead of the squared 3-momentum $|\vec{q}\,|^2$. For $Q^2=0$, one gets:
\begin{equation}
Q^2=0
\begin{Bmatrix}
G_{\text E}/e&=&\begin{matrix} 1 & \text{p} \\ 0 & \text{n} \end{matrix} \\
G_{\text M}/\mu_{\text N}&=&\begin{matrix} 2.79 & \text{p} \\ -1.91 & \text{n} \end{matrix} 
\end{Bmatrix}
\end{equation}
for the proton (p) and the neutron (n).
%
%
To separately determine  $G_{\text E}(Q^2)$ and $G_{\text M}(Q^2)$, the cross section has to be measured at fixed $Q^2$ and varying angles $\theta$ by changing the beam energy. If the results are plotted as in the right hand side of Fig.~\ref{fig:slac}  (the \emph{Rosenbluth separation}), $G_{\text M}(Q^2)$ is given by the slope and $G_{\text E}(Q^2)$ by the intercept on the $y$-axis. The proton charge radius can be determined (in analogy to the calcium radii above) by studying $\left.\text{d}G_{\text E}/\text{d}{Q^2}\right|_{Q^2=0}$ and is found to be around 1~fm. 

\subsection{DIS off the unpolarized proton}
\label{sec:disdis} 

To resolve structures below $\sim1~\mathrm{fm}=10^{-15}$~m, which is the size of a small nucleus, one needs to use projectiles of energy 1~GeV or larger. We are looking at lepton-proton scattering:
\begin{equation}
\ell \text{p}\rightarrow \ell^\prime \text{p}^\prime.
\label{eq:elp}
\end{equation}
When in the scattering process sufficient energy is transferred from the lepton to the proton, it becomes kinematically possible for the proton to be excited to a baryonic resonance state, for example, the $\Delta^+(1236)$.
The $\Delta^+$ is still very similar to the proton (mass 938~MeV). Only their quark spin alignments differ\footnote{see App.~\ref{sec:quarks} for more in-depth explanation of the baryon ordering scheme.}: p~$=|\text{u}^\uparrow \text{u}^\uparrow \text{d}^\downarrow\rangle$ and $\Delta^+=|\text{u}^\uparrow \text{u}^\uparrow \text{d}^\uparrow\rangle$.
Thus, in Eq.~\ref{eq:elp}, p$^\prime\neq$~p, and the invariant mass of the final state ($W$) is no longer identical to that of the initial-state proton ($M$). We have entered the \emph{inelastic regime}, which is categorized by 
\begin{equation}
W^2=M^2+\underbrace{2M\nu-Q^2}_{>0}.
\label{eq:inelastic}
\end{equation}
Note that for $2M\nu-Q^2=0$ (``elastic condition'' Eq.~\ref{eq:mainelastic}), the kinematics for elastic scattering are recovered. 

With further increasing energy transfer by the lepton, the proton will ``break up'', or \emph{fragment}, and \emph{hadronize}, i.e., it will produce a set of new color-saturated ``white'' hadrons (see Sec.~\ref{sec:hadrons}) while obeying quantum-number conservation laws (see Sec.~\ref{sec:conqua}). This process is called deep-inelastic scattering (DIS):
\begin{equation}
\ell {\text p} \rightarrow \ell^\prime \text{X}.
\label{eq:dis}
\end{equation}
The (a priori unknown) hadronic final state X with invariant mass $W$
is not the proton or a resonant state any longer. 

Throughout Sec.~\ref{sec:dis}, we probe proton structure by measuring the 4-momentum of the scattered lepton while ignoring the presence of any particles in state X. This experimental choice  is called \emph{inclusive measurement}. 

\begin{figure}
\begin{center}
\includegraphics[width=0.9\textwidth]{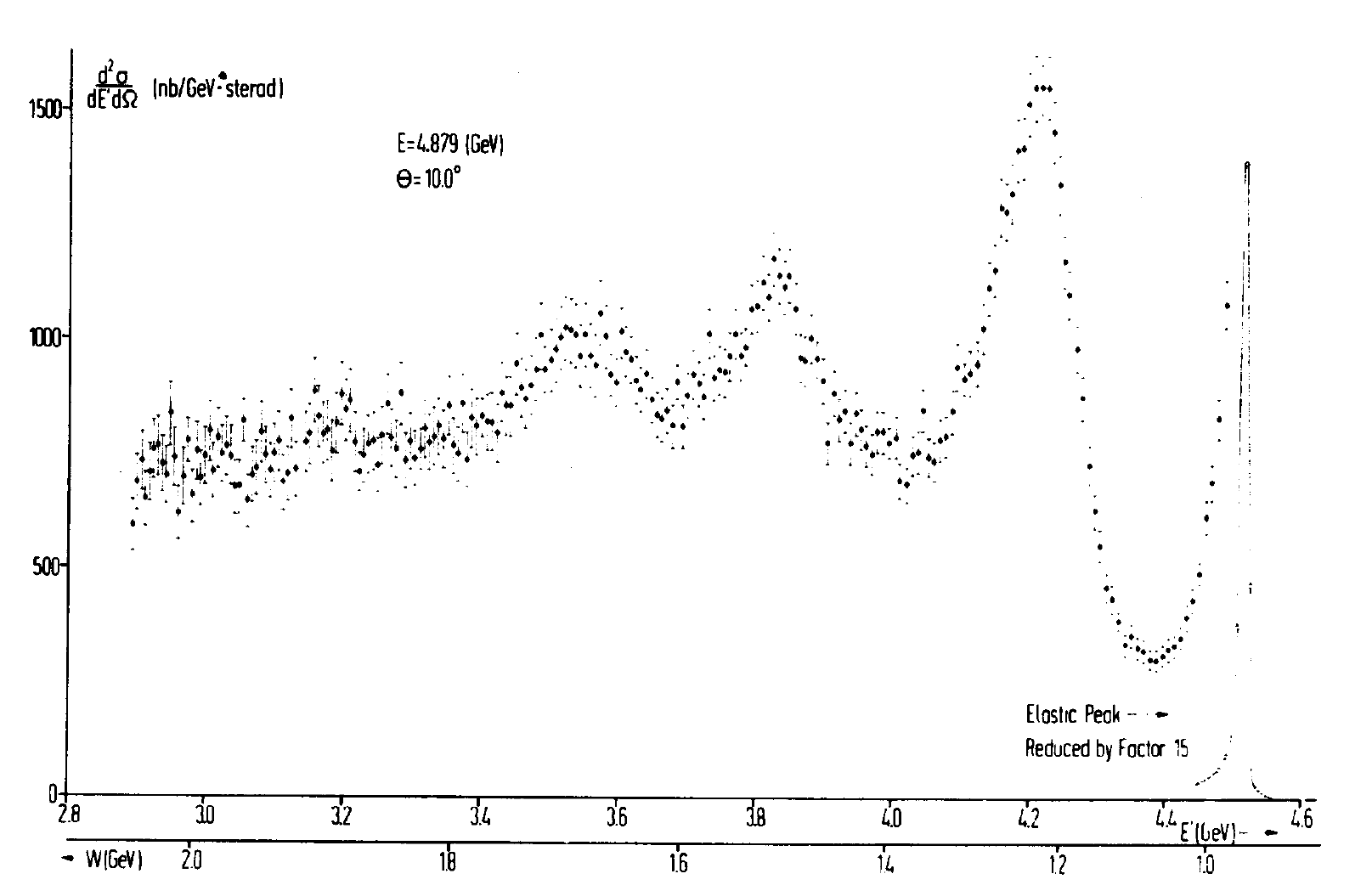}
\end{center}
\caption[Cross section of electron-proton scattering]{Cross section of electron-proton scattering at beam energy $E=5$~GeV vs.~energy of the scattered electron $E^\prime$, which increases to the right. The invariant mass of the photon-proton system $W$ increases to the left. The squared momentum transfer at the $\Delta^+$ peak ($W=1.236$~GeV) is 0.63~GeV$^2$. See text for discussion. Figure taken from Ref.~\cite{Ba68}.}
\label{fig:eldis}
\end{figure}

The transition from the elastic regime via the resonance region into the DIS regime is illustrated in Fig.~\ref{fig:eldis}, which allows to identify three distinctive regions: the sharp elastic peak (downscaled for better comparison) at the proton mass $W\approx0.938$~GeV, followed on the left by the resonance region with several broad inelastic nucleon excitations, and lastly the deep-inelastic continuum for values $W>2$~GeV.  The kinematic condition for the DIS regime is usually categorized by $Q^2>1$~GeV$^2$.

As in the case of elastic scattering (Eq.~\ref{eq:fefm}), there are two functions in DIS describing the structure of the target that is scattered of. In elastic scattering, the functions were called form factors, in DIS they are called \emph{structure functions}. The cross section for inclusive DIS on the proton is written in terms of two structure functions $W_2(\nu,Q^2)$ and $W_1(\nu,Q^2)$ describing the deep-inelastic process:
\begin{equation}
\frac{\mathrm{d}^2\sigma}{\mathrm{d}\Omega\mathrm{d}E^\prime}=\left(\frac{\mathrm{d}\sigma}{\mathrm{d}\Omega}\right)_{\mathrm{Mott}}^*\cdot\left[W_2(\nu,Q^2)+2\tan^2(\frac{\theta}{2})\cdot W_1(\nu,Q^2)\right].
\label{eq:xsectionw1w2}
\end{equation}
As in the case of elastic scattering, there is one function describing electric effects ($W_2$) and one describing magnetic effects ($W_1$). However, while the elastic structure functions depended on one kinematic variable, there are \emph{two} Lorentz invariants (see App.~\ref{sec:Linvariants}) that the deep-inelastic structure functions depend on because there are \emph{two} degrees of freedom in DIS. This can be understood by considering that now, unlike in elastic scattering, the energy of the hadronic system X (Eq.~\ref{eq:dis}) is not fixed by the scattering angle $\theta$ any longer, as it was the case in Eq.~\ref{eq:easlticond}. Here, we chose $\nu$, the energy of the virtual photon, and $Q^2$, the negative squared 4-momentum transfer by the virtual photon (see Tab.~\ref{tab:inclDISkine}). 

In the modern formulation of Eq.~\ref{eq:xsectionw1w2}, it is usual to use the dimensionless structure functions $F_1\equiv MW_1$ describing magnetic effects and $F_2\equiv\nu W_2$ electric effects:
\begin{equation}
\frac{\mathrm{d}^2\sigma}{\mathrm{d}\Omega\mathrm{d}E^\prime}=\underbrace{\underbrace{\frac{Z^2\alpha^2}{4E^2}\cdot\frac{1}{\sin^4(\frac{\theta}{2})}}_{\text{Rutherford}}\cdot\cos^2(\frac{\theta}{2})}_{\text{Mott}}\left[\frac{1}{\nu}\underbrace{F_2(\nu,Q^2)}_{\text{electric effects}}+\frac{2}{M}\underbrace{\tan^2(\frac{\theta}{2})\cdot F_1(\nu,Q^2)}_{\text{magnetic effects}}\right], 
\end{equation}
with $\alpha=e^2/(4\pi\epsilon_0\hbar c)$. For a spinless target, $F_1\equiv 0$ because there is no magnetic interaction (see Sec.~\ref{sec:magint}). 

It is common practice to express $F_1$ and $F_2$ as functions of the two Lorentz invariants $x$ and $Q^2$. The $x$-Bjorken, $x=Q^2/(2M\nu)$, provides a measure of the inelasticity of the lepton-proton scattering process:
\begin{eqnarray}
\mathrm{elastic:}&&\;\; 2M\nu-Q^2 = 0, \;\; W^2=Q^2, \;\; x=1 \\
\mathrm{inelastic:}&&\;\; 2M\nu-Q^2 > 0, \;\; W^2>Q^2, \;\; 0<x<1.
\end{eqnarray}
Section~\ref{sec:qpm} will introduce another, entirely independent interpretation of Bjor\-ken-$x$. 

\subsection{Lessons from first DIS experiments and quark-parton model}
\label{sec:qpm}

\begin{figure}
\begin{center}
\includegraphics[width=0.58\textwidth]{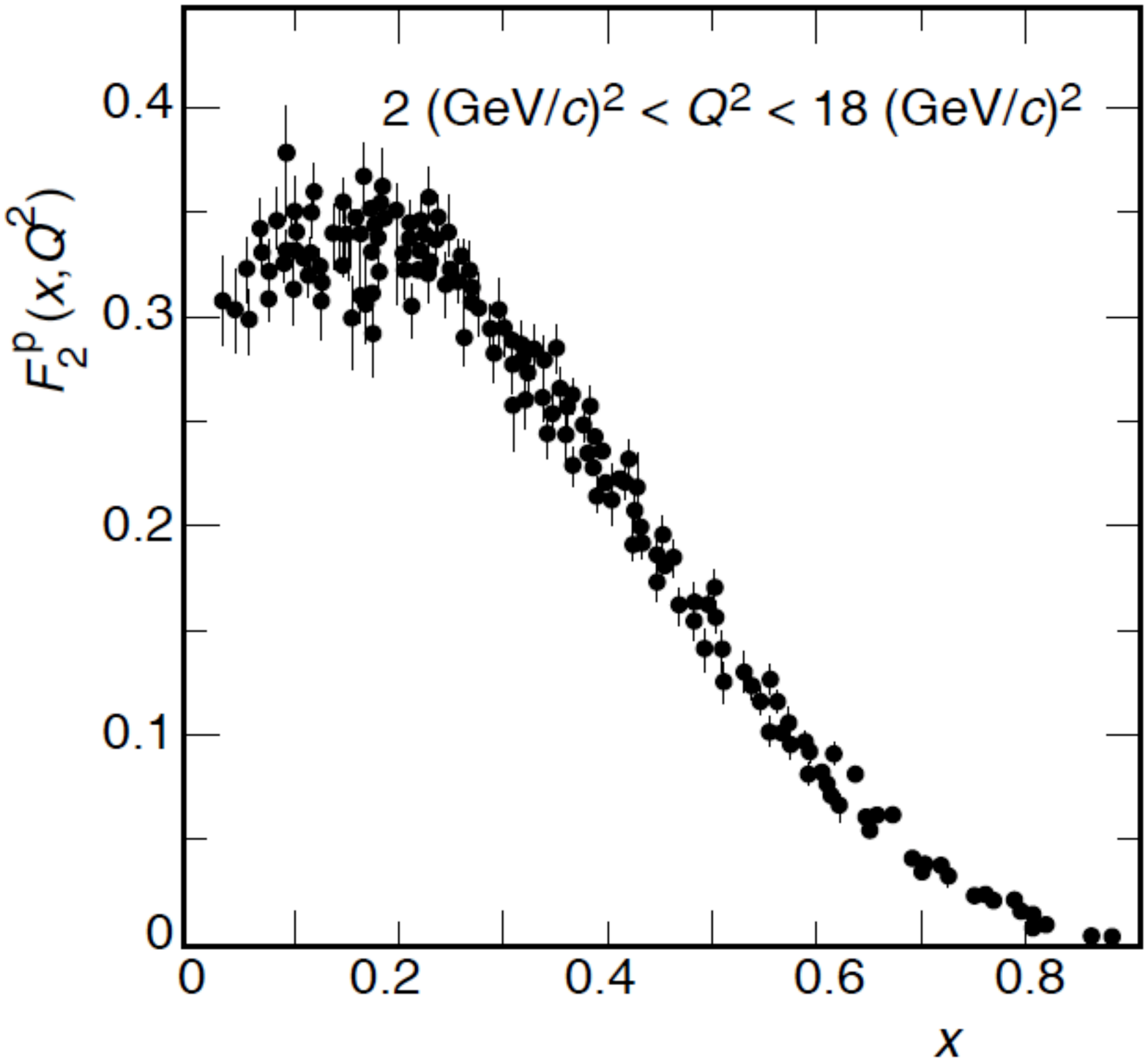}
\includegraphics[width=0.38\textwidth]{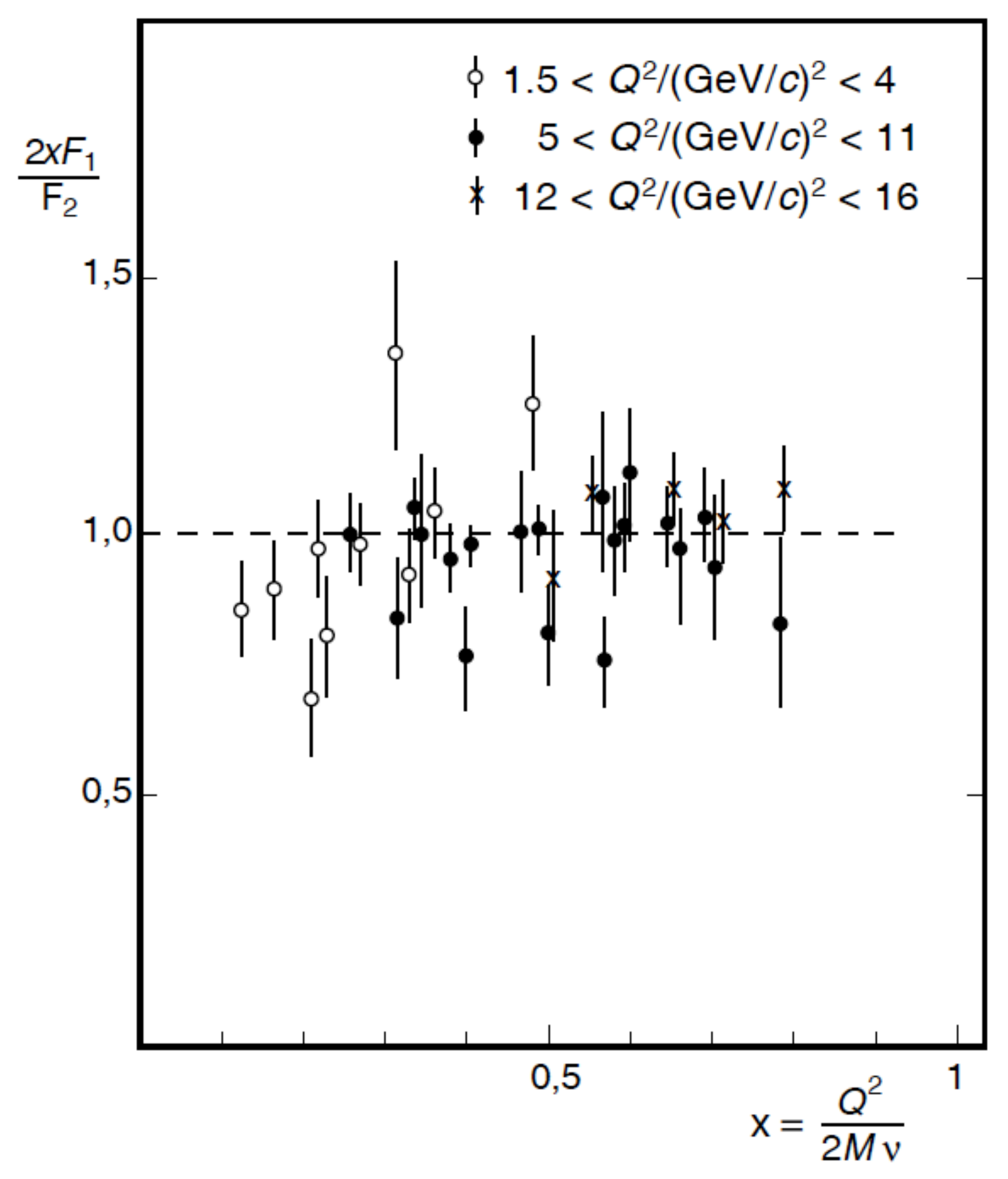}
\end{center}
\caption[Early DIS results]{Early DIS results from SLAC providing evidence for constituents inside the proton. Left: structure function $F(x,Q^2)$ at different values of $Q^2$. Figure taken from Ref.~\cite{At82}. Right: structure-function ratio $2xF_1/F_2$ to test the Callan-Gross relation. See Ref.~\cite{rith2006} and references therein.}
\label{fig:earlydis}
\end{figure}

Let's take a look back at Fig.~\ref{fig:eldis}. This electron-proton cross section was recorded for an electron beam energy of about 5~GeV and no structures are visible in the deep-inelastic region. What happens if the beam energy is increased? In the fall of 1967, a revolutionary set of experiments was launched at the Spectrometer Facility of the Stanford Linear Accelerator (SLAC) using an electron beam of 20~GeV energy and a liquid hydrogen target to provide a pure nuclear sample of protons. In short, the results are: 
\begin{description}
\item The structure function $F_2(x,Q^2)$ is in first order independent of $Q^2$, see Fig~\ref{fig:earlydis} left. This behavior is called (Bjorken) \emph{scaling}. From the discussion of form factors in Sec.~\ref{sec:formfactors}, we recall that a constant distribution in momentum space translates into a point-like structure in position space. We can conclude that \emph{nucleons have a substructure of point-like constituents.} 
\item The measured structure functions follow the Callan-Gross relation, see Fig~\ref{fig:earlydis} right,
\begin{equation}
2xF_1(x)=F_2(x),
\label{eq:callangross}
\end{equation}
which is expected for spin-1/2 Dirac particles. We can conclude that \emph{the point-like constituents of the proton have spin-1/2.}
\end{description}

Just a few years earlier, in 1964,  Murray Gell-Mann and George Zweig had independently proposed the quark model (see App.~\ref{sec:quarks}), but it had remained unclear whether those \emph{quarks} were real entities or just a mathematical construct. The early DIS experiments in the late 1960s provided the groundbreaking realization that there \emph{are} indeed constituents inside the proton - ``the peach does have a pit'' \cite{Taylor2000}, which seemed to be identical to the point-like constituents revealed in the experiments. This inspired Richard Feynman to formulate the quark-parton model (QPM) in the fall of 1968. 
``Parton'' is a generic term for ``constituent in the proton''.

\begin{figure}
\begin{center}
\includegraphics[width=0.6\textwidth]{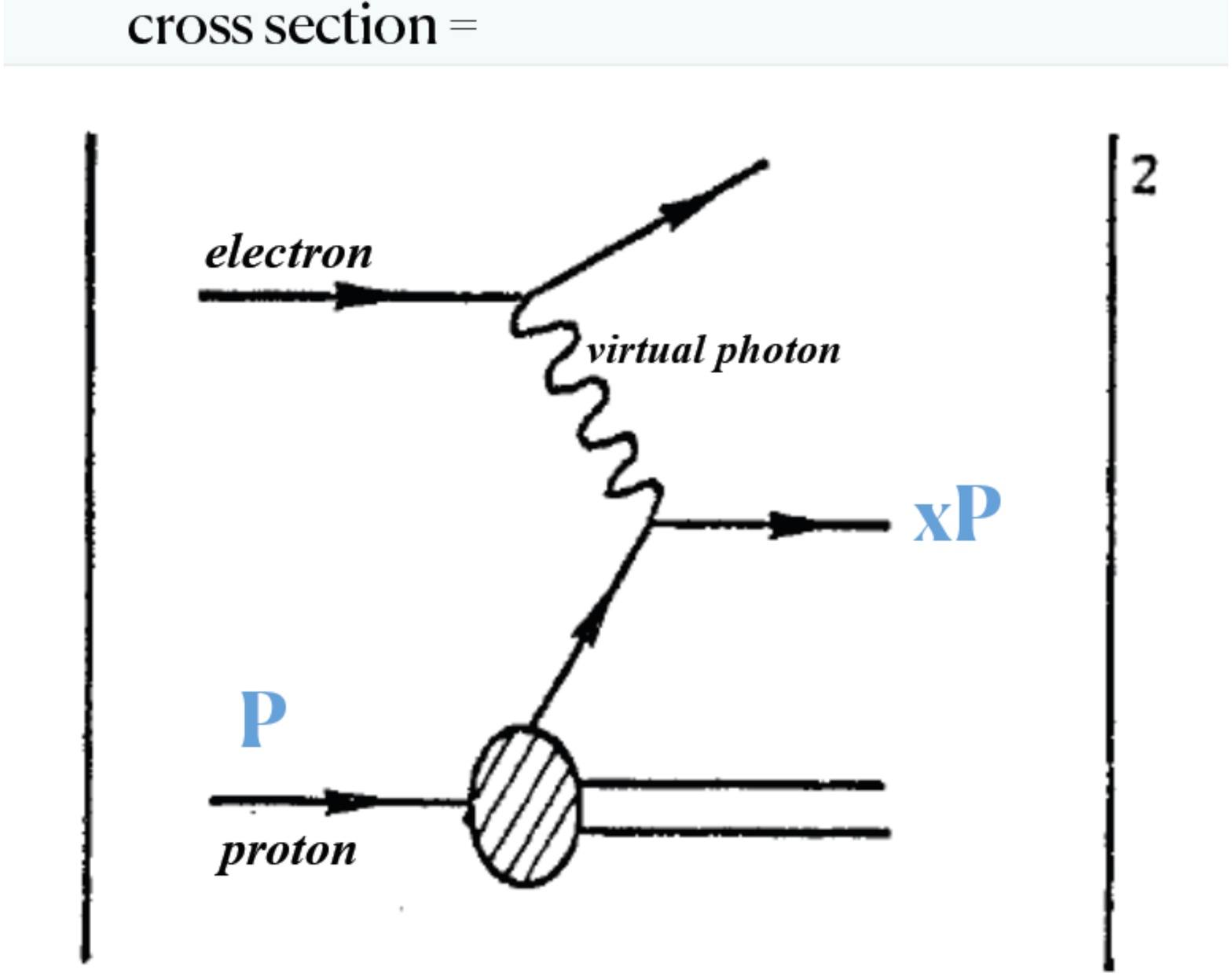}
\end{center}
\caption[Quark parton model]{Deep-inelastic scattering in the quark-parton model (QPM) under the impulse approximation, neglecting parton masses and  transverse momenta (\emph{collinear}), and $Q^2\gg M^2$: i.e., the
nucleon moves very fast in the longitudinal direction. A popular such choice is the Breit frame, where the photon transfers zero energy. Figure adapted from Ref.~\cite{close1979}.}
\label{fig:qpm}
\end{figure}

In the QPM, deep-inelastic scattering is viewed as the incoherent sum of elastic scattering processes of leptons off quasi-free point-like partons in the proton. This \emph{impulse approximation} is valid as long as the duration of the photon-parton interaction is so short that the interaction between the partons themselves can be safely neglected. This assumption is reasonable since the interaction of partons is weak over short distances (see App.~\ref{sec:qedqcd}). The $x$-Bjorken can then be interpreted as the \emph{longitudinal momentum fraction carried by the struck quark} in the DIS process, when the nucleon moves very fast in the longitudinal direction and other assumptions specified in the caption of Fig.~\ref{fig:qpm}.

This picture of the DIS process allows a probabilistic interpretation of $F_2(x)$ in the QPM, introducing the quark (longitudinal) momentum probability distributions $q(x)$ (sometimes referred to as \emph{quark densities}):
\begin{equation}
F_2(x)=x\cdot\sum_{q,\overline{q}}e_q^2\left(q(x)+\overline{q}(x)\right),
\label{eq:qpm}
\end{equation}
where the sum runs over all participating quark $q$ and anti-quark $\overline{q}$ flavors (see App.~\ref{sec:elferm}), and $e_q$ is the electrical charge of the (anti-) quark. That means, $q(x)\mathrm{d}x$ is the expectation value of the number of quarks of flavor $q$ in the proton whose momentum fraction lies within the interval $[x,x+\mathrm{d}x]$. 
Quarks of flavor $q$ carry momentum $p=xPq(x)\mathrm{d}x$ (with $P$ the proton momentum) and the probability to carry a momentum fraction $x$ is $p/P=xq(x)\mathrm{d}x$. Equation~\ref{eq:qpm} is valid for DIS with charged leptons and 1-photon (1$\gamma^*$) exchange (i.e., the DIS process is mediated by one virtual photon, not more, and not by bosons of the weak interaction as discussed in Sec.~\ref{sec:pdfs}). Why does $e_q^2$, the square of the parton's electrical charge, enter the sum? The cross section for electromagnetic (Coulomb) scattering on a point-like charged particle is proportional to the square of the charge, as we saw in the Rutherford formula Eq.~\ref{eq:rutherford}. This also explains why no gluons appear in the sum - they are electrically neutral and thus do not interact electromagnetically.

After introducing the concept of partons in the proton, we can conclude an important interpretation of $Q^2$: $\hbar/\sqrt{Q^2}$ is a measure of the resolution of the virtual photon probing the proton. The number of resolved partons grows with rising $Q^2$ \cite{rith2006}.

The close exchange between experimentalists and theorists, and the introduction of the concepts of \emph{asymptotic freedom} (at short quark distances) and \emph{confinement} (at long quark distances) led in the 1970s to the development of quantum chromo dynamics (QCD), the gauge theory of the strong interaction that describes how quarks and gluons interact with each other, and how the gluons as ``glue particles'' bind the quarks to form hadrons (see App.~\ref{sec:qedqcd}). A valuable review of the state-of-the art by the end of the 1970s is given in Ref.~\cite{close1979}. We will next look at some of the QCD effects in more detail.  

\subsection{Scaling violation and quantum chromo dynamics}
\label{sec:scalingviolation}

The reader may have noted that the $Q^2$-dependence was omitted in Eqs.~\ref{eq:callangross} and \ref{eq:qpm}. This was historically motivated by the observed scaling behavior in the early DIS experiments - $F_2(x=\mathrm{fixed},Q^2)$ appeared to be independent of $Q^2$ at first glimpse, which we interpreted as DIS is happening on point-like partons. Closer examination of Fig.~\ref{fig:earlydis} reveals however that the $F_2$ points for the same or similar $x$ do not lie on a thin line, but are rather scattered in a band. This \emph{scaling violation}, i.e., the finding that $F_2(x=\mathrm{fixed},Q^2)$ \emph{does} have a mild $Q^2$-dependence, becomes more obvious when increasing the lever arm of the lepton beam energy and thus the probed $Q^2$ range. 

\begin{figure}
\begin{center}
\includegraphics[width=0.48\textwidth]{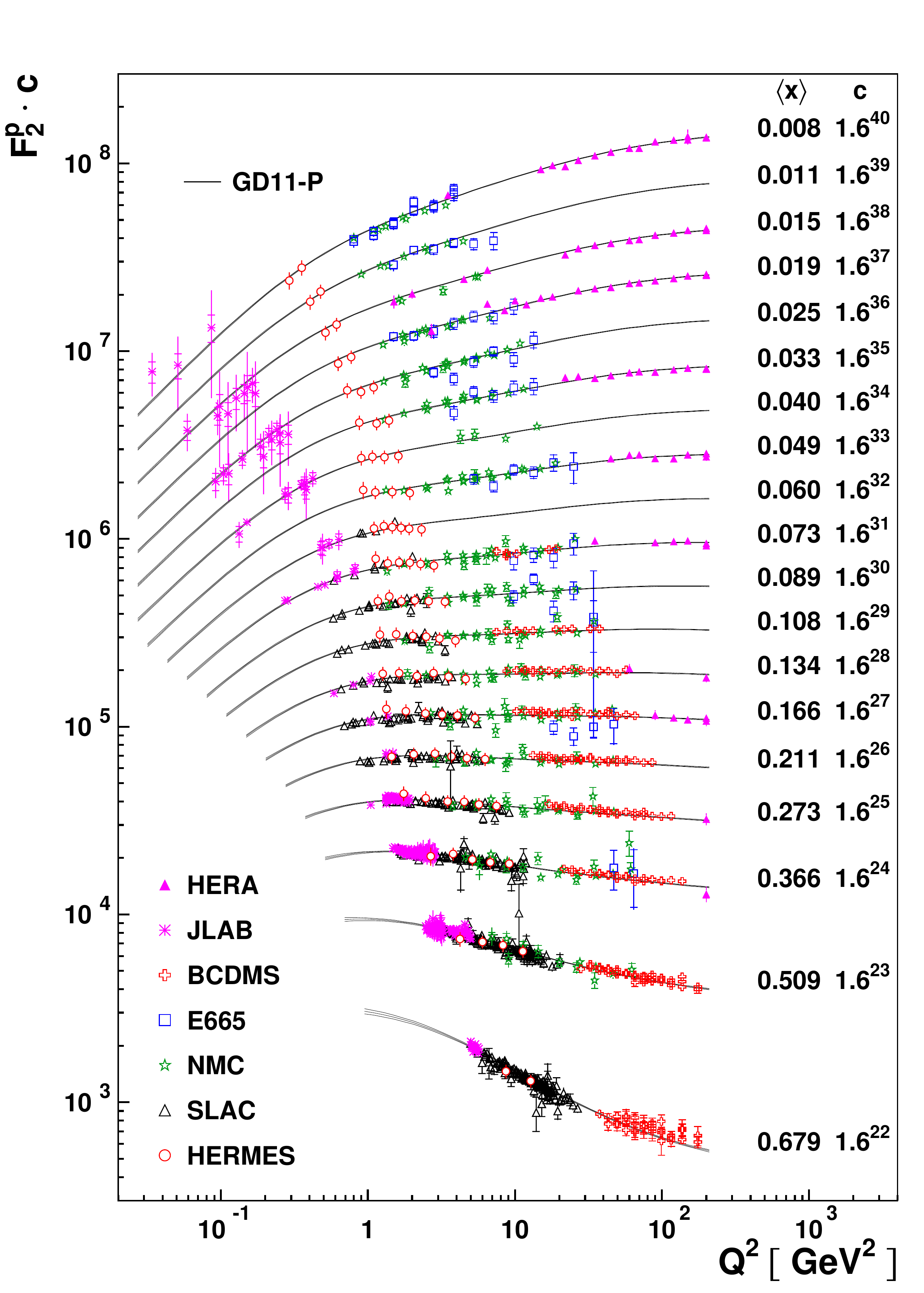}
\includegraphics[width=0.48\textwidth]{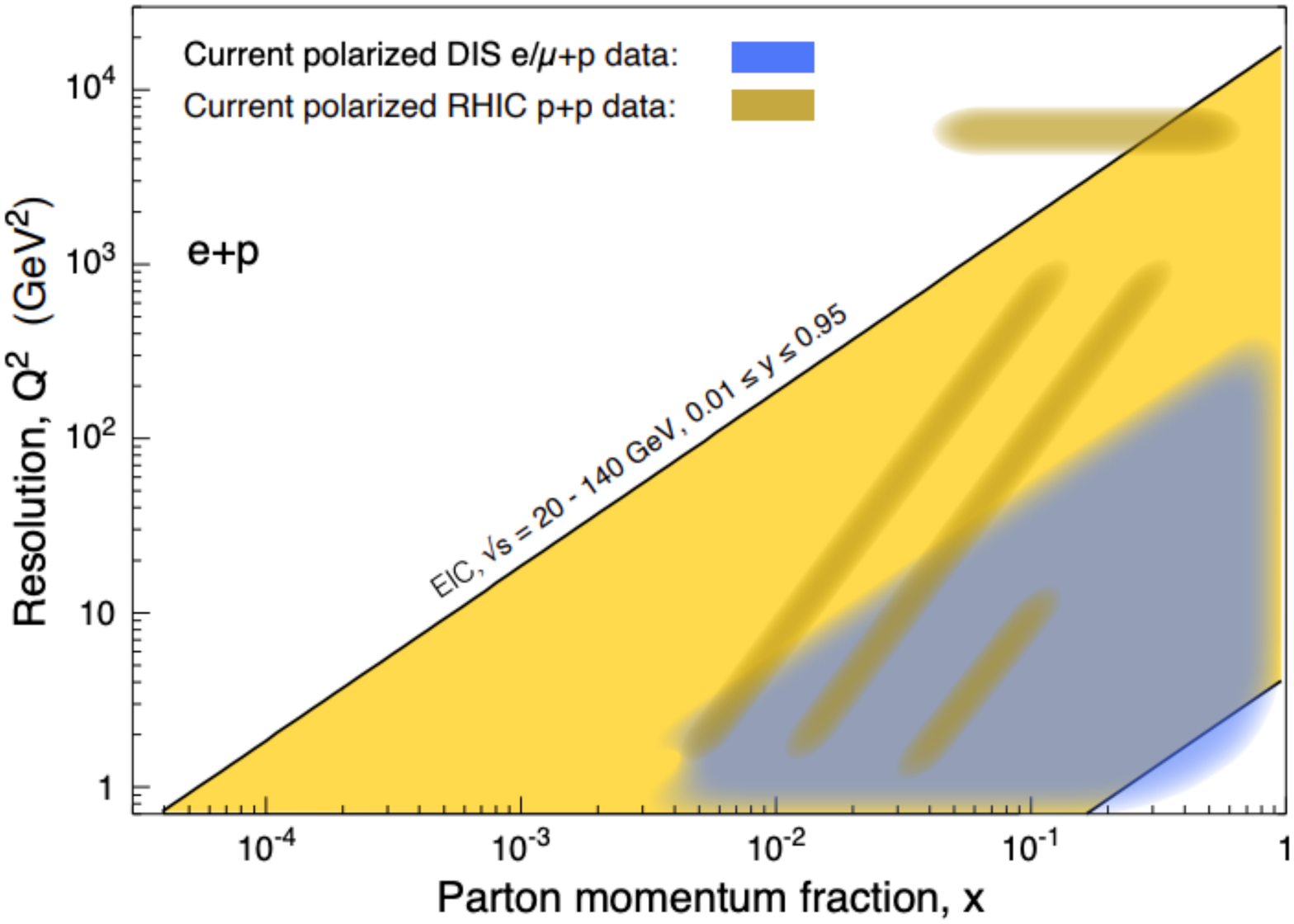}
\end{center}
\caption[Scaling of $F_2(x,Q^2)$ and $(x,Q^2)$ coverage]{Left: structure function $F_2(x,Q^2)$ from experiments at SLAC, CERN, Fermilab and DESY together with a phenomenological parameterization \cite{hermesf2}. Right: coverage of $(x,Q^2)$ at past and existing ep / $\mu$p DIS experiments at CERN, DESY, JLab, and SLAC, and pp experiments at BNL-RHIC, and the coverage of the future Electron-Ion Collider (EIC) at BNL in yellow. Figure from the 2021 EIC yellow report \cite{EICYellowReport2021}.}
\label{fig:f2}
\end{figure}

An overview including modern $F_2$ measurements at SLAC, CERN, Fermilab, Jefferson Lab, and DESY is shown in Fig.~\ref{fig:f2},
spanning a wide range in $(x,Q^2)$ phase space from the fixed-target experiments at JLab with electron beam energy of 6~GeV (center-of-mass energy $\sqrt{s}=\sqrt{2M_{\text p}E_{\text e}}\approx 3$~GeV) to the HERA collider experiments with proton beam energy of 920~GeV and electron beam energy of 27.6~GeV ($\sqrt{s}=\sqrt{4E_{\text p}E_{\text e}}\approx318$~GeV). The coverage in $x$ and $Q^2$ at past, current, and future DIS and pp facilities is also shown in Fig.~\ref{fig:f2}. The different ``branches'' of data points in the left hand side were recorded at different fixed values of $x$-Bjorken and are plotted in an exploded view for better visibility by multiplying $F_2$ by different constant factors $c$, as indicated in the Figure.  

\begin{figure}
\begin{center}
\includegraphics[width=.9\textwidth]{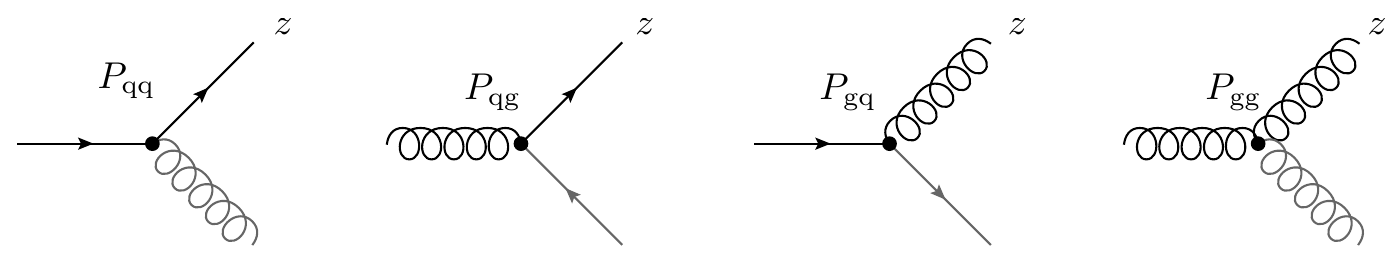}
\end{center}
\caption[QCD splitting graphs]{The QCD splitting functions in leading order QCD.}
\label{fig:qcd_processes}
\end{figure}

Clearly, there is some $Q^2$ dependence, or scaling violation, in $F_2(x=\mathrm{fixed},Q^2)$. This does however not mean that quarks are not point-like! What we observe here is a behavior dynamically generated by QCD. Partons constantly interact with each other via the exchange of gluons, thereby reshuffling momentum. The theoretical foundation for the description of these processes was laid in 1977 with the Dokshitzer–Gribov–Lipatov–Altarelli-Parisi, or DGLAP-, equations \cite{dglap}, a system of $(2n_{\text f}+1)$ coupled 
evolution equations (with $n_{\text f}$ the number of participating ``active'' quark flavors) which uses \emph{splitting functions} $P_{ij}$, $i,j\in \{{\text{q,p}}\}$ to describe the exchange of momentum between quarks q and gluons g. In other words, the splitting function represents the probability that
a daughter parton $i$ 
splits from a parent parton $j$. The different splitting graphs are shown in Fig.~\ref{fig:qcd_processes}. 
The processes shown in the figure, often referred to as \emph{QCD radiative effects}, can be seen as the first term of a perturbation series in powers of the strong coupling constant $\alpha_{\text S}$ (see App.~\ref{sec:qedqcd}). The perturbation series will be truncated at some order and we refer to this by saying leading (or lowest) order (LO), next-to-leading order (NLO), next-to-next-to-leading order (NNLO), and so on, \emph{pQCD} (perturbative QCD). 

The DGLAP equations can be divided into two branches: one describing the coupled \emph{singlet/gluon} evolution with the singlet quark distribution $q^\mathrm{S}$ and the gluon distribution $g$, and one describing the \emph{non-singlet} (NS) evolution with the non-singlet quark distributions $q_3$ and $q_8$. Setting  $n_{\text f}=3$ (thus only the lightest quark flavors u, d, s will participate), one has:
\begin{eqnarray}
q^{\mathrm{S}}\equiv a_0 &=&(u+\overline{u})+(d+\overline{d})+(s+\overline{s}),\\
a_3&=&(u+\overline{u})-(d+\overline{d}),\\
a_8&=&(u+\overline{u})+2(d+\overline{d})-(s+\overline{s}),\\
q^{\mathrm{NS}}&\equiv&a_3+a_8.
\end{eqnarray}
\label{page:dglap}
The $a_0$, $a_3$, and $a_8$ are called axial charges. In compact matrix notation, the DGLAP equations for the singlet/gluon evolution read:
\begin{equation}
\frac{\partial}{\partial\;\mathrm{ln}\mu^2}\begin{pmatrix}q^{\mathrm{S}}\\g\end{pmatrix}=\frac{\alpha_S}{2\pi}\begin{pmatrix}P_{\text{qq}}&2n_{\text f}P_{\text{qg}}\\P_{\text{gq}}&P_{\text{gg}}\end{pmatrix}\otimes\begin{pmatrix}q^{\mathrm{S}}\\g\end{pmatrix}.
\label{eq:dglap}
\end{equation}
The \emph{factorization scale} $\mu^2$ separates \emph{long} and \emph{short distance} physics (see also factorization theorem on page~\pageref{page:factorization}). Often one makes the simplifying assumption that the factorization scale is equal to the hard scale, $\mu^2=Q^2$. While we saw  in Eq.~\ref{eq:dglap} that the singlet distribution evolves coupled to the gluon distribution, the non-singlet distributions evolve independently of the gluon, and independently of each other. The dependence of the quark and gluon distributions is referred to as \emph{$Q^2$-evolution} of QCD.

We now return to the $F_2$ behavior in Fig.~\ref{fig:f2} and try to motivate the observations qualitatively:
\begin{description}
\item At small $x$, $F_2$ rises with rising $Q^2$. This is caused by \emph{gluon splitting} $\bf g\rightarrow q\overline{q}$ resulting in more quarks with small momentum being able to be resolved.
\item At large $x$, $F_2$ falls with rising $Q^2$. Due to gluon radiation $\bf q\rightarrow qg$, there are fewer quarks with large momentum fraction and more with smaller momentum fraction.
\end{description}
A ``QCD-improved'' QPM takes into account QCD radiative effects by introducing a $Q^2$-depen\-dence in the structure functions, e.g., $F_2(x)\rightarrow$  $F_2(x,Q^2)$. The Callan-Gross relation Eq.~\ref{eq:callangross} represents the lowest order QCD. The ``QCD-improved'' version reads:
\begin{equation}
2xF_1(x,Q^2)=\frac{1+\gamma^2}{1+R(x,Q^2)}F_2(x,Q^2)
\end{equation}
and includes the ratio $R$ of the longitudinal (L) over transverse (T) virtual-photon absorption cross section:
\begin{equation}
R(x,Q^2)\equiv\frac{\sigma_\text{L}}{\sigma_\text{T}}=(1+\gamma^2)\cdot\frac{F_2(x,Q^2)}{2xF_1(x,Q^2)},
\label{eq:r}
\end{equation}
with $\gamma=\sqrt{Q^2}/\nu$.

\subsection{Parton distribution functions}
\label{sec:pdfs}

We return to the quark longitudinal-momentum probability distributions from Eq.~\ref{eq:qpm}. These distributions are known under the name \emph{parton distribution functions} (PDFs).
\begin{figure}
\begin{center}
\includegraphics[width=0.30\textwidth]{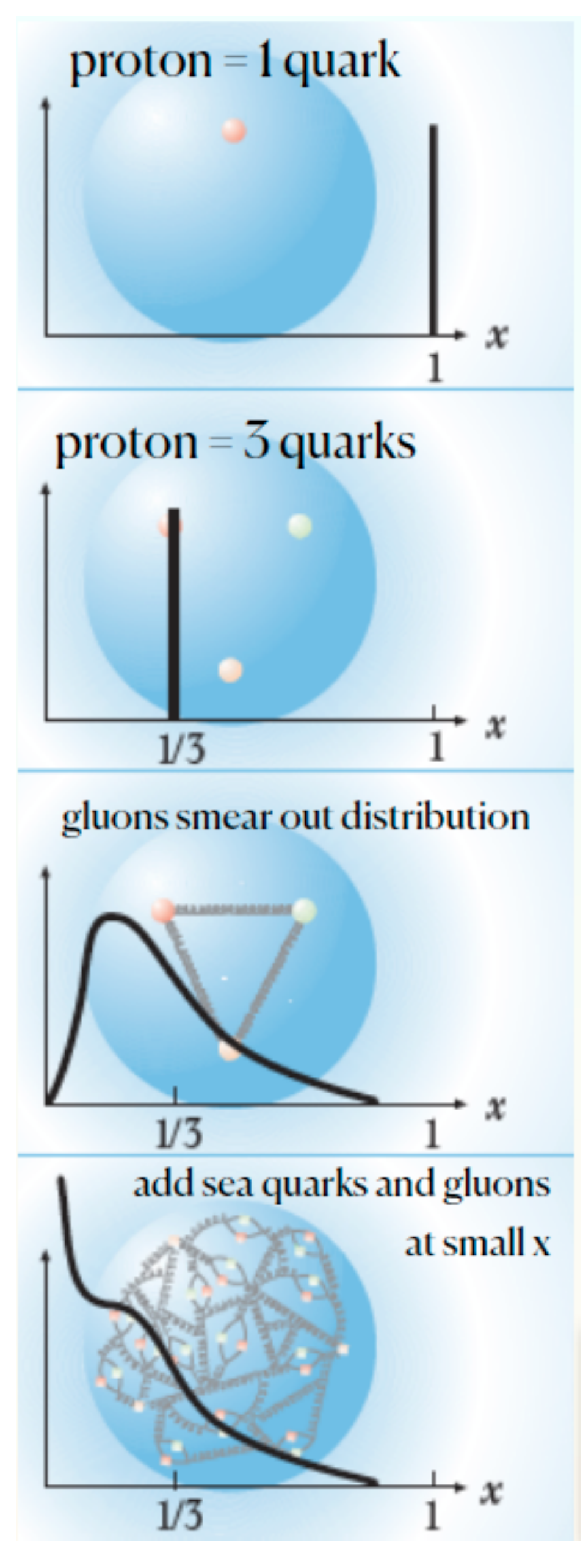}
\includegraphics[width=0.66\textwidth]{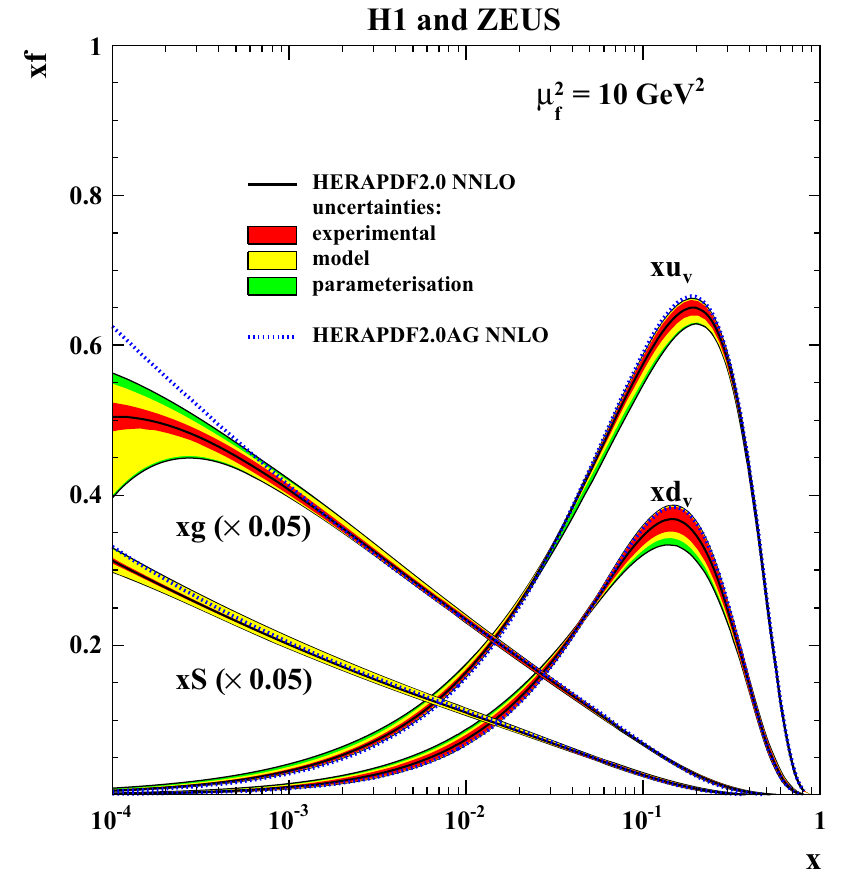}
\end{center}
\caption[Parton distribution functions]{Left: explanation of the shapes of parton distribution functions. Figure modified from Ref.~\cite{starkekraft}. Right: HERA PDFs \cite{herapdfs}.}
\label{fig:pdf}
\end{figure}
The expected PDF shape is motivated in Fig.~\ref{fig:pdf}. If the proton consisted of one quark only, the PDF would be a delta function at 1 since the one quark would have to make up for the entire nucleon's momentum. If there were three quarks in the proton, and nothing else, they would each carry 1/3 of the nucleon's momentum. Accounting for the presence of gluons mediating the interaction between the quarks, the quark PDFs get smeared out because gluons constantly exchange momentum with the quarks. A quark can receive extra momentum from gluons, or transfer momentum to gluons. The maximum of the PDF shifts to values smaller than 1/3 because gluons carry about 50\% of the proton's momentum, while at leading order QCD they do not contribute to DIS with charged leptons. 
If also sea quarks are considered, which are constantly produced and annihilated in the proton, {\bf g~$\rightarrow$~q$\overline{\text{q}}\rightarrow $~g}, there are sea-quark PDFs that rise with decreasing $x$. Sea quarks are seen in charged-lepton DIS because they have electrical charge.

The HERA PDFs represent a combined QCD analysis 
using data from the H1 and ZEUS experiments at the ep-collider HERA at DESY, which was operated from 1992 until 2007. An example from the most recent version called HERAPDF2.0 \cite{herapdfs} is shown in Fig.~\ref{fig:pdf}. Valence quarks (with subscript v) carry large fractions of the proton’s longitudinal momentum. The valence distribution has its maximum at around $x=0.2$. Sea quarks (s) and gluons (g) carry small fractions of the nucleon’s longitudinal momentum. At very large $x$, $F_2$ is very small. It is thus very unlikely that one quark alone carries the majority of longitudinal nucleon momentum.

At HERA and in general at collider center-of-mass energies, the DIS process can also be mediated by Z$^0$ and W$^\pm$ bosons of the weak interaction. The process {\bf ep~$\rightarrow$~eX} happens via the exchange of virtual photons ($\gamma^*$) and Z bosons\footnote{At low $Q^2$, i.e., $Q^2\ll M^2_{\text Z}$ and thus typical fixed-target experiments, the contribution of Z-boson exchange is negligible ($M_{\text Z}\approx 91.2$~GeV).}, collectively referred to as \emph{neutral current} (NC). So-far, we had only considered 1-photon exchange, which is a subcategory of NC. In addition to the structure function $F_2$, which is dominant at LO, there is an additional LO structure function $F_3$ arising from $\gamma^*/$Z interference. There is a parity violating (see App.~\ref{sec:conqua}) $\gamma^*/$Z contribution to the cross section, which changes its sign with the lepton charge. At NLO, there is is an additional ``longitudinal'' structure function $F_{\text L}$ that becomes important at low values of $x$, where gluon contributions grow in importance. 
At low $Q^2$ and low $y=Q^2/(sx)$, the cross section is driven by $F_2(x,Q^2)$ to a very good approximation. The \emph{charged-current} (CC) process $\bf ep\rightarrow \nu X$ mediated by W$^\pm$ bosons is sensitive to different quark flavor combinations if different electron beam charges\footnote{The symbol e in this section can both stand for electron and positron.} e$^\pm$ are used \cite{klein2008}. The exchange of the W$^{\pm}$ is selective on helicity and charge of the involved fermions and can therefore be used to separately determine quark- and anti-quark distributions in the nucleon.

Recently a PDF set has become available that includes in addition to the HERA legacy data diverse TeV-scale data in proton-proton collisions from the ATLAS experiment at CERN \cite{ATLASpdf21}. 

We finish this section with two important remarks. Firstly, the parton distribution functions cannot be derived perturbatively from first principles and therefore have to be determined experimentally. The QCD factorization theorem assumes that the measured inclusive DIS cross section $\sigma^{\mathrm{incl}\;\mathrm{DIS}}$ can be decomposed into a pQCD-calculable hard-scattering, short-distance part $\hat{\sigma}$ and the non-perturbative ``soft'', long-distance PDF:\\
``$\sigma^{\mathrm{incl}\;\mathrm{DIS}}=\hat{\sigma}(x,Q^2/\mu^2)\otimes\mathrm{PDF}(x,\mu^2)$'', where $\mu^2$ is again the factorization scale. The hard scattering part can then be calculated in pQCD and does not depend on the PDF. Secondly, on the other hand, this makes the PDFs universal - they do not depend on the process that they are probed with. We will return to this concept of universality later in this article. \label{page:factorization}

\subsection{Flavor (a)symmetry of the sea and neutrino-induced DIS}
\label{sec:neutrinodis}

In the early 1990s, the NMC experiment at CERN measured the $F_2$ structure functions of the proton and the neutron by scattering muon beams of 90~GeV and 280~GeV energy off hydrogen and deuterium targets \cite{nmc1992}. Deuterium targets are frequently used to overcome the unavailability (because instability) of free neutron targets. Following Eq.~\ref{eq:qpm}, we can write $F_2^{\ell,{\text p}}$ of the proton and $F_2^{\ell,{\text n}}$ of the neutron obtained from lepton ($\ell$)-induced DIS as:
\begin{eqnarray}
F_2^{\ell,{\text p}}(x)=x \left[\frac{1}{9}(d_{\text v}^{\text p}+d_{\text s}+\overline{d}_{\text s})+\frac{4}{9}(u_{\text v}^{\text p}+u_{\text s}+\overline{u}_{\text s})+\frac{1}{9}(s_{\text s}+\overline{s}_{\text s})\right], \label{eq:f2p}\\
F_2^{\ell,{\text n}}(x)=x \left[\frac{1}{9}(d_{\text v}^{\text n}+d_{\text s}+\overline{d}_{\text s})+\frac{4}{9}(u_{\text v}^{\text n}+u_{\text s}+\overline{u}_{\text s})+\frac{1}{9}(s_{\text s}+\overline{s}_{\text s})\right],
\label{eq:f2n}
\end{eqnarray}
with $u\equiv u(x)$, $d\equiv d(x)$, and $s\equiv s(x)$ the momentum distributions for up-, down-, and strange-quarks, respectively, and correspondingly for anti-quarks $\overline{u}$, $\overline{d}$, $\overline{s}$. The indices v and s stand for valence and sea quarks, respectively. In Eqs.~\ref{eq:f2p} and \ref{eq:f2n} we assumed that the sea quark distributions in the proton and neutron behave similarly by skipping the proton and neutron indices for sea quarks. This is justified a posteriori by the NMC's result of the structure function ratio: for small values of $x$, $F_2^{\ell,{\text p}}/F_2^{\ell,{\text n}}\approx 1$, which also tells us that sea quarks dominate over valence quarks at small $x$. For large values of $x$, valence quarks dominate, $F_2^{\ell,{\text p}}/F_2^{\ell,{\text n}}\approx 1/4=e_{\text d}^2/e_{\text u}^2$. Large momentum fractions in the proton are carried by the u-valence quarks and in the neutron by the d-valence quarks. 

The PDFs were introduced as probability distributions in longitudinal momentum space. Can we verify that we obtain correct quark numbers when we integrate the PDFs over $x$? Doing this for the proton, $\int_0^1\frac{\mathrm{d}x}{x}F_2^{\text p}(x)$, brings along the difficulty that the contributions from the sea quarks diverge. As we go closer to $x=0$, there is an infinite number of q$\overline{\text{q}}$ pairs coming into existence. We therefore use the trick of forming the difference of proton and neutron structure functions, thereby eliminating the sea-quark infinities. This is known as the \emph{Gottfried sum rule}:
\begin{eqnarray}
\int_0^1\frac{\mathrm{d}x}{x}\left[F_2^{\text p}(x)-F_2^{\text n}(x)\right]&\stackrel{\tiny \circled{1}}{=}&\int_0^1\frac{\mathrm{d}x}{x}\left[\frac{1}{9}d_{\text v}^{\text p}-\frac{1}{9}d_{\text v}^{\text n}+\frac{4}{9}u_{\text v}^{\text p}-\frac{4}{9}u_{\text v}^{\text n}\right]\nonumber\\
&=&\frac{1}{3}\left[\underbrace{\int_0^1\frac{\mathrm{d}x}{x}u_{\text v}^{\text p}}_{=2}-\underbrace{\int_0^1\frac{\mathrm{d}x}{x}d_{\text v}^{\text p}}_{=1}\right]=\frac{1}{3},
\label{eq:gsr}
\end{eqnarray}
where we have used that proton and neutron are isospin partners (App.~\ref{sec:isospin}), which arises from the isospin symmetry between u- and d-quarks. We therefore have $u_\text{v}^{\text p}=d_\text{v}^{\text n}$ and $d_\text{v}^{\text p}=u_\text{v}^{\text n}$. In $\tiny \circled{1}$, we assumed a flavor symmetric sea, i.e., $u_\text{s}^{\text p}=d_\text{s}^{\text p}=u_\text{s}^{\text n}=d_\text{s}^{\text n}$. We can then insert the number of u-quarks and the number of d-quarks in the proton. Analysis of the NMC data indicates that experimentally one measures a value that is not quite 1/3, but rather closer to 0.25 \cite{nmc1994}. One popular explanation for this finding is a flavor asymmetric sea:
\begin{equation}
\int_0^1\frac{\mathrm{d}x}{x}\left[F_2^p(x)-F_2^n(x)\right]=\frac{1}{3}-\frac{2}{3}\int_0^1\frac{\mathrm{d}x}{x}\left[\;\overline{d}(x)-\overline{u}(x)\right],
\label{eq:flaasy}
\end{equation}
which would reduce the measured value of the structure function difference. 
A review of the flavor asymmetry of sea quark distributions can be found in Ref.~\cite{JCP1999}.

\begin{figure}
\begin{center}
\includegraphics[width=0.7\textwidth]{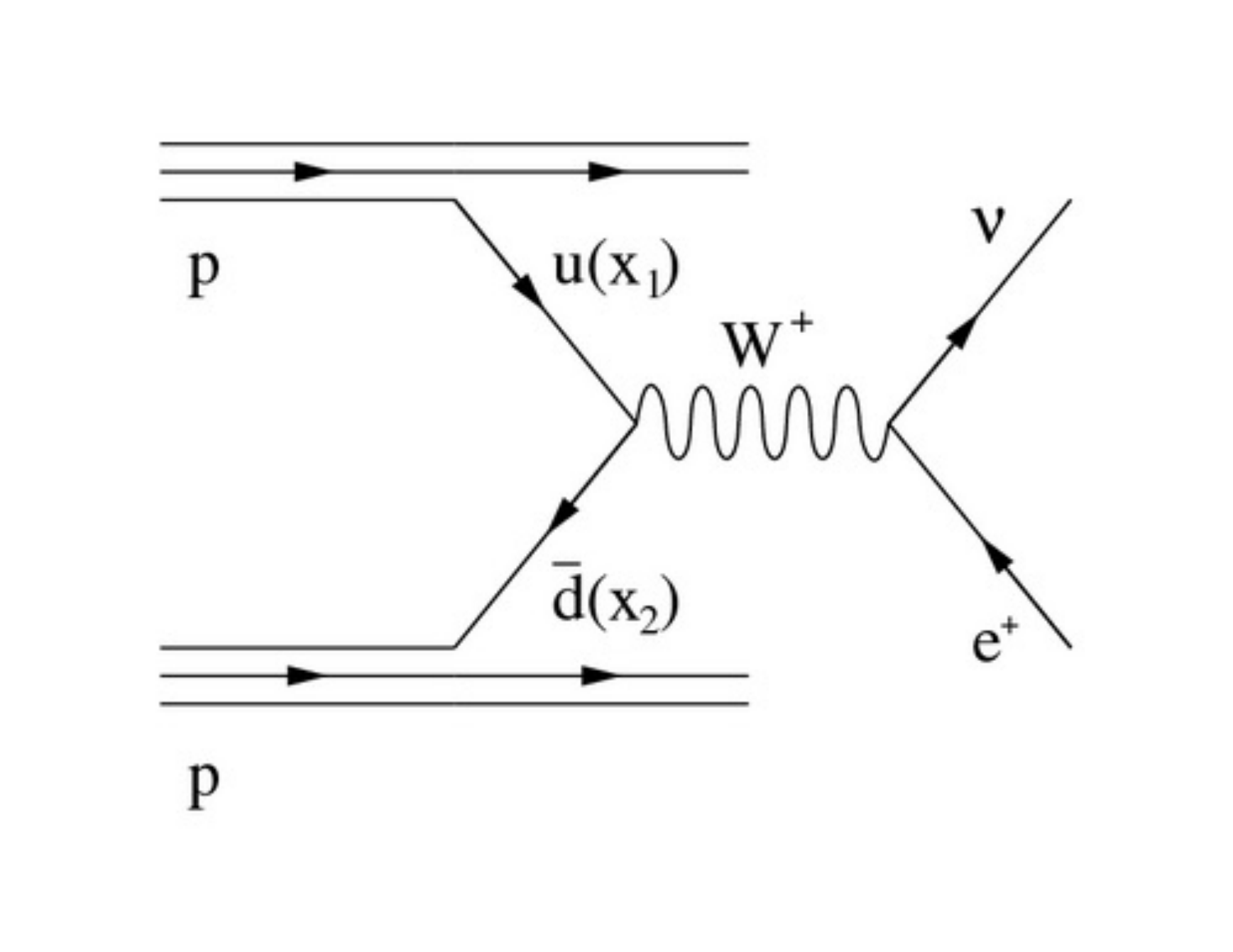}
\end{center}
\caption[W$^\pm$ production in proton-proton collisions]{W production in proton-proton collisions. Shown is the process pp$\rightarrow$W$^+(\rightarrow$e$^+\nu_{\text e})$X. There is the analogous process pp$\rightarrow$W$^-(\rightarrow$e$^-\overline{\nu}_{\text e})$X.}
\label{fig:wprod}
\end{figure}

New results on the flavor composition of sea quarks have recently been published by the SeaQuest experiment at Fermilab and the STAR experiment at BNL's proton-proton collider RHIC using two independent experimental approaches. Sea\-Quest measured the cross section of the proton-induced Drell-Yan process on hydrogen (pp~$\rightarrow\ell\overline{\ell}$~X) and deuterium (pd~$\rightarrow\ell\overline{\ell}$~X) targets in the sea-quark domain \cite{SeaQuest}. STAR measured the cross sections of weak-boson production pp~$\rightarrow$W$^\pm$($\rightarrow$~e$^\pm\stackrel{(-)}{\nu_{\text e}}$)~X (see Fig.~\ref{fig:wprod}) at the momentum scale of the weak-boson mass \cite{STAR:dbarubar}. The two experiments are complementary in their kinematic coverage and both find a flavor asymmetry in the sea: $\overline{d}(x)>\overline{u}(x)$.

Neutrino-induced DIS, $\nu\,{\text p}\rightarrow \ell$\,X, is a type of charged-current process we have not yet discussed. All CC features mentioned above in Sec.~\ref{sec:pdfs} apply also to DIS with neutrino beams, it for example allows the separation of valence and sea quark distributions. Here we will look at a result from the 1970s that provided important information about the gluon momentum distribution in the proton. A muon-neutrino beam was produced at CERN by sending the 26~GeV proton beam from the Proton Synchrotron (PS) on a beryllium target, thereby producing charged pions and kaons, which decay into neutrinos and muons: $\pi^\pm\rightarrow \stackrel{(-)}{\nu_\mu}\mu^\pm$ and K$^\pm\rightarrow \stackrel{(-)}{\nu_\mu}\mu^\pm$. The structure function $F_2^{\nu,{\text N}}(x)$ of the nucleon (N) from neutrino-induced DIS was then extracted from data taken with the Gargamelle bubble chamber. Since all quarks couple equally via their weak charge, there is \emph{no} $e_{\text q}^2$ in the sum of $F_2$ (Eq.~\ref{eq:qpm}) and we find:   
\begin{equation}
F_2^{\nu,{\text N}}(x)=\frac{18}{5}\cdot F_2^{\ell,{\text N}}(x),
\end{equation}
where $F_2^{\ell,{\text N}}$ is the averaged nucleon structure function of the proton and neutron from charged-lepton- ($\ell$-) induced DIS (assuming a flavor symmetric sea), and 5/18 is the mean square electrical charge of u- and d-quarks in the nucleon. The Gargamelle measurement yielded \cite{perkins:1972}:
\begin{equation}
\int_0^1\text{d}x\;F_2^{\nu,{\text N}}(x)\approx 0.5,
\end{equation}
which means that about half of the longitudinal proton’s momentum is carried by particles that have neither electric nor weak charge: the gluons. Historical reviews of the discovery of quarks are for example given in Jerome Friedman's Nobel lecture \cite{Friedman1990} and in Ref.~\cite{riordan:1992}.

\section{Longitudinal spin structure of the nucleon}
\label{sec:longitudinal}

\subsection{Polarizing targets and beams}
\label{sec:polarizing}

So far we have studied proton structure from scattering experiments on unpolarized protons. That means that in the used nuclear targets and lepton or hadron beams, the directions of the proton or lepton spins were randomly distributed, thus yielding effective zero polarization. To create a polarized sample, the spin projections of some or better most of the particles must point in the same direction. This is in many cases achieved by magnetic fields applied in a specific direction, which can be longitudinal ($\leftrightarrow$) or transverse ($\updownarrow$) to the beam direction. We then have the \emph{longitudinal vector polarization $P_z$} of a particle ensemble,
\begin{equation}
P_z=\frac{N^\rightarrow-N^\leftarrow}{N^\rightarrow+N^\leftarrow},
\label{eq:vectorpol}
\end{equation}
where $N^\rightarrow$ is the number of particles with spin projection along the beam direction, and $N^\leftarrow$ opposite to it. Analogously, a transverse polarization can be constructed from $N^\uparrow$ and $N^\downarrow$. 

As we will see in the following sections, scattering experiments using polarized samples, both nucleons and leptons, open the door to a rich set of phenomena and allow to probe proton structure at a deeper and higher-dimensional level. In this sense, this introductory subsection also applies to Secs.~\ref{sec:gpds} and \ref{sec:tmds}. The development of polarized targets and beams has therefore been, and continues to be, an intrinsically important aspect of spin physics. There are different principles of polarizing targets and beams. 

Beams of polarized particles are generated from polarized sources or via self-pol\-arization effects. At the Continuous Electron Beam Accelerator Facility (CEBAF) at Jefferson Lab, the polarized source is a gallium-arsenide photocathode emitting longitudinally polarized electrons \cite{cebaf2001}. The electrons are then accelerated to 6 or 11-12\,GeV and are led on fixed targets in the experimental Halls A, B and C.  Electron polarization values of over 80\% are achieved at high beam intensities. 

\begin{figure}
\begin{center}
\includegraphics[width=\textwidth]{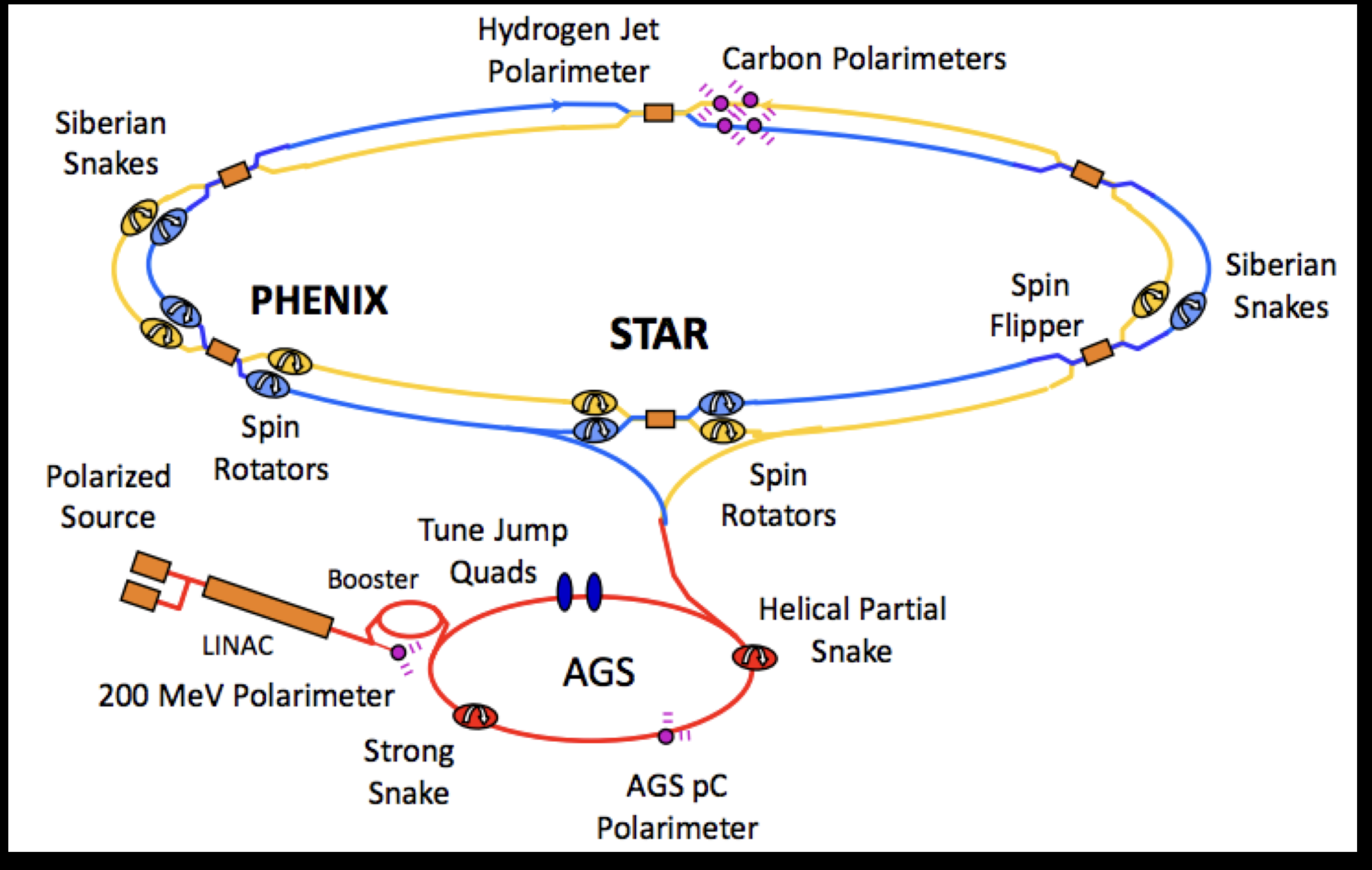}
\end{center}
\caption[RHIC pp collider]{The RHIC pp and heavy-ion collider at BNL (Long Island, NY, USA) with the locations of the STAR and (s)PHENIX experiments.}
\label{fig:rhic}
\end{figure}

The Relativistic Heavy Ion Collider (RHIC) at BNL uses an optically pumped polarized ion source (OPPIS) that transfers electron polarization to protons, which are then accelerated (see Fig.~\ref{fig:rhic}). To overcome the effects of depolarizing resonances in the circular accelerator, beam-line elements called \emph{Siberian Snakes} are introduced \cite{rhic2003}. RHIC with proton beams (pp) operates at typically $\sqrt{s}=200$\,GeV or 500\,GeV. \emph{Spin rotators} turn if desired the proton polarization, which is vertical (transverse) by default, into a (horizontal) longitudinal polarization at the collision points of the STAR, PHENIX (in the past), and sPHENIX experiments (in the nearest future). RHIC also collides heavy ions to study the quark-gluon plasma. 

\begin{figure}
\begin{center}
\includegraphics[width=\textwidth]{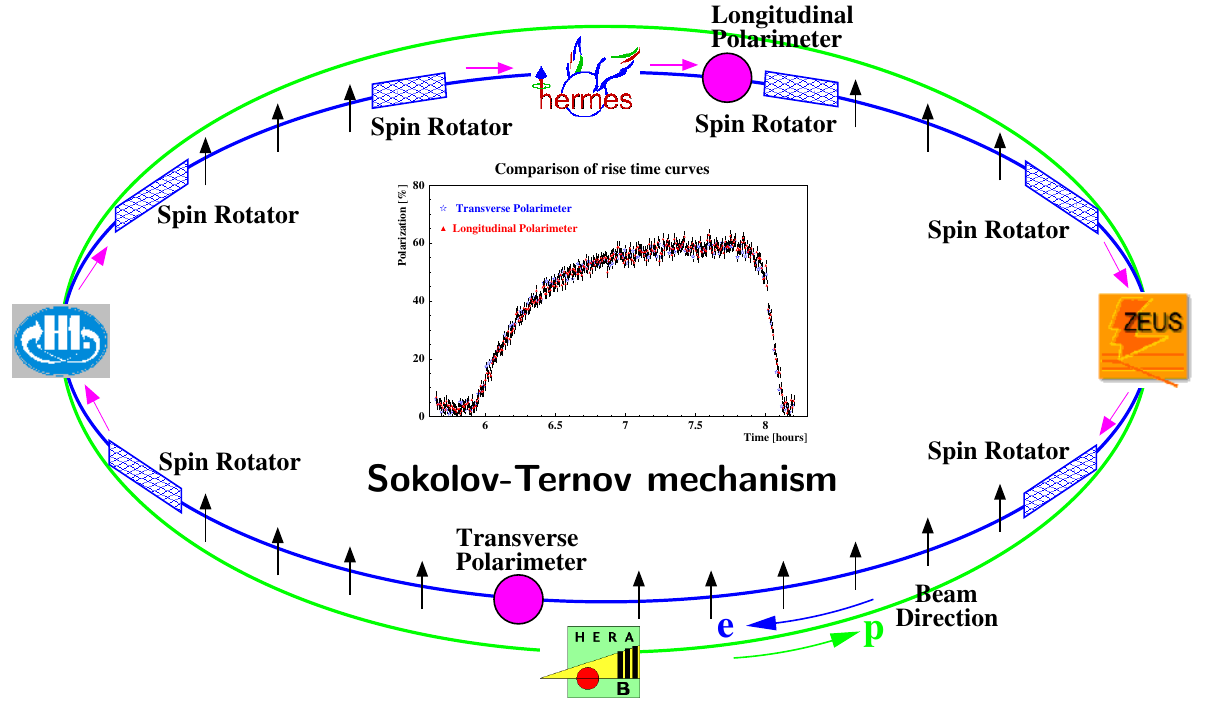}
\end{center}
\caption[HERA ep collider]{The HERA ep collider at DESY (Hamburg, Germany) with the locations of the H1 and ZEUS collider experiments and the HERMES and HERA-B fixed-target experiments (figure adapted from Ref.~\cite{hermes:deltaq}). Also shown is the build-up of electron polarization in the storage ring \cite{riedl:2009}.}
\label{fig:hera}
\end{figure}

The HERA electron-proton collider at DESY \cite{hera1993} operated with e$^\pm$ beams that became transversely self-polarized in the storage ring due to a small asymmetry in the emission of spin-flip synchrotron radiation (Sokolov-Ternov effect) \cite{hera1994}, see Fig.~\ref{fig:hera}. Once the e$^+$ or e$^-$ beam had been injected into HERA, it took about 40~minutes of rise time to reach the final electron polarization between 40\% and 60\%. Since the rise time scales with the mass of the circulating particle, it takes almost 2000 times longer for protons to become self-polarized, which is why other methods to polarize proton beams have been developed. The polarized electrons of energy 27.6\,GeV were brought in collision with unpolarized protons of typically 920\,GeV at the H1 and ZEUS experiments. The fixed-target experiments HERMES and HERA-B used the electron or proton beam only, respectively. Spin rotators up- and downstream of HERMES, and later also at H1 and ZEUS, supplied longitudinal electron beam polarization at the location of the experiments. HERA ran with the same electron beam charge over many months, while flipping the direction of the electron spin, i.e., helicity, every few months. Both beam charges e$^\pm$ were available for either beam helicity e$^{\substack{\rightarrow\\[-.05in]\leftarrow}}$. The Electron-Ion Collider (EIC) will be the first collider to provide both polarized electron and proton beams.

\begin{figure}
\begin{center}
\fbox{\includegraphics[width=\textwidth]{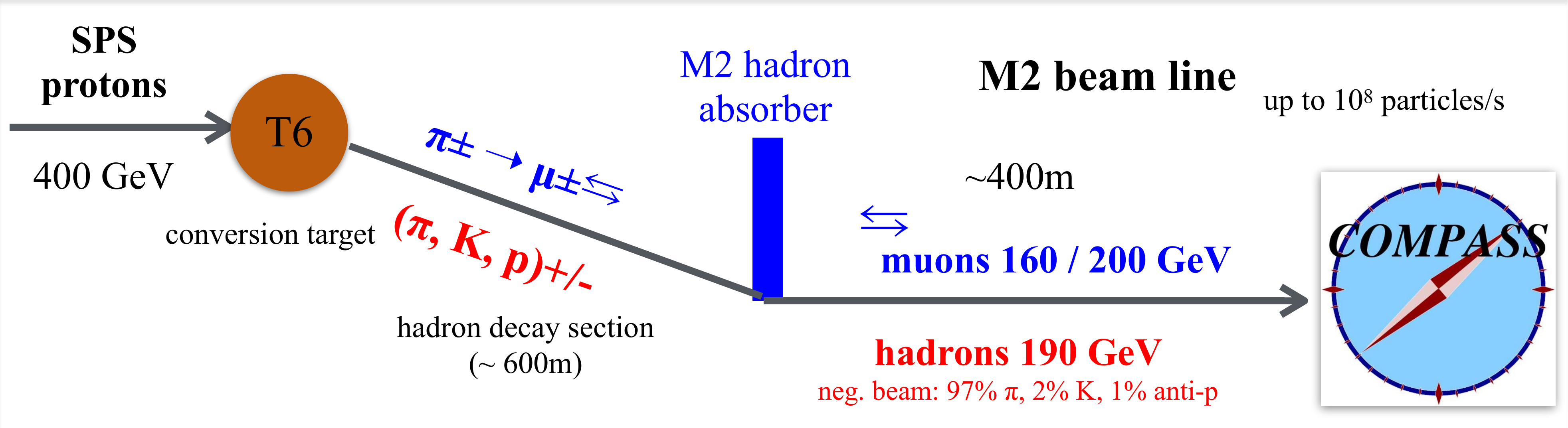}}
\includegraphics[width=0.35\textwidth]{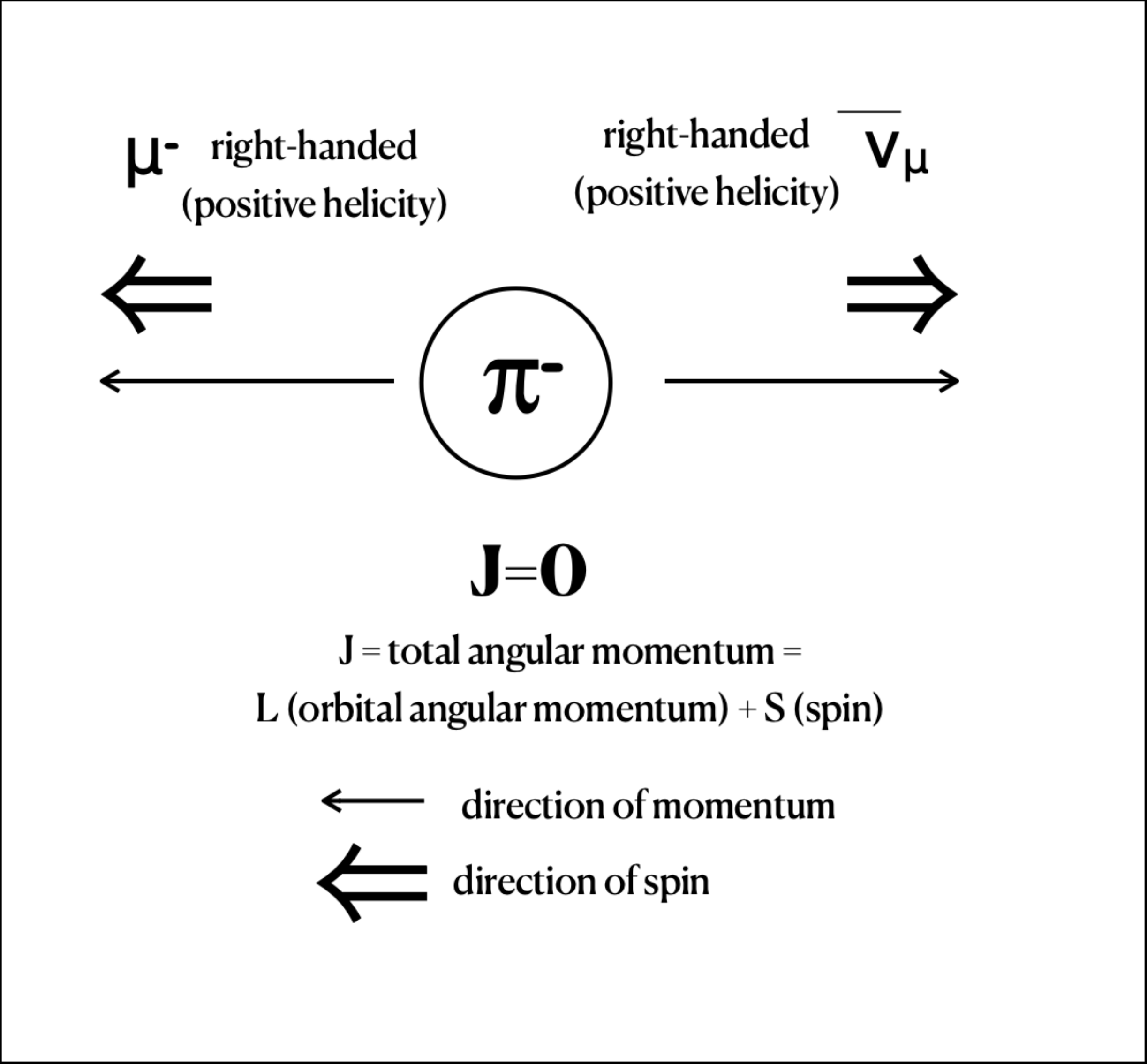}
\includegraphics[width=0.61\textwidth]{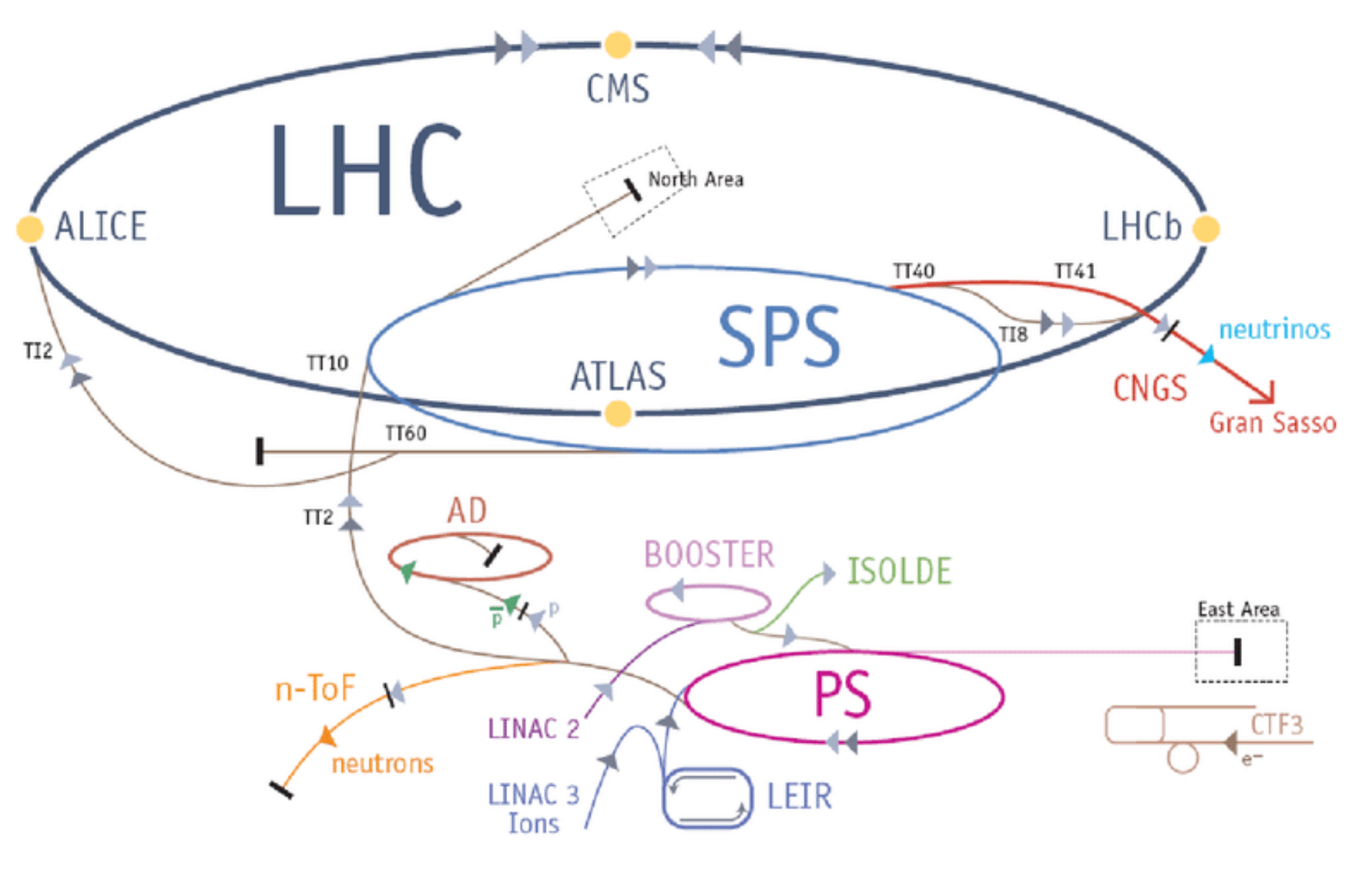}
\end{center}
\caption[CERN M2 beamline]{Top: the M2 beam line at the CERN SPS (CERN North Area), which can be operated with either muon (blue) or hadron beams (red).  Bottom left: the parity-violating pion decay yielding spin-polarized muons, and bottom right: the CERN accelerator complex with the Large Hadron Collider (LHC) of 27~km circumference at the Swiss-French border close to Geneva, Switzerland.}
\label{fig:m2}
\end{figure}

Starting in the mid-1970s and inspired by the DIS results from the SLAC experiments, there were a series of fixed-target experiments in the CERN North Area (NA) that made use of the high-energy and high-intensity muon beams from the M2 beamline of the Super Proton Synchrotron (SPS) in the experimental area EHN2 - EMC, BCDMS, NMC, SMC, and since 2002 the COMPASS experiment. The AMBER experiment and others will follow in the future \cite{PBC2019}. As shown in Fig.~\ref{fig:m2}, SPS (primary) protons of 400\,GeV are led on a beryllium production target, where they produce (secondary) mesons, mostly pions. The pions decay into neutrinos and (tertiary) muons, which are longitudinally polarized due to the parity-violating nature of the weak decay: the negatively charged muons have positive helicity, $\mu^{\substack{-\\[-.03in]\rightarrow}}$, and the positive muons negative helicity, $\mu^{\substack{+\\[-.03in]\leftarrow}}$. The remaining hadrons in the beam are filtered out by movable hadron absorbers. Muons of energies between 160 and 200\,GeV and a polarization of about 80\% are then led on the COMPASS target. If the hadron absorbers are driven out of the beamline, COMPASS can also receive pion beams of energies between 160 and 190\,GeV. 

\begin{figure}
\begin{center}
\includegraphics[width=.8\textwidth]{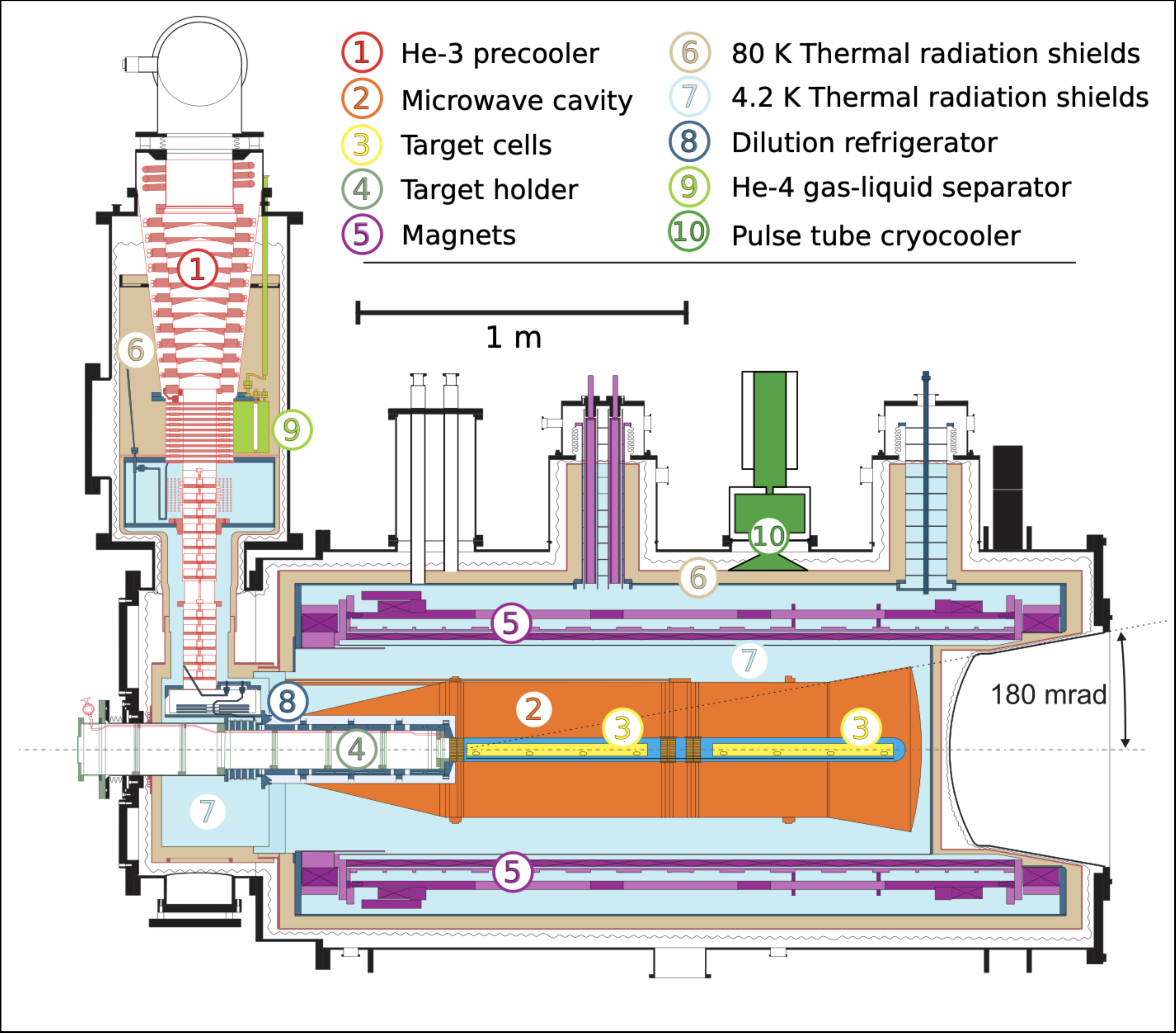}
\end{center}
\caption[COMPASS polarized target]{COMPASS polarized target \cite{compasstarget2016}.}
\label{fig:compasstarget}
\end{figure}

The polarized COMPASS target \cite{compasstarget2016} is shown in Fig.~\ref{fig:compasstarget}. It can provide longitudinal and transverse polarization of both protons and deuterons. The target material is immersed in liquid helium in a cylinder of 3-4~cm diameter and almost 1.5~m length along the beam axis and consists of frozen beads of ammonia (NH$_3$) for the proton measurements and of deuterated lithium ($^6$LiD) for the deuteron measurements. Polarization is built up via dynamic nuclear polarization (DNP) 
by irradiating the target material with microwaves to transfer polarization from the electrons to the protons or deuterons. The system is in a magnetic field of 2.5\,T (longitudinal solenoid) or 0.5\,T (transverse dipol). The target polarization is determined via the nuclear magnetic resonance (NMR) technique. Polarization values between 70\% and 90\% are achieved. An additional dilution factor to the polarization of 0.22 (for NH$_3$) or 0.5 (for $^6$LiD) has to be taken into account because of the presence of other, unpolarized nuclei in the target. To minimize systematic effects in spin asymmetry measurements, the target material is separated into two or three cells along the beamline with opposite-sign nuclear polarization, and the sign of the target polarization is reversed about every two weeks.
COMPASS has also used an unpolarized liquid hydrogen (LH$_2$) target and various solid nuclear targets. 

\begin{figure}
\begin{center}
\includegraphics[width=.8\textwidth]{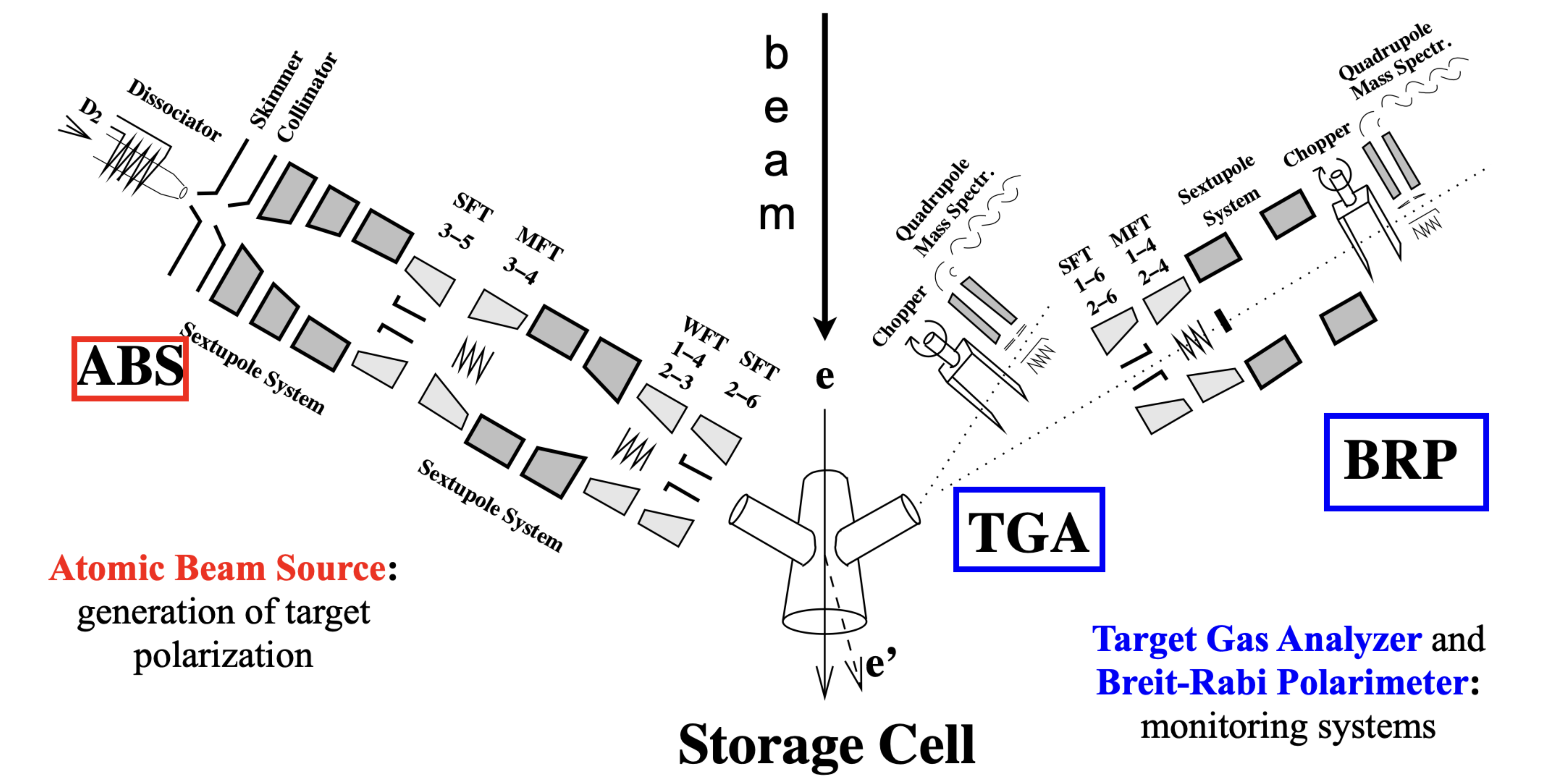}
\end{center}
\caption[HERMES polarized target]{HERMES polarized target \cite{hermespol2005}.}
\label{fig:hermes_target}
\end{figure}

HERMES used a novel technique that consisted of injecting pure nuclear-polarized gas into the HERA accelerator vacuum. The HERMES polarized target \cite{hermespol2005} shown in Fig.~\ref{fig:hermes_target}
consisted of an atomic beam source (ABS) to generate nuclear polarization, a Breit-Rabi polarimeter (BRP) to measure the polarization, and a target-gas analyzer (TGA) to analyze the atomic and the molecular content of the sample. In a magnetic field of about 300\,mT, the hyperfine states of hydrogen or deuterium gas undergo a Stern-Gerlach like separation. A series of sextupole magnets and high-frequency
transitions allowed to select the desired hyperfine states with nuclear polarization in a fast manner so that the direction of the target polarization could be reversed every 60 or 90~seconds, thereby minimizing possible systematic effects that might affect spin-asymmetry measurements. Longitudinal polarizations of about 85\% (proton and deuteron) and transverse polarizations of about 75\% (proton) were achieved. Because of the usage of pure gases, no dilution factor has to be taken into account. HERMES also used heavy gases as unpolarized nuclear targets. 

There are DNP-based polarized targets also at CLAS in Jefferson Lab's Hall B \cite{jlabfrost2012}. The SpinQuest experiment at Fermilab will use the unpolarized 120\,GeV proton beam from the Fermilab main injector to scatter off transversely polarized proton and deuteron targets similar to the COMPASS one \cite{spinquest2019}. The LHC-spin experiment will follow in HERMES' footsteps by using a fixed target of polarized gases at the upstream end of the LHCb detector \cite{lhcspin2020}. An overview of target polarization techniques is given in Ref.~\cite{targetoverview}.

\subsection{Inclusive DIS off longitudinally polarized protons}
\label{sec:dis_LL}

\begin{figure}
\begin{center}
\includegraphics[width=0.60\textwidth]{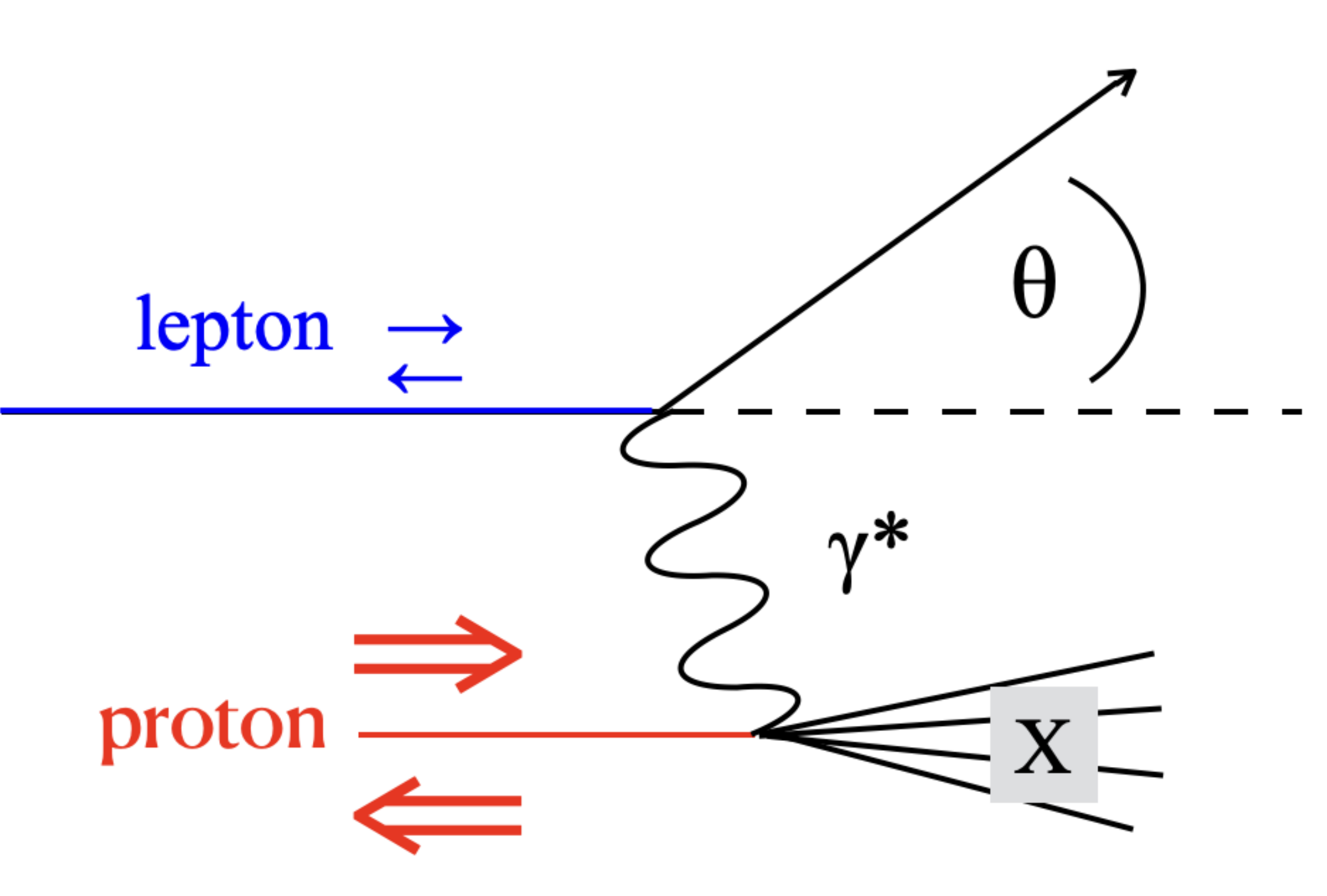}
\includegraphics[width=0.36\textwidth]{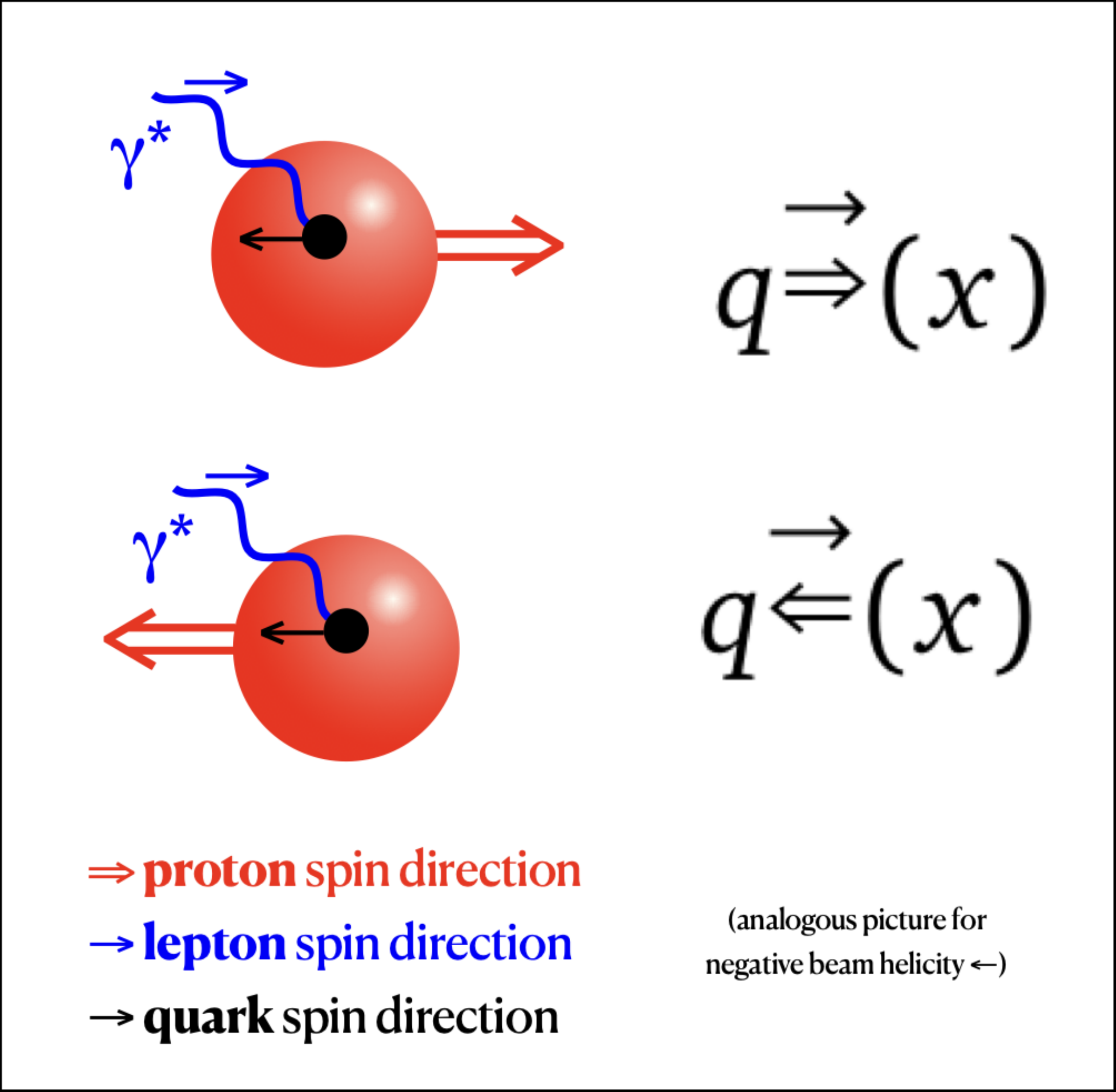}
\end{center}
\caption[DIS on longitudinally polarized protons]{The DIS process in $1\gamma^*$ exchange on longitudinally polarized protons. The virtual photon (spin 1) can only be absorbed by a quark (spin 1/2) with opposite spin orientation. The horizontal arrows indicate spin directions as detailed in the figure.}
\label{fig:DIS_LL}
\end{figure}

In Sec.~\ref{sec:disdis}, we saw that there are two structure functions
($F_1, F_2$) parameterizing the unpolarized spin-1/2 nucleon. The cross section $\mathrm{d}^2\sigma_{\mathrm{UU}}/\mathrm{d}x\mathrm{d}Q^2$ of scattering an unpolarized lepton beam (U) off an unpolarized proton (U) for $1\gamma^*$-exchange is written\footnote{We use here a different notation than in Sec.~\ref{sec:disdis} without changing the main message.}:
\begin{equation}
\frac{\mathrm{d}^2\sigma_{\mathrm{UU}}(x,Q^2)}{\mathrm{d}x\mathrm{d}Q^2}=\frac{4\pi\alpha^2}{Q^4}\left[F_1(x,Q^2)\cdot y^2+\frac{F_2(x,Q^2}{x}(1-y-\frac{y^2}{4}\gamma^2)\right], 
\label{eq:sigmaUU}
\end{equation}
with $y=\nu/E$, $\gamma^2=Q^2/\nu^2$, and $\alpha$ the electromagnetic fine structure constant. When both lepton and nucleon are longitudinally polarized (LL), there is an additional spin-dependent part $\mathrm{d}^2\sigma_{\text{LL}}(x,Q^2)/\mathrm{d}x\mathrm{d}Q^2$ in the cross section:
\begin{equation}
\frac{\mathrm{d}^2\sigma_{\text{LL}}(x,Q^2)}{\mathrm{d}x\mathrm{d}Q^2}=\frac{8\pi\alpha^2y}{Q^4}\cdot\left[\left(1-\frac{y}{2}-\frac{y^2}{4}\gamma^2\right)g_1(x,Q^2)-\frac{y}{2}\gamma^2g_2(x,Q^2)\right].
\label{eq:sigmaLL}
\end{equation}
Two \emph{spin structure functions} $g_1(x,Q^2)$ and $g_2(x,Q^2)$ are needed to describe the longitudinally polarized spin-1/2 nucleon in addition to the \emph{spin-independent structure functions} $F_1(x,Q^2)$ and $F_2(x,Q^2)$ \cite{hermes:g1long}. As was the case for the spin-independent structure functions (see Eq.~\ref{eq:qpm} with Callan-Gross relation Eq.~\ref{eq:callangross}), the spin-dependent structure function $g_1$ has a probabilistic interpretation in the simple (LO QCD) QPM formulated in terms of quark densities $q(x)$:
\begin{eqnarray}
g_1(x)&=&\frac{1}{2}\sum_qe_q^2\overbrace{\left( q^{\substack{\rightarrow \\ \Leftarrow}}(x)-q^{\substack{\rightarrow \\ \Rightarrow}}(x)\right)}^{\substack{\text{quark helicity distributions}\\\Delta q(x)}}\label{eq:g1}\\
F_1(x)&=&\frac{1}{2}\sum_qe_q^2\underbrace{\left(q^{\substack{\rightarrow \\ \Leftarrow}}(x)+q^{\substack{\rightarrow \\ \Rightarrow}}(x)\right)}_{\substack{q(x)\\ \text{quark momentum distributions}}}.\label{eq:f1}
\end{eqnarray}
As indicated in the left of Fig.~\ref{fig:DIS_LL}, the upper (single $\rightarrow$) arrow stands for the direction of the lepton polarization and the lower (double $\Rightarrow$) arrow for that of the nucleon. In Fig.~\ref{fig:DIS_LL} and Eqs.~\ref{eq:g1} and \ref{eq:f1}, we have for simplicity omitted the case of negative lepton helicity $\leftarrow$. What is important is the separate definition of anti-parallel ($\substack{\rightarrow \\ \Leftarrow}$ and $\substack{\leftarrow \\ \Rightarrow}$) and parallel ($\substack{\rightarrow \\ \Rightarrow}$ and $\substack{\leftarrow \\ \Leftarrow}$) spin-orientation cases. 
The previously introduced quark momentum distributions $q(x)$ are recovered by summing over parallel and anti-parallel quark densities. If we form their difference, we obtain \emph{quark helicity distributions} $\Delta q(x)$. By angular momentum conservation, a spin-1/2 quark in the nucleon can absorb a photon (with spin 1) only when their spin orientations are opposite, as shown in the right side of Fig.~\ref{fig:DIS_LL}. The DIS process on longitudinally polarized protons therefore probes \emph{quark helicities}. 

The structure function $g_2$ has no transparent interpretation in the QPM since it is related to transverse degrees of freedom.

\subsection{Quark spin contribution to the nucleon spin from inclusive DIS}
\label{sec:quarkspindis}

In order to filter out the additional information contained in the cross section when lepton and nucleon are polarized, it is usual to construct spin asymmetries of the type: 
\begin{equation}
A_{\|}=\frac{\sigma_{\text{LL}}}{\sigma_{\text{UU}}}=\frac{1}{P_{\text B}P_z}\cdot\frac{\sigma^{\substack{\rightarrow \\ \Leftarrow}}-\sigma^{\substack{\rightarrow \\ \Rightarrow}}}{\sigma^{\substack{\rightarrow \\ \Leftarrow}}+\sigma^{\substack{\rightarrow \\ \Rightarrow}}},
\label{eq:apara}
\end{equation}
with $P_{\text B}$ the polarization of the lepton beam and $P_z$ that of the nuclear target. These fractional polarization values enter because any imperfect polarization values $<~100\%$ will act as dilution to the number of polarized particles. If the target is an impure solid target (see Sec.~\ref{sec:polarizing}), an additional dilution factor $f$ will enter ($P_z\rightarrow fP_z$).

Generically, a cross section is experimentally determined by measuring the count rates $N$ of events divided by the luminosity $\mathcal{L}$, the latter of which ensures proper normalization. We can then write down the cross section for specific configurations of lepton and nucleon polarization (parallel or anti-parallel):
\begin{equation}
\sigma^{\substack{\rightarrow \\ \Rightarrow}}(x,Q^2)\equiv\frac{N^{\substack{\rightarrow \\ \Rightarrow}}(x,Q^2)}{\mathcal{L}^{\substack{\rightarrow \\ \Rightarrow}}} \;\;\;\;\; \text{or} \;\;\;\;\;
\sigma^{\substack{\rightarrow \\ \Leftarrow}}(x,Q^2)\equiv\frac{N^{\substack{\rightarrow \\ \Leftarrow}}(x,Q^2)}{\mathcal{L}^{\substack{\rightarrow \\ \Leftarrow}}},
\end{equation}
where we have again for brevity skipped the symmetric case of negative lepton helicity $\leftarrow$. The measured quantity is an asymmetry in the normalized count rates of inclusively detected scattered beam leptons for anti-parallel versus parallel spin configurations:
\begin{equation}
A_{\|}=\frac{1}{P_BP_z}\cdot\frac{\frac{N^{\substack{\rightarrow \\ \Leftarrow}}}{\mathcal{L}^{\substack{\rightarrow \\ \Leftarrow}}}-\frac{N^{\substack{\rightarrow \\ \Rightarrow}}}{\mathcal{L}^{\substack{\rightarrow \\ \Rightarrow}}}}{\frac{N^{\substack{\rightarrow \\ \Leftarrow}}}{\mathcal{L}^{\substack{\rightarrow \\ \Leftarrow}}}+\frac{N^{\substack{\rightarrow \\ \Rightarrow}}}{\mathcal{L}^{\substack{\rightarrow \\ \Rightarrow}}}}.
\end{equation}
Experimental access to $g_1(x,Q^2)$ is obtained by measuring the virtual-photon asymmetry $A_1(x)$,
\begin{eqnarray}
A_1(x)\approx\frac{A_{\|}(x,Q^2)}{D(y,Q^2)}&\approx&\frac{\color{magenta}{g_1(x,Q^2)}}{F_1(x,Q^2)}\label{eq:g1overf1}\\
&=&\frac{\sum_qe_q^2\color{magenta}{\Delta q(x,Q^2)}}{\sum_qe_q^2q(x,Q^2)},\label{eq:deltaq}
\end{eqnarray}
here neglecting the small contribution from the interference between transverse and virtual-photon longitudinal amplitudes (which is encoded in the virtual-photon asymmetry $A_2$). The asymmetry $A_1$ is a convenient representation when comparing different experiments because most of the $Q^2$-dependence of $A_{\|}$ is canceled due to the polarization transfer from the lepton to the virtual photon. This polarization transfer is described by the function $D(y,Q^2)$ and is often called depolarization factor. Equation~\ref{eq:deltaq} summarizes the relation between $g_1(x,Q^2)$ and the quark helicity distributions $\Delta q(x,Q^2)$.

Examination of Eq.~\ref{eq:g1overf1} reveals that for the extraction of the spin structure function $g_1$, it is, in addition to measuring $A_1$, necessary to use some input parameterizations from world data: the $F_2^{\text p}$ structure function of the proton; if scattering off a deuteron target, also the ratio of $F_2^{\text n}/F_2^{\text p}$; to get $F_1$ from $F_2$, we need $R$, the longitudinal over transverse virtual-photon absorption cross section (Eq.~\ref{eq:r}), and there are others.

\begin{figure}
\begin{center}
\includegraphics[width=0.48\textwidth]{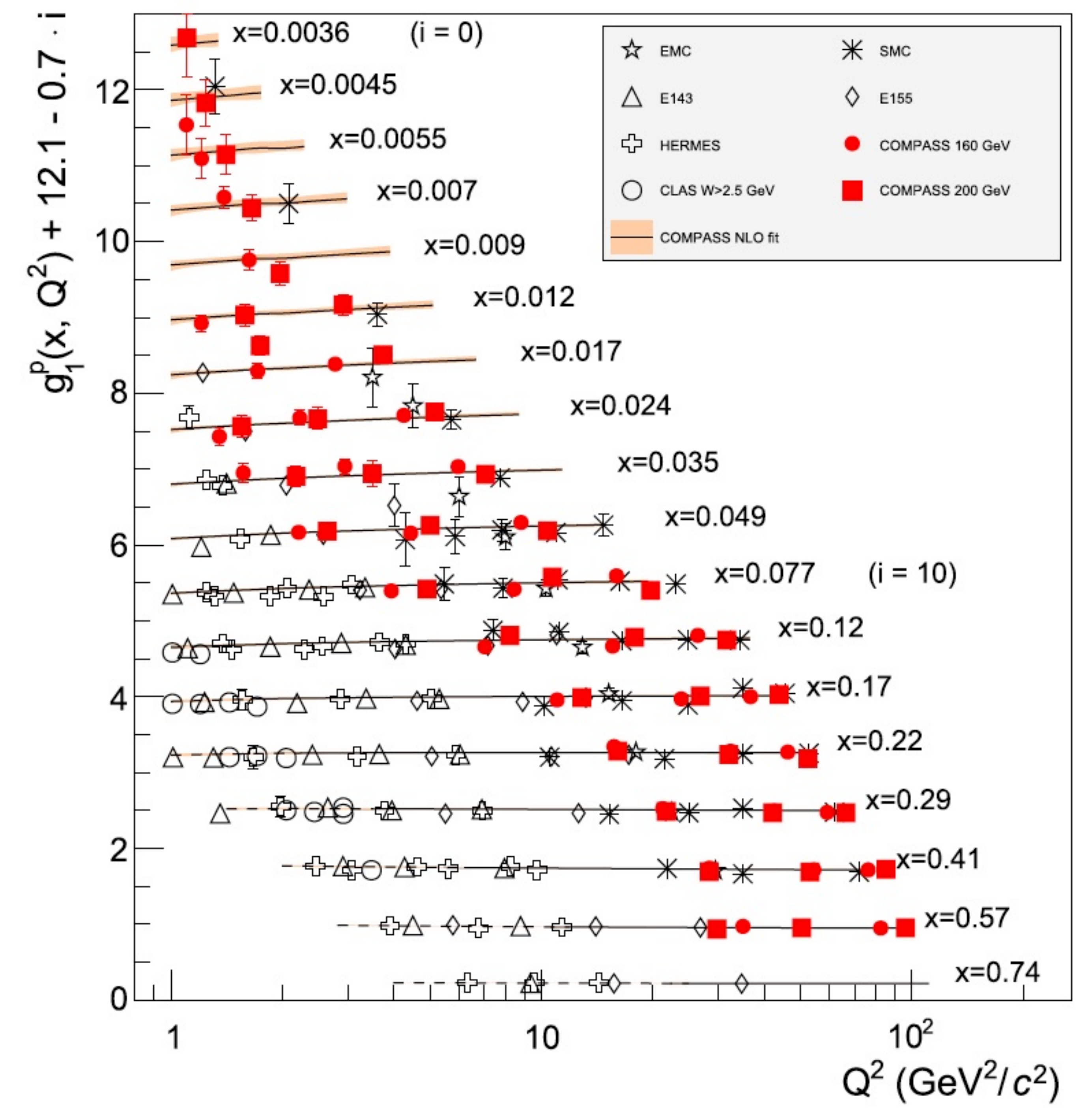}
\includegraphics[width=0.48\textwidth]{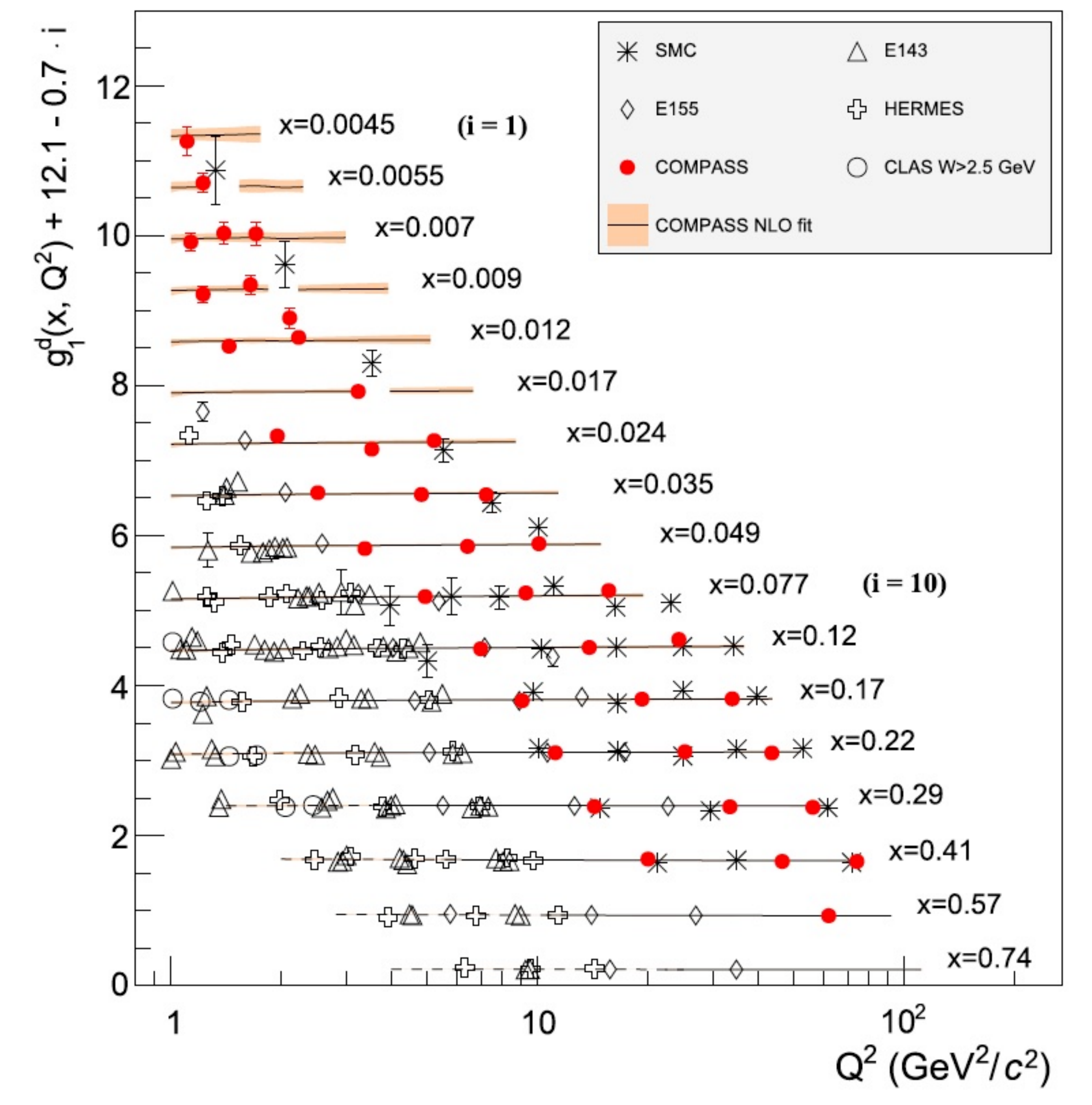}
\end{center}
\caption[Spin structure function $g_1$]{World data on the spin structure function $g_1$ as a function of $Q^2$ for various values of $x$, 
on the {\bf left} for the proton (Ref.~\cite{compass_g1p} and references therein) and on the {\bf right} for the deuteron (Ref.~\cite{compass_g1d} and references therein). The solid line represents a NLO QCD fit for $W^2>10$~GeV$^2$ and the dashed line an extension to lower values of $W^2$.}
\label{fig:g1_all}
\end{figure}

The spin structure function $g_1$ measured at experiments at CERN (COMPASS, SMC, EMC), DESY (HERMES), JLab (CLAS), and SLAC (E143, E155) is shown in Fig.~\ref{fig:g1_all} for longitudinally polarized proton and deuteron targets.
To obtain the contribution of quark spins to the nucleon spin, one has to integrate $g_1(x,Q^2)$ of the proton (p) or the deuteron (d) over $x$:
\begin{equation}
\Gamma_1^{\text{p,n}}=\int_0^1 \mathrm{d}x \;g_1^{\text{p,n}}(x,Q^2),
\label{fig:gamma1}
\end{equation}
thereby spanning as much as possible range in $x$. Because it is experimentally not possible to perform the measurement from $x=0$ to $x=1$, a saturation of the integral below a certain $x$-value is assumed. The contribution of the quarks spins $\Delta\Sigma=(\Delta u + \Delta \overline{u}) + (\Delta d + \Delta \overline{d}) + (\Delta s + \Delta \overline{s})$, with $\Delta q \equiv\int_0^1\mathrm{d}x\;\Delta q(x,Q^2)$, to the spin of the nucleon is related to the integral in Eq.~\ref{fig:gamma1} in the modified minimal subtraction, or $\overline{\mathrm{MS}}$, scheme\footnote{A particular renormalization scheme used in quantum field theory (QFT).},
\begin{equation}
\Delta\Sigma(Q^2)\stackrel{\overline{\mathrm{MS}}}{=}a_0(Q^2)=\frac{1}{\Delta C_\text{S}^{\overline{\mathrm{MS}}}}\left[\frac{9\Gamma_1^{\text d}}{(1-\frac{3}{2}\omega_{\text D})}-\frac{1}{4}a_8\Delta C_{\text{NS}}^{\overline{\mathrm{MS}}}\right],
\label{eq:deltasigma}
\end{equation}
here for the case of a deuteron target with integral $\Gamma_1^{\text d}$ and D-state admixture $\omega_{\text D}$. See page~\pageref{page:dglap} for $a_8$. The $\Delta C\equiv\Delta C(x,\alpha_{\text S}(Q^2))$ are Wilson coefficient functions for the flavor singlet (S) and non-singlet (NS) states. We omit further details of the NLO QCD analysis here and refer to the references instead. 

\begin{figure}
\includegraphics[width=.43\textwidth]{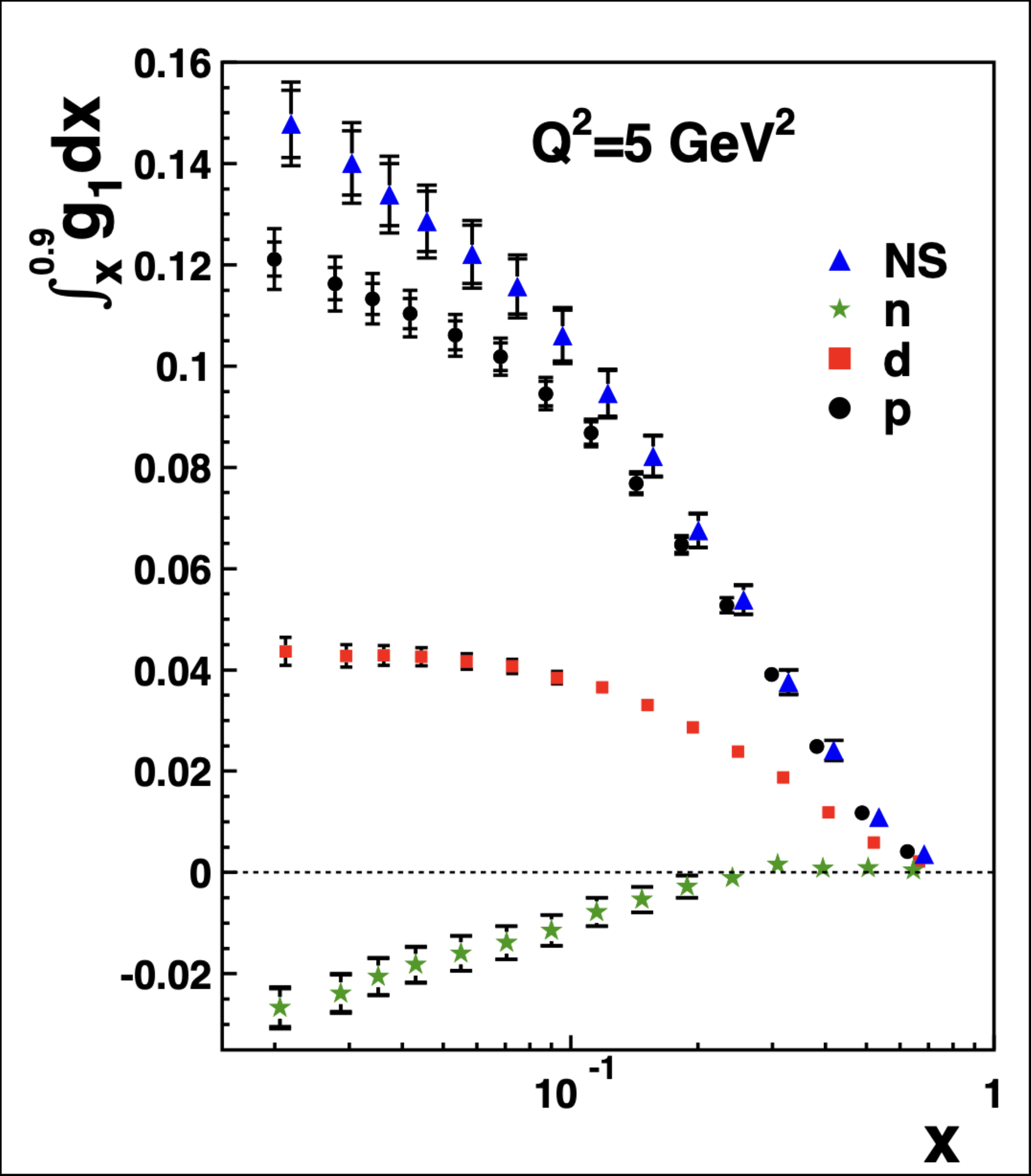}
\includegraphics[width=.53\textwidth]{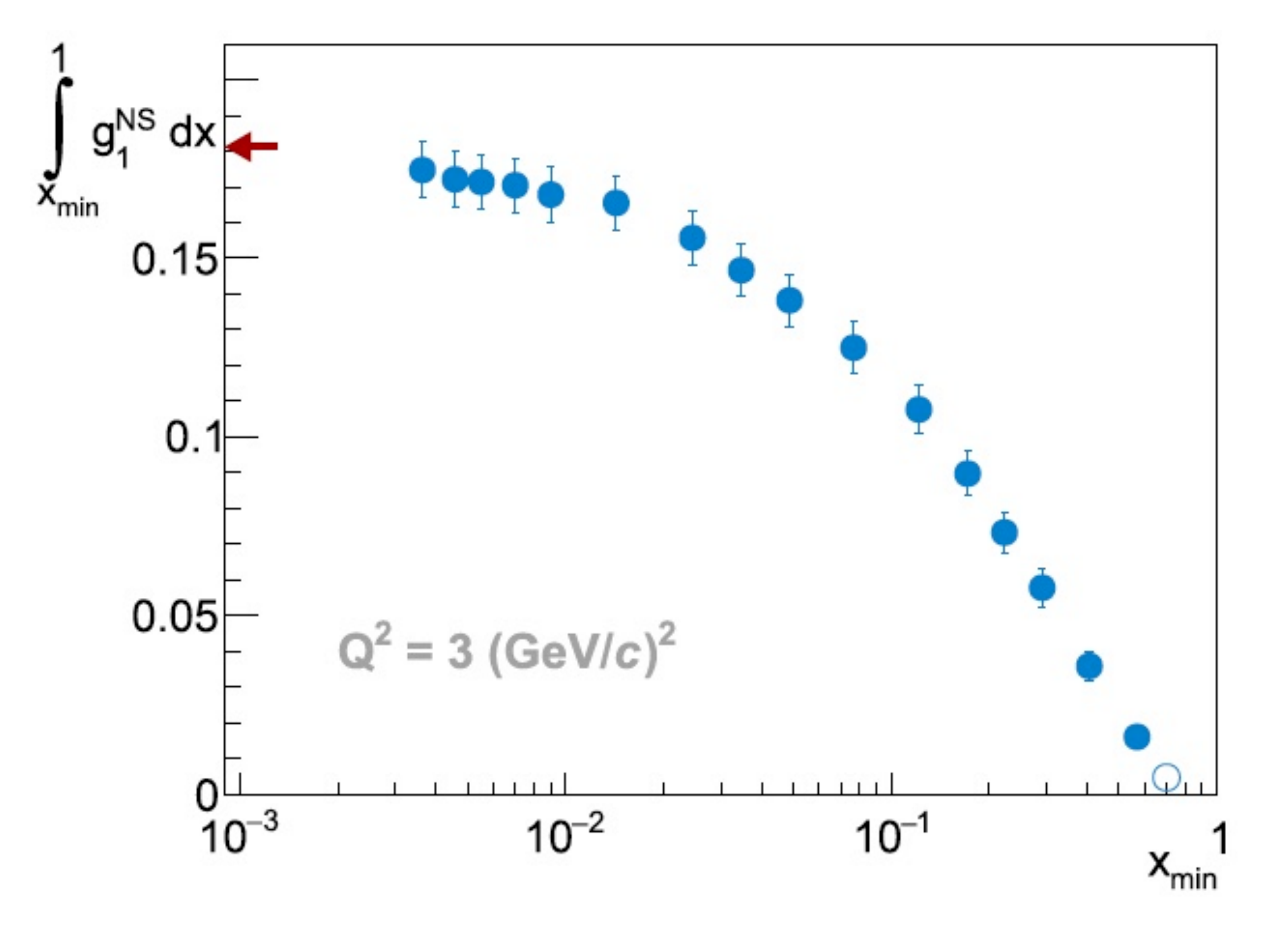}
\caption[Integral of $g_1$]{Spin structure function integrals at common values of $Q^2$. {\bf Left}: integrating the HERMES $g_1$ over $x$-Bjorken to obtain the contribution of quarks spins to the spin of the nucleon \cite{hermes:g1long}. {\bf Right}: integrating the non-singlet COMPASS $g_1^{\text{NS}}$ over $x$-Bjorken to test the Bjorken sum rule \cite{compass_g1p}.
}
\label{fig:g1_integral}
\end{figure}

Figure~\ref{fig:g1_integral} shows the integrals Eq.~\ref{fig:gamma1} from HERMES for different targets. A saturation for the deuteron integral is observed and therefore HERMES used the deuteron data to extract $\Delta\Sigma$. 
It was later demonstrated by COMPASS, whose $g_1$ results on the proton and deuteron target 
at very low values of $x$ were not yet available at the time of the HERMES publication in 2007,
that $g_1^{\text d}$ indeed does not contribute at even lower values of $x$. Due to the higher lepton-beam energy at COMPASS as compared to HERMES, COMPASS has a larger reach to small values of $x$. While the COMPASS proton data clearly show some spin effects at lowest $x$ \cite{compass_g1p_lowx}, the deuteron $g_1$ is compatible with zero in that range \cite{dis17badelek}, unlike indicated by older SMC measurements on the deuteron. 

The $g_1$ data on the proton (p) and neutron (n) also allow for a test of the fundamental \emph{Bjorken sum rule} \cite{bjorkensum}, which involves the non-singlet (NS) spin structure function $g_1^{\text{NS}}(x,Q^2)=g_1^{\text p}(x,Q^2)-g_1^{\text n}(x,Q^2)$:
\begin{equation}
\int_0^1\text{d}x\; g_1^{\text{NS}}(x,Q^2)=\frac{1}{6}\left|\frac{g_{\text A}}{g_{\text V}}\right|C_1^{\text{NS}}(Q^2).
\label{eq:bjorkensum}
\end{equation}
The non-singlet integral over $x$ at a given value of $Q^2$ (shown in Fig.~\ref{fig:g1_integral} for COMPASS data) is connected to the ratio $g_{\text A}/g_{\text V}$ of the axial (A) and vector (V) coupling constants, which is known from neutron beta decay. The COMPASS NLO analysis provides a validation of the Bjorken sum rule of Eq.~\ref{eq:bjorkensum} with an accuracy of 9\% and improves at NNLO level.

We return to the contribution from the quark spins to the spin of the nucleon, $\Delta\Sigma$. HERMES \cite{hermes:g1long} found in 2007:
\begin{equation}
\Delta\Sigma(Q^2=5\mathrm{GeV}^2)=0.330\pm0.011\;\text{(theo)}\pm0.025\;\text{(exp)}\pm\;0.028\;\text{(evol)},
\end{equation}
for the inclusive measurement on the deuteron evolved to a common $Q^2$ of 5~GeV$^2$. Also from the inclusive $g_1^{\text d}$ measurement, COMPASS \cite{compass_g1d} found in 2017 for a common $Q^2$ of 3~GeV$^2$:
\begin{equation}
\Delta\Sigma(Q^2=3~\mathrm{GeV}^2)= 0.32\pm0.02\;\text{(stat)} \pm0.04\;\text{(syst)} \pm0.05\;\text{(evol)}.
\end{equation}
These results have to be looked at in the historical context. In the late 1980s, the combined EMC and SLAC data had resulted in $\Delta\Sigma=0.12\pm0.16$ \cite{Jaffe1995}, i.e., a value that is compatible with \emph{no} quark spin contribution to the nucleon spin. This so-called ``spin crisis'' of 1988 resulted in the planning and construction of the SMC, HERMES, and COMPASS fixed-target experiments. With the data from these experiments, together with proton-proton collision data from PHENIX and STAR at RHIC, the crisis has become a spin puzzle that still awaits complete assembly. We recall the spin decomposition of the nucleon from  Eq.~\ref{eq:spinpuzle}, $1/2=1/2\Delta\Sigma + \Delta G + L$. Nowadays the contribution from quark spins to the spin of the nucleon, $\Delta\Sigma$, is from $g_1$ measurements and global QCD analyses at NLO \cite{dssv08} considered to be about one third, and we expect contributions from gluon spin, $\Delta G$, and parton orbital momenta, $L$, to fill up the missing budget. We will study these contributions later, after having taken a look at tensor-polarization effects in the deuteron and flavor-separated valence- and sea-quark helicities.

We finish this subsection with some remarks about (QED) radiative corrections, which can play an important role in the measurement and the interpretation of the results. In the context of charged-lepton DIS, we speak of \emph{internal bremsstrahlung} when a real photon is radiated off the initial lepton before it interacts with the nucleon, or off the final lepton before it is detected. \emph{External bremsstrahlung} is the emission of photons due to the interaction of the particles with the detector or target material, often referred to as ``detector smearing''. Both effects make the measured kinematics appear to be different than they actually would have been at the theoretical \emph{Born level}: radiative effects make events migrate from the Born bin to the measured bin and thus cause a ``smearing'' of the kinematic event distributions. For example, if the scattered DIS electron radiates a photon before it enters the experiment's calorimeter, its energy will be reconstructed at a too low value. The size of radiative effects depends greatly on the experimental context. For example, electrons are more affected than muons because of their smaller mass, and the radiative effects are smaller if one or more hadrons are detected in addition to the lepton. 

Many experiments use \emph{unfolding} techniques to separate smearing effects from the physical signal. In general, smearing processes occur in a statistical manner and can therefore not be corrected for on an event-by-event basis. They have to be calculated (in case e.g.~of QED radiative effects), simulated (detector smearing), or estimated from iterative experimental methods (SMC). For example, HERMES unfolded kinematic bin migrations from the measured inclusive asymmetries using a \emph{smearing matrix} obtained from simulated Monte-Carlo data that kept track about which kinematic bin on generated (Born) level the event migrated to on reconstructed (measured) level. The simulation included a full GEANT \cite{geant} model of the various detector materials, 
which is state-of-the-art for every modern particle-physics experiment. 

Care has to be taken when interpreting the experimental results corrected for radiative effects, as was for example demonstrated by the infamous \emph{HERMES effect} \cite{hermeseffect} created by overestimated radiative corrections. 

\subsection{Tensor structure of the deuteron}
\label{sec:b1}

\begin{figure}
\begin{center}
\includegraphics[width=.64\textwidth]{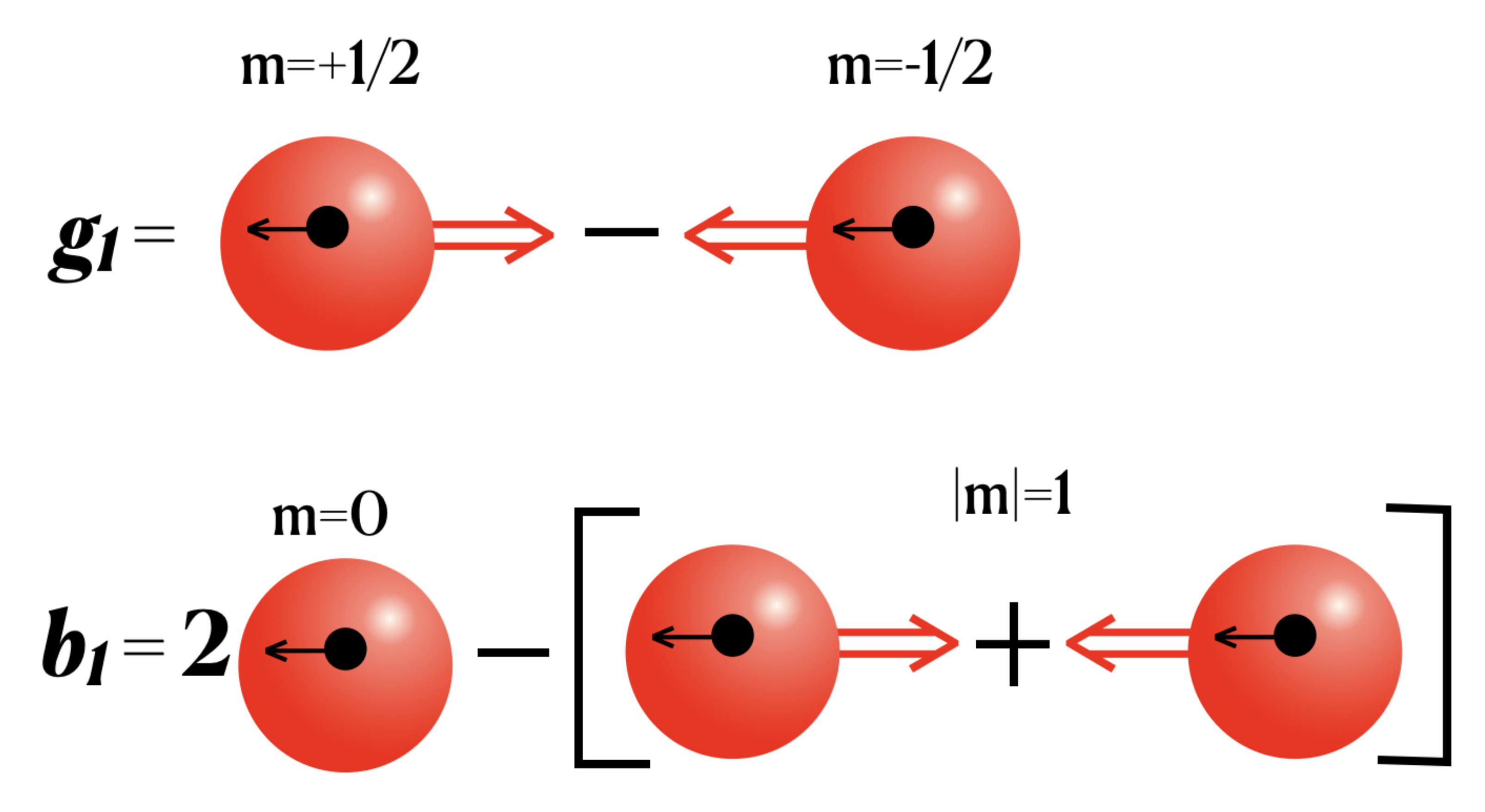}
\includegraphics[width=.32\textwidth]{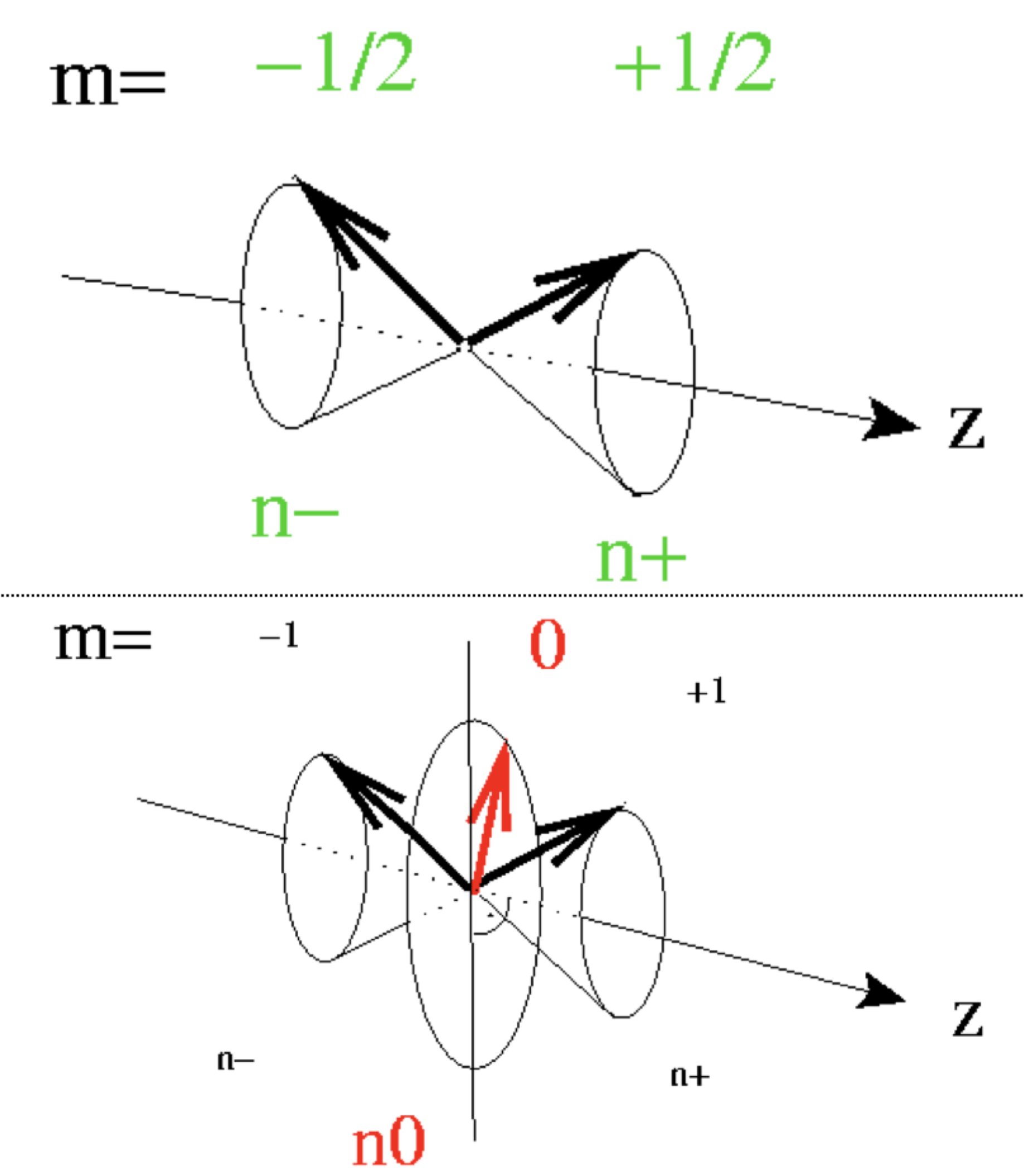}
\includegraphics[width=.80\textwidth]{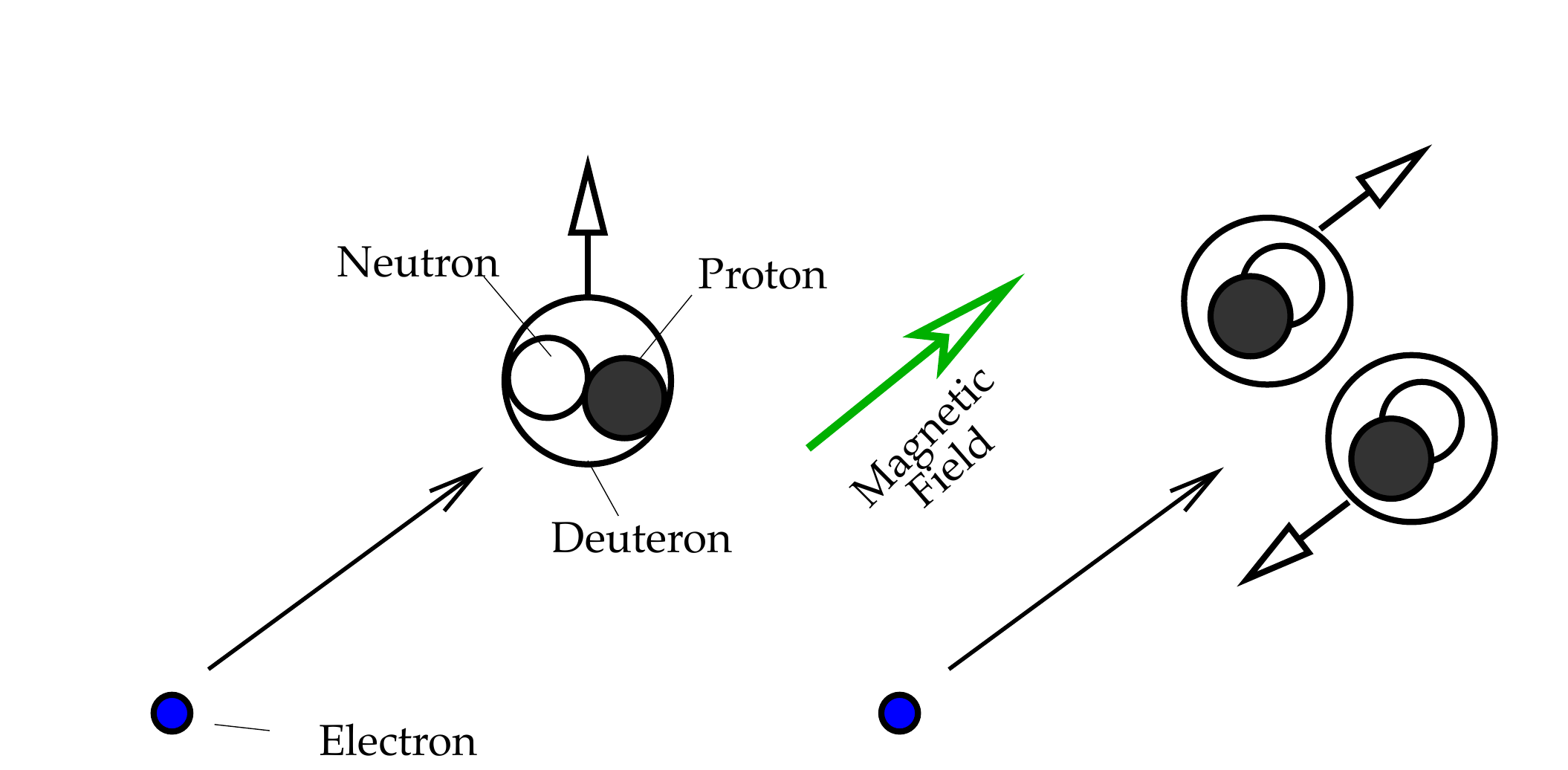}
\end{center}
\caption[Tensor effects in the deuteron]{Tensor effects on the deuteron. Top left: interpretation of $g_1$ (of the proton) and $b_1$ (of the deuteron) in the quark parton model. Top right: possible spin projections $m$ for spin-1/2 and spin-1 particles. Bottom: nuclear shadowing effects in the deuteron's tensor structure. When the deuteron is in the $m=0$ state, the scattering lepton sees the proton and neutron as ``side-by-side'' (doughnut shape), while for $m=\pm 1$, proton and neutron are in front of each other (dumbbell shape).}
\label{fig:tensor}
\end{figure}

We undertake a short excursion to an effect that occurs only for spin-1 hadrons such as the deuteron. 
For a spin-1 hadron, the parameterization of spin-dependent DIS cross section requires in addition to the spin structure functions $g_1$ and $g_2$ introduced in Eq.~\ref{eq:sigmaLL} another four structure functions usually labeled $b_1$, $b_2$, $\Delta$, and $b_3$ \cite{hoodbhoy_manohar1989}. We will here focus on the leading-twist\footnote{``Twist'' related to the exponent in operator production expansion (OPE) in QFT \cite{PeskinSchroeder}. Typically, twist-2 is leading twist.} \emph{tensor structure function $b_1(x,Q^2)$}, which has an interpretation in the simple QPM as shown in Fig.~\ref{fig:tensor}. While a spin-1/2 particle such as the proton has two possibilities for the spin quantum number $m=\pm 1/2$ (Eq.~\ref{eq:m}), the deuteron as spin-1 particle has three: $m=\pm 1, 0$. While the spin structure function $g_1$ is sensitive to the difference in quark densities between negatively and positively polarized quarks (the helicity distributions), the tensor structure function $b_1$ measures the difference in quark densities when the deuteron is in an $m=\pm 1$ state as compared to an $m=0$ state. It has been known already from elastic electron-deuteron scattering that the deuteron takes different shapes (dumbbell vs.~doughnut) depending on its polarization state. This is also illustrated in  Fig.~\ref{fig:tensor} - proton and neutron ``shadow'' each other for $m=\pm 1$, thereby decreasing the cross section for one of the two polarization states. But is this also true in DIS, where the parton level is probed and not the nucleon level?

\begin{figure}
\begin{center}
\includegraphics[width=.46\textwidth]{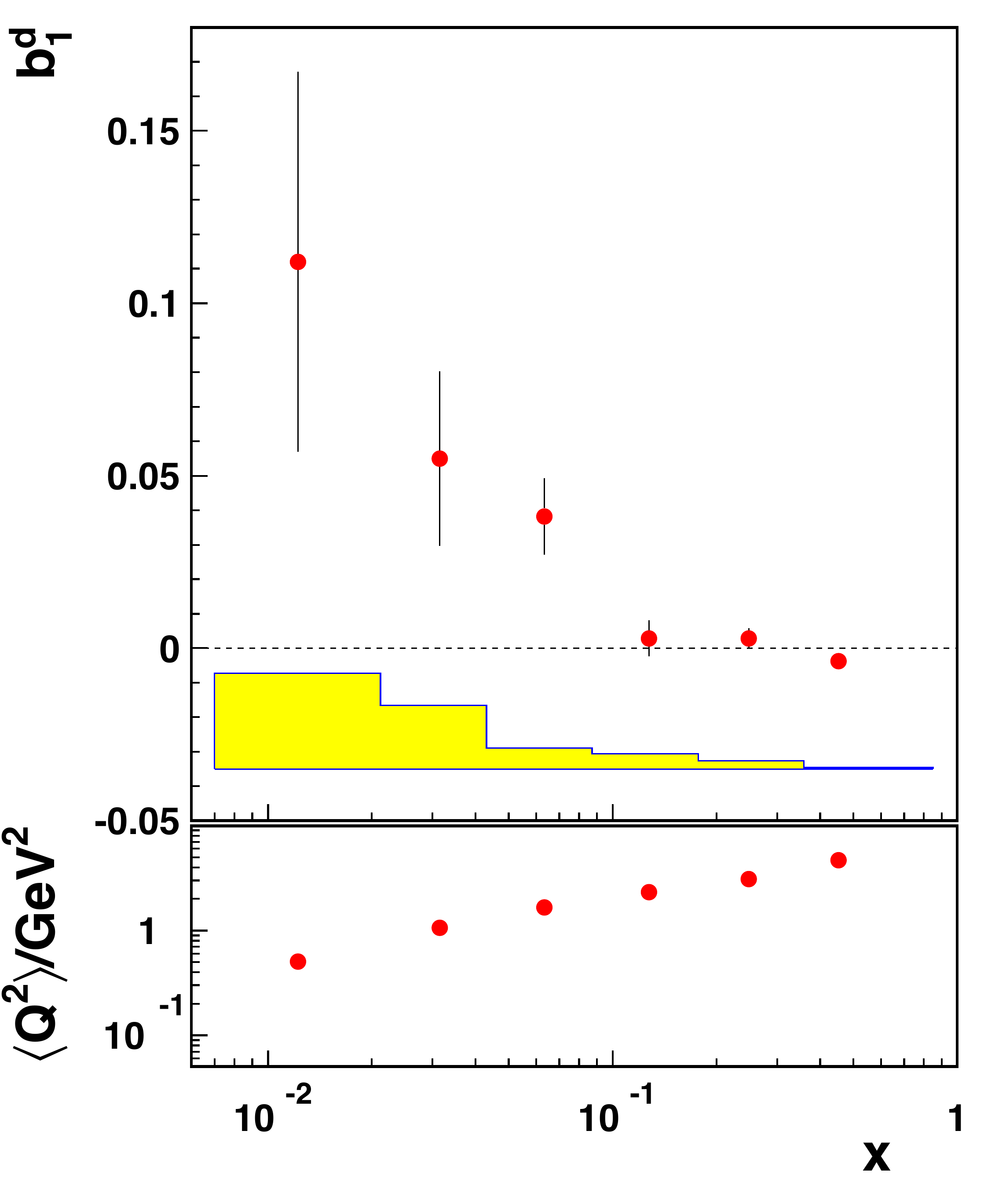}
\includegraphics[width=.50\textwidth]{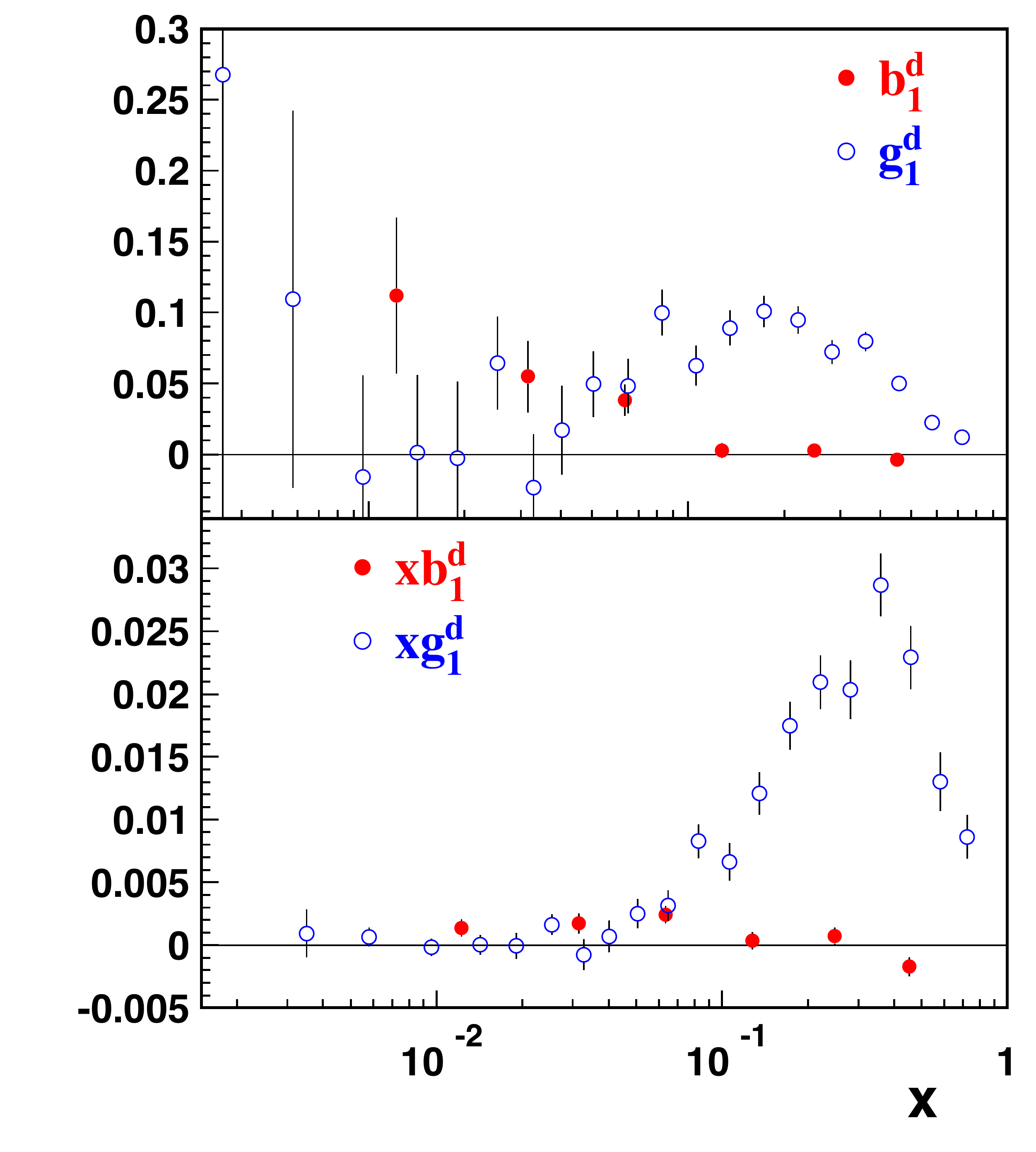}
\end{center}
\caption[Tensor structure function in DIS]{Tensor structure function $b_1$ of the deuteron measured at HERMES \cite{hermes:b1} and comparison to the HERMES spin structure function $g_1$ of the deuteron \cite{riedl:2005}.}
\label{fig:b1}
\end{figure}

To that end, HERMES created a pure sample of tensor-polarized deuter\-ons and scattered 27.6\,GeV positrons off them. Recalling the definition of vector polarization from Eq.~\ref{eq:vectorpol} with $|P_z|\le 1$, the tensor polarization $P_{zz}$ is defined as
\begin{equation}
P_{zz}=\frac{(N^1+N^{-1})-2N^0}{N^1+N^{-1}+N^0},
\label{eq:tensorpol}
\end{equation}
with $N^m$ the number of deuterons in state $m$, and $-2\le P_{zz}<1$. Very similar to $A_{\|}$ (Eq.~\ref{eq:apara}), the tensor asymmetry $A_{zz}$
\begin{equation}
A_{zz}=\frac{1}{P_{zz}}\cdot\frac{(\sigma^1+\sigma^{-1})-2\sigma^0}{\sigma^1+\sigma^{-1}+\sigma^0}
\end{equation}
was extracted from the data. The HERMES polarized gas target (Sec.~\ref{sec:polarizing}) allowed injecting the deuteron hyperfine state in the $m=0$ state resulting in high negative tensor polarization with at the same time very small vector polarization, which is important because otherwise the tensor asymmetry would receive non-negligible contributions from $A_{\|}$ (Eq.~\ref{eq:apara}) and it would be difficult to disentangle the vector and tensor parts. The HERMES results \cite{hermes:b1} for $b_1\sim A_{zz}F_1^{\text d}$ are shown in Fig.~\ref{fig:b1}.
Unlike the spin structure function $g_1^{\text d}(x)$ of the deuteron, $b_1^{\text d}(x)$ rises for small $x$ and is at the smallest measured $x$ values even larger than $g_1$. The observed behavior of $b_1$ for small $x$ is in qualitative agreement with coherent double scattering models, where the deuteron is probed as spin-1 object and not as compound of two spin-1/2 objects (which have no tensor effect). The tensor structure function $b_1$ is only accessible in DIS but also probes the spin of the embedding hadron. In this sense one could say that it crosses the border between nuclear and quark physics. 

A proposal to measure $b_1$ at JLab's Hall C was approved. 

\subsection{Flavor-separated valence-quark helicities from SIDIS }
\label{sec:deltaq}

\begin{figure}
\begin{center}
\includegraphics[width=.465\textwidth]{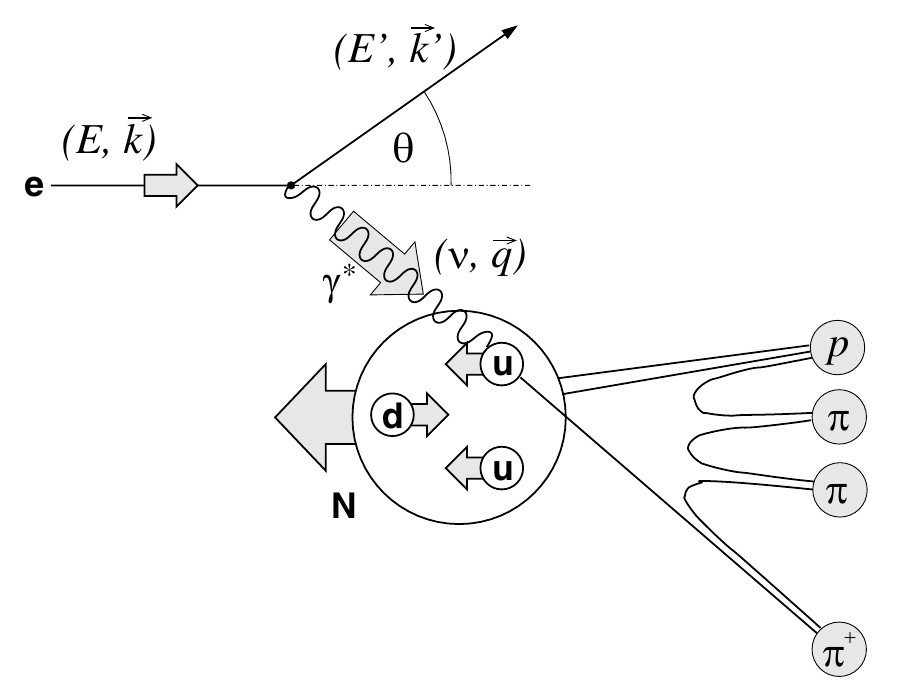}
\includegraphics[width=.495\textwidth]{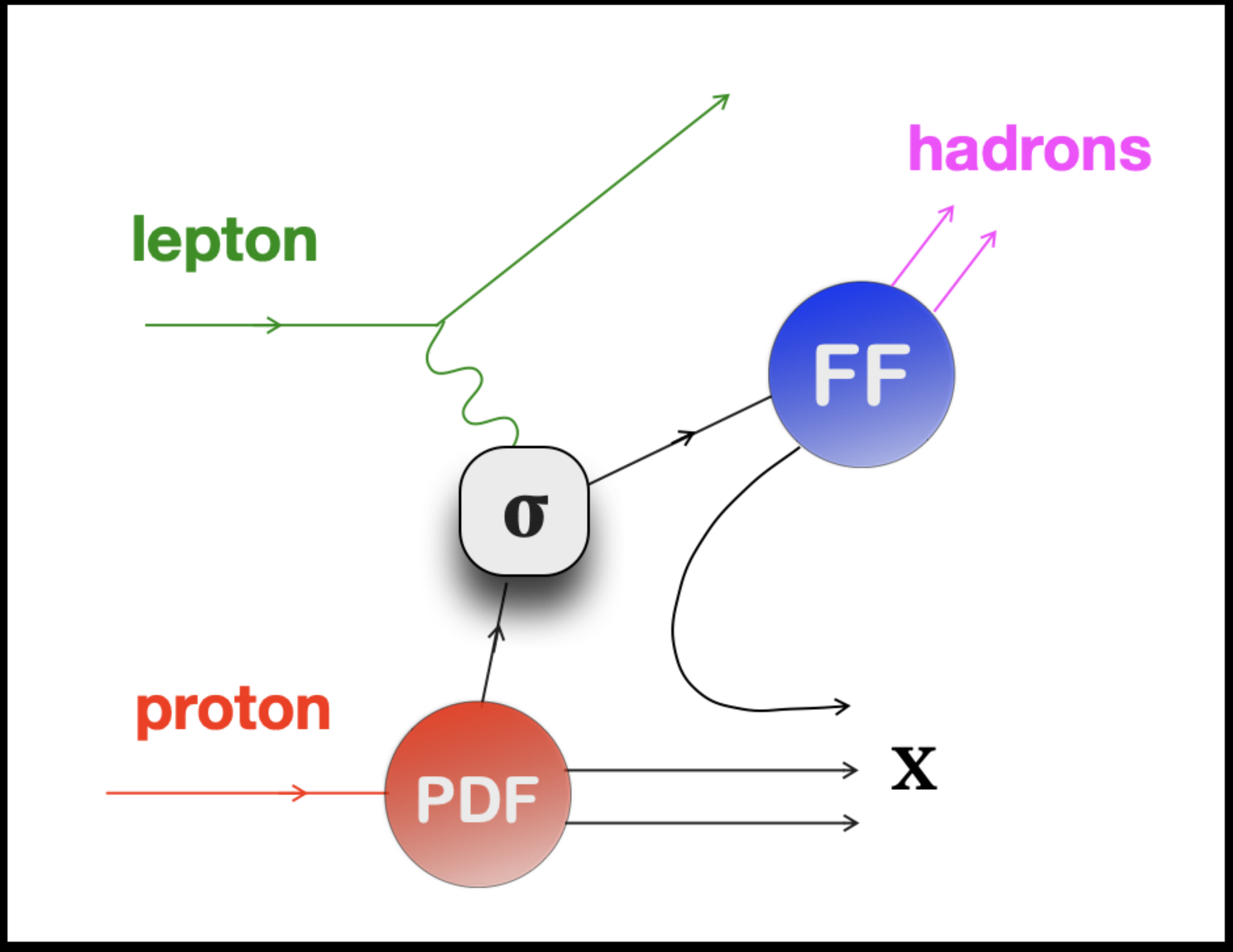}
\end{center}
\caption[Semi-inclusive DIS]{Left: semi-inclusive deep-inelastic scattering (SIDIS) on the proton with the scattered lepton and various hadrons in the final state. Right: decomposition of the SIDIS cross section.}
\label{fig:sidis}
\end{figure}

We return to the quark helicity distributions measured in DIS. To experimentally separate the various flavors of $\Delta q$ (Eq.~\ref{eq:deltaq}) it is necessary to detect in addition to the scattered beam lepton also one or more final-state hadrons created in the fragmentation process:
\begin{eqnarray}
\ell {\text p}&\rightarrow& \ell^\prime \text{hX}\;\;\mathrm{or} \nonumber\\
\ell \text{p}&\rightarrow& \ell^\prime \text{h}_1\text{h}_2\text{X}.
\label{eq:sidis}
\end{eqnarray}
This experimental choice is called a \emph{semi-inclusive} measurement, and correspondingly we speak of semi-inclusive DIS (SIDIS), see Fig.~\ref{fig:sidis} left. We will later (Sec.~\ref{sec:gpds}) also come across \emph{exclusive} measurements that aim to detect the entire final state. 

Experimentally, the discrimination of different types of hadrons, called hadron PID, is typically achieved with ring-imagining Cherenkov (RICH), or RICH-like, detectors.   

We recall the factorization theorem for inclusive DIS mentioned in Sec.~\ref{sec:pdfs} on page~\pageref{page:factorization} that allowed us to write the hard-scattering part $\hat{\sigma}$ separately from the non-perturbative parton distribution function. Also here in SIDIS, we assume that the measured cross section $\sigma$ can be decomposed, or factorized into parts that do not blend into each other: ``$\sigma^{\mathrm{SIDIS}}=\hat{\sigma}\otimes\mathrm{\color{red}{PDF}}\otimes\mathrm{\color{blue}{FF}}$'', where ``FF'' are the \emph{fragmentation functions} $D_q^\text{h}(z)$ representing the probability that a quark with flavor $q$ will fragment into a hadron h. Here, $z$ is the fractional energy of the final-state hadron, $z=E_{\mathrm{hadron}}/\nu$. The hadron with the largest momentum in the event, called \emph{leading hadron}, originates with large probability from the struck quark. With the FFs, we have introduced in addition to the PDFs a second class of objects that are universal, thus expected to not depend on the process they are measured with. 

\begin{figure}
\begin{center}
\includegraphics[width=.85\textwidth]{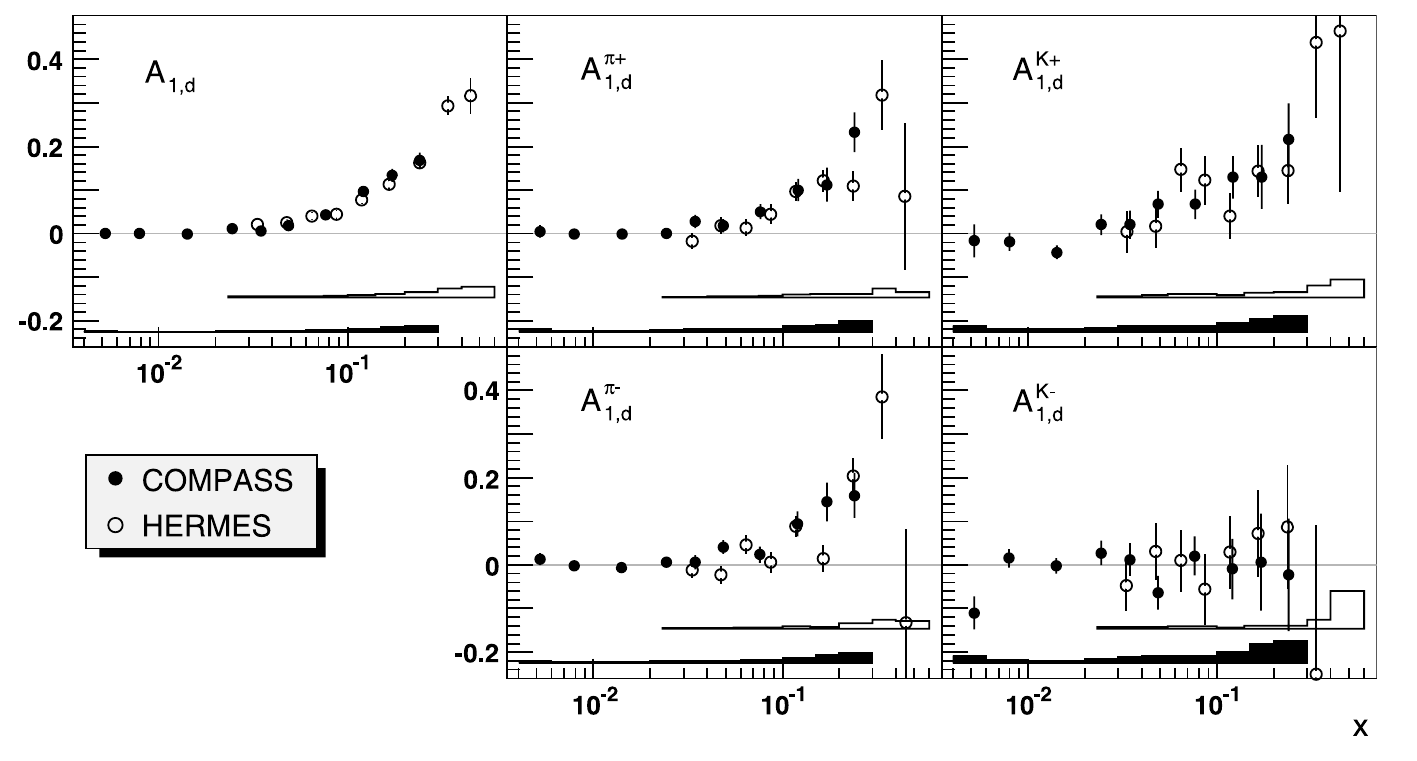}
\includegraphics[width=.55\textwidth]{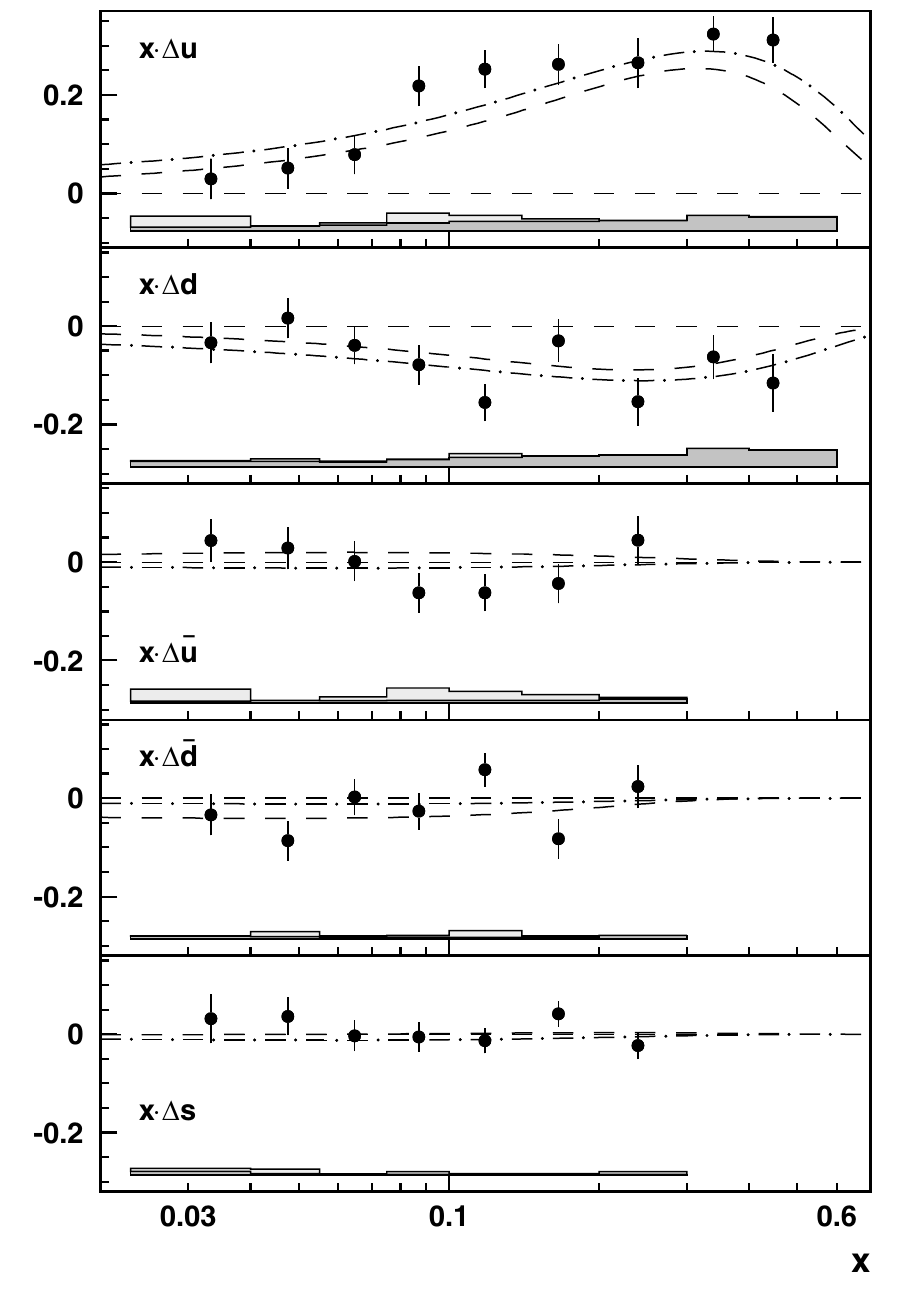}
\end{center}
\caption[SIDIS asymmetries on longitudinally polarized nucleons]{Top: virtual-photon asymmetries for single-hadron production in SIDIS on longitudinally polarized deuterons at COMPASS \cite{compass:deltaq} and longitudinally polarized protons at HERMES \cite{hermes:deltaq}. Bottom: flavor-separated quark-helicity distributions from HERMES at $Q^2=2.5$\,GeV$^2$.}
\label{fig:deltaq-flavor}
\end{figure}

The measured  virtual-photon asymmetry $A_1^\text{h}(x)$ for hadron h can then in leading order be approximated as \cite{klaus:2002}:
\begin{equation}
A_1^{\text{h}}(x,Q^2,z)\cong\frac{\sum_qe_q^2\color{red}{\Delta q(x,Q^2)}\color{blue}{D_{q}^{\text{h}}(z)}}{\sum_{q^\prime} e_{q^{\prime}}^2\color{red}{q^\prime(x,Q^2)}\color{blue}{D_{q^\prime}^{\text{h}}(z)}}.
\label{eq:a1-sidis}
\end{equation}
Both HERMES and COMPASS used the \emph{purity method} in their data analyses: 
\begin{equation}
A_1^{\text{h}}(x,Q^2,z)=\sum_q\mathcal{P}^{\text{h}}_q(x,Q^2,z)\cdot\frac{\Delta q(x,Q^2)}{q(x,Q^2)},
\end{equation}
where the purity $\mathcal{P}^{\text{h}}_q(x,Q^2,z)$ is the conditional probability that a hadron of type h observed in the final state originated from a struck quark of flavor $q$. The $A_1^\text{h}$ measurements are shown in Fig.~\ref{fig:deltaq-flavor}. 

The flavor-separated helicity distributions of the nucleon were extracted from the measured $A_1^\text{h}$  in a LO QCD analysis. For the a priori unknown fragmentation functions from Eq.~\ref{eq:a1-sidis}, the LUND model describing the fragmentation process was tuned to the specific kinematic phase space of the fixed-target experiment (in the case of the HERMES analysis \cite{hermes:deltaq}), or parameterizations from a global fit to world data from e$^+$e$^-$ annihilation, SIDIS, and pp data \cite{dssv:ff2007} were used (in the case of the later COMPASS analysis \cite{compass:deltaq}, at which point more reliable FF parameterizations had become available). 

In a recent publication \cite{hermes:deltaq2019}, HERMES analyzed \emph{hadron charge-difference} longitudinal double-spin asymmetries $A_{\mathrm{LL}}$, which provide a direct way to extract valence-quark helicities under isospin-symmetry assumptions of the fragmentation functions. The observed transverse-momentum dependence is found to be weak and consistent with findings by COMPASS and the CLAS experiment at JLab \cite{clas:deltaq2010}. 

\subsection{Sea-quark helicities from $\vec{\text{p}}\text{p}$}
\label{sec:deltaq-pp}

Quark helicity distributions $\Delta q$ from longitudinally polarized protons were determined at fixed-targets experiments for the valence-quark region (Sec.~\ref{sec:deltaq}). For the sea-quark region, the STAR \cite{star:deltaq} and PHENIX \cite{phenix:deltaq} experiments at RHIC measured $\Delta q$ in W$^\pm$ production from proton-proton collisions at a center-of-mass energy of $\sqrt{s}=510$\,GeV (see Fig.~\ref{fig:wprod} on page~\pageref{fig:wprod}). 

\begin{figure}
\begin{center}
\includegraphics[width=.48\textwidth]{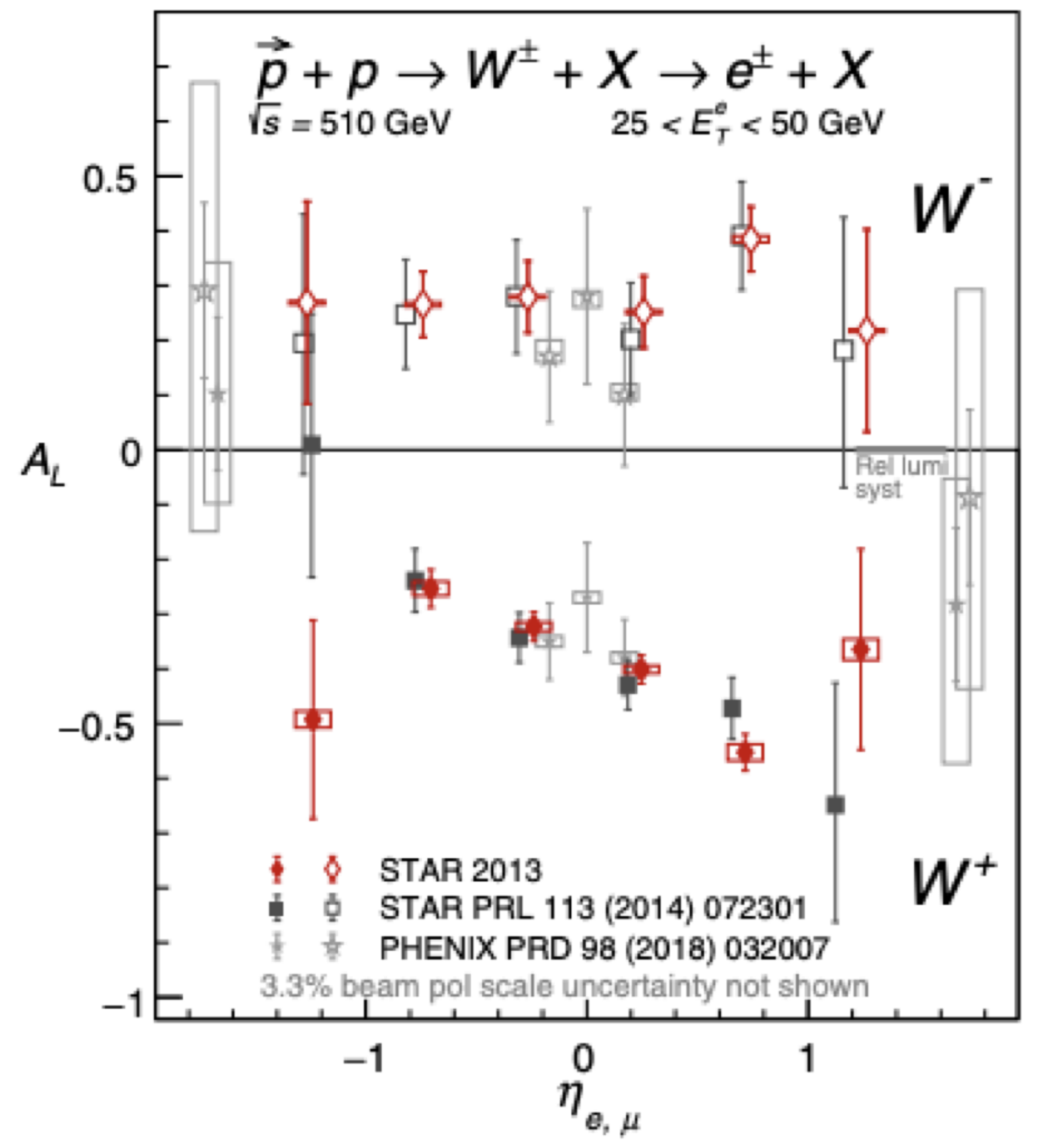}
\includegraphics[width=.48\textwidth]{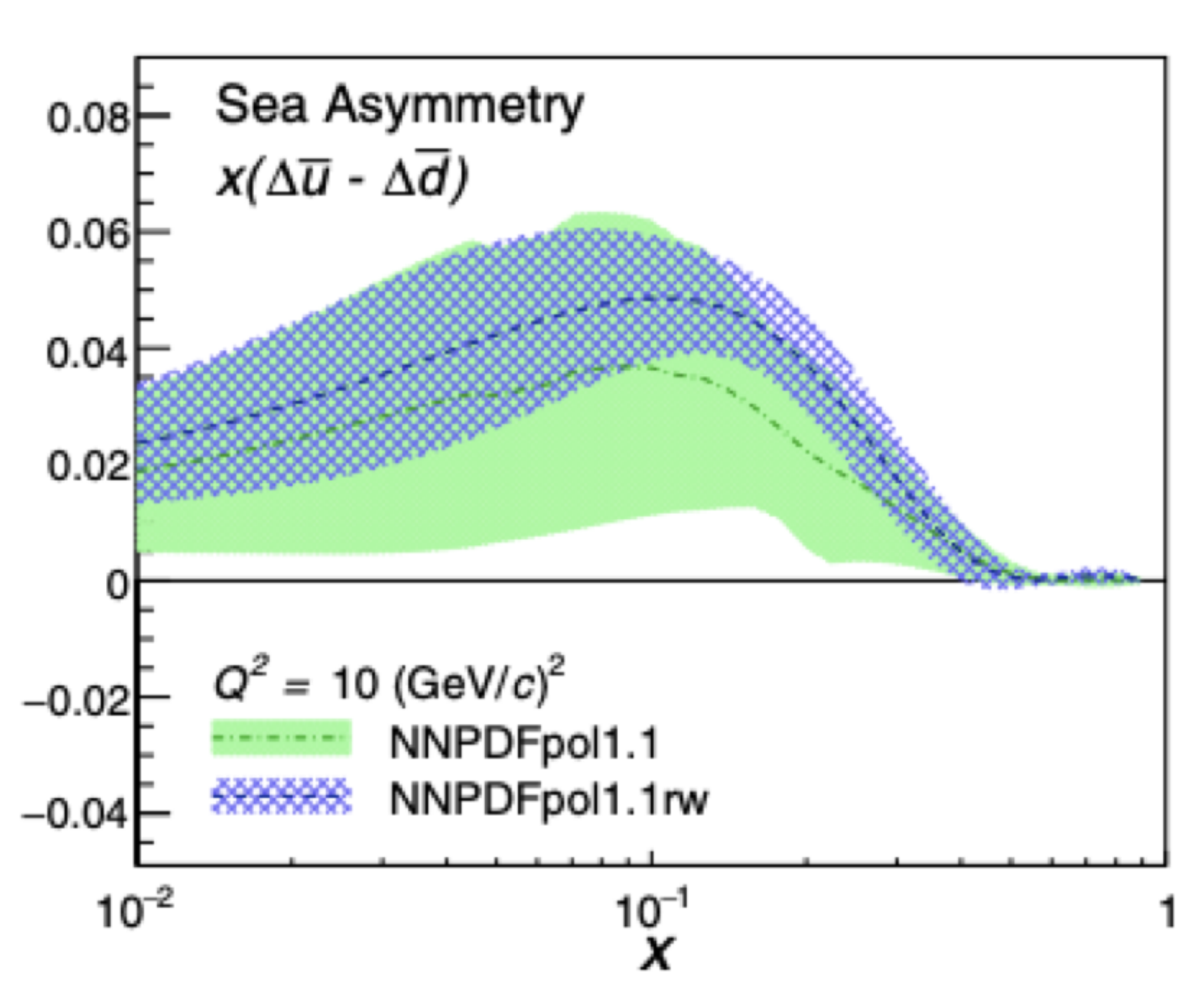}
\end{center}
\caption[Single-spin asymmetries in W$^\pm$ production from $\vec{\text{p}}\vec{\text{p}}$]{Left: single-spin asymmetries in W$^\pm$ production from collisions of longitudinally polarized protons at STAR \cite{star:deltaq} and PHENIX \cite{phenix:deltaq}. Right: asymmetry in the sea quark helicity distributions from the NNPDF global analysis \cite{nnpdf}. The blue band includes the recent results from the STAR 2013 data.}
\label{fid:deltaq_sea_rhic}
\end{figure}

Considering the longitudinal polarization of one of the RHIC proton beams, a single-spin asymmetry $A_{\text{L}}$ is extracted from the data:
\begin{equation}
A_{\text{L}}=\frac{\sigma^+-\sigma^-}{\sigma^++\sigma^-},
\end{equation}
where + (-) stands for proton beams with positive (negative) helicity with the single arrow indicating the momentum direction and the double arrow the spin direction of the beam: 
\begin{eqnarray}
+ & \substack{\Rightarrow\\\rightarrow} \;\;\text{or}\;\; \substack{\Leftarrow\\\leftarrow} & \text{positive helicity},\nonumber\\
- & \substack{\Rightarrow\\\leftarrow} \;\;\text{or}\;\; \substack{\Leftarrow\\\rightarrow}& \text{negative helicity}. \label{eq:rhicLhelicity}
\end{eqnarray}
The single-spin asymmetries $A_{\mathrm{L}}^{\text{W}^\pm}$ measured by STAR and PHENIX are shown in the left of Fig.~\ref{fid:deltaq_sea_rhic}. Separately studying W$^+$ and W$^-$ production allows to separate different combinations of quark flavors. If in addition the produced W bosons are requested to be detected in very different ranges of rapidity\footnote{Rapidity is defined with respect to the direction of the polarized particle, i.e. here, in the direction of the longitudinally polarized proton beam.} $y$,
\begin{equation}
y=\frac{1}{2}\text{ln}\frac{1+\beta\cos(\theta)}{1-\beta\cos(\theta)},
\label{eq:rapiditymain}
\end{equation}
related to the polar angle $\theta$ of the W boson (see Fig.~\ref{fig:rapidity}), the longitudinal asymmetries in W$^\pm$-boson production can be approximated as follows:
\begin{eqnarray}
A_{\text{L}}^{\text{W}^+}=\frac{1}{2}\left(\frac{\Delta \overline{d}}{\overline{d}}\right.&-&\left.\frac{\Delta u}{u}\right)\\
A_{\text{L}}^{\text{W}^-}=\frac{1}{2}\underbrace{\left(\frac{\Delta \overline{u}}{\overline{u}}\right.}_{\substack{\text{very}\\\text{backward}\\y^\text{W}\ll0}}&-&\underbrace{\left.\frac{\Delta d}{d}\right)}_{\substack{\text{very}\\\text{forward}\\y^\text{W}\gg0}}.
\end{eqnarray}
The ``very backward'' ($y^\text{W}\ll 0$) W$^+$ bosons are sensitive to $\Delta \overline{d}/\overline{d}$, while the very backward W$^-$ are sensitive to $\Delta \overline{u}/\overline{u}$. A global analysis called NNPDFpol1.1 \cite{nnpdf} provides evidence for flavor-symmetry breaking in the ``polarized sea sector'' (which is slang for sea-quark helicity distributions):
\begin{equation}
\Delta \overline{u}(x,Q^2)>\Delta \overline{d}(x,Q^2),
\end{equation}
as shown in the right of in Fig.~\ref{fid:deltaq_sea_rhic}. The NNPDF fits incorporate a Neural-Network-based Monte-Carlo approach and are obtained from a global set of DIS and collider pp data both including longitudinal proton polarization. The flavor-symmetry breaking in the polarized sea sector is opposite to that in the ``unpolarized sector'' (Sec.~\ref{sec:neutrinodis}) from SeaQuest and RHIC data, where we found $\overline{d}(x,Q^2)>\overline{u}(x,Q^2)$. 

\subsection{Gluon helicity from $\vec{\text{p}}\vec{\text{p}}$ and fixed-target}
\label{sec:deltag}

The reader may have wondered why the spin structure function $g_1(x,Q^2)$ does not also provide some information about the contribution of the gluon spin $\Delta G$ to the spin of the proton? After all, we had seen on page~\pageref{page:dglap} in Sec.~\ref{sec:scalingviolation} that the singlet quark and gluon distributions do not evolve independently of each other: even though the gluons do not contribute to charged-lepton DIS at leading order QCD, they get mixed with the quark singlet distribution through QCD evolution (Eq.~\ref{eq:dglap}). It is thus true that $g_1$ offers an indirect way to access gluon spin, however is the current kinematic reach of the DIS experiments with proton polarization not sufficient to adequately constrain $\Delta G$. 

\begin{figure}
\begin{center}
\includegraphics[width=.48\textwidth]{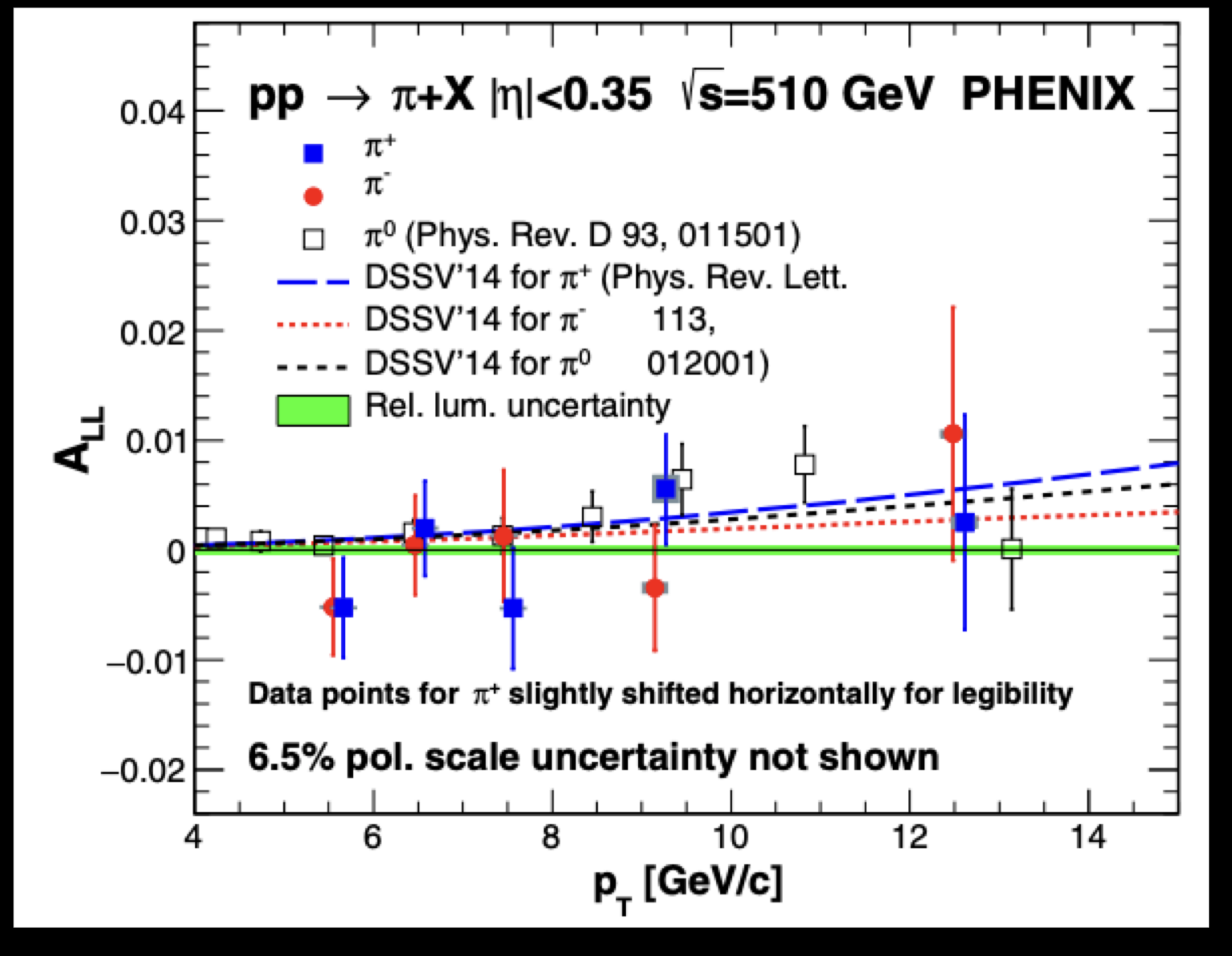}
\includegraphics[width=.48\textwidth]{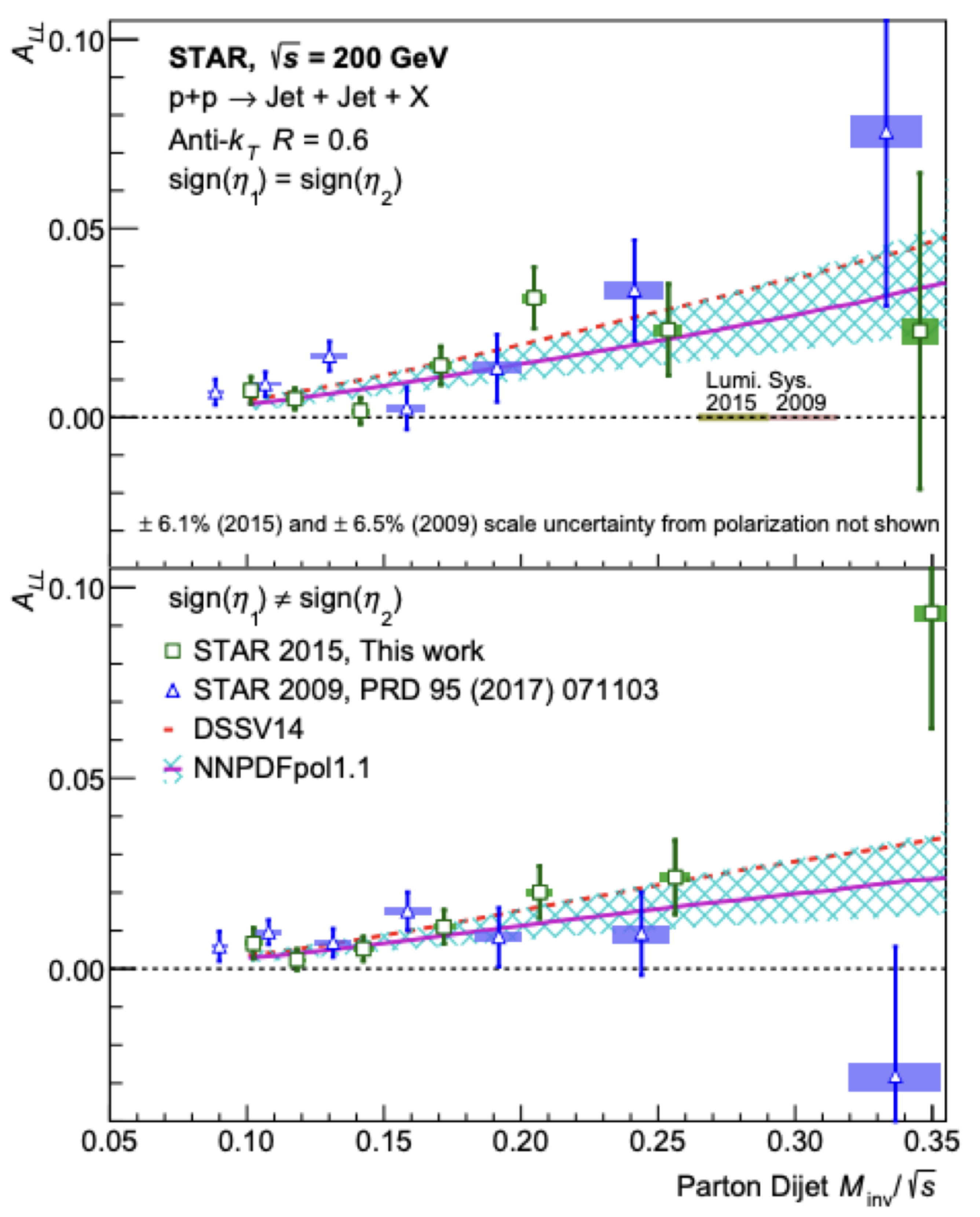}
\end{center}
\caption[Double-spin asymmetries from $\vec{\text{p}}\vec{\text{p}}$]{PHENIX \cite{phenix:all} and STAR \cite{star:all} measurements of longitudinal double-spin asymmetries in proton-proton collisions.}
\label{fig:all_rhic}
\end{figure}

On the other hand, RHIC's proton-proton collisions with doubly longitudinally (LL) polarized beams at $\sqrt{s}=200$ or 500\,GeV provide direct access to the helicity distributions of gluons in the proton, or in other words, the contribution by the gluon spin to the spin of the proton, $\Delta G$. At RHIC energies, gluon information can be probed at leading order QCD via quark-gluon ({\bf qg}) and gluon-gluon ({\bf gg}) hard scattering. Possible production channels are for example charged pions, isolated direct photons, inclusive jets, and dijets (see page~\pageref{page:jets} for jets) and a double-spin (LL) asymmetry is measured in these channels:
\begin{equation}
A_{\text{LL}}=\frac{\sigma^{++}-\sigma^{+-}}{\sigma^{++}+\sigma^{+-}},
\end{equation}
where the superscripts follow the same helicity convention as in Eq.~\ref{eq:rhicLhelicity}, only that two longitudinally polarized proton beams are considered now. Recent high-precision results from PHENIX  in charged-pion production \cite{phenix:all} and from STAR in di-jet and inclusive-jet production \cite{star:all} are shown in Fig.~\ref{fig:all_rhic}.
With ``mid-rapdity'' we refer to small absolute values of pseudo-rapidity $\eta$, 
\begin{equation}
\eta=-\ln\tan\frac{\theta}{2},
\label{pseudorapmain}
\end{equation}
typically $|\eta|<0.35$. Global QCD fits (NNPDFpol1.1 \cite{nnpdf}) indicate non-zero positive gluon-spin contribution $\Delta G=\int\text{d}x\;\Delta g(x,Q^2)$ to the proton spin in the RHIC kinematic range $(0.05 < x < 0.2)$.

\begin{figure}
\begin{center}
\includegraphics[width=.7\textwidth]{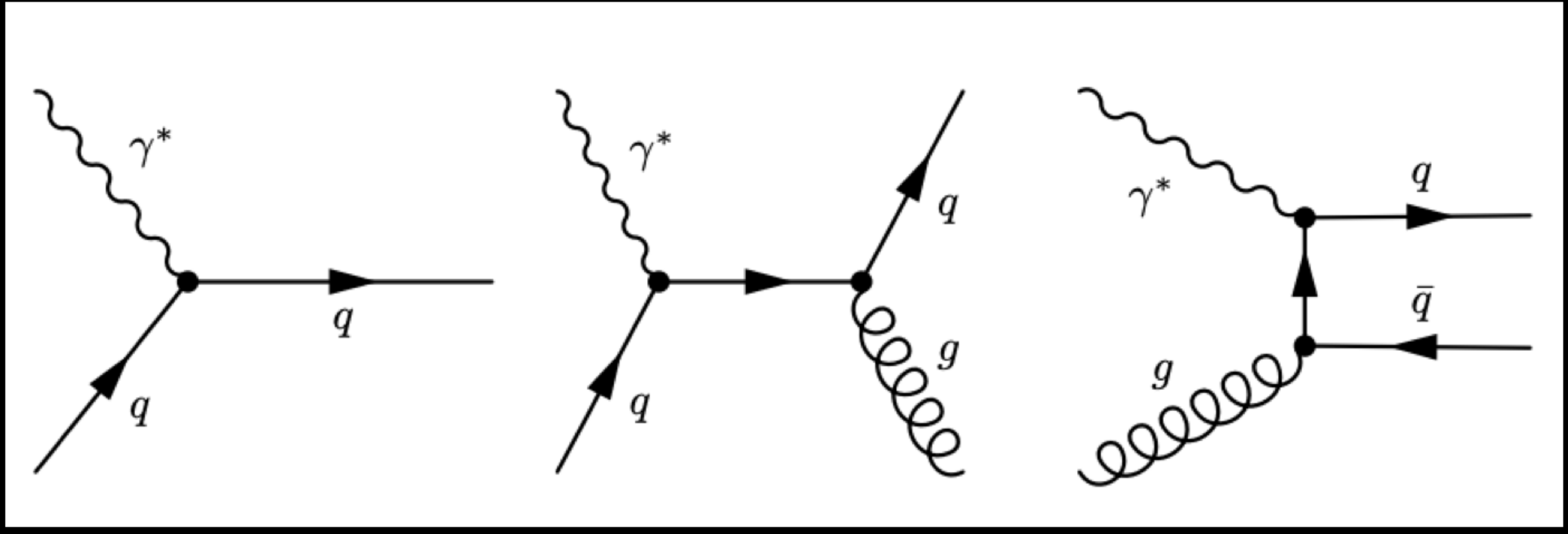}
\end{center}
\caption[QCD processes at fixed-target DIS]{QCD processes at fixed-target DIS. Left: DIS leading order (LO) not sensitive to $\Delta G$; middle: QCD-Compton (NLO) not sensitive to $\Delta G$; right: photon-gluon-fusion (NLO) sensitive to $\Delta G$.}
\label{fig:deltaq_ft}
\end{figure}

\begin{figure}
\begin{center}
\includegraphics[width=.55\textwidth]{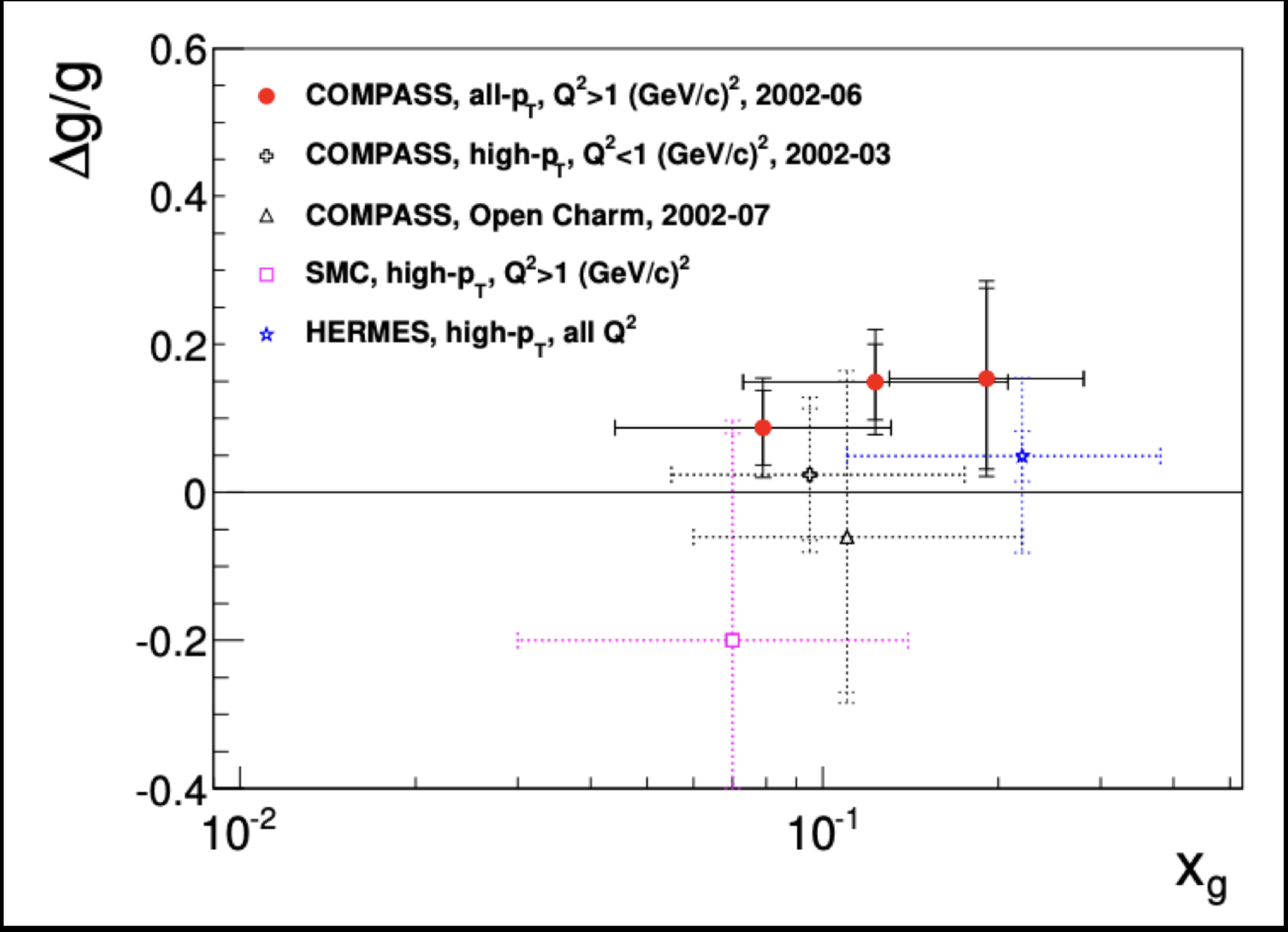}
\includegraphics[width=.41\textwidth]{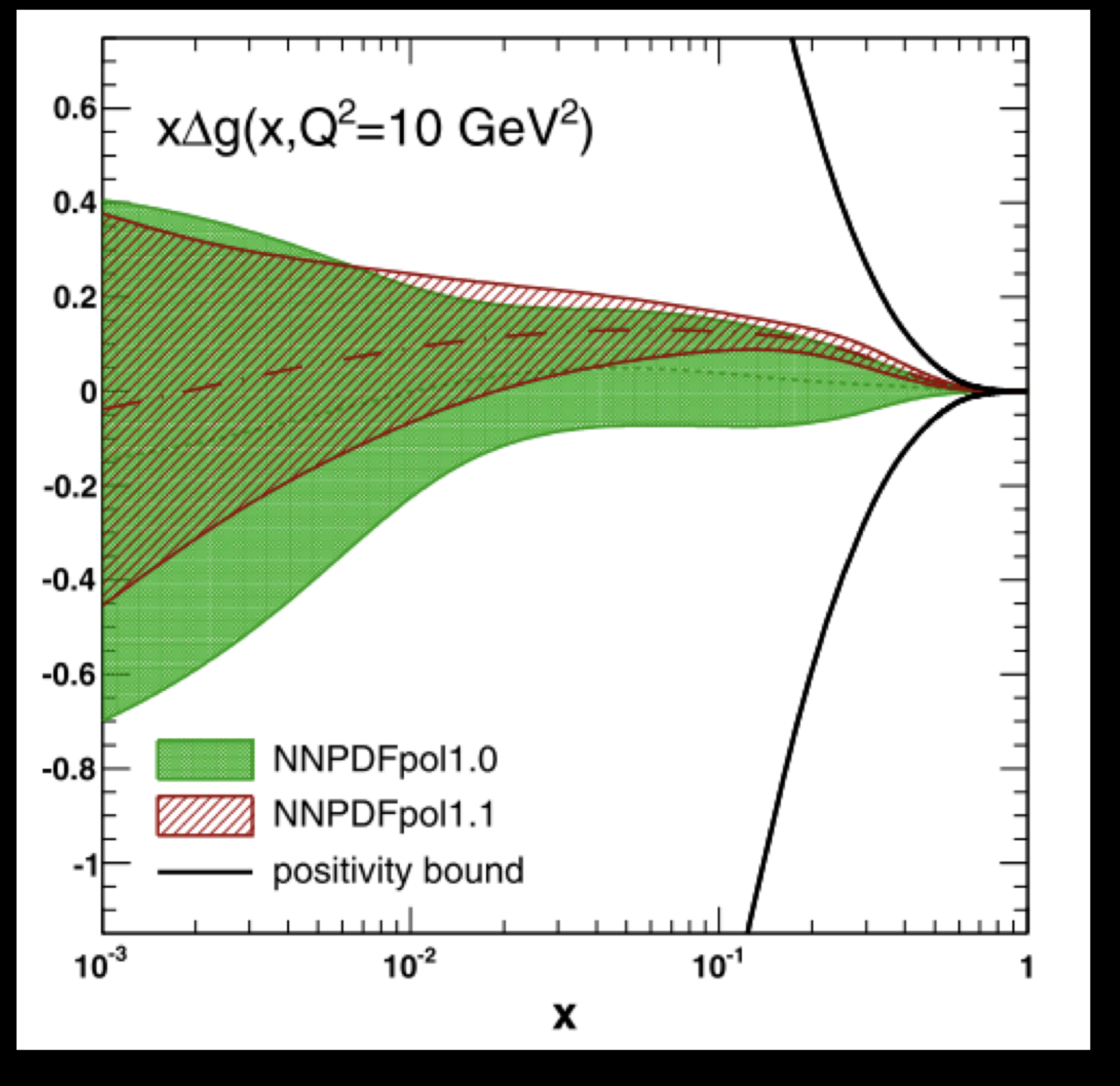}
\end{center}
\caption[Constraints on gluon helicity]{Left: HERMES \cite{hermes:deltag} and COMPASS \cite{compass:deltag} extractions of $\Delta G$ from the detection of high-$p_\text{T}$ hadrons originating from the NLO gluon-fusion process. Right: gluon helicity from a global NNPDF analysis \cite{nnpdf} including RHIC collider and fixed-target experiments.}
\label{fig:deltaq_nnpdf}
\end{figure}

Gluon helicity distributions can also be accessed at fixed-target DIS experiments, however only at QCD next-to-leading order and involving complex pQCD analyses. Only one of the processes involved at fixed-target DIS experiments shown in Fig.~\ref{fig:deltaq_ft} is sensitive to $\Delta G$ - the NLO photon-gluon fusion process. 
HERMES \cite{hermes:deltag} and COMPASS \cite{compass:deltag} selected hadrons of high transverse momenta $p_\text{T}$ to enhance the processes involving gluons. The results including also an older SMC analysis are shown in Fig.~\ref{fig:deltaq_nnpdf}.
Combining all currently available RHIC collider data and data from the fixed-target experiments, the NNPDFpol1.1 global QCD analysis  \cite{nnpdf} yields for the contribution of the gluon spin to the spin of the proton:
\begin{equation}
\int_{0.05}^{0.5}\text{d}x\;\Delta g(x,Q^2)=0.23\pm0.07\;\;\text{at}\;Q^2=10\;\text{GeV}^2,
\end{equation}
in the currently covered kinematic range. 

{\bf Recapitulation.}
We close the chapter \emph{Longitudinal Structure} by taking a step back and looking again at the spin puzzle from Eq.~\ref{eq:spinpuzle}, $1/2=1/2\,\Delta\Sigma + \Delta G + L$. The experimental results from inclusive and semi-inclusive DIS and pp experiments give:
\begin{description}
\item The quark spins contribute about 1/3 to the spin of the proton.
\item The gluon spins contribute some positive amount in the currently covered range. 
\end{description}
One may ask if there is there a substantial contribution from $\Delta G$ or $\Delta q$ at very low values of $x$-Bjorken? We expect the Electron Ion Collider to give us the answer. 
Lastly, one must ask: \emph{where is the remaining spin coming from?} Do quarks and gluons have orbital angular momentum? We will take a look at that next. 
\section{Generalized Parton Distributions}
\label{sec:gpds}

Over the past 20 years there has been substantial interest in generalized parton distributions (GPDs) and transverse momentum dependent (TMD) PDFs both in experimental and theoretical nuclear physics. The experimental challenges to measure quantities sensitive to GPDs and TMDs include setting up detector systems that allow measuring the entire final state, developing transversely polarized targets, and recording high-statistics data samples in a wide kinematic range that allow a fine slicing (binning) of the kinematic variables, just to name a few. The theoretical efforts include creating a unified picture of the nucleon by first combining the concepts of elastic form factors and deep-inelastic PDFs, and eventually unifying these two separate concepts of \emph{nucleon tomography} in the framework of the \emph{Wigner functions} \cite{diehl_GPDTMD}. 

\subsection{Hard exclusive reactions}
\label{sec:gpdintro}

\begin{figure}
\begin{center}
\includegraphics[width=.7\textwidth]{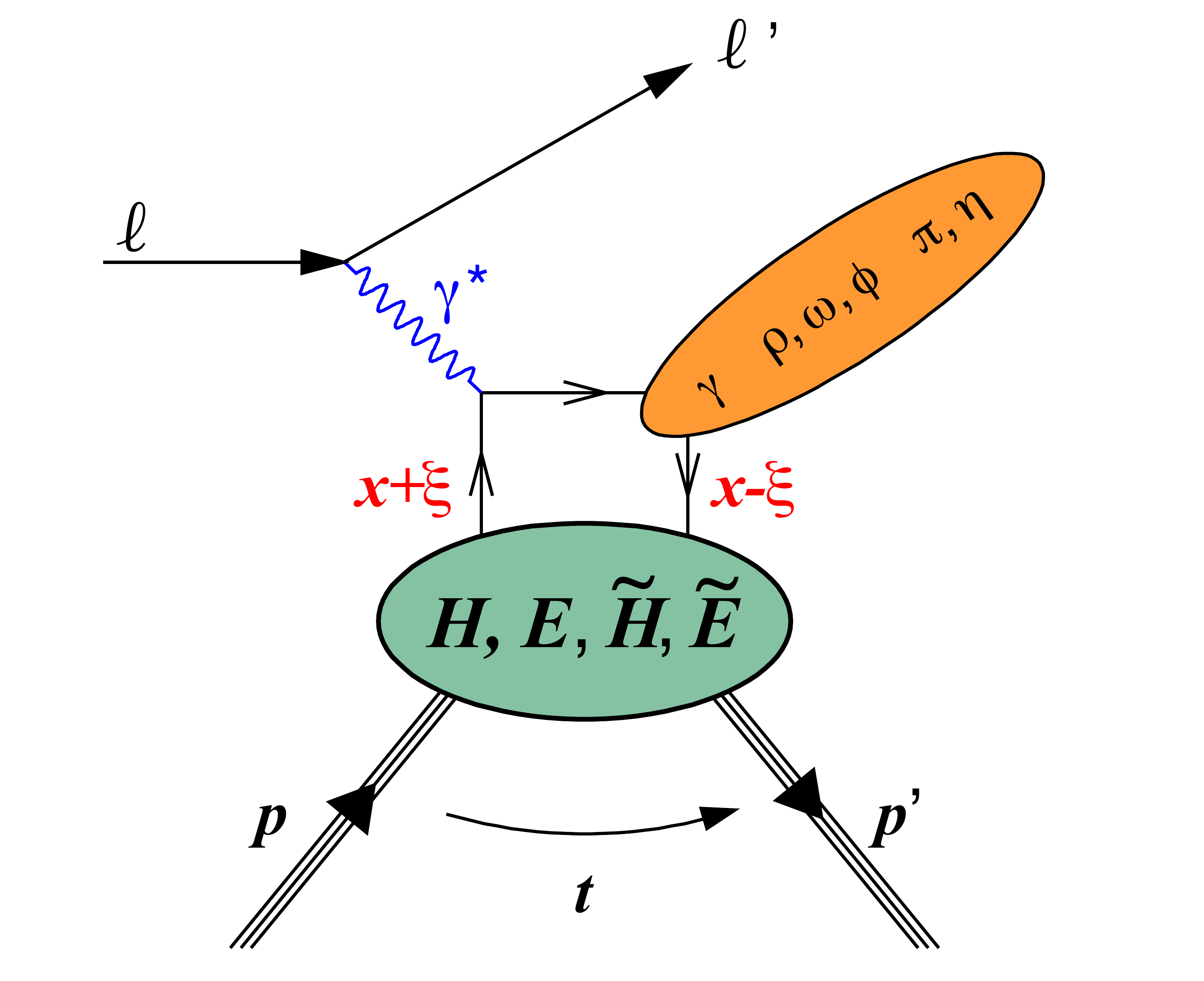}
\end{center}
\caption[Handbag diagram in DVCS and DVMP]{Deeply virtual Compton scattering (for a real photon $\gamma$ in the final state) or deeply virtual meson production (for a meson in the final state). The soft part of the hard exclusive reaction is described by generalized parton distributions (GPDs), which depend on the kinematic variables $t$, $x$, and $\xi$, where $x$ is not identical to $x$-Bjorken (which is labeled $x_\text{B}$ throughout Sec.~\ref{sec:gpds}).}
\label{fig:handbag}
\end{figure}

\begin{figure}
\begin{center}
\includegraphics[width=.95\textwidth]{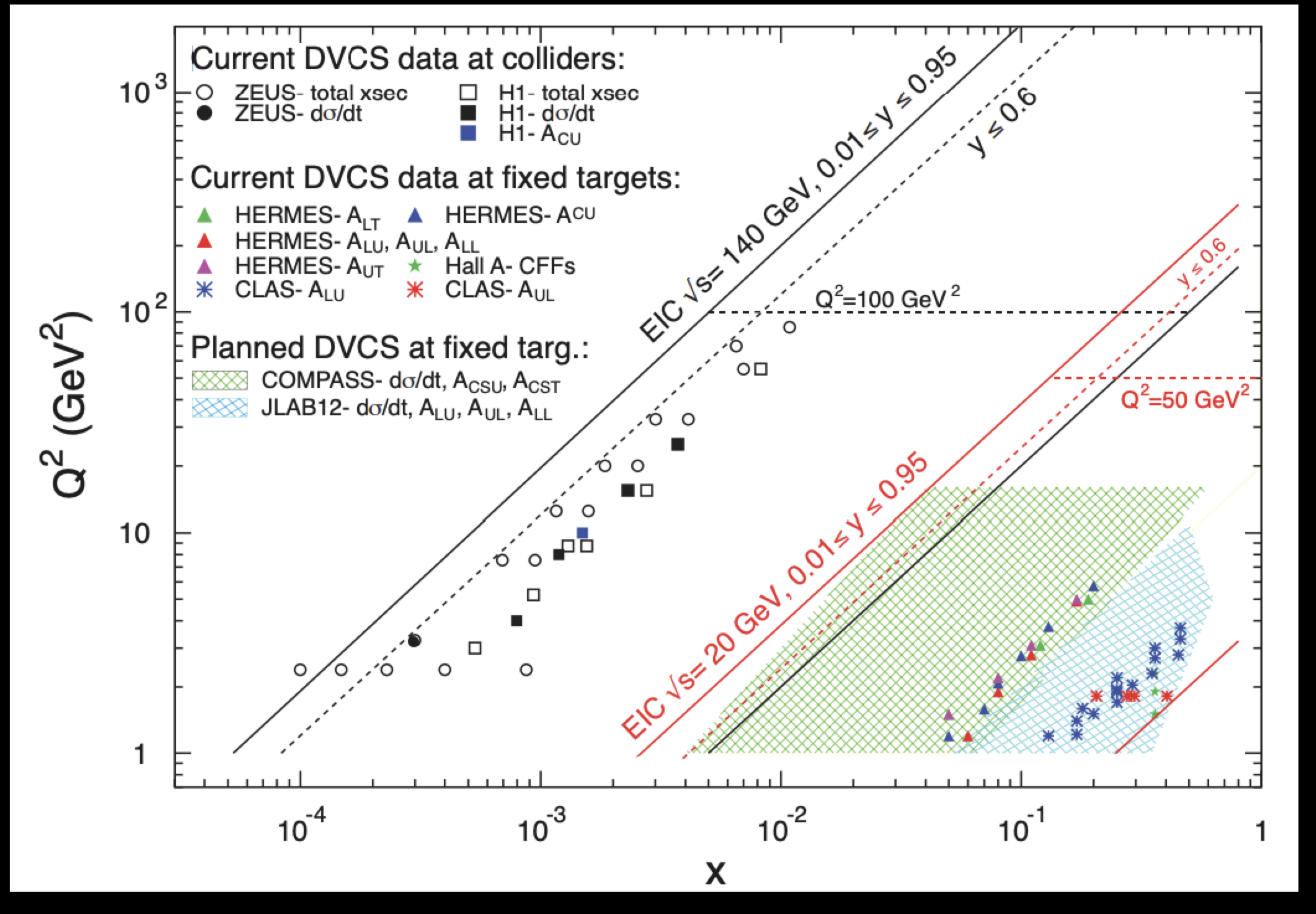}
\end{center}
\caption[DVCS kinematic coverage]{Kinematic coverage $(x_\text{B},Q^2)$ of DVCS experiments at HERMES, JLab, HERA, COMPASS, and the future Electron-Ion Collider (EIC). The COMPASS data have been collected and are being analyzed. The JLab12 data are in the process of being collected and/or analyzed. Figure from Ref.~\cite{GPDsEIC}.}
\label{fig:dvcskinecoverage}
\end{figure}

We will now study reactions in lepton ($\ell$) - proton (p) deep-inelastic scattering such as depicted in Fig.~\ref{fig:handbag}, 
\begin{equation}
\ell \text{p}\rightarrow \ell^\prime \text{p}^\prime \gamma.
\label{eq:dvcs}
\end{equation}
The process in Eq.~\ref{eq:dvcs} is called deeply virtual Compton scattering (DVCS). The wording \emph{virtual Compton} scattering is in resemblance of the traditional Compton effect, which is the elastic scattering of a real photon off an electron, while here we scatter \emph{virtual photons} off the proton, thereby producing a real photon $\gamma$. The 4-momentum transfer of the virtual photon $Q^2$ is sometimes also called \emph{photon virtuality}. Since DVCS is a subcategory of DIS, we speak of a \emph{deeply virtual} process: the $Q^2$ of the process is very different from zero, which is the 4-momentum squared of a real photon. If there is not a real photon in the final state, but a meson M, 
\begin{equation}
\ell \text{p}\rightarrow \ell^\prime \text{p}^\prime \text{M},
\label{eq:dvmp}
\end{equation}
the process is called deeply virtual meson production (DVMP). Both DVCS and DVMP have been measured at a variety of nuclear-physics experiments. The kinematic coverage of past, present and future DVCS experiments is shown in Fig.~\ref{fig:dvcskinecoverage}. 

The reactions depicted in Fig.~\ref{fig:handbag} are in general called hard exclusive reactions. The \emph{hard} refers, similarly to deeply virtual, to the $Q^2$, which is well beyond the elastic domain. The \emph{exclusive} refers to the experimental measurement technique. In order to be able to categorize the processes in Eq.~\ref{eq:dvcs} or \ref{eq:dvmp}, we need to fully assess the final state. 

\begin{figure}
\begin{center}
\includegraphics[width=.65\textwidth]{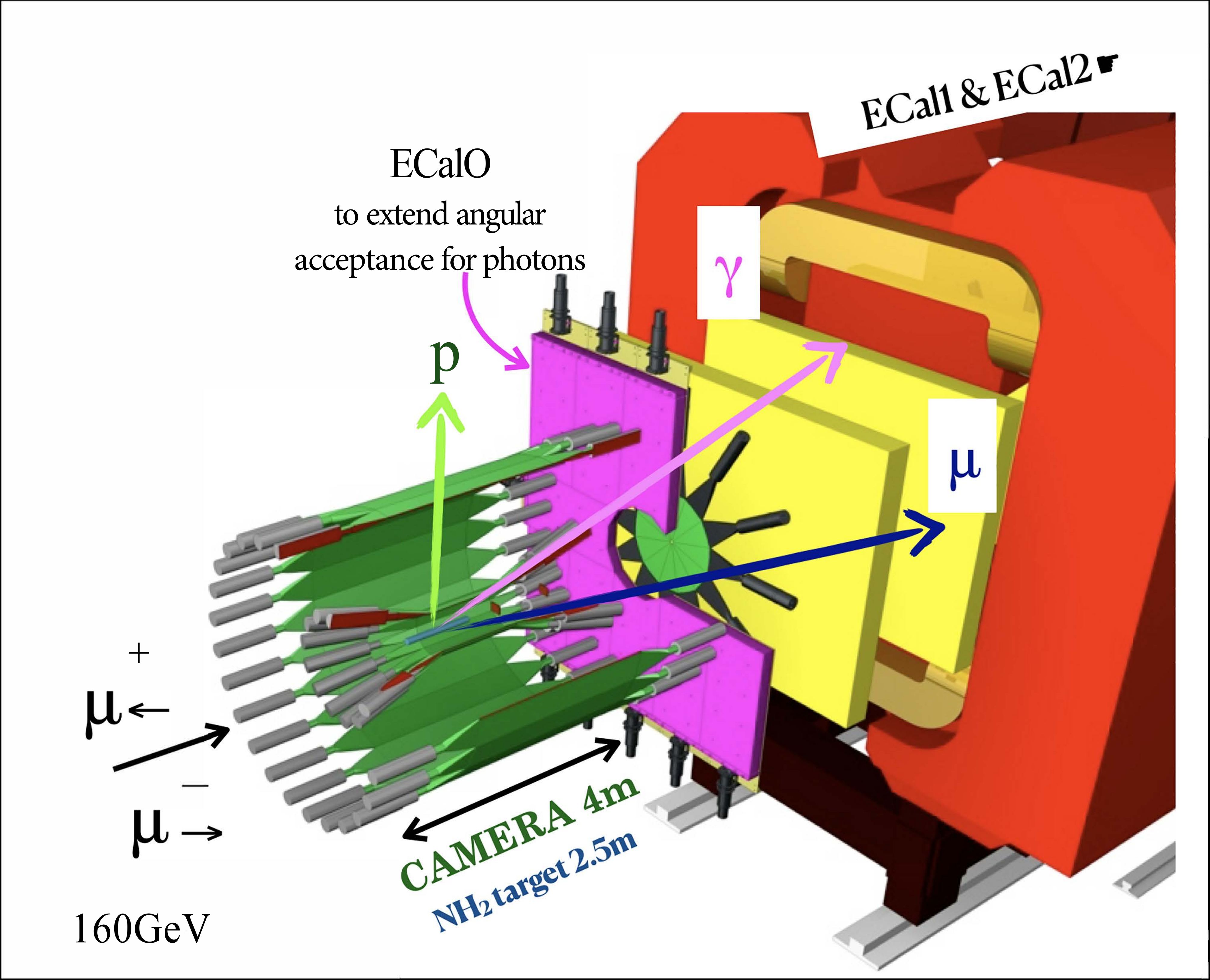}
\end{center}
\caption[COMPASS GPD setup]{COMPASS GPD setup with unpolarized hydrogen target surrounded by a time-of-flight recoil proton detector. The example of a DVCS event is shown, with muon-photon-proton detection. The forward going scattered beam lepton and real photon are detected by the standard COMPASS spectrometer supplied with an extra electromagnetic calorimeter.}
\label{fig:dvcscompass}
\end{figure}

\begin{figure}
\begin{center}
\includegraphics[width=.95\textwidth]{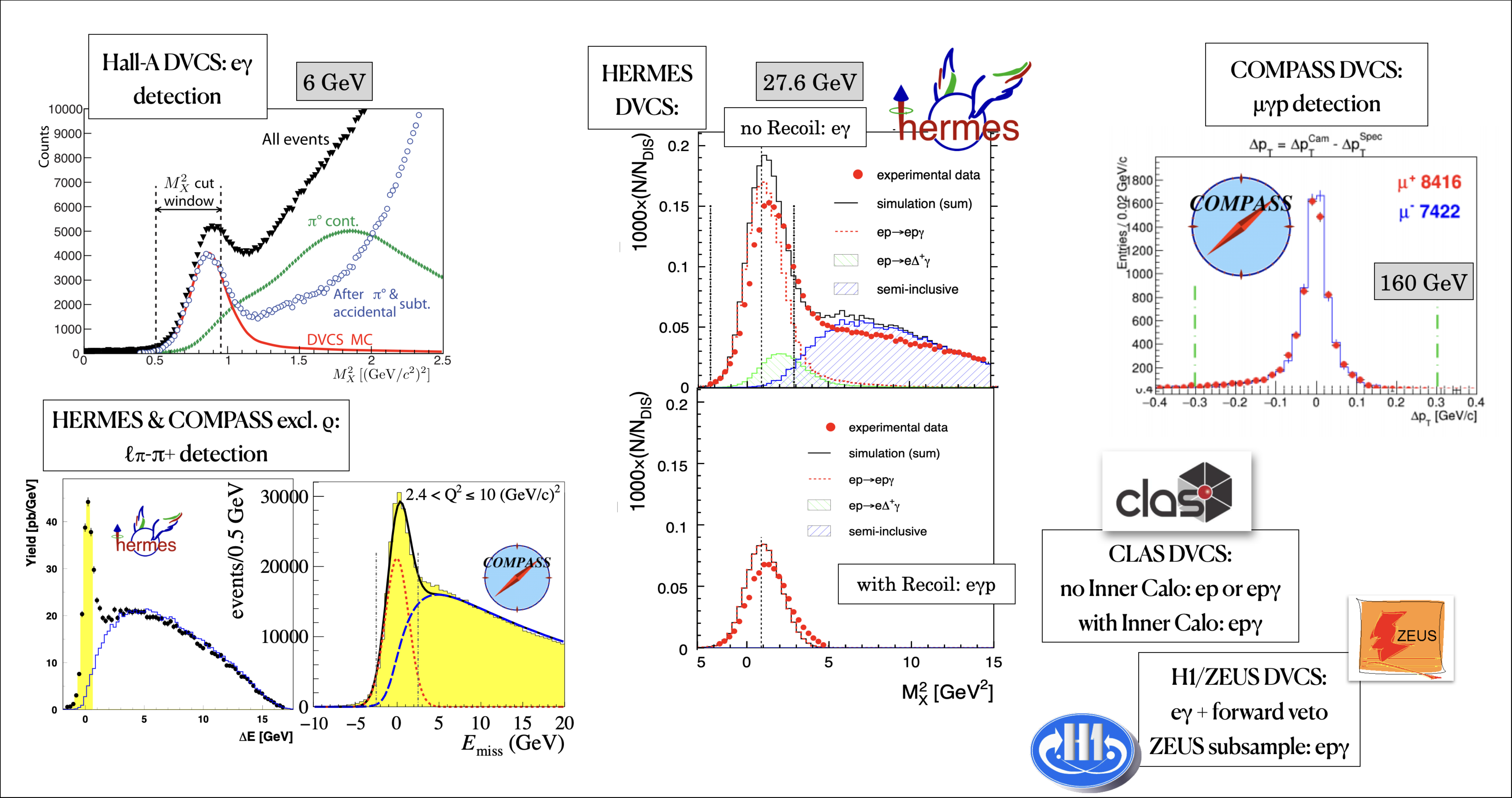}
\end{center}
\caption[Exclusive event selection]{Examples of exclusive event selections using missing-energy or missing-mass techniques, or kinematic event fitting.}
\label{fig:dvcsexp}
\end{figure}

Note that there is no unmeasured state ``X'' anymore unlike in the cases of inclusive (Eq.~\ref{eq:dis}) or semi-inclusive DIS (Eq.~\ref{eq:sidis}). We claim to fully reconstruct or at least assume to know the final state, and if we make mistakes, there will be contamination by SIDIS events. If the detector coverage is not sufficient, we will (depending on the center-of-mass energy of the experiment) either not be able to measure certain exclusive reactions at all (as was the case for DVCS at COMPASS without recoil proton detector), or we will have to make assumptions about the unmeasured recoil proton at the expense of including in the exclusive event sample unwanted resonant excitations of the proton (as was the case for DVCS at HERMES without recoil detector). At fixed-target geometry, the recoiling protons p$^\prime$ scatter under large polar angles and not into the existing forward spectrometer, making it necessary to build a recoil detector around the target (see Fig.~\ref{fig:dvcscompass}). 
This is the reason why both HERMES and COMPASS have not performed exclusive measurements with \emph{both} polarized target and recoil-proton detection. Both the HERMES and COMPASS polarized targets are quite bulky and did in their default designs not allow the simultaneous installation of a detector around them, not to speak of the extra materials that those low-energy protons would have to traverse. Some exclusive event selections at different experiments are shown in Fig.~\ref{fig:dvcsexp}. Missing-energy or missing-mass techniques are applied whenever one of the particles cannot be detected (usually the proton). If all particles of the event can be detected, the best choice of tagging exclusive events is kinematic event fitting: under the assumption of 4-momentum conservation, a $\chi^2$ that includes measured and fitted kinematic parameters is minimized. 

As is the case for the longitudinal spin observables discussed earlier, we are looking to the EIC \cite{GPDsEIC} to add missing data and extend the kinematic reach, in particular for the observables related to total angular momentum of partons. 

\subsection{Generalized parton distributions}

The DVCS and DVMP processes represent the standard experimental channels to access GPDs, which describe the soft, non-perturbative part of the scattering process. For large $Q^2$ and small $t$, the DVCS cross section will factorize into a hard-scattering kernel and the GPDs. Factorization also holds for exclusive production of mesons from longitudinally polarized virtual photons. The GPDs depend on four kinematic variables: 
\begin{itemize}
\item $Q^2$, the squared 4-momentum transfer by the lepton (introduced in Sec.~\ref{sec:elascat}). 
\item The Mandelstam variable $t$, which is the squared 4-momentum transfer to the target proton. 
\item The two longitudinal momentum fractions $x$ and $\xi$ of the probed quark. Here $x$ represents the average of the quark momentum fractions of initial and final state, and $\xi$ half of their difference. \emph{Note that $x$ is not equal to $x$-Bjorken} introduced in Sec.~\ref{sec:disdis}, the latter of which we will refer to as $x_{\text B}$ throughout Sec.~\ref{sec:gpds}. 
\end{itemize}
In the \emph{Bjorken limit} (large $Q^2$ and $x_{\text B}$ at fixed $t$), $\xi$ can be approximated as $\xi\approx x_{\text B}/(2-x_{\text B})$ and is referred to as \emph{skewness}. No information about the (``mute'') variable $x$ is contained in DVCS and DVMP, it can however be directly accessed in double DVCS (DDVCS), which allows mapping of GPDs in $(x,\xi)$ space. The virtuality of the final photon provides an additional lever arm,
which allows to vary $x$ and $\xi$ independently. DDVCS measurements are planned at CLAS12. 

We can now describe a hard exclusive reaction such as in Fig.~\ref{fig:handbag} as follows: a quark with longitudinal momentum fraction $x+\xi$ is taken out of the proton and is returned with different longitudinal momentum fraction $x-\xi$. In that process, a particle (photon or meson) is emitted carrying away information about proton structure, and the proton stays intact after having received the (typically small) squared 4-momentum transfer $t$.

Remarkably, the GPDs contain both elastic form factors (Sec.~\ref{sec:formfactors}) and ordinary PDFs (Sec.~\ref{sec:pdfs}) as limiting cases in certain kinematic limits, or when integrating over one of the kinematic variables: 
\begin{itemize}
\item Form factors: the form factor $F^q(t)$ is recovered when integrating over $x$:
\begin{equation}
\int_{-1}^1\text{d}x\;H^q(x,\xi,t)=F^q(t),
\end{equation}
where the superscript $q$ denotes the quark flavor. Note that the form factor depended on $Q^2$ in Sec.~\ref{sec:formfactors}, while now they depend on $t$, which is also a squared momentum.
\item PDFs: in the \emph{forward limit} $\xi\rightarrow 0$ and $t\rightarrow 0$, the GPD $H^q(x,\xi,t)$ reduces to the collinear PDF $q(x_{\text B})$:
\begin{equation}
H^q(x,0,0)=q(x_{\text B}).
\end{equation}
\end{itemize}
This means that GPDs encode simultaneously information about transverse parton positions (their form factor character) and longitudinal momenta (their PDF character) - see also Fig.~\ref{fig:gpdmapping} on page~\pageref{fig:gpdmapping}. Theoretical GPD overviews are given in Ref.~\cite{diehl_GPDTMD}, and references therein. 

At leading twist, there are four chiral-even\footnote{quark-helicity conserving} GPDs describing the proton (or generically, any spin-1/2 hadron), as summarized in Tab.~\ref{tab:gpds}. As was the case for collinear PDFs, there is a larger set of GPDs needed to describe hadrons of higher spin (e.g., spin-1). 
\begin{table}
\begin{center}
\begin{tabular}{c|c|c|c|c}
& \small flips & \small conserves &  \multicolumn{2}{c}{} \\
GPD &  \small nucleon  & \small nucleon  &\multicolumn{2}{c}{experimental} \\
& \small helicity & \small helicity & \multicolumn{2}{c}{channel} 
\\\hline
&&&&\\
\small does not depend & \multirow{2}{*}{$E$} & \multirow{2}{*}{$H$} & $J^{\mathcal{P}}=1^-$&  $J^\mathcal{P}=1^-$ \\
\small on quark helicity & & &\small vector mesons&\small DVCS\\\hline
&&&&\\
\small depends on & \multirow{2}{*}{$\widetilde{E}$} & \multirow{2}{*}{$\widetilde{H}$} & $J^{\mathcal{P}}=0^-$& $J^\mathcal{P}=1^-$\\
\small quark helicity & & &\small pseudo-scalar mesons&\small DVCS\\
\end{tabular}
\caption[Chiral-even GPDs for spin-1/2 hadron]{The four chiral-even GPDs for a spin-1/2 hadron. In the forward limit, GPD $H(x,\xi,t)$ reduces to the spin-independent structure function $F_1(x_{\text B})$ and $\widetilde{H}(x,\xi,t)$ to the spin structure function $g_1(x_{\text B})$ in DIS.}
\label{tab:gpds}
\end{center}
\end{table}
In hard exclusive production of vector mesons, GPDs $E$ and $H$ can be accessed, while the production of pseudo-scalar mesons allows studying GPDs $\widetilde{H}$ and $\widetilde{E}$. The DVCS process allows access to all four GPDs. Because of that and because DVCS does not involve an unknown meson amplitude, it is sometimes referred to as \emph{golden channel} to access GPDs. 

There are in addition four chiral-odd\footnote{quark-helicity flipping} GPDs, some of which are related to transverse-momentum-dependent PDFs (TMDs, see Sec.~\ref{sec:tmds}). The chiral-odd GPDs are distinguished from the chiral-even GPDs via a subscript $T$. 

The  GPD $E$ (and the Sivers TMD discussed below) involve a switch of nucleon helicity and are sensitive to spin-orbit correlations. They are thus closely connected with orbital angular momentum of partons.

\subsection{DVCS amplitude and Compton form factors}

\begin{figure}
\begin{center}
\includegraphics[width=.95\textwidth]{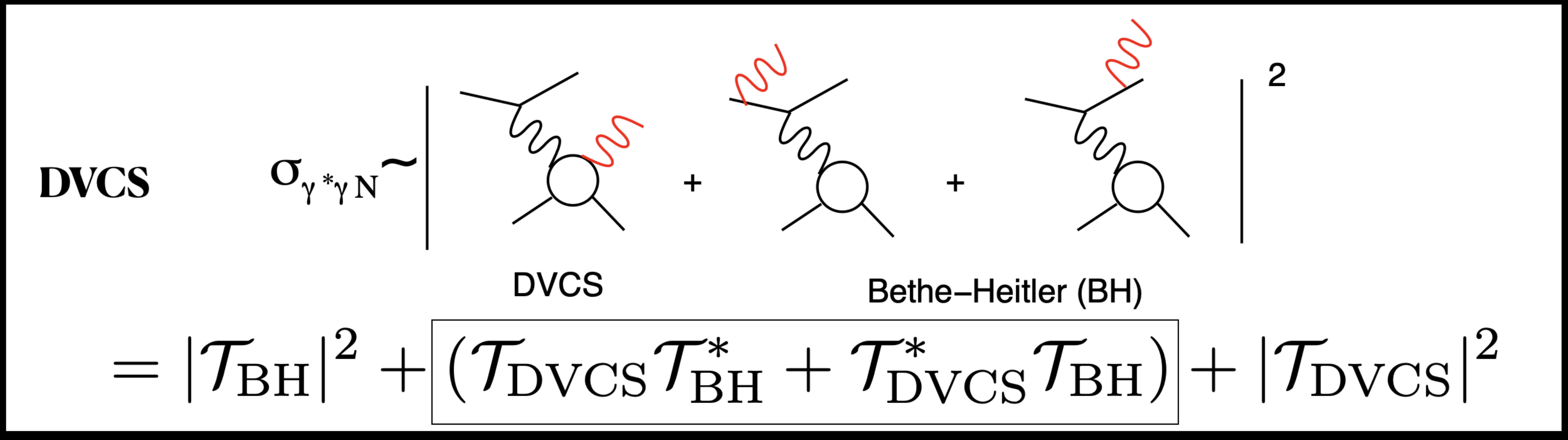}
\end{center}
\caption[DVCS / Bethe-Heitler interference]{Terms in the cross section (proportional to the square of the complex scattering amplitude $\mathcal{T}$) for the production of a single real photon in deep-inelastic lepton-proton scattering: there are pure Bethe-Heitler (BH) and pure DVCS contributions, and a contribution arising from the interference of DVCS and the BH process (boxed). The starred amplitude is the complex conjugate.}
\label{fig:dvcsbh}
\end{figure}

There is a process that has the same initial and final state as the DVCS process (Eq.~\ref{eq:dvcs}) - the Bethe-Heitler (BH) process, which is the emission of a real photon off the initial- or final-state lepton. Because of their identical initial and final states, the DVCS and BH processes interfere, as shown in Fig.~\ref{fig:dvcsbh}. There is no experimental detection technique to distinguish the two processes - the experimental signature is the coincident detection of the scattered beam lepton, a real photon, and the recoil proton. The presence of the BH process is however a lucky coincidence, and not only an unwanted dilution of the desired signal: in general, a cross section is proportional to the square of the scattering amplitude $|\mathcal{T}|^2=R^2$ with $\mathcal{T}=R\cdot \mathrm{e}^{i\varphi}$ (with $R$ the real part and the phase $\varphi$ the imaginary part), so that the phase information is usually lost in scattering experiments. The interference term between DVCS and BH, on the other hand, allows to determine both magnitude (real part) and phase (imaginary part) of the DVCS amplitude since the DVCS amplitude enters only linearly.  This is sometimes referred to as \emph{holographic principle} \cite{BeMu2002}. In other words, the  DVCS-BH interference term allows to disentangle the real and imaginary parts of the DVCS amplitude. 

While the BH contribution is exactly calculable in QED with the input of the elastic form factors of the nucleon, the DVCS amplitude is a priori unknown. At low center-of-mass energies (typical for HERMES and JLab), the BH amplitude dominates the DVCS amplitude: $|\mathcal{T}_{\mathrm{DVCS}}|^2\ll |\mathcal{T}_{\mathrm{BH}}|^2$, while at higher energies (at HERA and COMPASS), the magnitude of the two amplitudes become comparable: $|\mathcal{T}_{\mathrm{DVCS}}|^2\approx |\mathcal{T}_{\mathrm{BH}}|^2$.  

\begin{figure}
\begin{center}
\includegraphics[width=.75\textwidth]{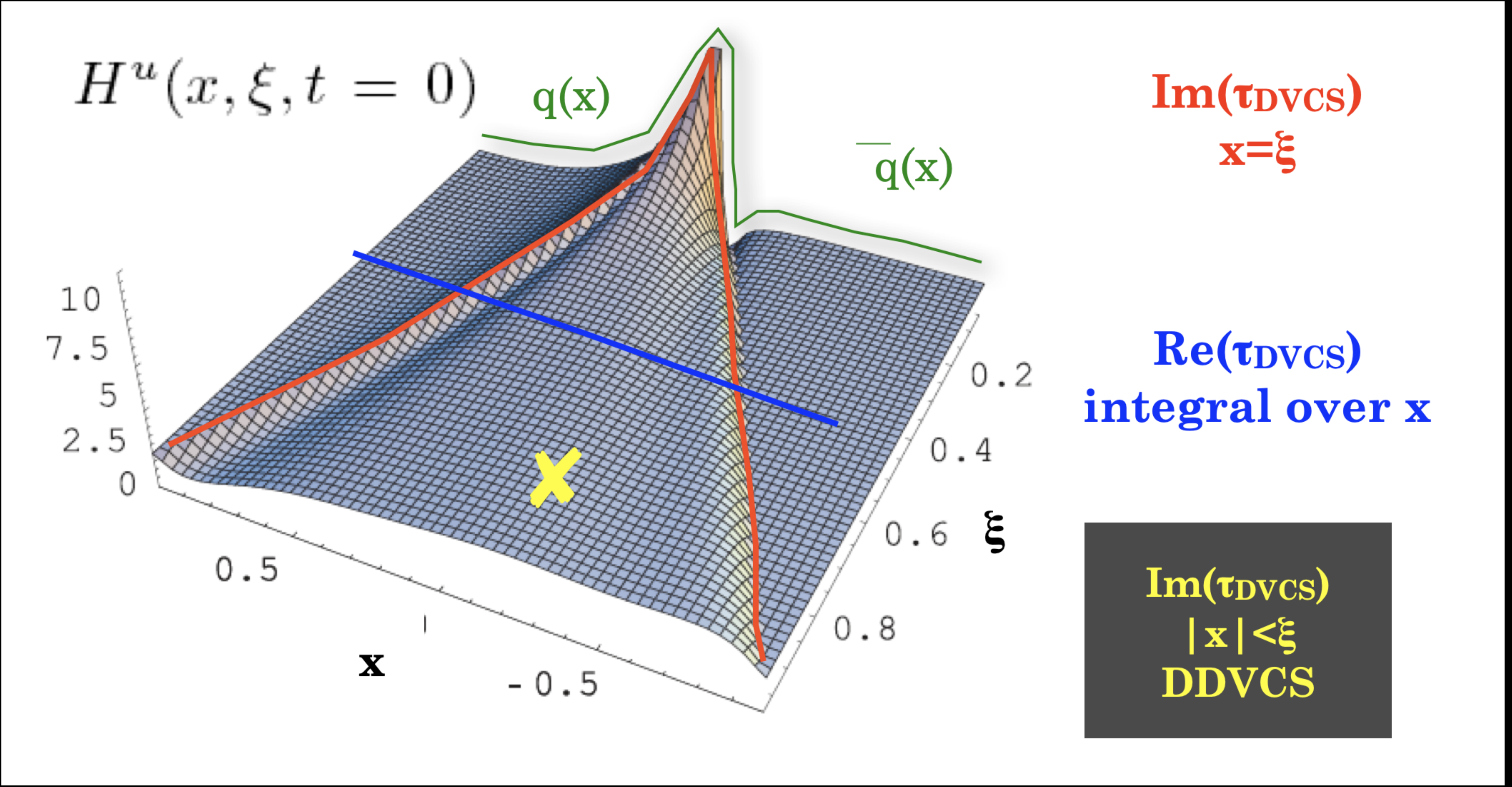}
\end{center}
\caption[GPD $H$]{Relations between GPD $H$ and DVCS amplitude. See text for details.}
\label{fig:gpdh}
\end{figure}

The experimentally accessed quantity is a complex Compton form factor (CFF) $\mathcal{H}$, usually written as calligraphic letter to distinguish it from the GPD $H$:
\begin{equation}
\mathcal{H}(\xi,t)=\textcolor{blue}{\mathcal{P}\int_{-1}^{+1}\mathrm{d}x\;\frac{H(x,\xi,t}{x-\xi}}-i\textcolor{red}{H(\xi,\xi,t)},
\label{eq:cff}
\end{equation}
where $\mathcal{P}$ is the principal value. Similarly, there is a CFF $\mathcal{E}$ for GPD $E$, and so forth. A graphical representation of GPD $H^{\text u}$ (for the u-quark) is given in Fig.~\ref{fig:gpdh}. Both in the figure and in Eq.~\ref{eq:cff}, the red color indicates the imaginary part of the DVCS scattering amplitude (integral over $x$) and the blue color its real part (GPD at $x=\xi$). The green line in Fig.~\ref{fig:gpdh} indicates the PDF $q(x_{\text B})$ in the forward limit ($\xi\rightarrow 0$ and $t\rightarrow 0$). The region $|x|<\xi$ is accessible only in double DVCS. 

The dispersion relation between the real and imaginary parts of the CFF includes a \emph{$D$-term} $D(t)$,
\begin{equation}
\textcolor{blue}{{\mathrm{\bf Re}}\;\mathcal{H}(\xi,t)}=\mathcal{P}\int_{-1}^{+1}\mathrm{d}x\;\frac{\textcolor{red}{\mathrm{\bf Im}\mathcal{H}(x,t)}}{x-\xi}+D(t),
\label{eq:dterm}
\end{equation}
which is interestingly related to shear forces and radial distribution of pressure inside the nucleon (see also Fig.~\ref{fig:gpdglobal}). 

\subsection{Experimental access to GPDs via DVCS}
\label{sec:expdvcs}

\begin{figure}
\begin{center}
\includegraphics[width=.75\textwidth]{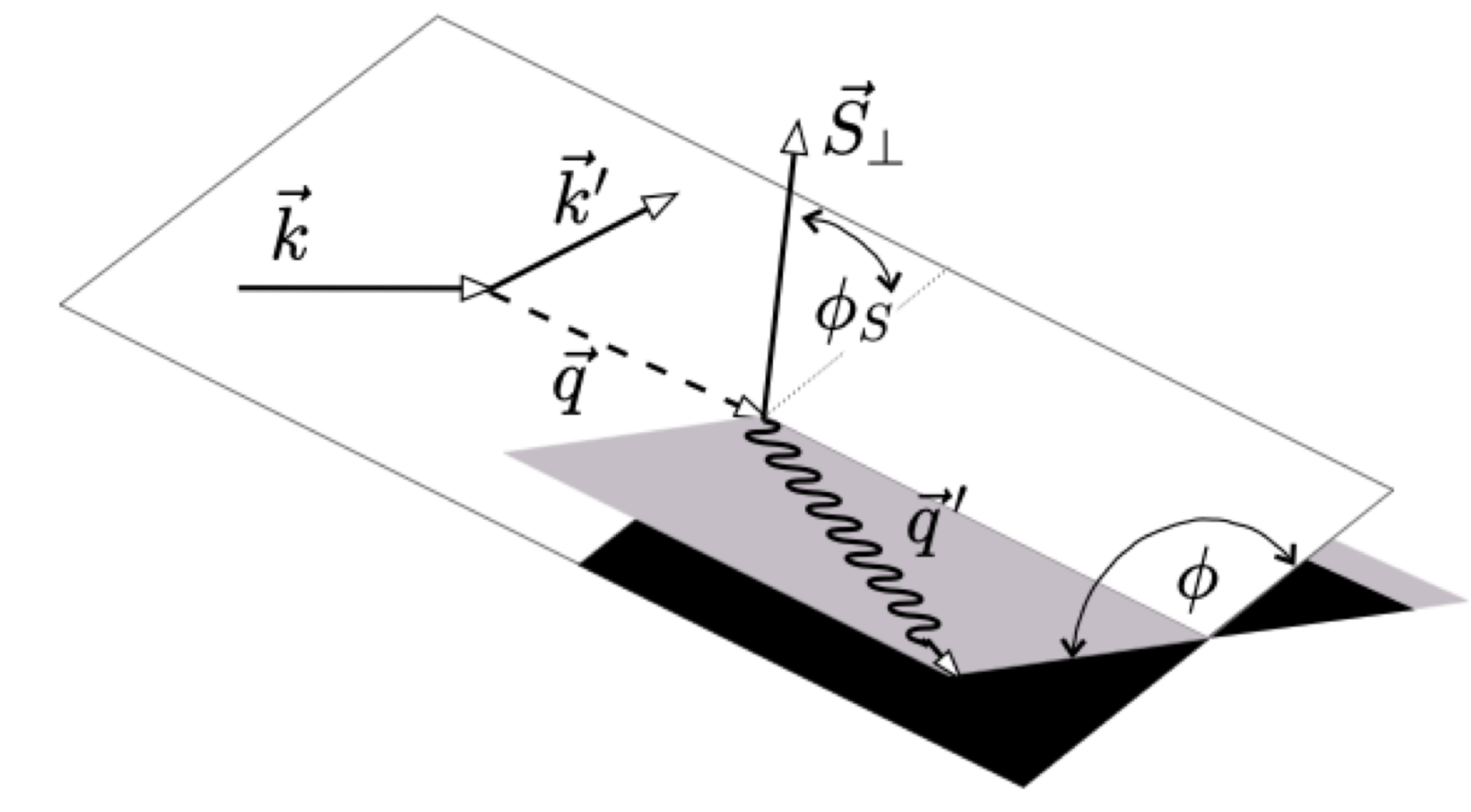}
\end{center}
\caption[Azimuthal angle definition in DVCS at fixed target]{Definition of 3-momentum vectors and angles for DVCS at a fixed target: $\vec{k}$ ($\vec{k}^\prime$) is the momentum of the incoming (scattered) lepton, $\vec{q}$ that of the virtual photon, $\vec{q}^{\;\prime}$ that of the real photon, and $\vec{S}_\perp$ is the vector pointing in the direction of the transverse target polarization (if given). The angle $\phi$ is the angle between the lepton scattering plane and the and the plane spanned by the virtual and real photons. The angle $\phi_\text{S}$ is the angle between the lepton scattering plane and the target-polarization vector.}
\label{fig:dvcsangles}
\end{figure}

The 4-fold differential cross section for exclusive single-photon production (Eq.~\ref{eq:dvcs}) on the  unpolarized proton is given by: 
\begin{equation}
\frac{\mathrm{d}^4\sigma}{\mathrm{d}Q^2\;\mathrm{d}x_{\text B}\;\mathrm{d}t\;\mathrm{d}\phi}=\frac{x_{\text B}e^6}{32(2\pi)^4Q^4\sqrt{1+\epsilon^2}}|\mathcal{T}_{\ell p\rightarrow \ell^\prime\text{p}^\prime\gamma}|^2,
\label{eq:dvcsxsec}
\end{equation}
with $\epsilon=2x_{\text B}M/Q$ and $\mathcal{T}_{\ell\text{p}\rightarrow \ell^\prime\text{p}^\prime\gamma}$ the scattering amplitude of the process, $|\mathcal{T}_{\ell\text{p}\rightarrow \ell^\prime\text{p}^\prime\gamma}|^2=|\mathcal{T}_{\mathrm{DVCS}}|^2+|\mathcal{T}_{\mathrm{BH}}|^2+\mathcal{I}$, and $\mathcal{I}$ the DVCS-BH interference term. Each of these three contributions in Fig.~\ref{fig:dvcsbh} can be written in terms of a harmonic series with respect to the azimuthal angle $\phi$ (see Fig.~\ref{fig:dvcsangles}). If the proton is transversely polarized, there is in addition the angle $\phi_\text{S}$. 

\begin{figure}
\begin{center}
\includegraphics[width=.96\textwidth]{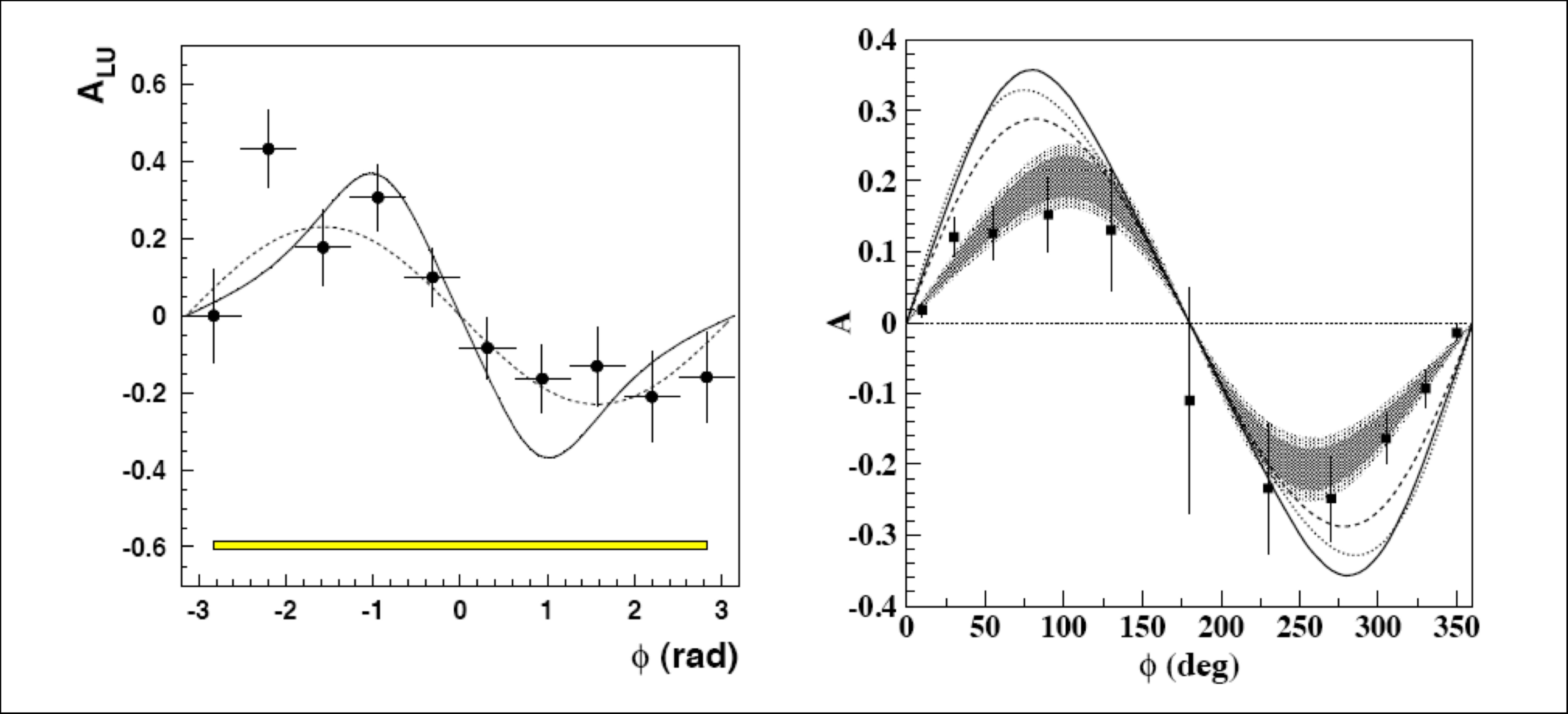}
\end{center}
\caption[Early DVCS results]{First DVCS beam-helicity asymmetries $\mathcal{A}_{\mathrm{LU}}(\phi)$ on unpolarized protons and longitudinally polarized leptons for HERMES \cite{dvcshermes2001} (left) from a 27.6~GeV positron beam and CLAS \cite{dvcsclas2001} (right) from a 6~GeV electron beam. Note the different locations of $\phi=0$ and the opposite signs of the $\sin\phi$ amplitudes, as expected from the dependence on the lepton beam charges, which are opposite at HERMES and CLAS for these measurements.}
\label{fig:dvcsfirst}
\end{figure}

At HERMES and the JLab experiments, information about the DVCS process is accessed experimentally by measuring an azimuthal asymmetry with respect to (for example) beam helicity:
\begin{equation}
\mathcal{A}_{\text{LU}}(\phi)=\frac{\sigma^\rightarrow-\sigma^\leftarrow}{\sigma^\rightarrow+\sigma^\leftarrow},
\label{eq:dvcsalu}
\end{equation}
where the arrow $\rightarrow$ ($\leftarrow$) indicates positive (negative) beam helicity of the longitudinally polarized lepton beam. The \emph{beam-helicity asymmetry} from Eq.~\ref{eq:dvcsalu} is often also referred to as beam-spin asymmetry. Here and in the following, we use the double-subscript notation $\mathcal{A}_{[\mathrm{beam}\;\mathrm{target}]}(\phi)$ indicating the polarization of beam and target: U for unpolarized, L for longitudinally polarized, and T for transversely polarized (the exception being the beam-charge asymmetry $\mathcal{A}_{\mathrm{C}}$ with single index). The single-photon production cross section from Eq.~\ref{eq:dvcsxsec} can then be written in terms of the beam-helicity asymmetry of Eq.~\ref{eq:dvcsalu}:
\begin{equation}
\left.\frac{\mathrm{d}^4\sigma}{\mathrm{d}Q^2\;\mathrm{d}x_{B}\;\mathrm{d}t\;\mathrm{d}\phi}\right|_{\lambda,e_\ell}=\sigma_\mathrm{UU}(\phi,e_\ell)\left[1+\lambda\mathcal{A}_{\text{LU}}(\phi,e_\ell)\right],
\end{equation}
where $\sigma_\mathrm{UU}$ is the helicity-(spin-)averaged cross section, $\lambda$ the lepton-beam helicity, and $e_\ell$ the lepton charge. Note that we here indicate the dependence on the beam charge because the sign of the beam-helicity asymmetry is dependent on the sign of the beam charge. The beam-helicity asymmetry can then be approximated by a truncated Fourier series:
\begin{equation}
\mathcal{A}_{\mathrm{LU}}(\phi,e_\ell)\simeq A_{\mathrm{LU}}^{\sin\phi}\sin\phi+A_{\mathrm{LU}}^{\sin(2\phi)}\sin(2\phi).
\label{eq:aludev}
\end{equation}
Note the difference between the azimuthal asymmetries (written as calligraphic $\mathcal{A}$) and the Fourier coefficients (written as standard letter $A$), often called \emph{azimuthal asymmetry amplitudes}. The first DVCS beam-helicity asymmetries $\mathcal{A}_{\mathrm{LU}}$ were measured at HERMES and CLAS at JLab in 2001 and published back-to-back in the same journal edition. As shown in Fig.~\ref{fig:dvcsfirst}, the asymmetries exhibit the expected sinusoidal behavior in the azimuthal angle.  

\begin{figure}
\begin{center}
\includegraphics[width=.98\textwidth]{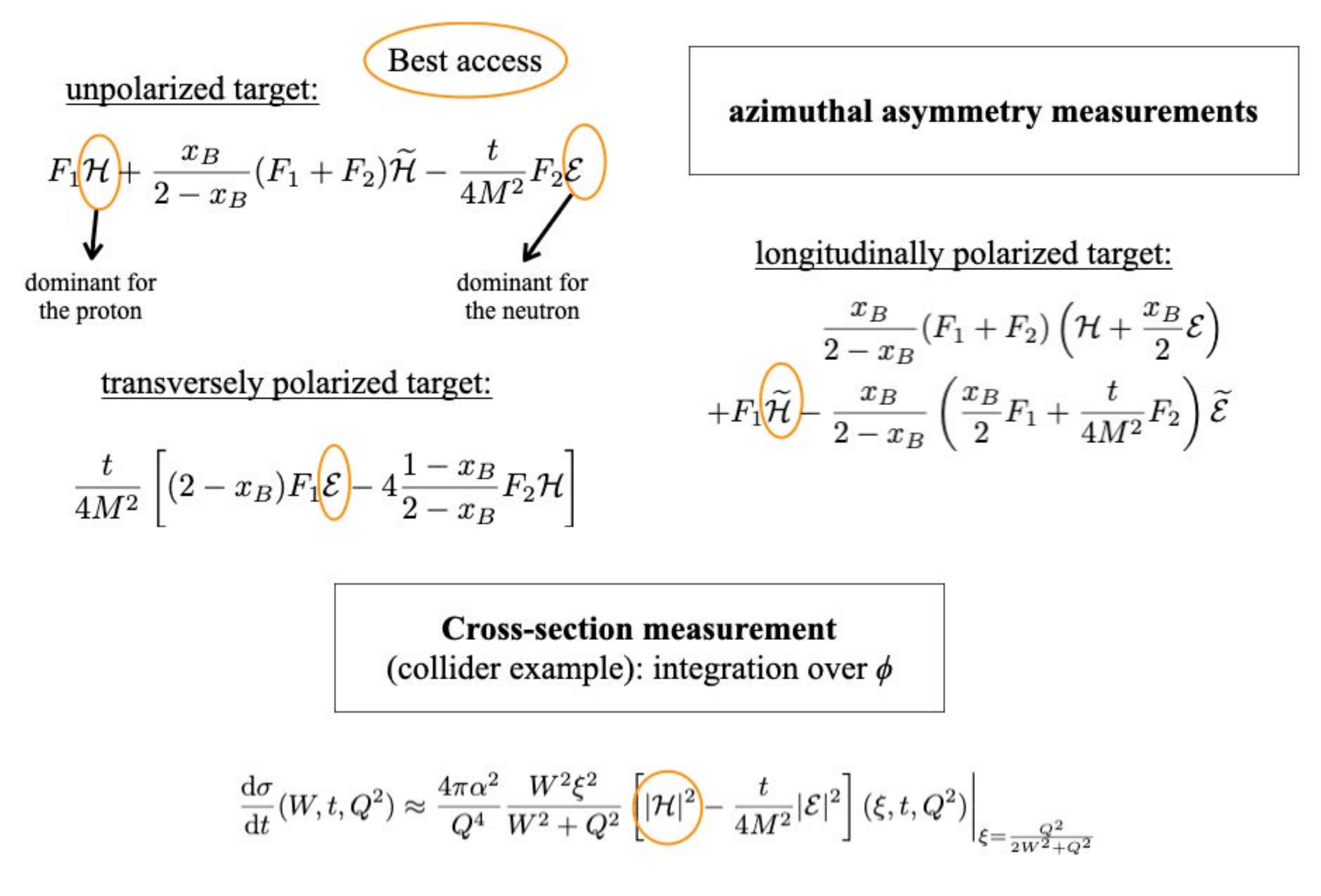}
\end{center}
\caption[GPD access via DVCS]{Different experimental configurations with respect to beam helicity, beam charge, longitudinal or transverse target polarization, and used target nucleon allow access to different combinations of GPDs / CFFs.}
\label{fig:dvcsaccess}
\end{figure}

The asymmetry amplitudes (Fourier coefficients) are extracted from the experimental data typically using maximum likelihood techniques.  
If both beam charges are available, and both beam helicities for either beam charge, additional Fourier coefficients for the beam-helicity asymmetry can be accessed. Harmonic analyses are also performed with respect to beam charge and longitudinal or transverse target polarization. The more experimental configurations are available, the more Fourier coefficients can be extracted from DVCS data. For example, the beam-helicity asymmetry explicitly discussed here provides information about GPD $H$ / CFF $\mathcal{H}$, while a transverse-target spin asymmetry on the proton is sensitive to GPD $E$ / CFF $\mathcal{E}$. An overview is given in Fig.~\ref{fig:dvcsaccess}.

\begin{figure}
\begin{center}
\includegraphics[width=.95\textwidth]{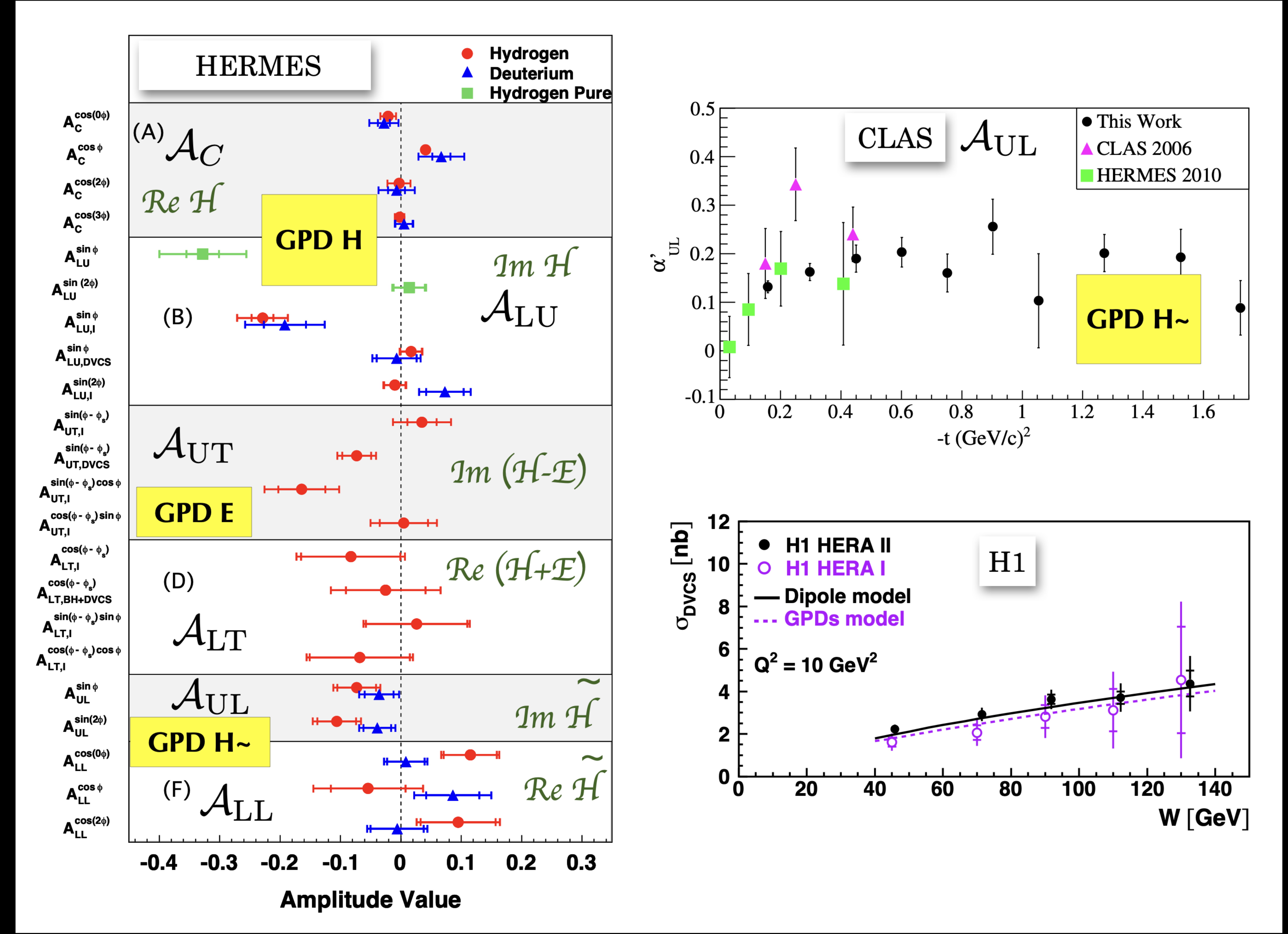}
\end{center}
\caption[DVCS early experimental highlights]{Some DVCS highlights until the year 2012 from HERMES \cite{dvcshermes:unpol} \cite{dvcshermes:unpold} \cite{dvcshermes:rd} \cite{hermesdvcs:aut} \cite{hermesdvcs:aut2} \cite{hermesdvcs:aul} \cite{hermesdvcs:auld} including the sensitivity to various CFFs (left), CLAS \cite{dvcsclas:aul} \cite{dvcsclas:aul2} (right top) and H1 at HERA \cite{dvcsh1} (right bottom). The HERMES and CLAS results are azimuthal asymmetry measurements, while the H1 collider experiment measured the cross section.}
\label{fig:dvcshighlights}
\end{figure}

The beam-helicity asymmetry shown in Fig.~\ref{fig:dvcsfirst} was only the beginning of numerous experimental campaigns to measure DVCS. A multitude of measurements was performed following the year 2001, enabling the start of global GPD analyses (see Sec.~\ref{sec:gpduniversal}). Some of the early experimental highlights are shown in Fig.~\ref{fig:dvcshighlights}. 
Campaigns to measure DVCS are currently pursued at JLab12 providing high-luminosity measurements and COMPASS providing measurements in the kinematic domain between the other fixed-target and the collider experiments.

\begin{figure}
\begin{center}
\includegraphics[width=.98\textwidth]{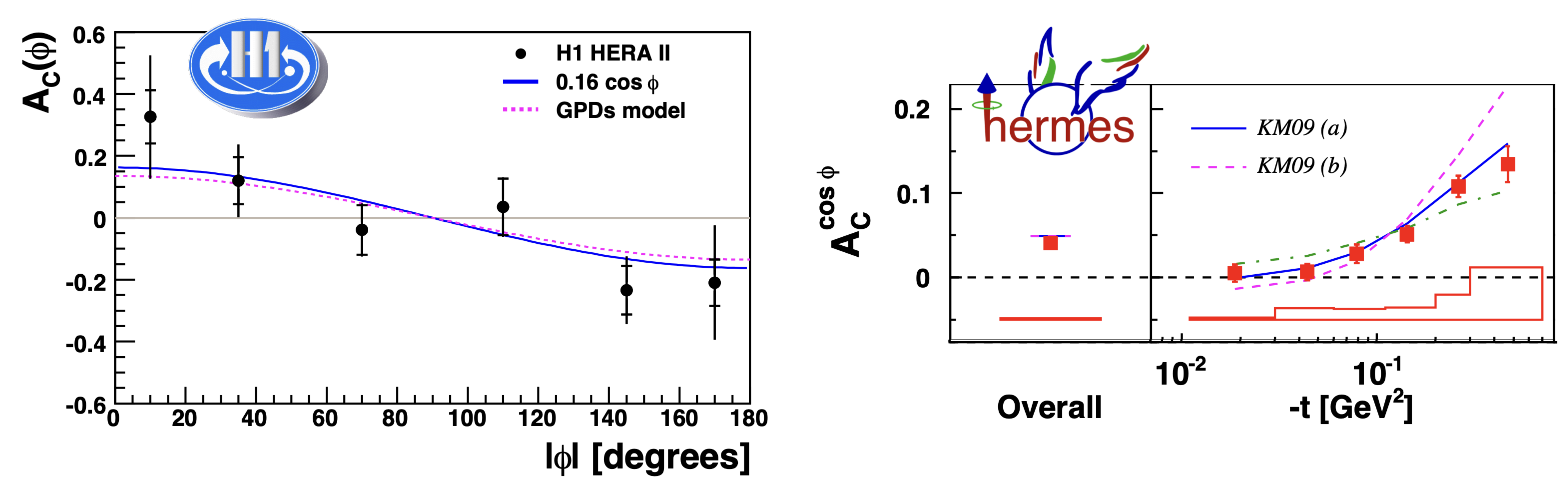}
\end{center}
\caption[DVCS beam-charge asymmetry]{DVCS beam-charge asymmetry $\mathcal{A_{\mathrm{C}}}$ measured at H1 \cite{dvcsh1} (left) and beam-charge asymmetry $\cos\phi$ amplitude $A_{\mathrm{C}}^{\cos\phi}$ measured at HERMES \cite{dvcshermes:unpol} (right).}
\label{fig:dvcsac}
\end{figure}

Of particular interest are the beam-charge asymmetries measured at HERMES and the collider experiment H1. The signs of the measured beam-charge asymmetry $\cos\phi$ amplitudes (see Fig.~\ref{fig:dvcsac}) indicate that the real part of the DVCS scattering amplitude switches sign in the kinematic domain between the two experiments and one can ask the question: \emph{where is the zero crossing?} The gap may be filled by the future COMPASS results at intermediate energies. 
COMPASS studies helicity dependent \textcolor{red}{$\sigma(\rightarrow)-\sigma(\leftarrow)$} or helicity-averaged cross sections \textcolor{blue}{$\sigma(\rightarrow)+\sigma(\leftarrow)$} (often called combined beam-charge and -spin asymmetries). Both in this case and in case of the azimuthal-asymmetry analysis at HERMES and JLab, the respectively \textcolor{red}{imaginary} and \textcolor{blue}{real} parts of the DVCS amplitude are probed independently. The harmonic analysis of DVCS COMPASS data is ongoing. 

\begin{figure}
\begin{center}
\includegraphics[width=.98\textwidth]{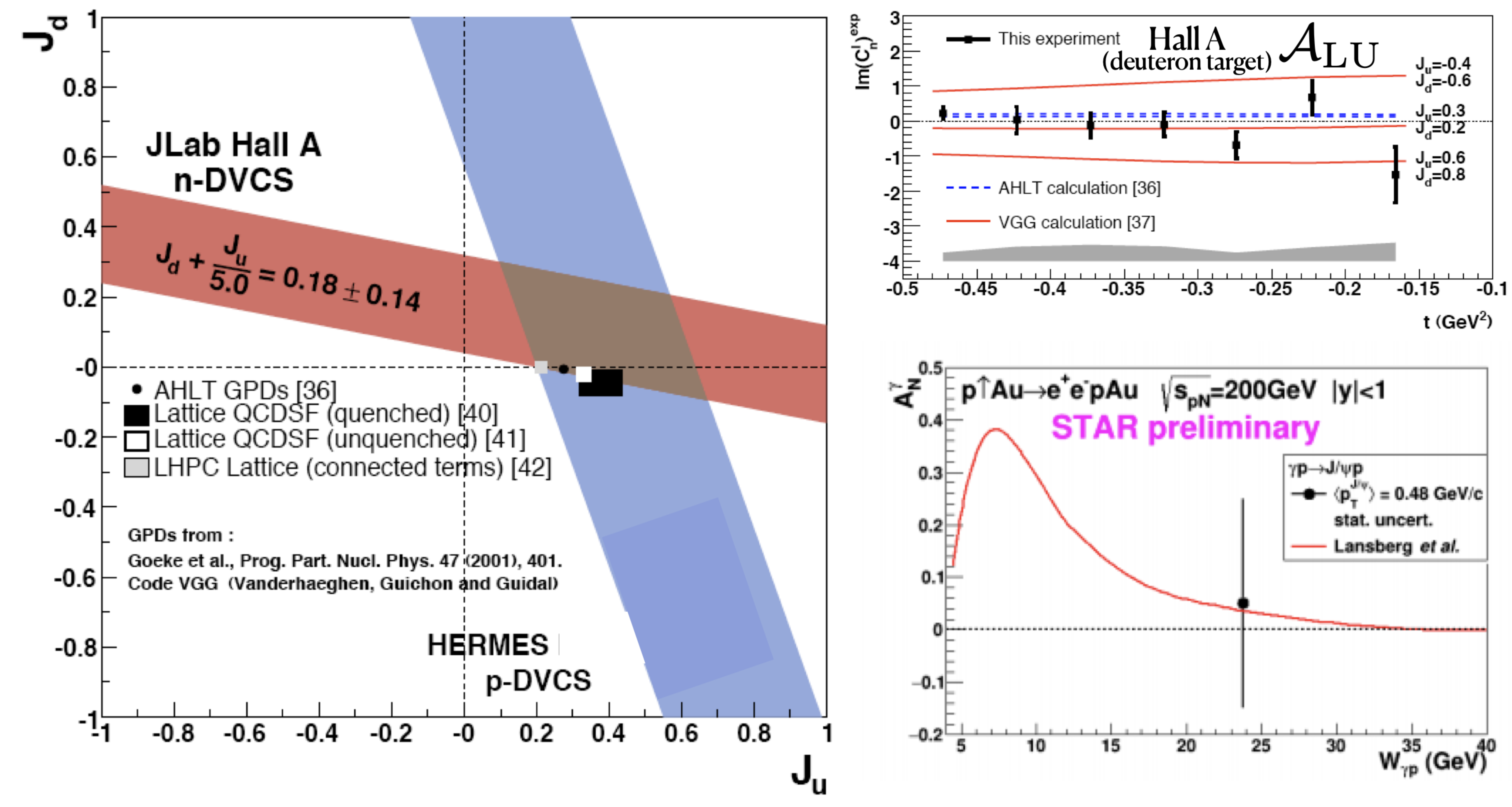}
\end{center}
\caption[GPD $E$ and parton orbital angular momentum]{Right: exclusive measurements sensitive to GPD $E$; top right: DVCS beam-helicity asymmetry on the neutron at JLab Hall A \cite{dvcsclas:neutronalu}, bottom right: transverse target spin asymmetry in exclusive J$/\psi$ production in proton-gold collisions at STAR (BNL) \cite{star:excljpsi}. Left: constraints on the total angular momentum of d-quarks $J_{\text d}$ versus that of u-quarks $J_{\text u}$ from HERMES DVCS data on the transversely polarized proton \cite{hermesdvcs:aut} \cite{hermesdvcs:aut2} and JLab Hall A DVCS on the neutron \cite{dvcsclas:neutronalu}.}
\label{fig:gpde}
\end{figure}

While plenty of experimental information on GPD $H$ is nowadays available, data on GPD $E$ are rather sparse. Measurements sensitive to GPD $E$ allow (in principle) access to the total angular momentum of partons $J_q$ via the \emph{Ji sum rule} for the nucleon \cite{Jisumrule}:
\begin{equation}
J_q=\frac{1}{2}\lim_{t\rightarrow 0}\int_{-1}^{+1}\mathrm{d}x\; x[H^q(x,\xi,t)+E^q(x,\xi,t)].
\label{eq:jisum}
\end{equation}
The experimental access to GPD $E$ requires either transversely polarized protons or a neutron target with longitudinal polarization of the lepton beam. HERMES measured the DVCS transverse target spin asymmetry \cite{hermesdvcs:aut} \cite{hermesdvcs:aut2} and CLAS the DVCS beam-helicity asymmetry on the deuteron \cite{dvcsclas:neutronalu} (which provides effectively a neutron target), shown in the top right of Fig.~\ref{fig:gpde}. Also shown is a constraint from these two data sets on the total angular momentum of quarks in the nucleon. A caveat is in place here: this constraint is very model-dependent. 
New CLAS12 data on the beam-helicity asymmetries on the neutron and a transversely polarized proton target will contribute to our knowledge of GPD $E$. 

All so-far discussed GPDs are \emph{quark} GPDs. For the first time, data sensitive to the GPD $E$ of the \emph{gluon} were collected by STAR in exclusive J$/\psi$ production in ultra-peripheral pAu collisions at RHIC \cite{star:excljpsi}, see Fig.~\ref{fig:gpde}. Significant improvements in precision for that channel are expected with the future STAR data to be taken with the just finished instrumentation upgrades.

\begin{figure}        
\begin{center}
\includegraphics[width=0.75\textwidth]{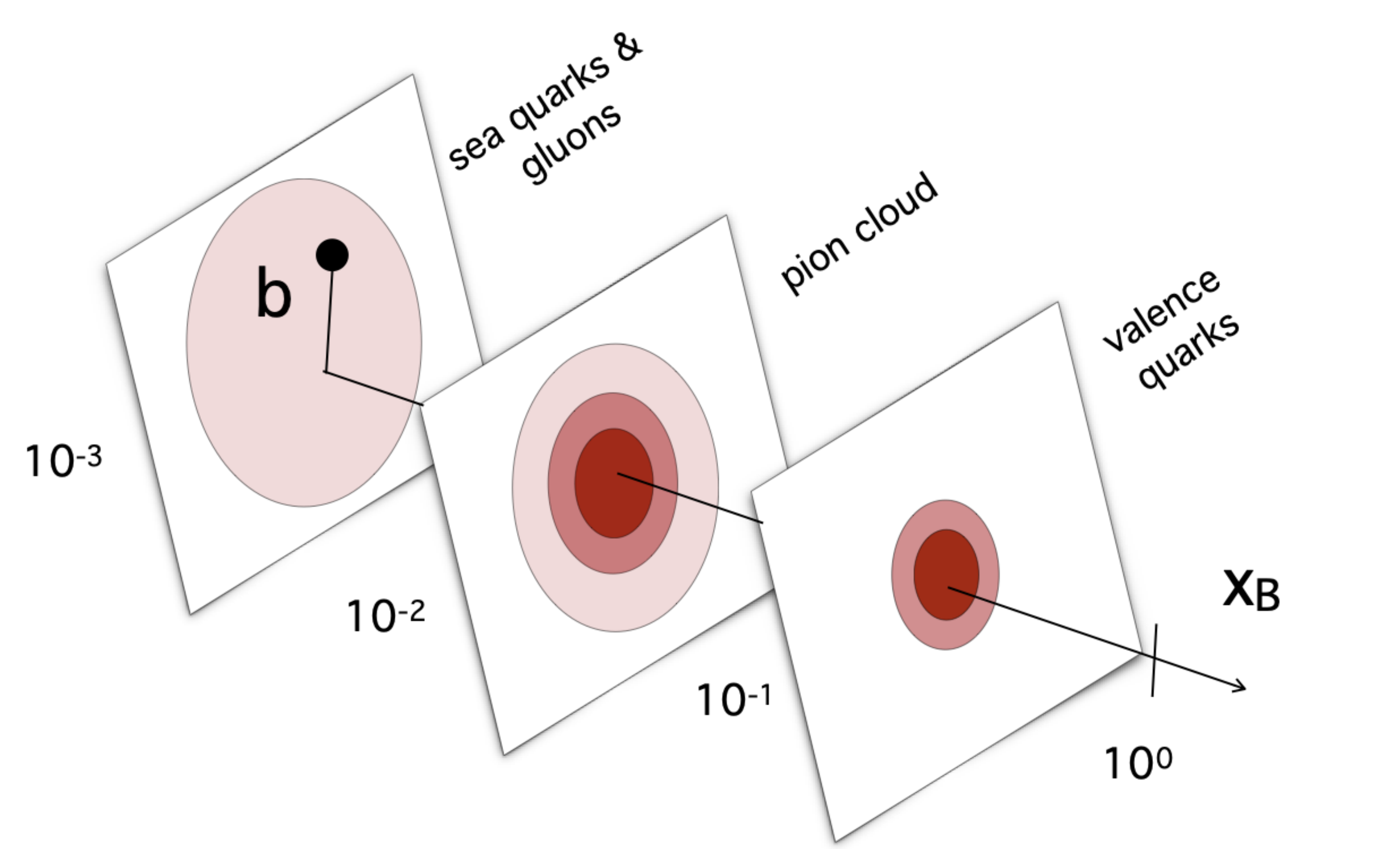}
\end{center}
\caption[Nucleon tomography with GPDs]{Mapping of spatial densities while scanning longitudinal momentum $x_\text{B}$. }
\label{fig:gpdmapping}
\end{figure}

The HERA experiments and COMPASS with the higher center-of-mass energies are able to measure DVCS cross sections without being dominated by the BH contribution. Of the three contributions in Fig.~\ref{fig:dvcsbh}, $\mathcal{T}_\mathrm{BH}$ dominates for small $x_{\text B}$ 
and allows to determine a BH reference yield. The region of medium $x_{\text B}$
is dominated by the DVCS-BH interference term  and is used to perform the harmonic analysis ($\phi$ modulation in the cross section) described earlier. The region of large $x_\text{B}$ 
receives contributions by $|\mathcal{T}_\mathrm{BH}|$ and $|\mathcal{T}_\mathrm{DVCS}|$ of similar size. The pure DVCS yield is analyzed after subtraction of the BH reference yield. 
The DVCS cross section differential in $t$
\begin{equation}
\frac{\mathrm{d}\sigma_{\mathrm{DVCS}}}{\mathrm{d}t}\propto e^{-b|t|}
\label{eq:tslope}
\end{equation}
falls exponentially with $|t|$ following the \emph{t-slope} $b$. It allows to study the \emph{transverse extension of partons in the nucleon} (``transverse imaging''), as becomes apparent by the \emph{impact parameter} ($b_\perp$) representation \cite{burkardt:2003}:
\begin{equation}
q^f(x,b_\perp)=\int\frac{\mathrm{d}^2\Delta_\perp}{(2\pi)^2}\;e^{-b_\perp\Delta_\perp} H^f(x,0,-\Delta_\perp^2),
\label{eq:ipr}
\end{equation}
where $\Delta_\perp$ is the difference of initial and final transverse momenta, and $\Delta_\perp^2$ is related to the Mandelstam $t$. As demonstrated in Fig.~\ref{fig:gpdmapping}, the GPDs allow a \emph{tomography of the nucleon} in transverse position (impact parameter) space. 

\begin{figure}
\begin{center}
\includegraphics[width=0.75\textwidth]{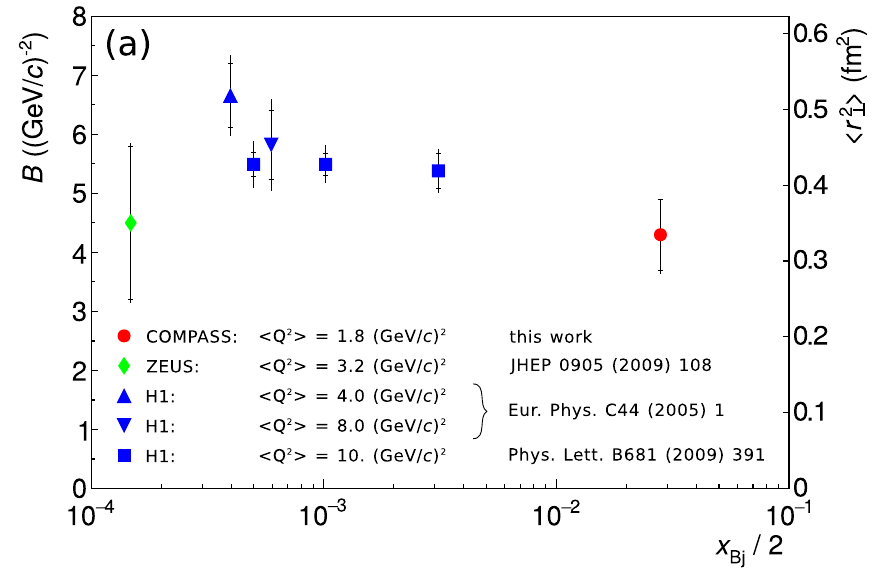}
\end{center}
\caption[$t$-slope from DVCS data]{COMPASS DVCS $t$-slope and comparison to HERA ep-collider (H1 and ZEUS) data \cite{compass:dvcs}. }
\label{fig:compassdvcs}
\end{figure}

The DVCS $t$-slope from the COMPASS differential DVCS cross section 
\cite{compass:dvcs} is shown in Fig.~\ref{fig:compassdvcs} in comparison with HERA collider results at lower $x_\text{B}$.
The measurements allow determination of the transverse extension of partons (``transverse size of the nucleon'') in the $x_\text{B}$ domain between valence quarks and gluons.
COMPASS finds $\sqrt{\langle r_\perp^2\rangle}=(0.58\pm0.04_{\mathrm{stat}}\substack{+0.01\\-0.02}|_{\mathrm{sys}}$ $\pm0.04_{\mathrm{model}})$~fm. More COMPASS DVCS data are being analyzed. 

For the first time, a DVCS beam-spin asymmetry was measured to be larger in the coherent (nucleus) channel than in the incoherent (proton) channel through detecting helium recoils using a radial TPC in CLAS \cite{clas:coherentdvcs}. Coherent DVCS allows studying if the DVCS amplitude rises with atomic mass number and if there is a ``generalized'' EMC effect, i.e, the effect of nuclear modification and shadowing when the nucleon is not free (as proton), but embedded in a nucleus. CLAS12 studies time-like Compton scattering (TCS), which is the time-reversal conjugate process of DVCS and sensitive to the real part of CFF $\mathcal{H}$ \cite{CLAS12:TCS}.

\subsection{Hard exclusive meson production}

\begin{figure}
\begin{center}
\includegraphics[width=0.98\textwidth]{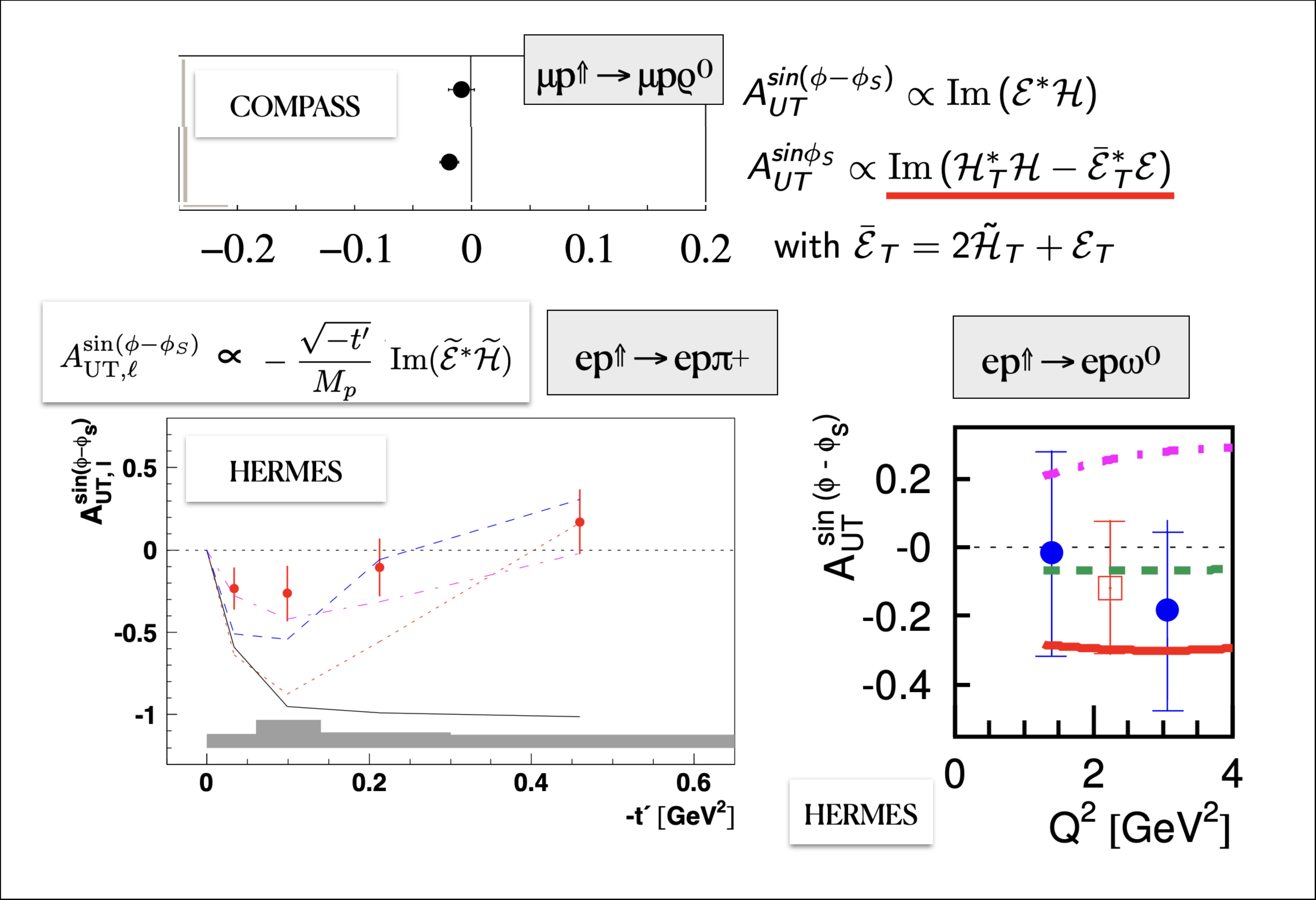}
\end{center}
\caption[Transverse target spin asymmetry in DVMP]{Transverse target spin asymmetries in deeply virtual production of mesons - top: $\rho^0$ at COMPASS \cite{compass:autrho}, bottom: $\pi^+$ \cite{hermes:autpiplus} and $\omega$ \cite{hermes:automega} at HERMES.}
\label{fig:exclaut}
\end{figure}

Selecting either a vector or a pseudo-scalar meson in deeply virtual meson production (DVMP) provides sensitivity to different sets of GPDs, as indicated in Tab.~\ref{tab:gpds}. For example, HERMES measured the transverse target spin asymmetry in exclusive $\pi^+$ production. The result (shown in Fig.~\ref{fig:exclaut}) is sensitive to both GPDs $\widetilde{H}$ and $\widetilde{E}$. Exclusive vector meson production is sensitive to GPD $E$. The results of the $\sin(\phi-\phi_\text{S})$ amplitude of the transverse target spin asymmetries for exclusive $\rho^0$ and $\omega$ production measured at COMPASS and HERMES, respectively, are also shown in Fig.~\ref{fig:exclaut}. Different mesons filter different quark flavors (u, d) and have different sensitivity to gluon (g) GPDs:
\begin{eqnarray}
E^{\rho^0}&=&\frac{1}{\sqrt{2}}\left(\frac{2}{3}E^\text{u}+\frac{1}{3}E^\text{d}+\frac{3}{4}E^\text{g}\right),\\
E^{\omega}&=&\frac{1}{\sqrt{2}}\left(\frac{2}{3}E^\text{u}-\frac{1}{3}E^\text{d}+\frac{3}{4}E^\text{g}\right),\\
E^{\phi}&=&-\frac{1}{3}E^\text{s}+\frac{1}{8}E^\text{g}.
\end{eqnarray}
All so-far discussed GPDs are chiral-even. In recent years, there have been multiple experimental campaigns at the fixed-target experiments to access the chiral-odd GPDs $H_T$ and $E_T$ via DVMP. The $\sin\phi_\text{S}$ component of the transverse target spin asymmetry for $\rho^0$ mesons shown in Fig.~\ref{fig:exclaut} carries sensitivity to GPDs $H_T$ and $E_T$. Further measurements sensitive to chiral-odd GPDs include the cross section for exclusive neutral pions at COMPASS \cite{compass:pizero}. The analyses on beam-spin asymmetries for neutral and positively charged pions at CLAS12 are in progress. 

\begin{figure}
\begin{center}
\includegraphics[width=0.48\textwidth]{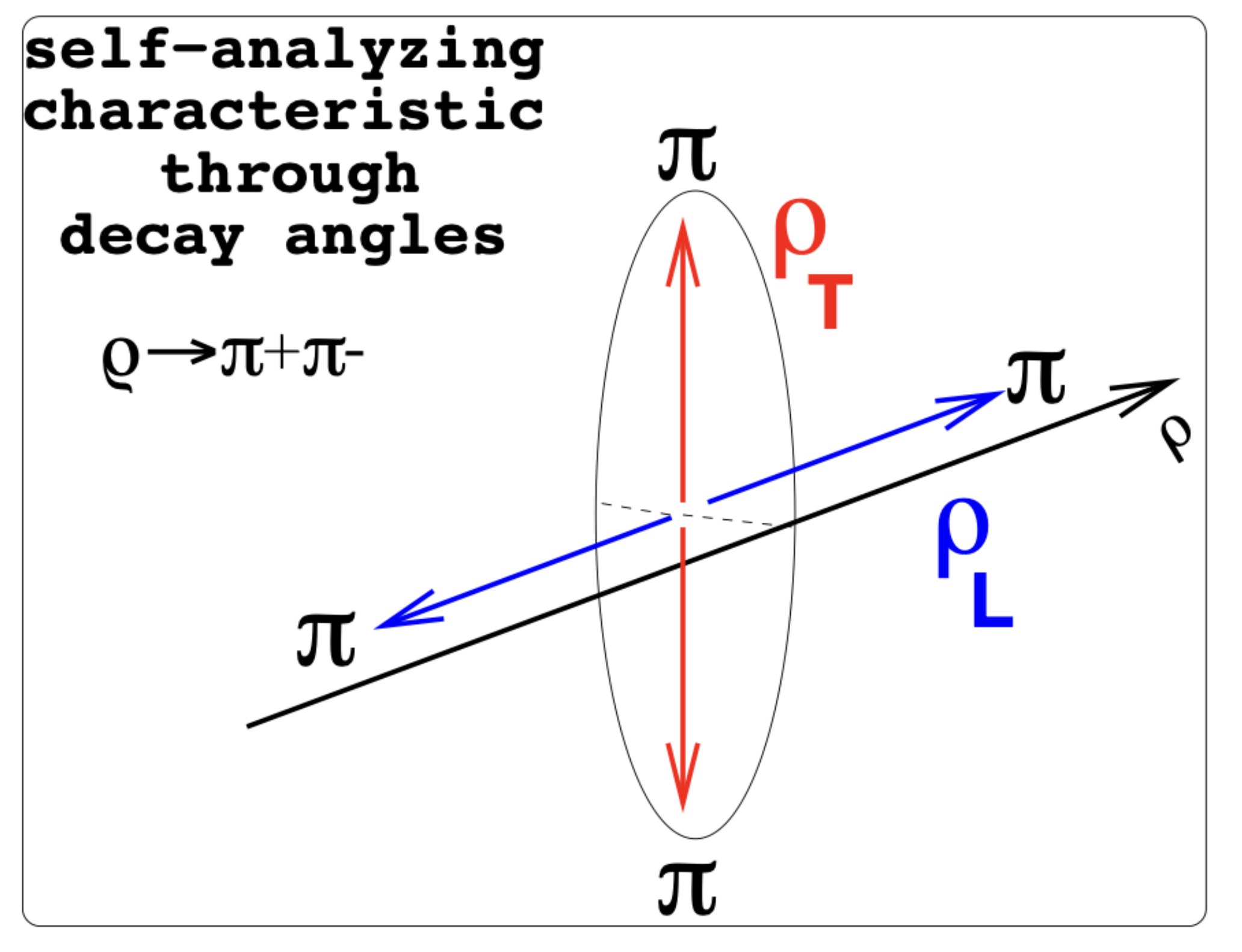}
\includegraphics[width=0.48\textwidth]{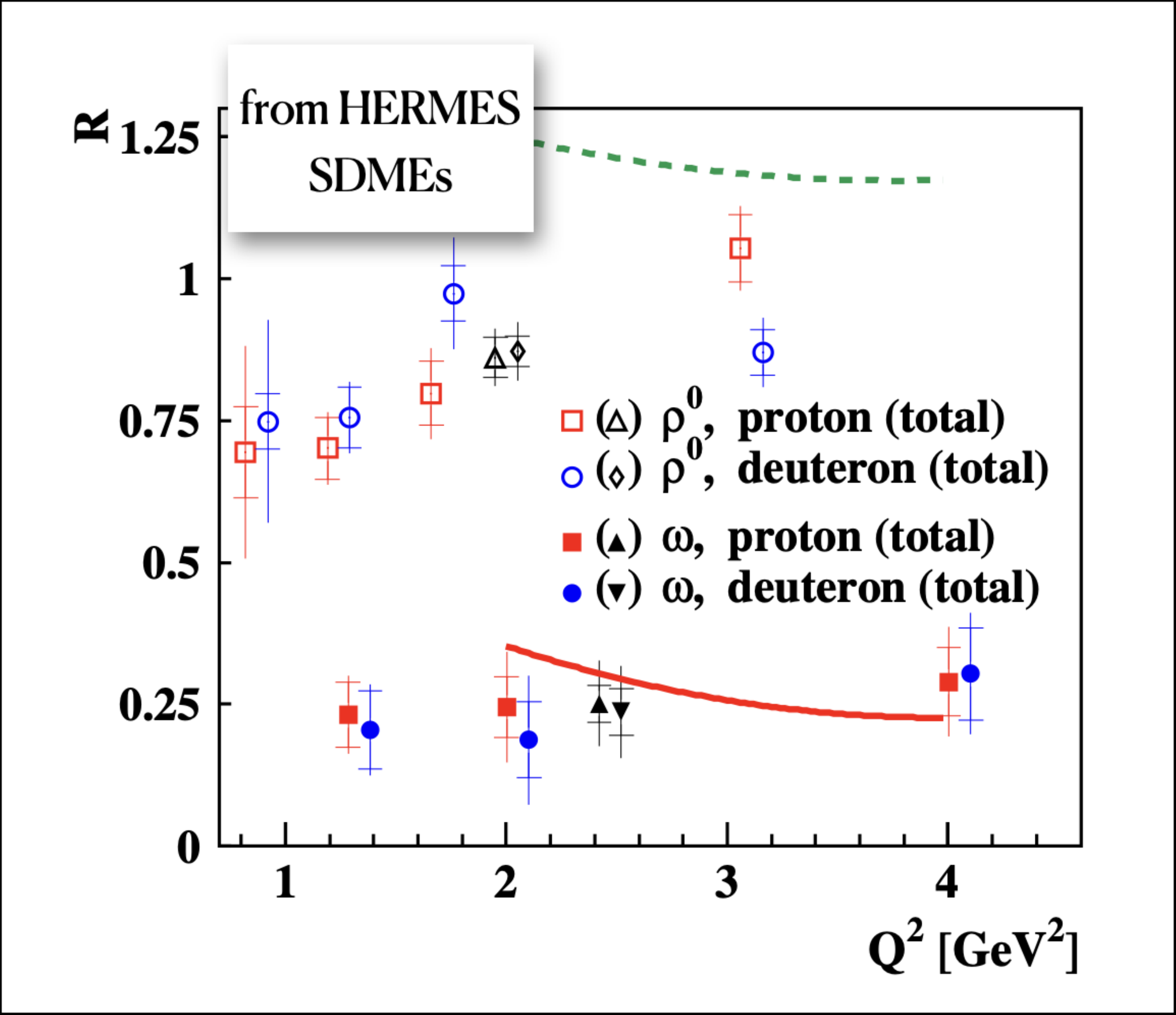}
\includegraphics[width=0.60\textwidth]{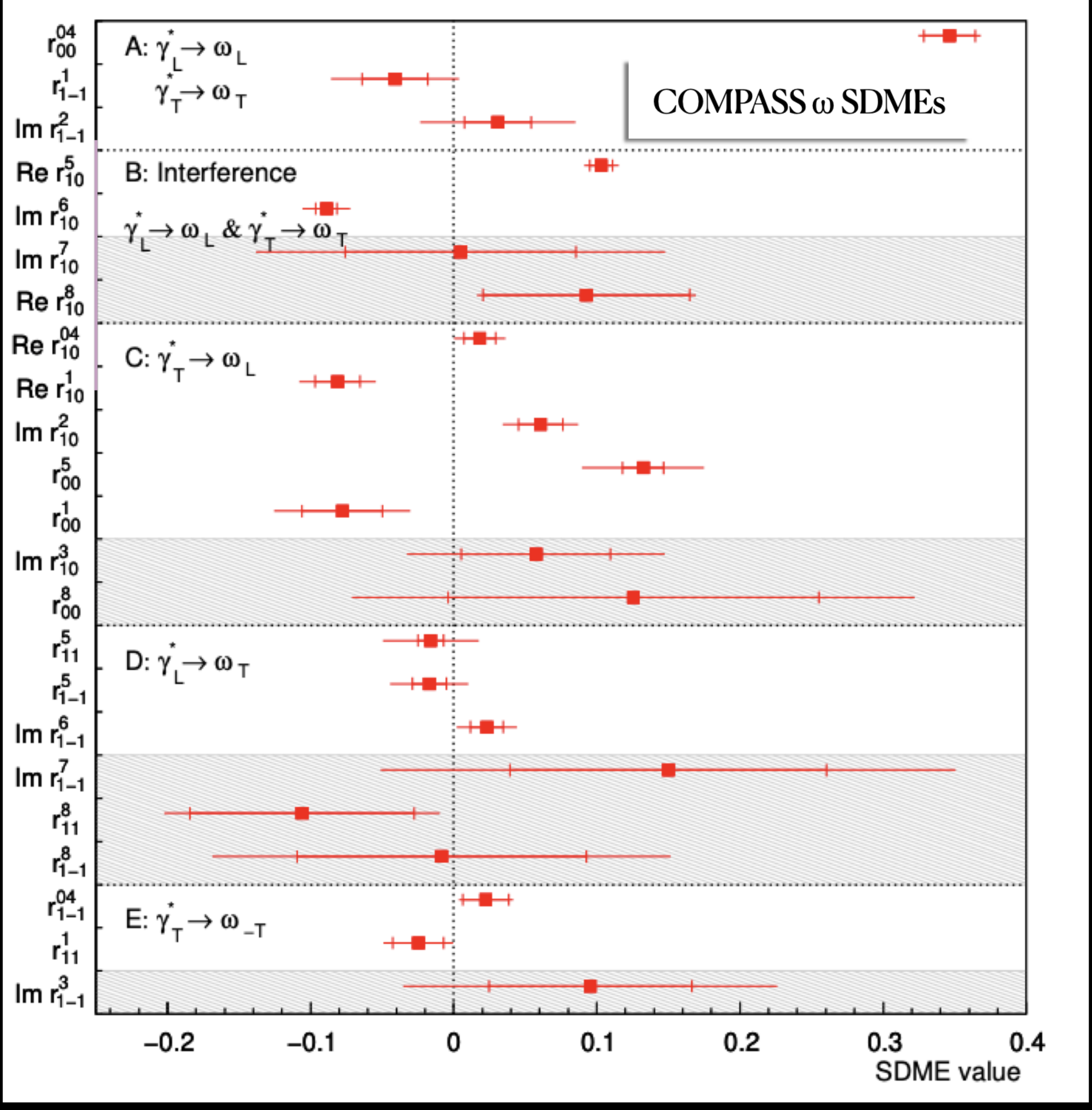}
\end{center}
\caption[Spin density matrix elements]{Top left: decay of vector meson. Top right: ratio $R$ (longitudinal to transverse vector mesons) at HERMES \cite{hermes:sdmes}, in comparison to calculations with (red solid) and without (green dashed) pion pole (Goloskokov-Kroll model \cite{gk:2014}). Bottom: COMPASS SDMEs \cite{compass:omega} describing the spin transfer from the virtual photon $\gamma^*$ to the vector meson (here: $\omega$), both of which can be either longitudinally (L) or transversely (T) polarized. The SDMEs are grouped in five categories A through E.}
\label{fig:sdmes}
\end{figure}

Using unpolarized nuclear targets, HERMES \cite{hermes:sdmes} and COMPASS \cite{compass:omega} determined \emph{spin density matrix elements} (SDMEs), which describe how the spin components of the virtual photon are transferred to the created vector meson. The SDMEs parameterize the function $\mathcal{W}(W,Q^2,t,\Phi,\phi,\cos\Theta)$ appearing in the cross section for vector-meson production. The decay angles $\Phi$, $\phi$, and $ \Theta$ are defined in the references. In the analysis, one studies the angles of the decay products, for example two pions, as shown in Fig.~\ref{fig:sdmes} and deduces the original polarization of the vector meson (longitudinal or transverse) from the decay-angle distributions. One of the SDMEs is related to the cross-section ratio $R$ of longitudinally- to transversely polarized vector mesons. The results from HERMES for $\rho^0$ and $\omega$ mesons for both proton and deuteron targets are shown in Fig.~\ref{fig:sdmes}. The SDMEs measurements provide further constraints on GPD parameterizations beyond the cross-section and spin-asymmetry measurements. They also provide sensitivity to chiral-odd GPDs.

CLAS has recently extracted exclusive $\pi^+$ beam-spin asymmetries  for backward scattering angles \cite{clas:backward}, which allows studying nucleon-to-pion baryonic \emph{transition distribution amplitudes} (TDAs), a further generalization of the GPD concept.

\subsection{GPD universality and advanced extractions}
\label{sec:gpduniversal}

\begin{figure}
\begin{center}
\includegraphics[width=0.98\textwidth]{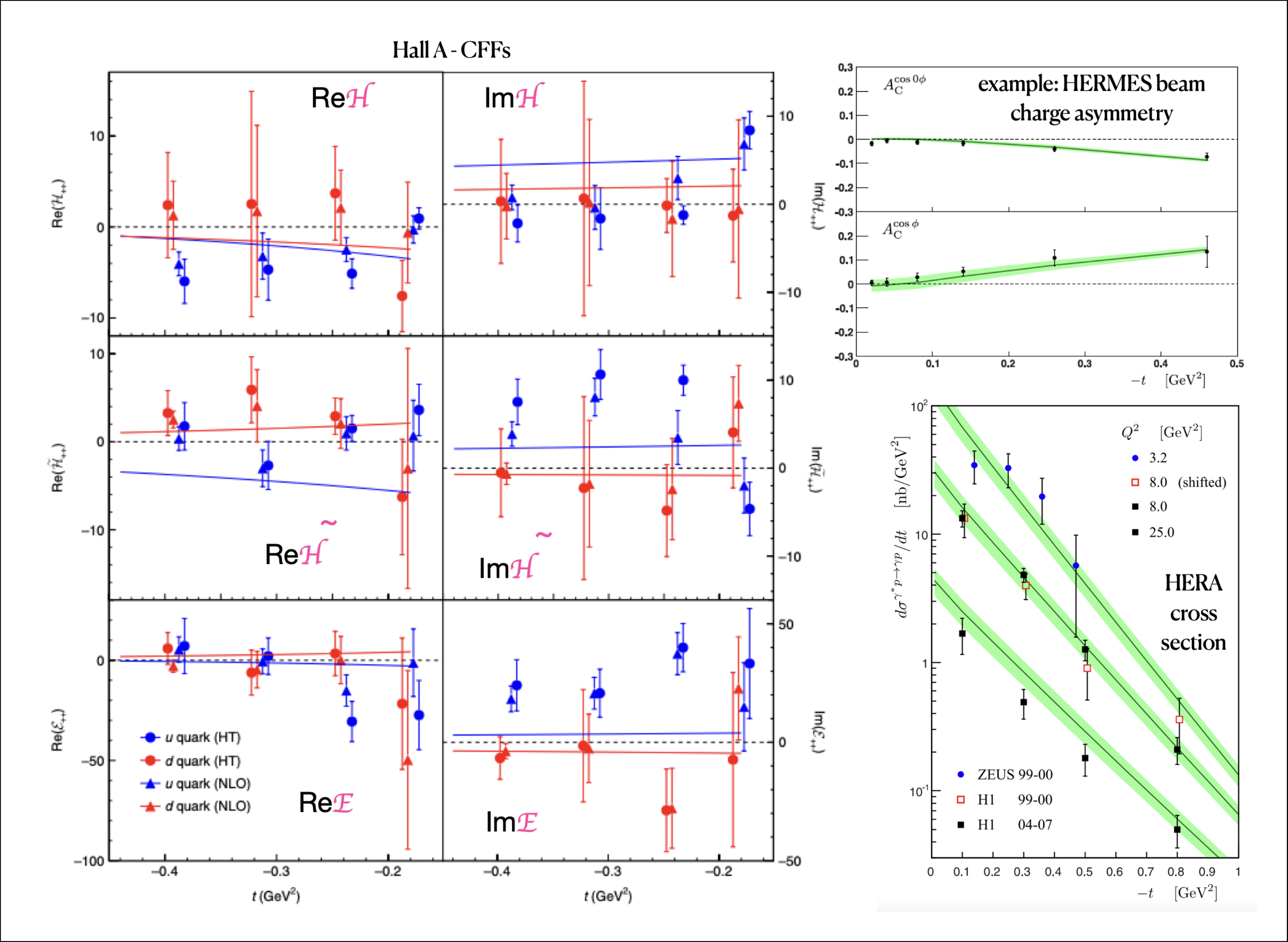}
\end{center}
\caption[Global analysis of GPD data]{Left: JLab Hall A - real and imaginary parts of various Compton form factors for u- and d-quarks \cite{halla:dvcs}. Right: DVCS observables extracted from global DVMP data and comparison to measurement \cite{kms:2013}.}
\label{fig:gpdglobal}
\end{figure}

The global analysis of GPD data and analyses that aim at the extraction of physical quantities have become a very active field. We will look at only a few here. With Jefferson Lab Hall A DVCS on the neutron, a flavor separation of CFFs was performed \cite{halla:dvcs}, see Fig.~\ref{fig:gpdglobal}. The same figure also shows theory predictions for DVCS observables (asymmetries and cross sections) obtained by constraining GPD parameters from global DVMP data. The good comparison demonstrates that GPDs indeed represent universal entities that do not depend on the process they are measured with - DVCS or DVMP. 

\begin{figure}
\begin{center}
\includegraphics[width=0.98\textwidth]{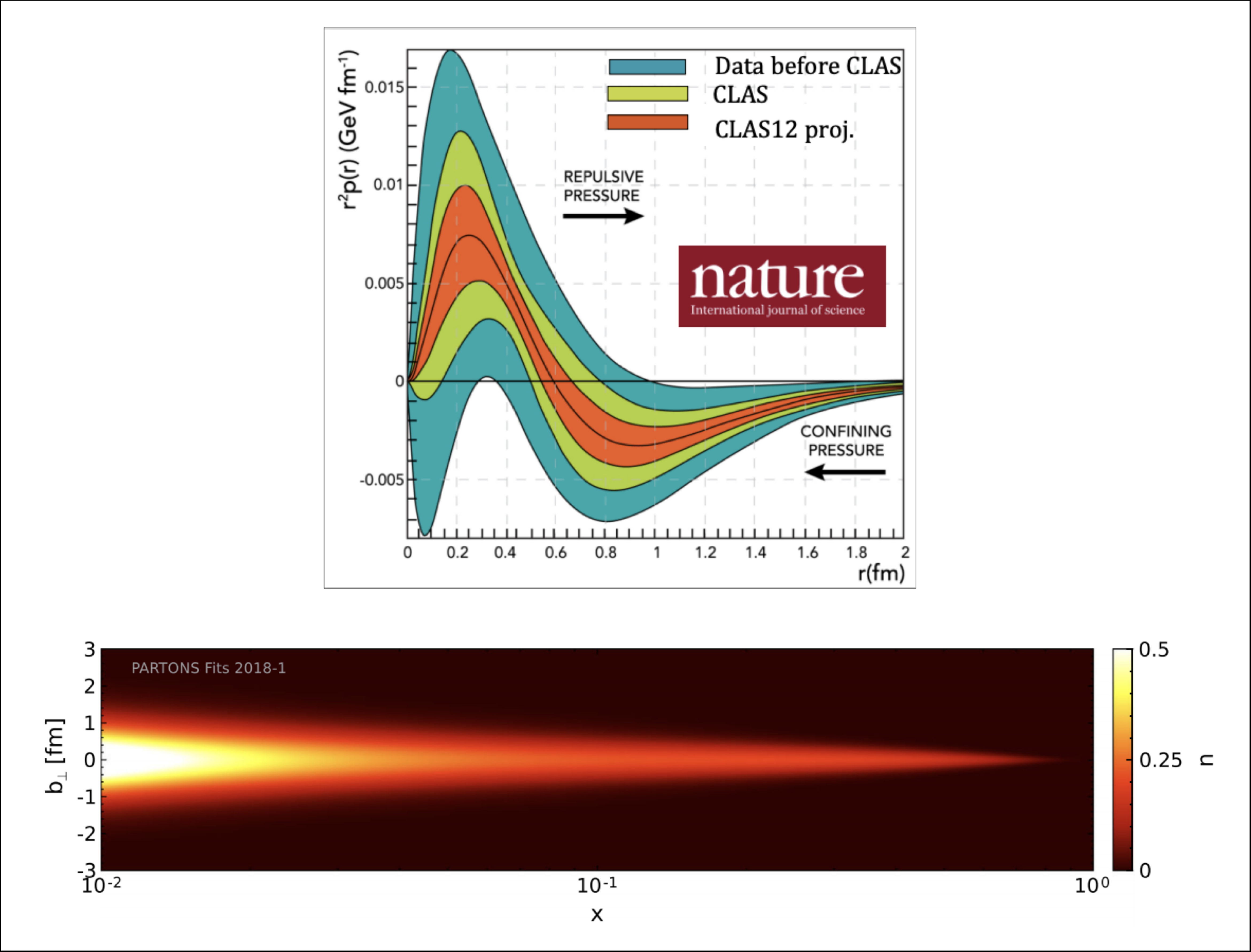}
\end{center}
\caption[Advanced extractions from GPD data]{Top: radial pressure distribution from CLAS DVCS data \cite{clas:shear}. Bottom: position of up quarks from PARTONS fits 2018-1 using world data of elastic form factors and DVCS proton data from HERMES, CLAS, Hall A and COMPASS \cite{msw:2018}.}
\label{fig:gpdadvanced}
\end{figure}

The radial pressure distribution inside the nucleon is shown in Fig.~\ref{fig:gpdadvanced} as extracted from CLAS data. The expected impact from the new CLAS12 data is also indicated. The figure also shows the position of u-quarks in impact-parameter space $(b_\perp, x)$ in the unpolarized nucleon extracted from a global fit to DVCS world data. 

%
\clearpage
\section{Transverse Momentum Dependent functions}
\label{sec:tmds}

\subsection{Nucleon tomography}
\begin{figure}
\begin{center} 
\includegraphics[width=.98\textwidth]{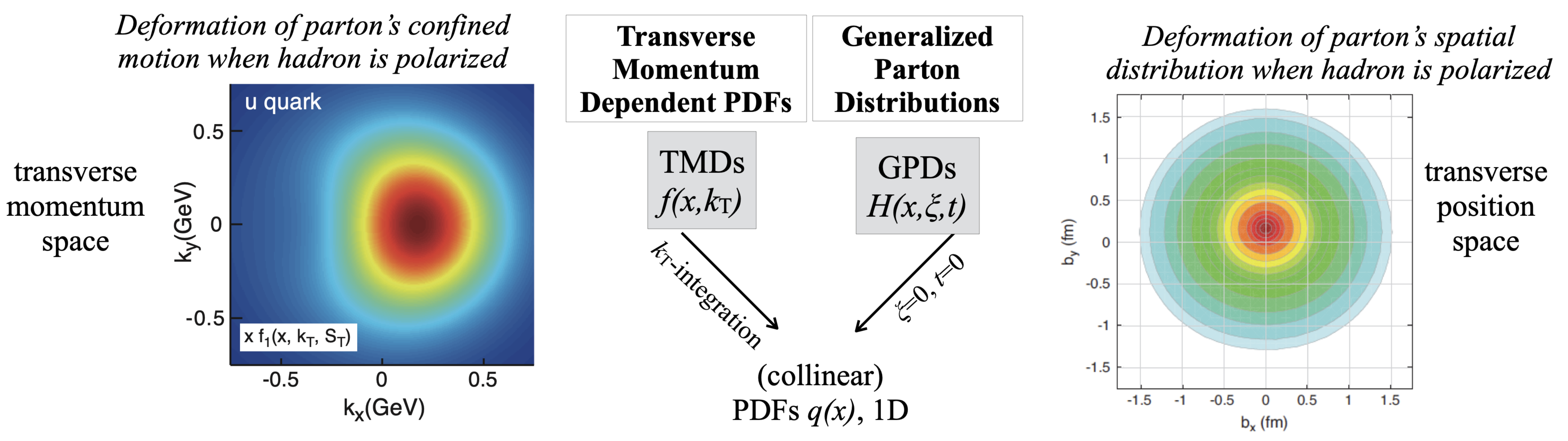}
\end{center}
\caption[Nucleon tomography with TMDs and GPDs]{Nucleon tomography with TMDs in transverse momentum space (left \cite{eic:white} based on \cite{anselmino:2011}) and with GPDs in transverse position space (right \cite{GPDsEIC}). Both frameworks contain the collinear PDFs as special cases.}
\label{fig:tomography}
\end{figure}

Both transverse momentum dependent PDFs (TMDs) and generalized parton distributions (GPDs) explore the proton beyond the collinear approximation, which considers the partons to move ``very fast'' in the longitudinal direction in absence of transverse parton momenta. Including transverse parton position in case of the GPDs and transverse parton momentum in case of the TMDs allows for the first time to model a multi-dimensional picture of the proton, often referred to as ``nucleon tomography'' (see Fig.~\ref{fig:tomography}). 
Because of the mutually perpendicular momentum- and position vectors involved, this nucleon picture includes orbital angular momenta via the non-zero cross product $\vec{r}\times\vec{p}$.
The TMDs assess the question, \emph{How is the proton spin correlated with the motion of quarks and gluons?}, encoding the deformation of the parton's distribution of {\bf transverse momentum} when the proton is polarized. On the other side, with GPDs it is investigated, \emph{How does the proton spin influence the spatial distribution of partons?}, encoding the deformation of the parton's {\bf spatial} distribution when the proton is polarized. 

The TMDs and GPDs are not unconnected: for example is the chiral-odd GPD $H_T$ related to the transversity TMD. Both TMDs and GPDs hold the collinear PDFs as special cases: the TMDs by integrating over parton transverse momentum, and the GPDs in a certain kinematic (forward) limit.

\subsection{Transverse momentum dependent PDFs}
\begin{figure}
\begin{center}
\includegraphics[width=0.98\textwidth]{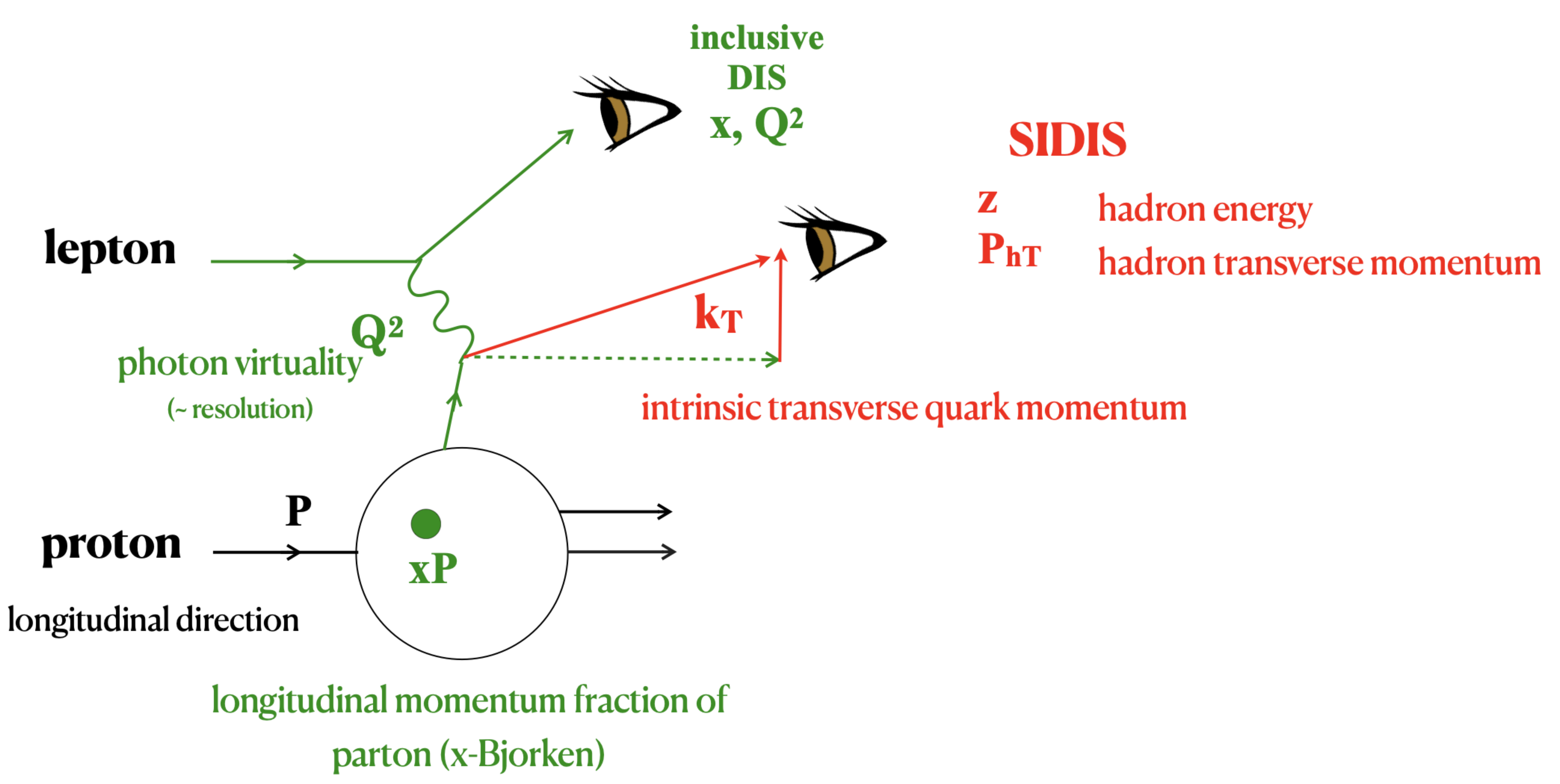}
\end{center}
\caption[SIDIS with transverse momentum]{Semi-inclusive deep-inelastic scattering (SIDIS) with measurement of the transverse momentum of a hadron from the fragmentation process.
}
\label{fig:sidis_withpt}
\end{figure}

We have seen in  Sec.~\ref{sec:pdfs} that the DIS ($\ell \text{N}\rightarrow\ell^\prime \text{X}$) cross section can be expressed in terms of non-perturbative parton distribution functions (PDFs), which encode information about the moment\-um-depen\-dent distribution of quarks and gluons inside the proton.
If one or more hadrons are detected (semi-inclusive DIS or SIDIS),  $\ell \text{N}\rightarrow\ell^\prime \text{hX}$ or $\ell \text{N}\rightarrow\ell^\prime \text{h}_1\text{h}_2\text{X}$, in addition to the PDFs also fragmentation functions (FF) enter the cross section, as we have learned in Sec.~\ref{sec:deltaq}. The \emph{intrinsic transverse momentum of partons} in the proton, $k_\text{T}$, can be accessed indirectly by measuring the \emph{transverse momentum} $P_\text{hT}$ of one or more hadrons emerging from the fragmentation process, as demonstrated in Fig.~\ref{fig:sidis_withpt}. 
In fact, additional information is gained by measuring the transverse momentum of one or more of the produced hadrons. In  Sec.~\ref{sec:deltaq}, we did not take into account the $k_\text{T}$ or $P_\text{hT}$ explicitly (by integrating over it) and we did not consider semi-inclusive DIS (SIDIS) with transversely polarized targets. From now on, we will do both. We will study the QCD-based framework built on TMD PDFs 
- a framework that has emerged from intense experimental and theoretical efforts over the past almost three decades. For an exhaustive recent experimental and theoretical TMD overview the reader is referred to Ref.~\cite{Vossen2020}.

\begin{figure}
\begin{center}
\includegraphics[width=0.98\textwidth]{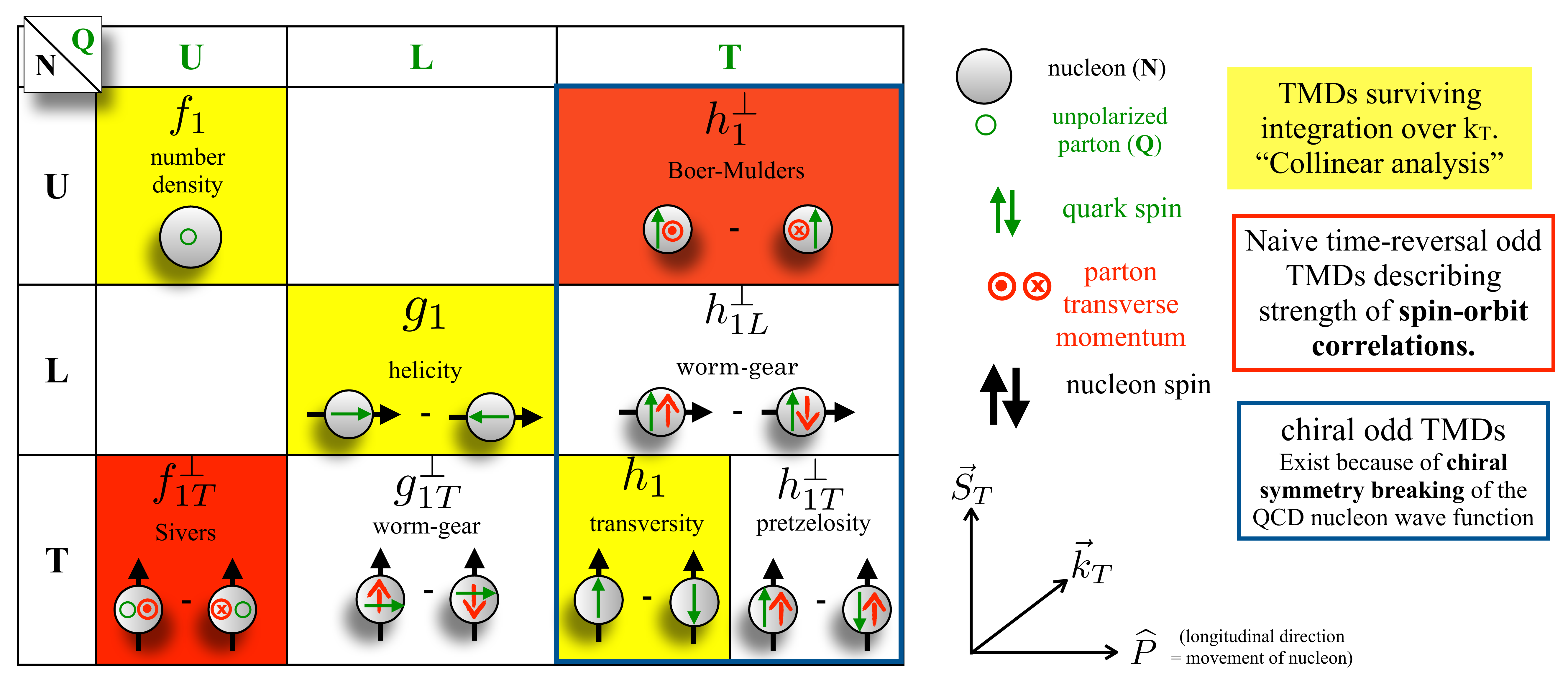}
\end{center}
\caption[TMD PDFs]{TMD PDFs at leading twist for unpolarized (U), longitudinally (L) and transversely (T) polarized partons (Q) and nucleons (N). The transverse-target polarization vector is labeled $\vec{S}_\text{T}$, the parton transverse-momentum vector $\vec{k}_\text{T}$, and the momentum unit vector in the longitudinal direction $\hat{P}$.
}
\label{fig:tmdtable}
\end{figure}
\begin{figure}
\begin{center}
\includegraphics[width=0.8\textwidth]{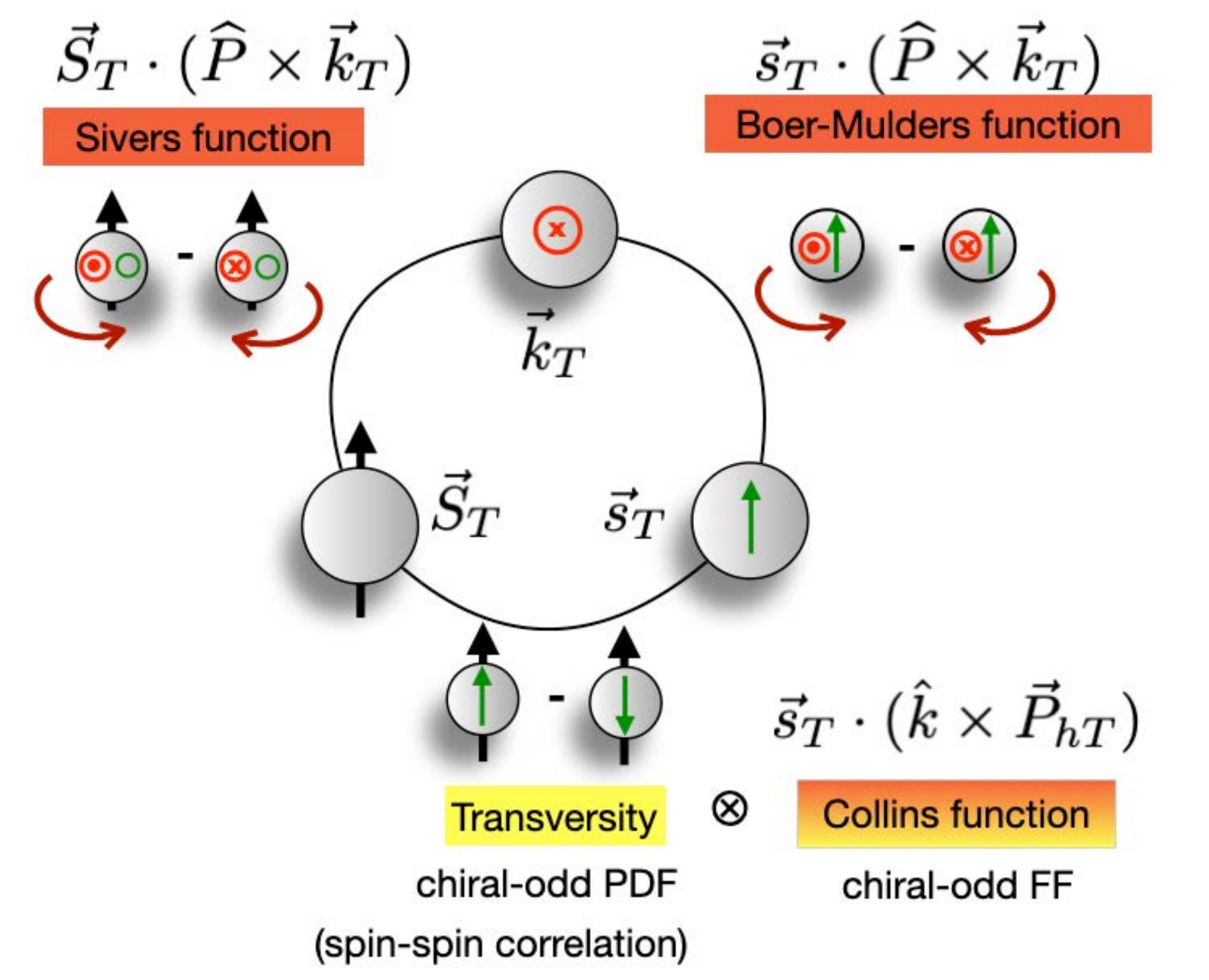}
\end{center}
\caption[$T$-odd TMDs]{Naive time-reversal odd TMDs and the spin-orbit correlations they describe: $\vec{S}_\text{T}$ is the transverse proton polarization, $\vec{s}_\text{T}$ the transverse parton polarization, and $\vec{k}_\text{T}$ the transverse parton momentum.}
\label{fig:spin-orbit-correlations}
\end{figure}

In the SIDIS cross section, each TMD PDF is convoluted with a fragmentation function (FF) encoding the fragmentation of a quark into a final-state hadron, and appears with a specific azimuthal modulation, for example,
\begin{equation}
\sin(\phi-\phi_{\mathrm{S}})\cdot (\mathrm{Sivers}\;\mathrm{TMD} \;\mathrm{PDF}) \otimes \mathrm{(FF)}.
\label{eq:sidisconvolution}
\end{equation}
Figure~\ref{fig:tmdtable} presents the set of TMD PDFs at leading twist organized by quark polarization (columns) and proton polarization (rows). The analog table for the FFs (capital letters except for UU=D$_1$) is not shown. Note that longitudinal proton direction $\hat{P}$, transverse parton momentum $\vec{k}_\text{T}$, and transverse proton polarization $\vec{S}_\text{T}$ are mutually perpendicular. The parton spin vector is labeled $\vec{s}_\text{T}$ in the following.
\begin{description}
\item The \emph{collinear TMDs} (yellow in Fig.~\ref{fig:tmdtable}) do not vanish when integrated over the intrinisc transverse parton momentum $k_T$. In Sec.~\ref{sec:dis}, we got to know the spin-independent PDF (here labeled $f_1$) and the spin-structure function $g_1$ for longitudinally polarized protons. If the proton is transversely polarized, there is a new PDF appearing in the cross section - the \emph{transversity TMD} $h_1$, which reflects the quark transverse polarization in a transversely polarized nucleon and is related to the \emph{tensor charge of the nucleon}. All other TMD PDFs ``do not survive'' integration over $k_T$.
\item The \emph{naive time-reversal ($T$-)odd}\footnote{
Naive time reversal is the
time-reversal operation without interchange of initial and final states, i.e., reversal of momentum and spin vectors only.} \emph{TMDs} (red in Fig.~\ref{fig:tmdtable}) describe the strength of \emph{spin-orbit correlations} in the proton, as summarized in Fig.~\ref{fig:spin-orbit-correlations} and in the following:
\begin{itemize}
\item The existence of the naive $T$-odd Sivers TMD $f_{1T}^\perp$ is known as  the \emph{Sivers effect}, which describes the correlation between the nucleon transverse-spin direction and the parton transverse momentum in the transversely polarized nucleon, relating the motion of unpolarized quarks and gluons to the nucleon spin:\\ 
{\bf Sivers~$\sim \vec{S}_\text{T}\cdot(\hat{P}\times\vec{k}_\text{T})$.}
\item The \emph{Boer-Mulders} (BM) TMD PDF $h_1^\perp$  describes the correlation between parton transverse spin and parton transverse momentum: {\bf BM$\sim \vec{s}_\text{T}\cdot(\hat{P}\times\vec{k}_\text{T})$.}
\item The existence of a naive $T$-odd (and chiral-odd) Collins FF $H_1^\perp$ is known as the \emph{Collins effect}, which describes the fragmentation of a transversely polarized parton into a final-state hadron: \\
{\bf Collins~$\sim \vec{s}_\text{T}\cdot(\hat{k}\times\vec{P}_\text{hT})$.}
\end{itemize}
\item The \emph{chiral-odd TMDs} (boxed blue in Fig.~\ref{fig:tmdtable}) can only be measured in conjunction with another chiral-odd object (to fulfill the requirement that the terms appearing in the cross section are chiral-even). They cannot be measured in inclusive DIS, where such matching chiral-odd partner does not exist. In SIDIS, they can be partnered with a chiral-odd fragmentation function, allowing them to be accessed.
\begin{itemize}
\item The \emph{transversity TMD} is chiral-odd and can be accessed in SIDIS because it is convoluted with the chiral-odd Collins FF:\\ transversity-TMD~$\otimes$~Collins-FF
\item Similarly are the other chiral-odd TMD PDFs \emph{Boer-Mulders, worm-gear (Kotzinian-Mulders), and pretzelosity} convoluted with the chiral-odd Collins FF in the SIDIS cross section.
\end{itemize}
\end{description}

What if we measure the observables related to TMDs describing spin-orbit correlations to be non-zero? This would indicate that partons possess orbital angular momentum (OAM). There is no quantitative relation between TMDs and OAM available yet.

\subsection{SIDIS with transverse degrees of freedom}

We look again at Fig.~\ref{fig:sidis_withpt}: in SIDIS, we measure the transverse momentum of one or more of the produced hadrons. The cross section contains convolutions of TMD PDFs and FFs (Eq.~\ref{eq:sidisconvolution}) together with a harmonic modulation in some azimuthal angle. Experimentally, the magnitude of a given convolution is at fixed-target experiments usually determined by measuring an azimuth-dependent, transverse \emph{single-spin asymmetry} in suitably normalized count rates $N$ for one proton-spin orientation ($\uparrow$) versus that for the other ($\downarrow$),
\begin{equation}
A_{\mathrm{T}}(\varphi)=\frac{1}{S_{\mathrm{T}}}\cdot\frac{N^\uparrow(\varphi)-N^\downarrow(\varphi)}{N^\uparrow(\varphi)+N^\downarrow(\varphi)},  
\label{eq:asy}
\end{equation}
with $S_{\mathrm{T}}$ the transverse proton polarization and $\varphi$ some azimuthal angle (or linear combination of azimuthal angles). Equation~\ref{eq:asy} is more complex in reality and was simplified here to illustrate a typical transverse spin observable. The full expression of the SIDIS cross section with various TMD-FF convolutions can be found in Ref.~\cite{bacchetta:2007} in a generic form containing contributions from unpolarized, longitudinally polarized, and transversely polarized protons. 
The analysis of azimuthal asymmetries to access harmonic amplitudes appearing in the SIDIS cross section proceeds similarly to the procedure described for GPDs in Sec.~\ref{sec:expdvcs}. Figure~\ref{fig:dvcsangles} can be used to define the angles $\phi$ and $\phi_\text{S}$ in SIDIS if the outgoing real DVCS photon is replaced by the outgoing final-state hadron(s).
While fixed-target experimenters often choose to extract the amplitudes of the various harmonic modulations using a simultaneous maximum-likelihood fit to the data, the typical collider spin observable is $A_\mathrm{N}$, a left-right asymmetry of normalized counts with respect to the proton-spin orientation. The two observables are related. 

\subsection{Sivers TMD}
\label{sec:sivers}

\begin{figure}
\begin{center}
\includegraphics[width=0.9\textwidth]{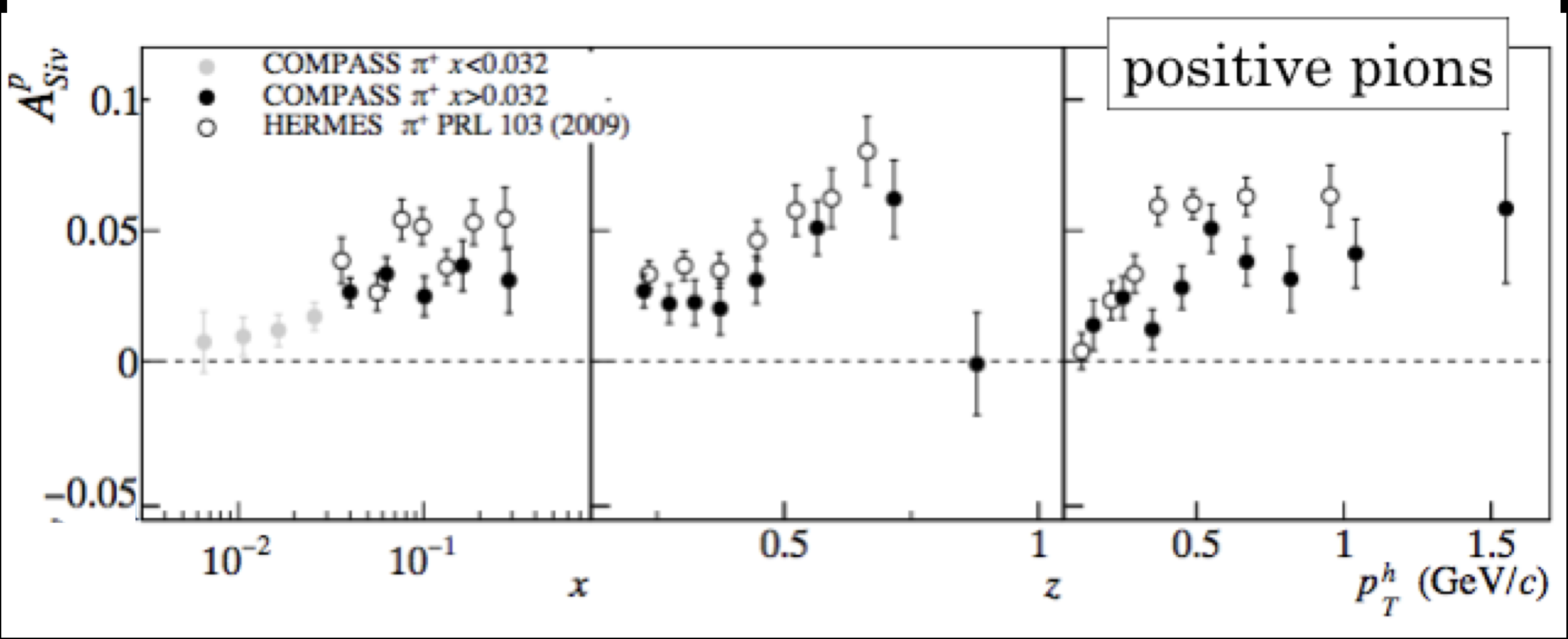}
\end{center}
\caption[Sivers SIDIS asymmetries at HERMES and COMPASS]{Sivers $\sin(\phi-\phi_\text{S})$ asymmetry amplitudes (here called $A^\text{p}_{\mathrm{Siv}}$) from HERMES \cite{hermes:sivers} and COMPASS \cite{compass:siverssidis} SIDIS data for positively charged pions.}
\label{fig:sivers-compass-hermes}
\end{figure}

The Sivers TMD PDF, often simply called Sivers function, was originally thought to vanish \cite{Collins:1993}. About 20 years ago, a non-zero single-spin asymmetry and thus Sivers function was shown to be allowed due to QCD final-state interactions (soft gluon exchange) in SIDIS between the outgoing quark and the target remnant \cite{Brodsky:2002}, see also Fig.~\ref{fig:DYvsSIDIS}. Indeed, significant non-zero Sivers asymmetries were measured in SIDIS for the first time by HERMES  \cite{hermes:sivers}  in 2004 and by COMPASS \cite{compass:siverssidis} in 2010 on transversely polarized protons. As shown in Fig.~\ref{fig:sivers-compass-hermes}, the Sivers signal for positively charged pions produced in SIDIS (``SIDIS $\pi^+$'') is measured to be smaller at COMPASS than at HERMES with its lower lepton-beam energy (HERMES: 27.6\,GeV, COMPASS: 160\,GeV), which may be interpreted as an effect of \emph{TMD evolution} - which is different to the standard $Q^2$ evolution of QCD (Eq.~\ref{eq:dglap}). 
The theoretical formalism tells us that the
\begin{equation}
\mathrm{SIDIS}\;\pi^+ \;\mathrm{Sivers}\;\mathrm{amplitude} \simeq-\textcolor{red}{f_{1T}^\perp(x,P_\text{hT}^2)}\otimes \textcolor{blue}{D_1^{\text{u}\rightarrow\pi^+}(z,k_\text{T}^2)}, 
\label{eq:sidissiverspiplus}
\end{equation}
where the red part is the \textcolor{red}{Sivers TMD PDF} and the blue part the spin-independent \textcolor{blue}{fragmentation function (FF)} for a u-quark fragmenting into a $\pi^+$ (see also Sec.~\ref{sec:deltaq} - and $x$ is again $x$-Bjorken). Since the Sivers asymmetry amplitude is measured to be positive, we can conclude that the Sivers function for the u-quark must be negative (note the minus sign in Eq.~\ref{eq:sidissiverspiplus}). 

\begin{figure}
\begin{center}
\includegraphics[width=0.8\textwidth]{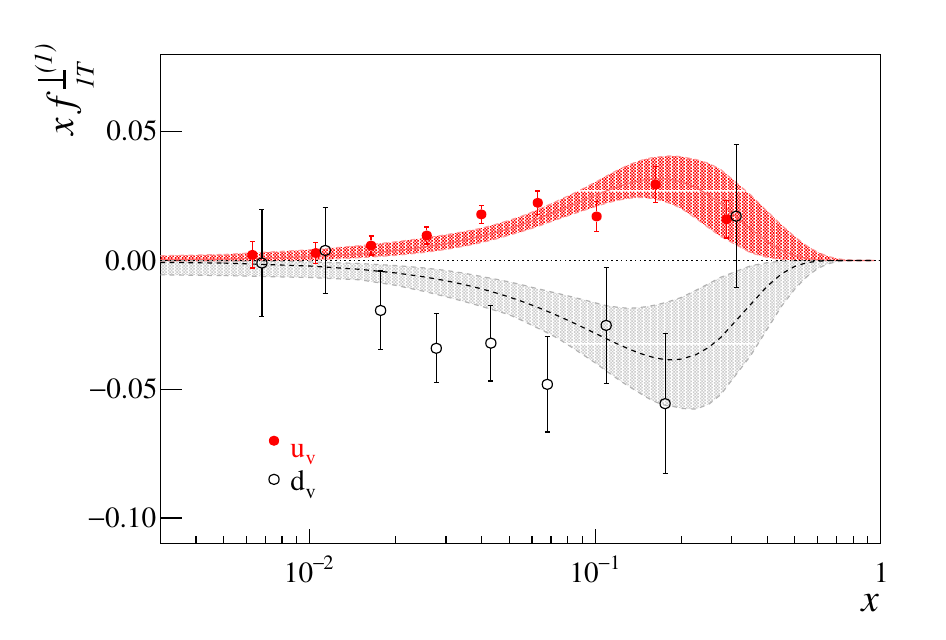}
\end{center}
\caption[Sivers moments for SIDIS u- and d-quarks]{First Sivers moments for valence u- and d-quarks from COMPASS $P_\text{hT}$-weighted asymmetries in SIDIS \cite{compass:ptweighted}, together with a fit to SIDIS world data \cite{anselmino:2012}.}
\label{fig:sivers_ptweighted}
\end{figure}

The Sivers amplitudes for SIDIS $\pi^-$ off a transversely polarized proton, and for SIDIS $\pi^+$ off a transversely polarized deuteron target \cite{compass:siversdeuteron} are measured to be compatible with zero, hinting to cancellation effects between u- and d-quarks. Indeed are the Sivers functions for u- and d-quarks known to have different signs. This flavor dependence of the Sivers function is demonstrated for example in a recent COMPASS publication for SIDIS data \cite{compass:ptweighted}, see Fig.~\ref{fig:sivers_ptweighted}. 
This analysis with $P_\text{hT}$-weighted asymmetry amplitudes provides a direct measurement of TMD $k_\text{T}^2$ moments while avoiding assumptions on the shape of the $k_\text{T}$ distribution.  

\begin{figure}
\begin{center}
\includegraphics[width=0.9\textwidth]{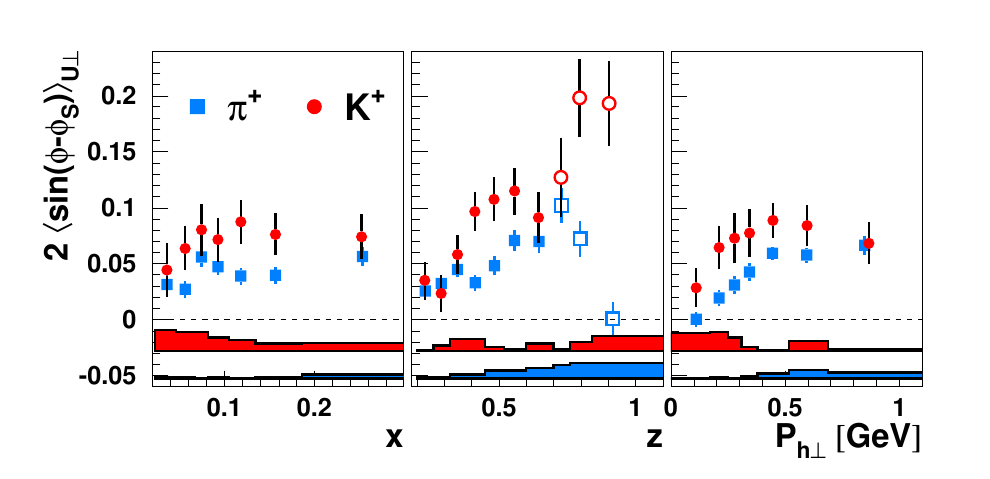}
\end{center}
\caption[Sivers SIDIS asymmetry for pions and kaons]{Sivers $\sin(\phi-\phi_\text{S})$ asymmetry amplitudes (here called $2\langle\sin(\phi-\phi_\text{S})\rangle_{\text{U}\perp} $) from HERMES SIDIS data for positively charged pions and kaons \cite{hermes:tmdbible}. The hadron transverse momentum is labeled $P_{\text{h}\perp}$ here.}
\label{fig:sivers_hermeskaons}
\end{figure}
The final compendium of HERMES TMD results from SIDIS with refined analysis, multi-dimensional binnings, and first (anti-)proton measurements has recently been published \cite{hermes:tmdbible}. Remarkably, as seen in Fig.~\ref{fig:sivers_hermeskaons}, the Sivers amplitudes for kaons (K~$=(\text{us})$) are larger than those for pions ($\pi=(\text{ud})$), which is unexpected if u-quark scattering dominates and which may point to a role of sea quarks.  

We conclude this subsection by summarizing: significant Sivers amplitudes were measured in SIDIS on the transversely polarized proton. Next we will widen our horizon to the global situation - TMDs are not only accessible in SIDIS, but in a variety of processes.

\subsection{Experimental TMD probes and TMD (modified) universality}
\label{sec:tmdglobal}

\begin{figure}
\begin{center}
\includegraphics[width=0.9\textwidth]{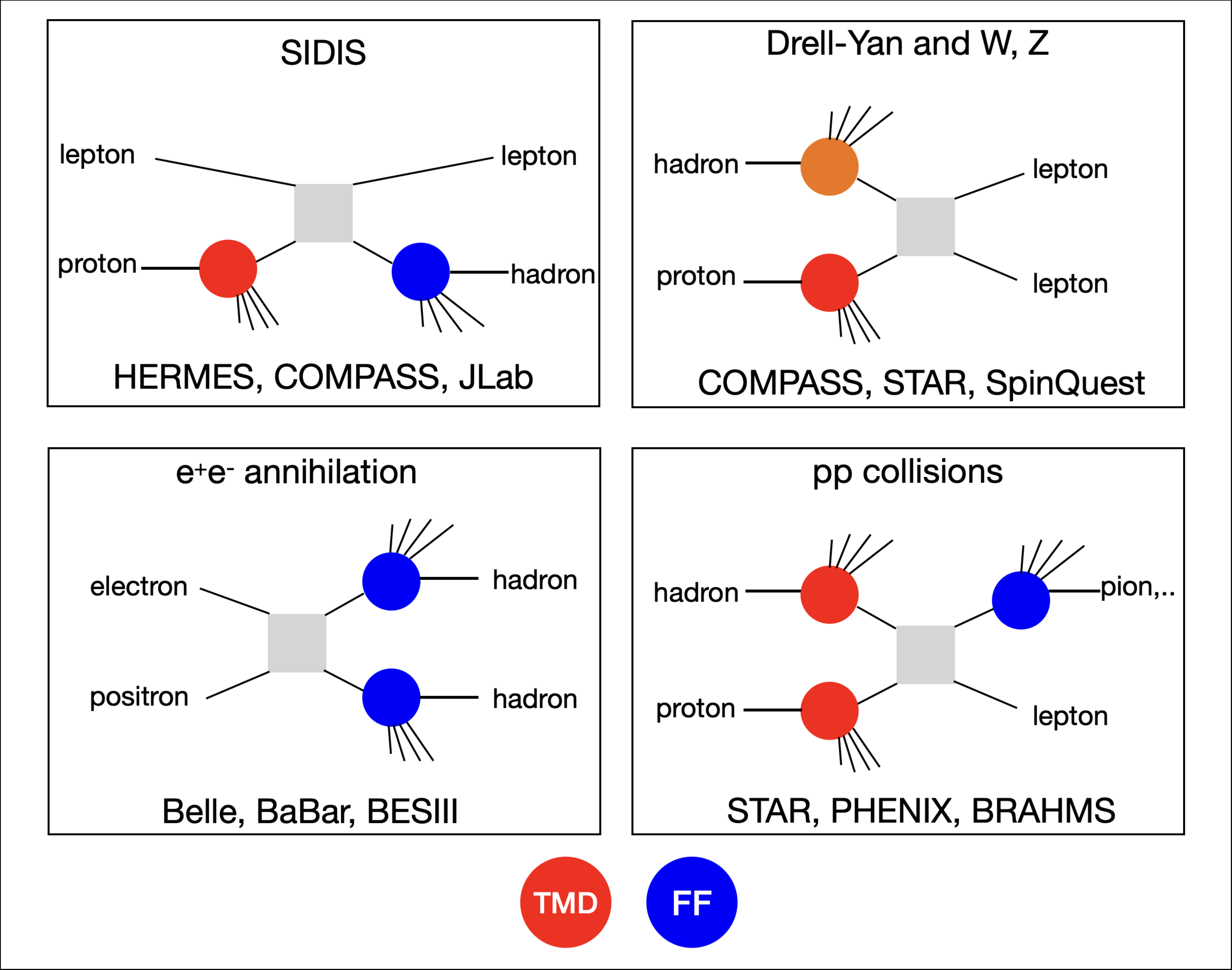}
\end{center}
\caption[Experimental TMD probes]{Experimental TMD probes. The red (or orange) circles are TMD PDFs, the blue circles fragmentation functions, and the grey boxes represent the hard (pQCD) part of the cross section. Figure adapted from Ref.~\cite{commonorigin}.}
\label{fig:TMD_exps}
\end{figure}

\begin{figure}
\begin{center}
\includegraphics[width=0.9\textwidth]{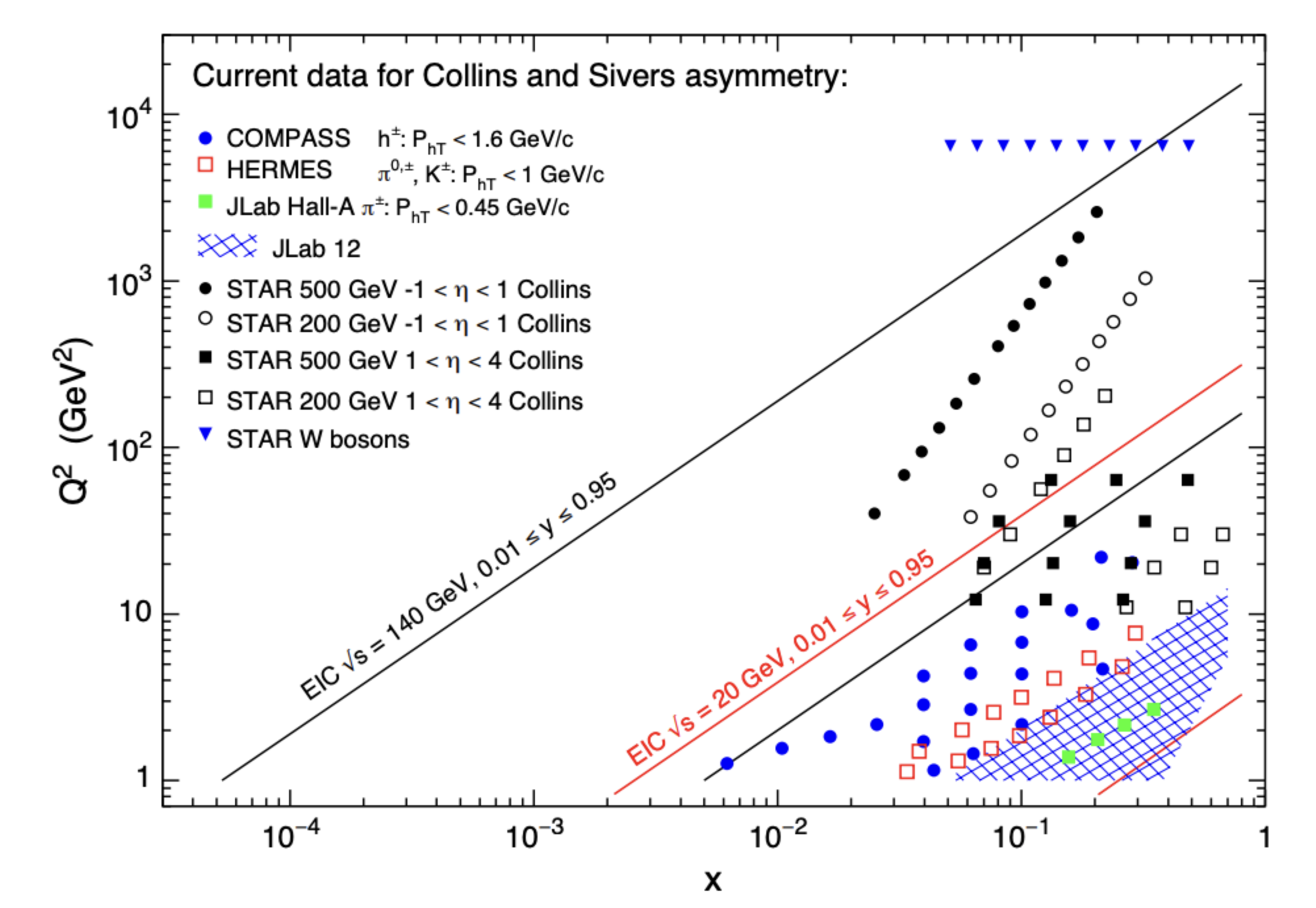}
\end{center}
\caption[Global TMD-related measurements]{Global TMD-related measurements at DESY, CERN, JLab and BNL and the coverage by the future EIC. Figure from Ref.~\cite{EICYellowReport2021}.}
\label{fig:TMD_globalmeas}
\end{figure}

There is a worldwide effort to access proton structure and in particular TMD-related observables using different types of scattering processes. Experimental TMD probes are not limited to semi-inclusive DIS: they include apart from SIDIS also the Drell-Yan process, proton-proton collisions, and electron-positron annihilation, as shown systematically in Fig.~\ref{fig:TMD_exps}. 
We had discussed proton-proton collisions earlier in Secs.~\ref{sec:neutrinodis} (flavor asymmetry in the sea from unpolarized protons), \ref{sec:deltaq-pp} (sea quark helicities from longitudinally polarized protons), and \ref{sec:deltag} (gluon helicity from longitudinally polarized protons). RHIC can also provide transversely polarized proton beams, which are used to study TMDs. Fragmentation functions can cleanly only be accessed in electron-positron experiments such as the Belle experiment at KEK \cite{belle:2019}. The hadron-hadron Drell-Yan process (DY),
\begin{equation}
\text{h}_1\text{h}_2\rightarrow \ell\overline{\ell}\text{X},
\label{eq:dy}
\end{equation}
has the experimental signature of two oppositely charged leptons $\ell$ (electrons or muons). It was first theoretically described in 1971 \cite{dy:1970} and has been probed in dedicated experiments at BNL, CERN and Fermilab since then (and is a source of background at any hadron-hadron collider). Upon again examining Fig.~\ref{fig:TMD_exps}, one realizes the complementarity between DY and SIDIS - the DY diagram looks similar to the SIDIS diagram after rotation by 90 degrees. The major apparent difference is the presence of two TMD PDFs in DY (one for each colliding hadron species), while there is a TMD PDF and an FF in SIDIS. Because time is running to the right in the diagrams, we can conclude that the virtual photon in DY is time-like (running to the right, with positive virtuality $m^2>0$), while it is space-like in SIDIS (running vertically, with negative virtuality $m^2<0$). The kinematic landscape for the TMD measurements is shown in Fig.~\ref{fig:TMD_globalmeas}.

One may expect the TMD PDFs and FFs to not depend on the process they are measured with - after all, we are probing fundamental intrinsic features of the nucleon and the answer that nature gives us should be independent on the specific experiment we carry out? Measuring TMD observables in different scattering processes allows to probe TMD universality. Looking closer at the QCD-based TMD theory, we find a peculiar prediction. The naive time-reversal odd TMD PDFs - the Sivers and Boer-Mulders TMDs - are expected to switch sign when measured in SIDIS or in DY. This is strikingly different from previously studied quark-momentum and quark-spin distributions in the nucleon.

\begin{figure}
\begin{center}
\includegraphics[width=0.98\textwidth]{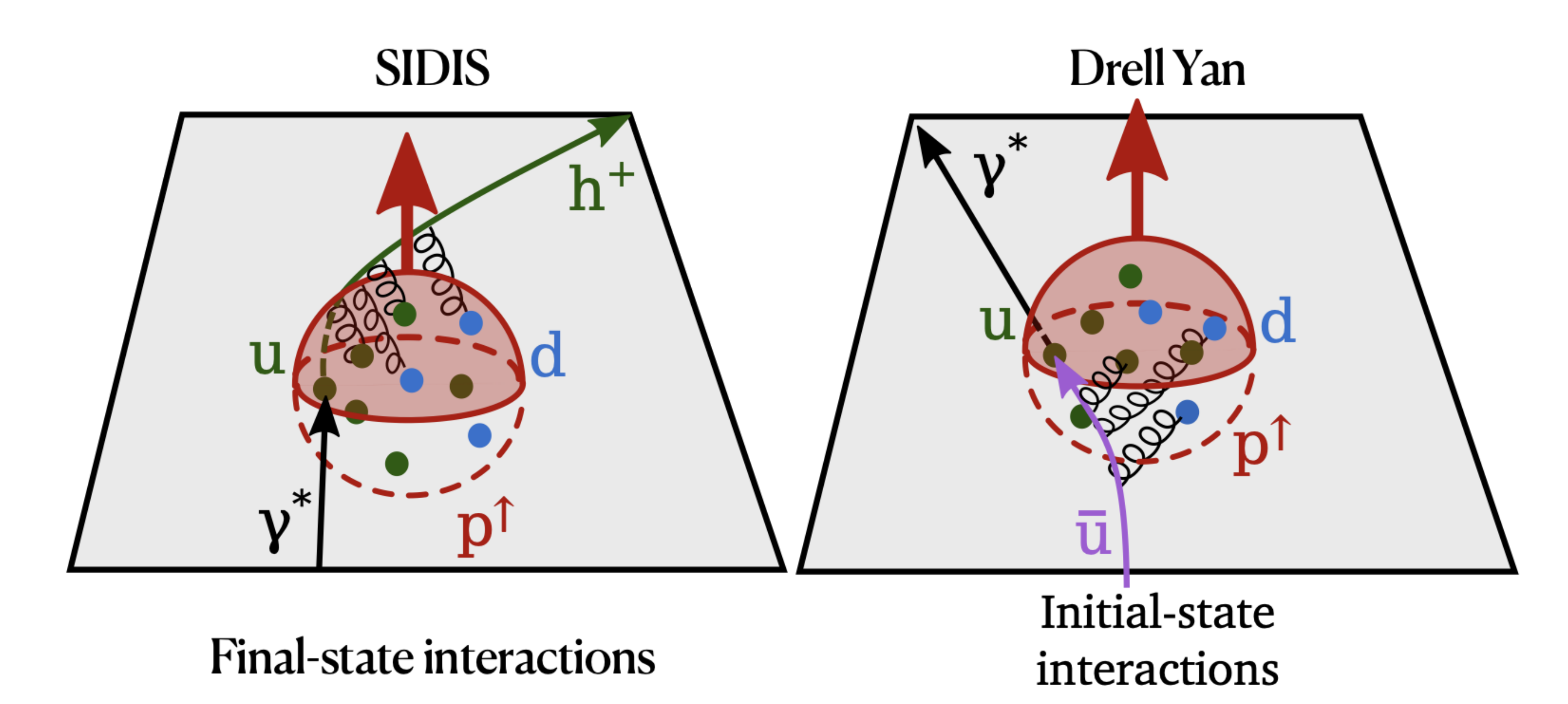}
\end{center}
\caption[Initial- / final-state state interactions in DY / SIDIS]{The expected sign switch of the naive time-reversal odd TMD PDFs (Sivers and BM) when measured in SIDIS (left) or Drell-Yan (right) can be explained by final-state resp.~initial-state interactions via the exchange of soft gluons. Figure courtesy J.~Matousek (COMPASS).}
\label{fig:DYvsSIDIS}
\end{figure}

Why is that? The path of the Wilson lines (objects in quantum field theory) depends on the space-time structure of the process in that the TMDs PDFs are embedded. The Wilson lines required for the Drell-Yan process point to the past, whereas those appearing in the TMD PDFs for SIDIS point to the future. If it were not for the gluon exchange represented by the Wilson line, the Sivers modulation would be zero \cite{diehl_GPDTMD}. Rephrasing this in the language we have used above in Sec.~\ref{sec:sivers}, final-state SIDIS interactions translate into DY initial-state interactions (see Fig.~\ref{fig:DYvsSIDIS}), changing the direction of gauge link integrals and thereby also the sign of the Sivers and Boer-Mulders TMD PDFs.

The experimental quest for the sign switch of the Sivers function and thus the \emph{modified universality concept} of the Sivers and Boer-Mulders TMDs has started a few years ago. The published COMPASS and STAR data tend to support the Sivers sign switch, albeit still within large experimental uncertainties. More data for the same channels are being analyzed by both collaborations. 

\begin{figure}
\begin{center}
\includegraphics[width=0.98\textwidth]{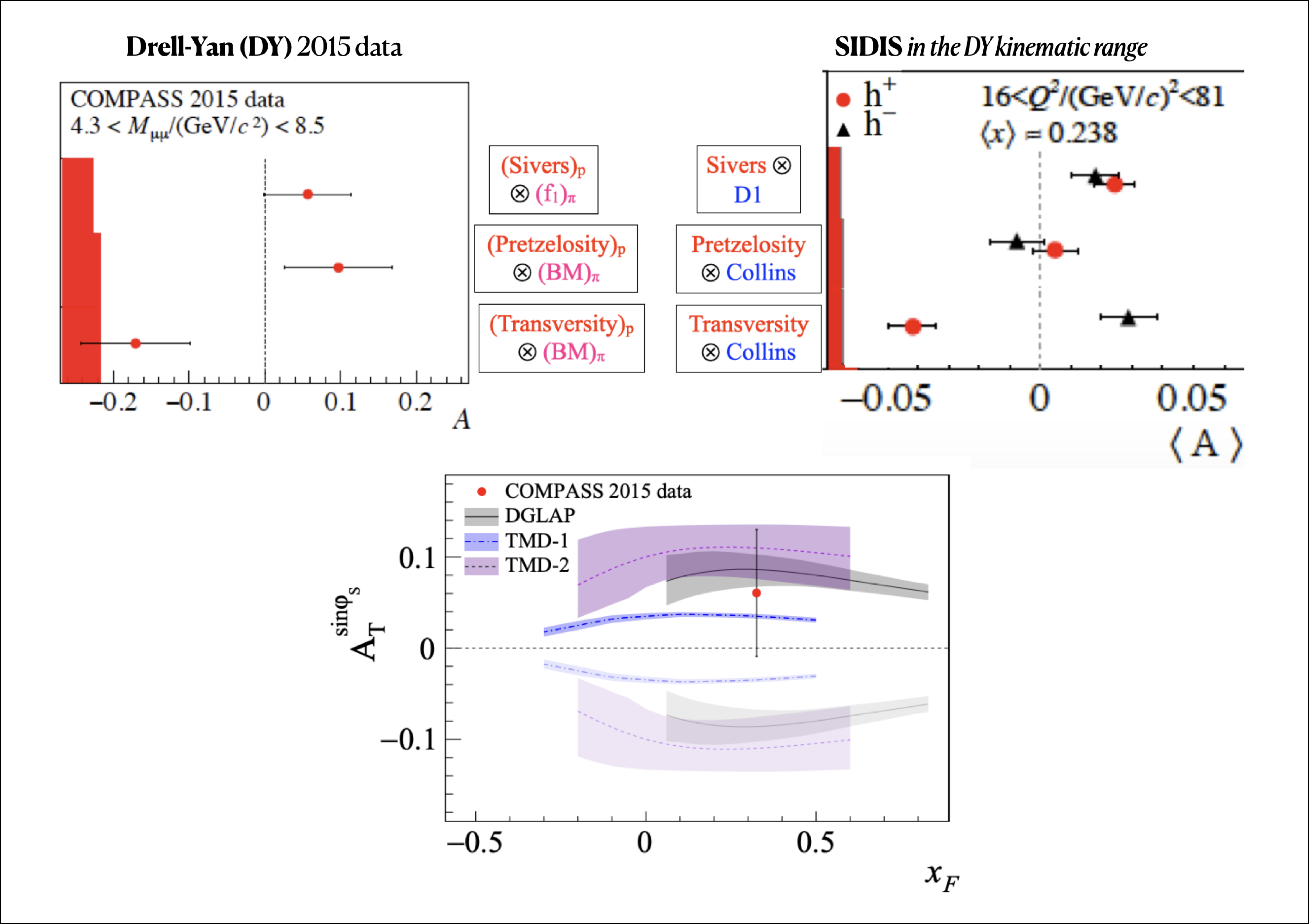}
\end{center}
\caption[COMPASS Sivers from SIDIS and DY and Sivers sign switch]{Top: COMPASS Sivers and other TMD asymmetry amplitudes from pion-induced DY \cite{compass:sivers_dy} (left) and SIDIS \cite{compass:sivers_sidis_dykine} (right) on a transversely polarized proton target. Bottom: COMPASS Sivers asymmetry amplitude in DY. The dark-shaded bands (DGLAP \cite{DGLAP}, TMD-1 \cite{TMD1}, TMD-2 \cite{TMD2}) belong to fits to SIDIS world data that \emph{include} the Sivers-function sign switch. The mirrored lighter versions of the bands are \emph{without} the sign switch. Note that the Sivers DY and SIDIS asymmetry amplitudes in the top have the same sign because of the used azimuthal-angle definitions. The sign switch is only revealed in the bottom plot. }
\label{fig:sivers_signswitch}
\end{figure}

COMPASS measured Sivers asymmetry amplitudes in SIDIS \cite{compass:sivers_sidis_dykine} and pion-induced DY \cite{compass:sivers_dy} with almost the same apparatus and in overlapping kinematic domains. The results are shown in Fig.~\ref{fig:sivers_signswitch} including other TMD-related asymmetry amplitudes such as those for the transversity and pretzelosity TMDs. The respective probed combinations of TMD-PDF~$\otimes$~TMD-PDF (for DY) and TMD-PDF~$\otimes$~FF (for SIDIS) are also shown in the figure for each of the TMD asymmetry amplitudes. The spin-dependent DY data were collected in 2015 and 2018 with a negatively charged pion beam scattering off a transversely polarized proton target, $\pi^-\text{p}^\uparrow\rightarrow \mu^+\mu^-\text{X}$. The dominant {\bf $\text{q}\overline{\text{q}}$} process is the $\overline{\text{u}}$ from the beam $\pi^-$ annihilating with a valence u-quark in the target proton, making the COMPASS DY experiment sensitive to the u-quark Sivers function in the valence-quark region, where the Sivers function has its largest magnitude. Figure~\ref{fig:sivers_signswitch} also shows the predicted DY asymmetry amplitude in case of a sign switch of the Sivers function. Note that the Feynman-$x$ variable $x_\text{F}$ is the difference in $x$-Bjorken of the participating partons in the two hadrons, $x_\text{F} = x_{\mathrm{beam}}-x_{\mathrm{target}}$. 

\begin{figure}
\begin{center}
\includegraphics[width=0.98\textwidth]{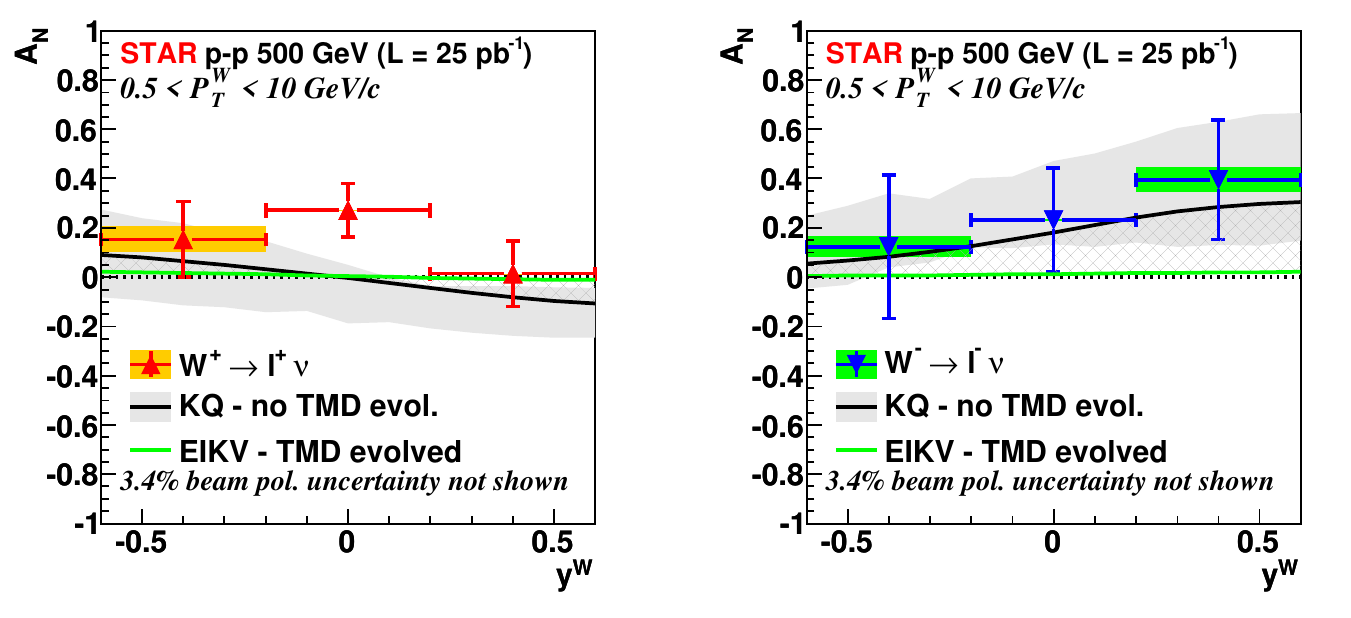}
\end{center}
\caption[STAR $A_\text{N}$ in W$^\pm$ and Sivers sign switch]{Left-right asymmetry $A_\text{N}$ in W-boson production from $\text{p}^\uparrow \text{p}$ collisions at STAR \cite{star:sivers}.}
\label{fig:sivers_starW}
\end{figure}

\begin{figure}
\begin{center}
\includegraphics[width=0.98\textwidth]{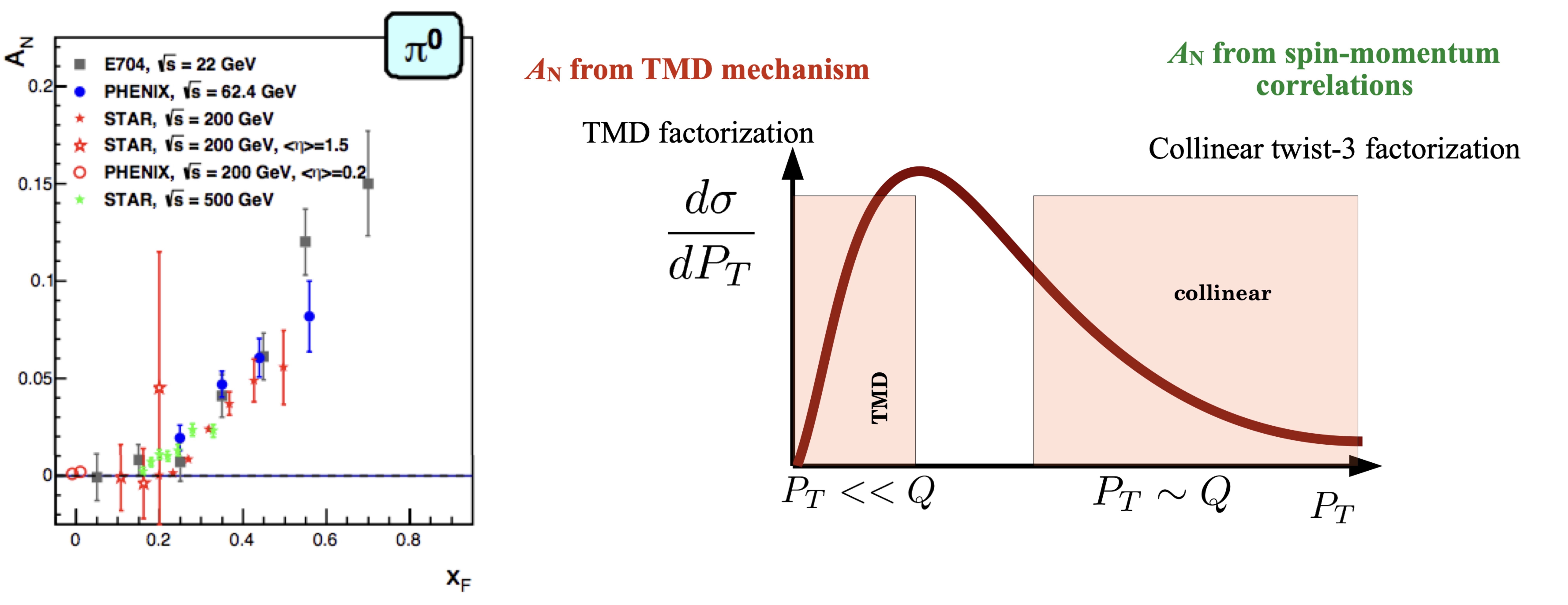}
\end{center}
\caption[Left-right asymmetries in $\text{p}^\uparrow \text{p}$]{Left: left-right asymmetries $A_\text{N}$ in $\text{p}^\uparrow \text{p}\rightarrow\pi^0\text{X}$ \cite{Aschenauer:2016our}. Right: The two factorization schemes - TMD factorization and collinear twist-3 factorization - are related and equivalent in the overlapping kinematics region \cite{Ji2006}.}
\label{fig:AN}
\end{figure}

STAR measured left-right asymmetries in W- and Z-boson production from $\text{p}^\uparrow \text{p}$ collisions at $\sqrt{s}=500$\,GeV RHIC energies \cite{star:sivers}.  
The results together with predictions from model calculations including the Sivers-function sign switch are shown in Fig.~\ref{fig:sivers_starW}.
The left-right asymmetry $A_\text{N}$ from Fig.~\ref{fig:sivers_starW} is a typical collider observable in $\text{p}^\uparrow \text{p}$ collisions:
\begin{equation}
A_\text{N}=\frac{\sigma_\text{L}-\sigma_\text{R}}{\sigma_\text{L}+\sigma_\text{R}},
\label{eq:an}
\end{equation}
essentially comparing left (L) and right (R) counts in a certain experimental channel. Because only the transverse polarization of one of the proton beams is made use of,  these $A_\text{N}$ asymmetries are often referred to as \emph{transverse single-spin asymmetries}. As shown in Fig.~\ref{fig:AN}, the $A_\text{N}$ are measured to be large, rising with $x_\text{F}$, and independent of the center-of-mass energy $\sqrt{s}$ over two orders of magnitude. 

The theoretical efforts over the past few years have demonstrated that it is possible to simultaneously describe left-right asymmetries across multiple collision species \cite{commonorigin}, including the single-spin asymmetries in SIDIS. This indicates that all $A_\text{N}$ have a common origin that is related to multi-parton correlations. Different but related factorization schemes come into play at different scales \cite{Ji2006}, see also Fig.~\ref{fig:AN}: \emph{TMD factorization} applies to processes with one hard (e.g., $Q^2$) and one soft (e.g., $p_\text{T}$) scale - SIDIS, DY, W/Z production, di-jets, hadrons in jets, ... -, with $p_\text{T}\ll Q$. For single-scale ($p_\text{T}\sim Q$) hard processes in pp such as single inclusive production (particle or jet $p_\text{T}$), \emph{collinear twist-3 factorization} applies. In the latter case, $A_\text{N}$ is thought not to arise from the TMD mechanism but from spin-momentum correlations ({\bf qgq} or tri-gluon {\bf ggg}). The two factorization schemes are equivalent in the overlapping kinematic region. For example, the $k_\text{T}$ moment of the Sivers TMD PDF is related to the twist-3 Efremov-Teryaev-Qiu-Sterman (ETQS) function.

\begin{figure}
\begin{center}
\includegraphics[width=0.8\textwidth]{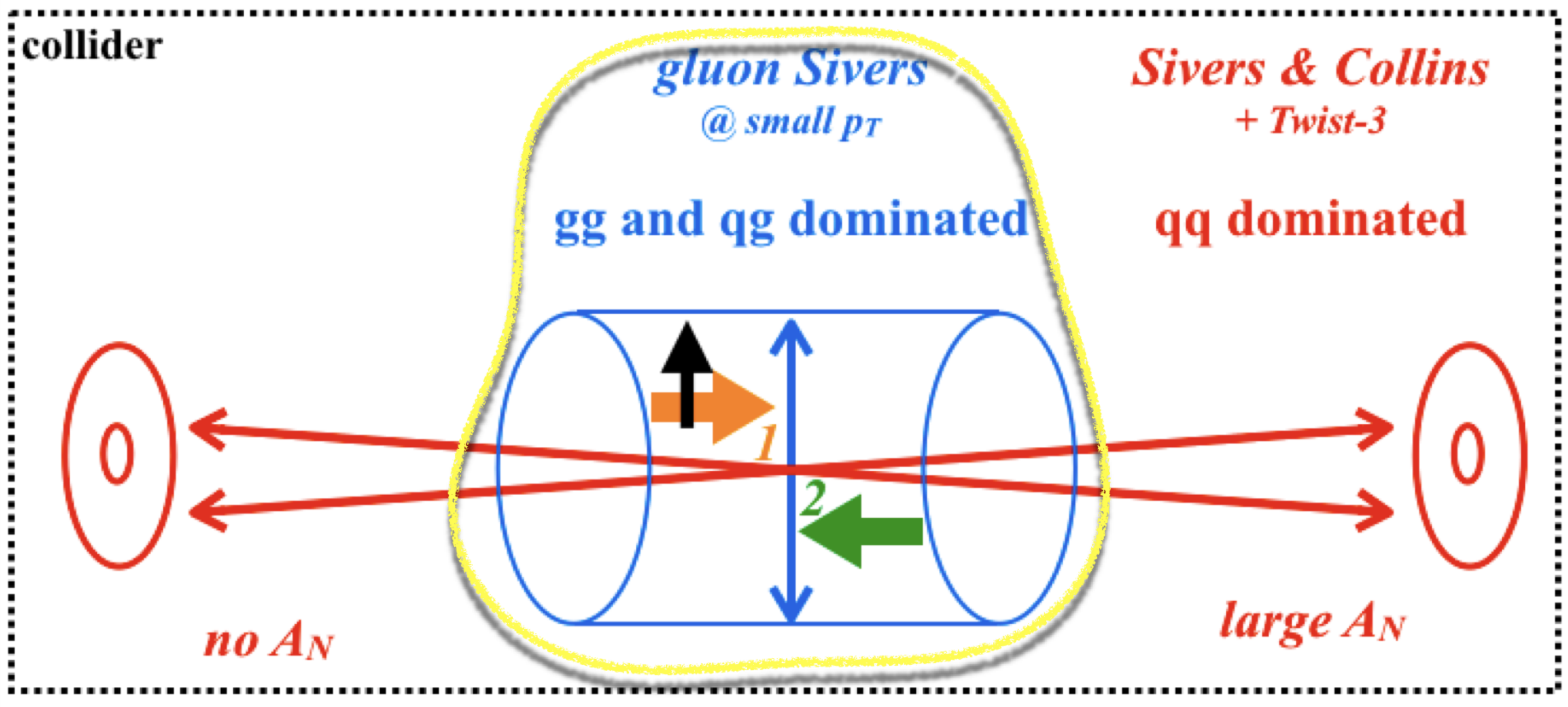}
\end{center}
\caption[Sensitivity to TMDs in collider geometry]{Sensitivity to TMDs in collider geometry. The blue barrel indicates mid-rapidity, which is expected to be dominated by gluonic processes and thus sensitive to gluon TMDs. To the right (defined by the direction of the transversely polarized proton beam) is the region of high positive rapidity (``forward'') with sensitivity to quark TMDs.}
\label{fig:TMD_collider}
\end{figure}

\begin{figure}
\begin{center}
\includegraphics[width=0.98\textwidth]{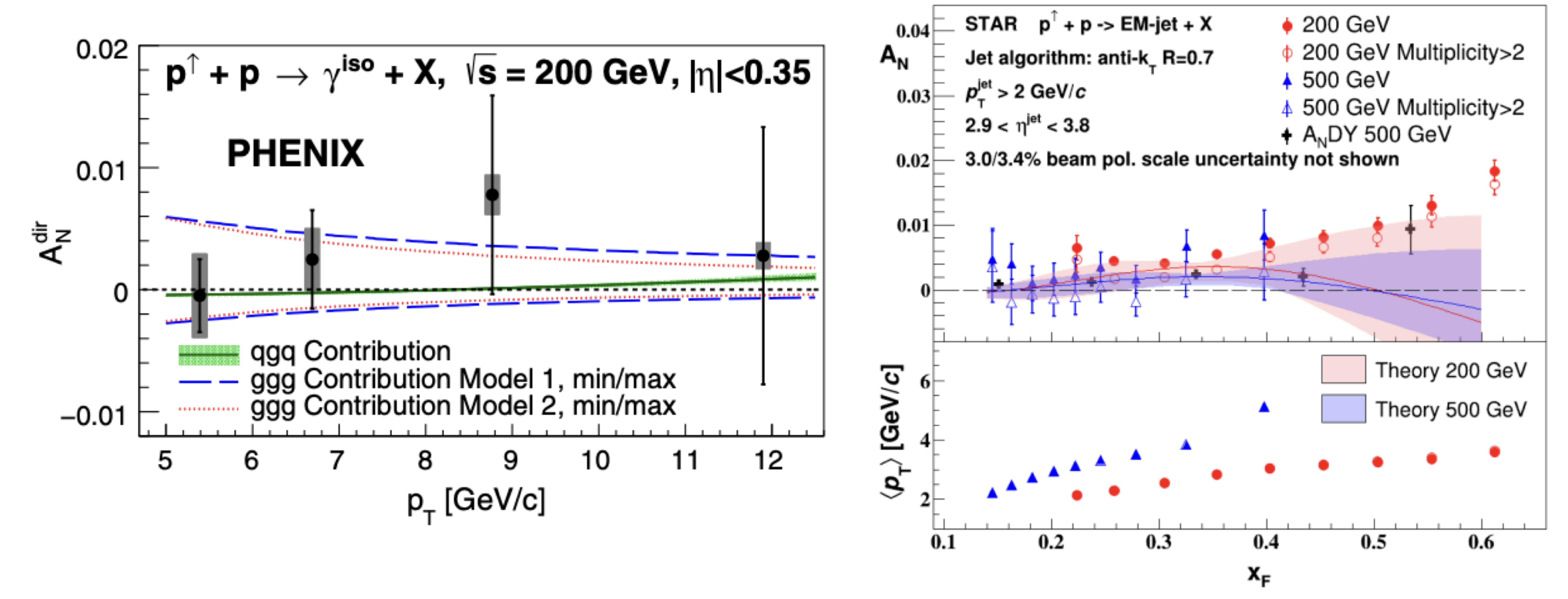}
\end{center}
\caption[RHIC p$^\uparrow$p $A_{\text N}$ in mid-rapidity and the forward]{RHIC p$^\uparrow$p $A_{\text N}$  at $\sqrt{s}=200$~GeV. Left: PHENIX in isolated direct-photon production at mid-rapidity \cite{phenix:directphoton}, right: STAR in electromagnetic jets in the forward \cite{star:forward}.}
\label{fig:AN_rap}
\end{figure}

As shown in Fig.~\ref{fig:TMD_collider}, the $A_\text{N}$ is expected to be dominated by {\bf gg} and {\bf qg} processes at mid-rapidity (small polar angles) and small transverse momenta $p_\text{T}$. This means that measurements in this kinematic domain are expected to be sensitive to tri-gluon twist-3 correlation functions, which are related to the \emph{Sivers effect of the gluon} \cite{anselmino:2006} \cite{DAlesio:2016}. 
PHENIX measured an $A_\text{N}$ in $\text{p}^\uparrow \text{p}\rightarrow \pi^0 \text{X}$ at mid-rapidity compatible with zero at high precision down to very low transverse momenta $p_\text{T}$ of the detected pions \cite{phenix:an2014}. Also recent results published by PHENIX revealed no signal at mid-rapidity: for both the isolated direct-photon $A_\text{N}$ \cite{phenix:directphoton} (see Fig.~\ref{fig:AN_rap}) and the $A_\text{N}$ in $\pi^0$ and $\eta$ production \cite{phenix:pieta}, no signals were found at high precision.
On the other hand, COMPASS found a non-zero result at the 2.5-sigma level for the Sivers asymmetry in photon-gluon fusion from SIDIS data \cite{compass:sivers_highpt}. Collecting more experimental information about the gluon Sivers TMD from other experimental channels is therefore important and analysis is currently in progress of heavy-flavor production at PHENIX \cite{phenix:heavy-flavor} and of J$/\psi$ production in pion-proton collisions at COMPASS.

While transverse single-spin asymmetries are measured to be extremely small at mid-rapidity, they grow substantially at forward rapidity for various observables. A plethora of new RHIC results in p$^\uparrow$p has recently become available. Using a calorimeter 18\,m away from the STAR interaction region, the RHICf collaboration measured $A_\text{N}$ from $\pi^0$ in electromagnetic jets \cite{rhicf:2020} for very forward rapidity $2.4<\eta<4$. PHENIX measured $A_\text{N}$ by detecting very forward neutrons \cite{phenix:forward}, and STAR by detecting $\pi^0$ and electromagnetic jets \cite{star:forward} (see Fig.~\ref{fig:AN_rap}). The left-right asymmetry increases with $p_\text{T}$, forwardness, $\pi^0$ isolation, and $\gamma$ multiplicity (STAR), suggesting it may in this domain arise from soft processes such as diffractive scattering.

\subsection{Transversity TMD and spin-dependent fragmentation}
\label{sec:collins}

\begin{figure}
\begin{center}
\includegraphics[width=0.71\textwidth]{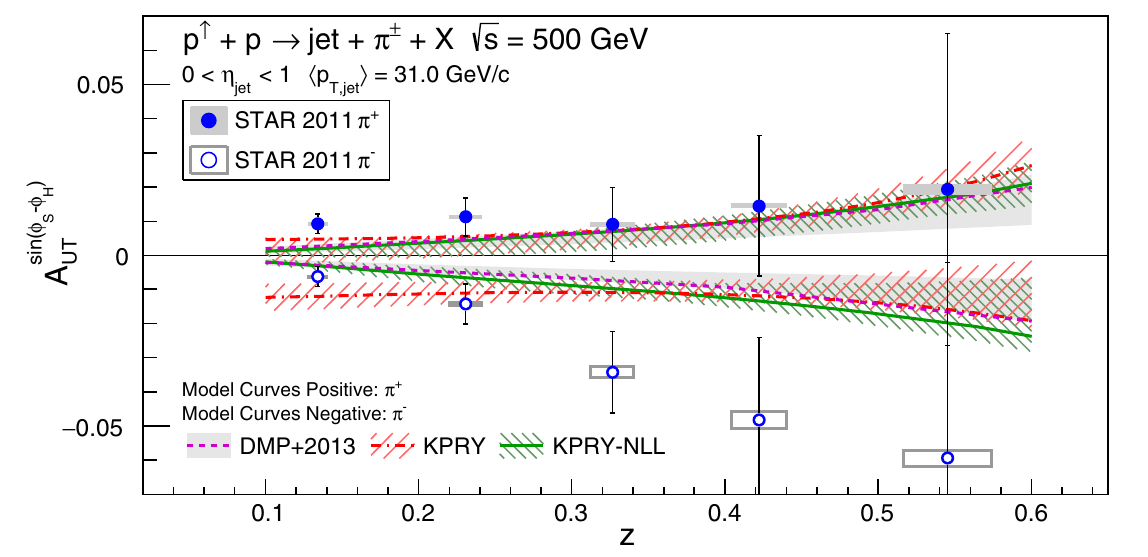}
\includegraphics[width=0.25\textwidth]{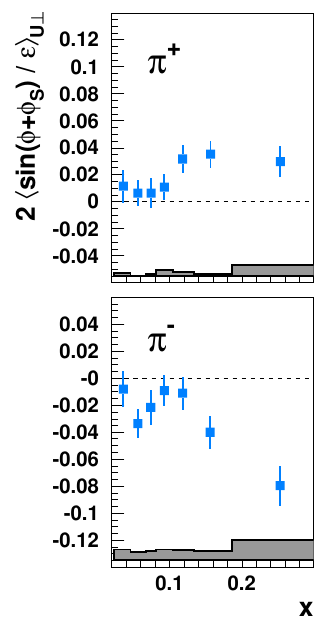}
\end{center}
\caption[Collins asymmetries]{Left: STAR Collins asymmetries from charged pions in jets in p$^\uparrow$p  \cite{star:collins_pions}. Right: HERMES $\sin(\phi+\phi_\text{S})$ Collins asymmetry amplitudes for charged pions in SIDIS  \cite{hermes:tmdbible}.}
\label{fig:collins}
\end{figure}

The Collins effect is the fragmentation of a transversely polarized parton into a final-state hadron. The Collins fragmentation function, $\sim\vec{s}_\text{T}\cdot(\widehat{k} \times \vec{P}_\text{hT})$, appears in the SIDIS cross section convoluted with the transversity TMD (which represents a spin-spin correlation). Both are chiral-odd. The coupling of the transversity TMD and the Collins FF in p$^\uparrow $p leads to azimuthal modulations of charged-hadron yields around the jet axis, and the involvement of two momentum scales (jet and hadron transverse momenta) allows for an interpretation within the TMD framework.   
STAR measured Collins asymmetries for charged pions in jets \cite{star:collins_pions} to be different from zero above the 5-sigma level at higher jet transverse momenta and with different signs for the two pion charges. Figure~\ref{fig:collins} shows the results together with model curves \cite{collins:theo1} \cite{collins:theo2}, which are based on calculations using SIDIS and ee data. The good comparison between model and data confirms the expected \emph{universality of the Collins FFs}.

STAR also reported the first constraint on the Collins-like asymmetry \cite{star:collins_pions}, which is sensitive to linear gluon polarization. STAR has also measured a significant Collins-dihadron interference-fragmentation asymmetry \cite{star:dihadronIFF} with the expected prominent enhancement at the $\rho$-meson mass. More STAR data are being analyzed including kaons and protons in jets \cite{talk:pokhrel}.

\begin{figure}
\begin{center}
\includegraphics[width=0.9\textwidth]{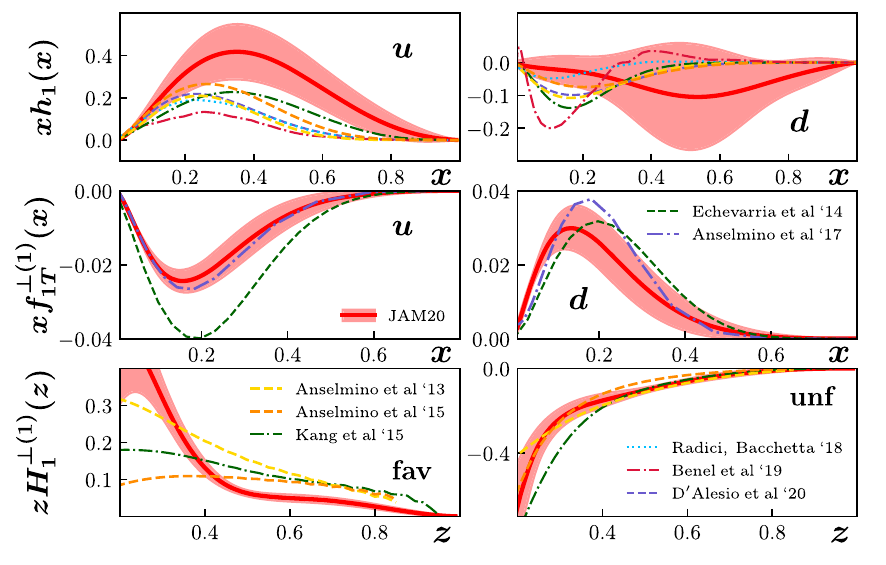}
\end{center}
\caption[JAM global TMD f its]{JAM global fits for transversity TMD (top), Sivers TMD (middle), Collins FF (bottom) \cite{commonorigin}. }
\label{fig:jam}
\end{figure}

The final results of the HERMES SIDIS Collins asymmetries \cite{hermes:tmdbible} are shown in the right side of Fig.~\ref{fig:collins}. The HERMES and COMPASS SIDIS Collins asymmetries agree well for the common kinematics \cite{compass:siverssidis} (not shown). There is a mirror symmetry for $\pi^+$ and $\pi^-$, which indicates that u- ($\delta_\text{u}$) and d-quark transversity ($\delta_\text{d}$) have approximately equal magnitude but opposite signs. 
Figure~\ref{fig:jam} presents universal fits to TMD data by the JAM collaboration \cite{commonorigin}, which demonstrate that $\delta_\text{d}$ is poorer constrained than $\delta_\text{u}$, given the u-quark dominance of many of the processes used in the global fits. It is therefore important to collect more data sensitive to $\delta_\text{d}$. The 2022 COMPASS run on the transversely polarized deuteron is expected to double the experimental precision on the proton’s tensor charge $g_T=\delta_\text{u}-\delta_\text{d}$ \cite{compass2-proposal}.

Already with existing data, it will be possible to test the predicted genuine universality for the transversity and pretzelosity TMDs by comparing extractions from SIDIS data and the new COMPASS DY data.  An alternative method of accessing transversity, as recently made available by COMPASS \cite{compass:lambda} and STAR \cite{star:hyperon}, is the measurement of hyperon transverse polarization, which may have been transferred from the struck quark in the transversely polarized proton.

Novel spin-dependent fragmentation effects are studied by COMPASS for $\rho$-meson production on the transversely polarized proton \cite{albitalk}, investigating the difference in the Collins mechanism of spin-1 vector mesons versus that for pseudoscalar mesons (ordinary Collins FF).
CLAS12 published a higher-twist di-hadron beam-spin asymmetry \cite{hayward}. The measurement establishes the first empirical evidence of a non-zero helicity-dependent di-pion fragmentation function, which encodes spin-momentum correlations in hadronization and is equivalent to the Collins FF for two pions. CLAS published a non-zero higher-twist di-hadron beam-spin asymmetry \cite{mirazita}. Sizeable higher-twist beam-spin asymmetries in SIDIS were reported by CLAS12 \cite{diehlclas12} and HERMES \cite{hermes:alu}. These asymmetries provide access to so-far poorly known sub-leading twist-3 TMD PDFs and FFs containing information about quark-gluon correlations in the proton and in the hadronization process.

%
\section{Summary and outlook}

Since the \emph{spin crisis} triggered by the findings of the EMC collaboration in 1988, the proton spin structure has undergone extensive experimental mapping campaigns and various novel theoretical frameworks describing proton and hadron structure have emerged. 
The experimental results from inclusive and semi-inclusive DIS and pp experiments indicate that the quark spins contribute a third to the spin of the proton and the gluon spins contribute some positive amount in the currently covered range. Only the future Electron-Ion Collider (EIC) \cite{EICYellowReport2021} will reveal if there is a significant contribution from the gluon or sea-quark spins at very low values of $x$-Bjorken. The question remains, where is the remaining spin coming from? Do quarks and gluons have orbital angular momentum? The experimental data sensitive to GPDs and TMDs support this scenario, but quantitative relations yet have to be developed and measurements be performed in extended kinematic regions, in particular at small $x$-Bjorken. The EIC is expected to deliver the final answer to the question of parton orbital angular momentum.  

Rich future experimental programs related to nucleon spin structure are planned prior to the EIC: until 2025, these are on the fixed-target side the just started final COMPASS d-quark transversity run \cite{compass2-proposal}, the continuation of the JLab12 program \cite{clas12talk} \cite{solidtalk}, the SpinQuest / E1039 TMD campaigns with polarized targets \cite{spinquest} to access sea-quark TMDs, AMBER \cite{amber} as COMPASS successor at the SPS M2 beamline in the CERN North Area, and LHCspin \cite{lhcspin} at CERN, which will use transversely polarized gas targets and LHCb as forward spectrometer. On the RHIC-collider side, the cold QCD programs will be continued at STAR with several upgrades \cite{starupgrade} and the new sPHENIX experiment \cite{sphenix}, for which assembly has started in summer 2021.  

\vspace{.5in}
{\bf Acknowledgements.} The author thanks Prof.~Barbara Badelek for the fruitful exchanges. 

\begin{appendices}
\newpage
\section{Introductory references and conventions}
\label{sec:conventions}

Some recommendations for further reading are given in the following. An excellent \emph{very short} introduction to particle physics is given in Ref.~\cite{Close2004}. It is useful to keep in mind that 
\begin{eqnarray}
hc &\approx& 10^{-6}~\mathrm{eV\cdot m},\\
\hbar c &\approx& 200~\mathrm{MeV\cdot fm}, \\
10^4~\mathrm{K}\rightleftharpoons 1~\mathrm{eV}&\rightleftharpoons& 10^{-6}~\mathrm{m},
\label{eq:evm}
\end{eqnarray}
and that light travels about the length of your foot in one nanosecond. A fascinating summary of the history of particle physics in the 20th century until 1979 is given in Ref.~\cite{Trefil1979}. Very good textbooks are (for example) Refs.~\cite{rith2006} and \cite{Thomson2013} on the experimental side and Ref.~\cite{PeskinSchroeder} on the theory side (quantum field theory).

A useful source of up-to-date information on particle physics including the summary of the concepts related to theory, instrumentation and data analysis techniques is the Particle Data Group (PDG) \cite{Zyla:2020zbs}, which can also be accessed online.  

Throughout this paper, the convention $c=\hbar=1$ is used. When written out, momenta have units of GeV/$c$ and masses GeV/$c^2$. For the definition of azimuthal angles, the community tries to follow the so-called Trento conventions from 2004 \cite{trento:conventions}.

\section{Kinematics}

\subsection{Relativistic energy-momentum relation}
\label{sec:relativistickine}

We recall the relativistic energy-momentum relation,
\begin{equation}
E=\sqrt{|\vec{p}\,|^2+m^2},
\label{eq:relep}
\end{equation}
with $p$ the general energy-momentum 4-vector (4-momentum vector) with energy $E$ and 3-momentum $\vec{p}$:
\begin{equation}
p=(E,\vec{p}\,).
\label{eq:vecp}
\end{equation}
From Eqs.~\ref{eq:relep} and \ref{eq:vecp}, and using the space-time metric of special relativity, follows the Lorentz-invariant squared mass, or \emph{short invariant mass}:
\begin{equation}
p^2=m^2
\label{eq:invmass}
\end{equation}
by explicitly squaring the 4-momentum vector:
\begin{equation}
p^2=(E,\vec{p}\,)\begin{pmatrix}E\\-\vec{p}\end{pmatrix}=E^2-|\vec{p}\,|^2\equiv m^2.
\end{equation}
For a real photon ("light-like"), one has $p^2=m^2=0$; for a space-like virtual photon $p^2=m^2<0$ (e.g., DIS); and for a time-like virtual photon $p^2=m^2>0$ (e.g., Drell-Yan). 

\subsection{Elastic scattering - 4-momentum conservation}
\label{sec:elastic}

\begin{figure}[h]
\begin{center}
\includegraphics[width=0.75\textwidth]{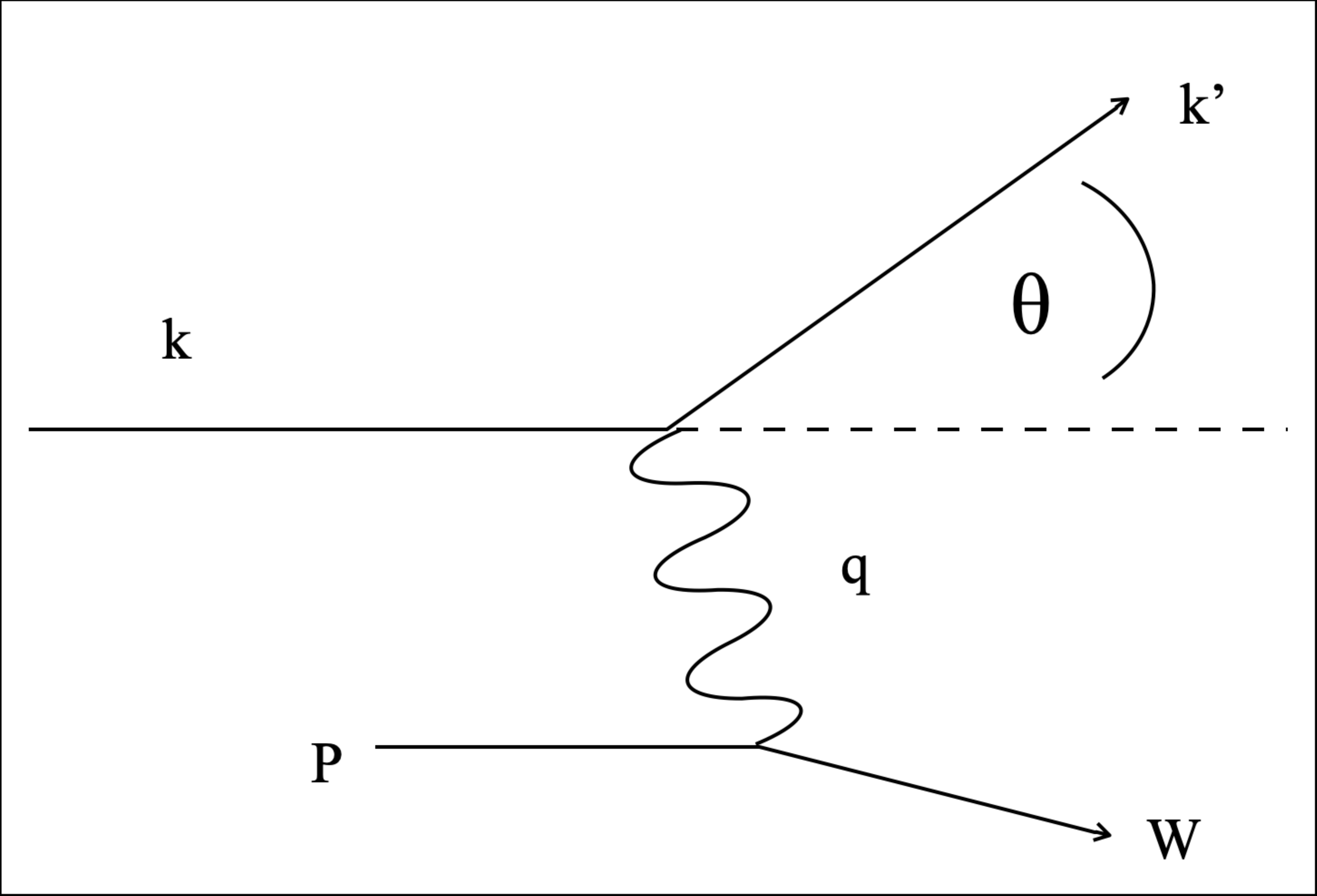}
\end{center}
\caption[Lepton-proton scattering]{Lepton-proton scattering via the exchange of a virtual photon indicated by the wavy line. The 4-momenta, as indicated in the figure, are defined in Tab.~\ref{tab:elscat}.}
\label{fig:elscat}
\end{figure}

We derive the kinematic relations and definitions for elastic lepton-proton scattering,
\begin{equation}
\ell \text{p} \rightarrow \ell^\prime \text{p}^\prime.
\end{equation}
The lepton is an electron or muon of either electrical charge. 
The diagram for the elastic process is shown in Fig.~\ref{fig:elscat} and the 4-momenta of the involved particles are defined in Tab.~\ref{tab:elscat}.
\begin{table}[h]
\begin{tabular}{c|c|c|c}
&initial (incoming) & exchange & final (scattered) \\\hline
&&&\\
lepton $\ell$ & $k=(E,\vec{k})$ & & $k^\prime=(E^\prime,\vec{k^\prime})$ \\
proton p & $P=(E_\text{p},\vec{P})$ & & $W=(E^\prime_\text{p},\vec{P^\prime})$ \\
virtual photon $\gamma^*$ & & $q=(E-E^\prime,\vec{k}-\vec{k^\prime})$ & \\\hline
\end{tabular}
\caption[4-momenta for lepton-proton scattering]{Declaration of 4-momenta for lepton-proton scattering.
The 4-momentum of the hadronic final state is labeled as $W$ and that of the virtual exchange boson (virtual photon) as $q$.}
\label{tab:elscat}
\end{table}

For the following calculation, we will assume that the proton is at rest, $P=(E_\text{p},\vec{P})\equiv(M,\vec{0})$, an assumption that will show explicitly from Eq.~\ref{eq:elscat2} on. The more generic expression is given at the end in Eq.~\ref{eq:elastic_pbeam}. We start with the conservation of 4-momenta:
\begin{equation}
k+P=k^\prime+W
\label{eq:4momcons}
\end{equation}
and rearranging Eq.~\ref{eq:4momcons}:
\begin{equation}
k-k^\prime=W-P.
\label{eq:kmkprime}
\end{equation}
After squaring both sides of Eq.~\ref{eq:kmkprime}, we obtain:
\begin{equation}
\underbrace{k^2}_{\circled{1}\approx 0}+\underbrace{k^{\prime 2}}_{\circled{1}\approx 0}-2kk^\prime=\underbrace{W^2}_{\circled{3}=M^2}+P^2-2WP.
\label{eq:squared}
\end{equation}
The following approximations, substitutions and assumptions are made. \hspace{.5in} $\circled{1}$ With Eq.~\ref{eq:invmass} and neglecting the lepton mass $m_\ell$ with respect to the nucleon mass:
\begin{equation}
k^2=k^{\prime 2}\equiv m_\ell^2\approx 0.
\label{eq:me}
\end{equation}
$\circled{2}$ Then follows with Eqs.~\ref{eq:relep} and \ref{eq:me}: \begin{equation}
E=|\vec{k}|, E^\prime=|\vec{k^\prime}|.
\label{eq:em}
\end{equation}
$\circled{3}$ Elastic scattering - the invariant mass is unchanged, the proton is not excited:
\begin{equation}
W^2=M^2.
\label{eq:noex}
\end{equation}
$\circled{4}$ Elastic scattering - energy conservation: 
\begin{equation}
E^\prime_\text{p}=E-E^\prime+M.
\label{eq:mew}
\end{equation}
We now return to Eq.~\ref{eq:squared} and make approximation $\circled{1}$ (Eq.~\ref{eq:me}), substitution $\circled{2}$ (Eq.~\ref{eq:em}), and elastic assumptions $\circled{3}$ and $\circled{4}$ (Eqs.~\ref{eq:noex} and \ref{eq:mew}):
\begin{equation}
-2EE^\prime+2\underbrace{|\vec{k}||\vec{k^\prime}|}_{\circled{2}=EE^\prime}\cos\theta=2M^2-2\underbrace{E^\prime_\text{p}}_{\circled{4}=E-E^\prime+M}M
\label{eq:elscat2}
\end{equation}
Equation~\ref{eq:elscat2} can be rewritten as in Eq.~\ref{eq:elastic} to obtain the relation between the energy of the scattered lepton $E^\prime$ and scattering angle $\theta$, at fixed incident beam energy $E$:
\begin{eqnarray}
-2EE^\prime+2EE^\prime\cos\theta&=&2M^2-2ME+2ME^\prime-2M^2\nonumber\\
-2EE^\prime(1-\cos\theta)&=&-2ME+2ME^\prime\nonumber\\
E^\prime\left(2E(1-cos\theta)+2M\right)&=&2ME\nonumber\\
E^\prime&=&\frac{2ME}{2E(1-\cos\theta)+2M}\nonumber\\
E^\prime&=&\frac{E}{1+\frac{E}{M}(1-\cos\theta)}.
\label{eq:elastic}
\end{eqnarray}
In case the proton is not at rest, Eq.~\ref{eq:elastic} becomes
\begin{equation}
E^\prime=\frac{E+E_\text{p}-M}{1+\frac{E}{M}(1-\cos\theta)}.
\label{eq:elastic_pbeam}
\end{equation}
We summarize: (a) in elastic scattering, $E^\prime$ and $\theta$ are uniquely correlated; (b) at fixed lepton beam energy $E$, there is exactly one free parameter; (c) the invariant masses of the proton before and after elastic scattering are \emph{unchanged}, i.e., the \emph{proton is not excited}.

Lastly, we introduce notation that will also be used for the deep-inelastic case. Using Eq.~\ref{eq:noex} and introducing $\nu$ (the energy transfer by the lepton) and $Q^2$ (the squared momentum transfer of the virtual photon), one gets:
\begin{eqnarray}
(P+q)^2&=&P^2 \nonumber\\
P^2+q^2+2Pq&=&P^2\nonumber\\
2M(E-E^\prime)+q^2&=&0\nonumber\\
2M\nu-Q^2&=&0.
\label{eq:nuq2el}
\end{eqnarray}
Equation~\ref{eq:nuq2el} is the \emph{elastic condition} and represents an alternative formulation to Eq.~\ref{eq:elastic}, with
\begin{eqnarray}
\nu&=&E-E^\prime,\label{eq:nulab}\\
Q^2&=&-q^2.
\label{eq:q2}
\end{eqnarray}
Note that lines 3 and 4 of Eq.~\ref{eq:nuq2el} and Eq.~\ref{eq:nulab} only hold for the case that the proton is at rest (target rest frame). The frame-independent definition of $\nu$ can be found in Tab.~\ref{tab:inclDISkine}.

\subsection{Deep-inelastic scattering - properties}
\label{sec:disprops}

In deep-inelastic lepton-proton scattering (DIS),
\begin{equation}
\ell \text{p} \rightarrow \ell^\prime \text{X},
\label{eq:appdis}
\end{equation}
the invariant mass of the final state is no longer identical to that of the initial-state proton. The (a priori unknown) hadronic final state X with invariant mass $W$ is not the proton or a resonant state any longer. Now we take again a look at the 4-momentum conservation and derive Eq.~\ref{eq:discond} as DIS condition formulated in terms of $Q^2$ and $\nu$:
\begin{eqnarray}
k+P&=&k^\prime+W\nonumber\\
k-k^\prime +P&=&W\nonumber\\
q+P&=&W\nonumber\\
(q+P)^2&=&W^2\nonumber\\
q^2+P^2+2Pq&=&W^2\nonumber\\
-Q^2+M^2+2M\nu&=&W^2\nonumber\\
W^2&=&M^2+\underbrace{2M\nu-Q^2}_{>0}.
\label{eq:discond}
\end{eqnarray}
In the second-to-last and last lines of Eq.~\ref{eq:discond}, the proton is again assumed to be at rest.

\subsection{Lorentz invariant quantities and other kinematic variables}
\label{sec:Linvariants}
Several kinematic quantities are derived from the 4-momentum $P$ of the nucleon and the 4-momenta $k$ and $k^{\prime}$ of the lepton (with initial energy $E$), see Tab.~\ref{tab:elscat}. We have previously learned about $Q^2$, the negative squared momentum transfer by the lepton; $\nu$, the energy transfer by the lepton, and $W^2$, the squared mass of the final hadronic state. Since these quantities are linear combinations of products of 4-momenta, they are independent of the reference frame and thus Lorentz invariant.

In Tab.~\ref{tab:inclDISkine}, the standard kinematic variables in inclusive DIS are compiled. In DIS, each two Lorentz invariants are independent of each other. Care has to be taken to distinguish frame-dependent and frame-independent definitions. For the sake of simplicity, kinematic variables in the analysis of fixed-target data are often calculated from quantities in the lab frame assuming the proton is at rest, which is not the case in, e.g., the EIC collider case. 
\begin{table}
\begin{center}
\begin{tabular}{ll}
$W^2:=(P+q)^2=M^2+2M\nu-Q^2$ & Invariant squared mass of the\\
&final hadronic state \\
& \\
$Q^2\equiv -q^2:=(k-k^{\prime})^2$ & Negative squared 4-momentum transfer\\
\hspace{0.22in}$\stackrel{\mathrm{lab}}{\approx}4EE^{\prime}\sin^2(\theta/2)$ &  lepton $\rightarrow$ virtual photon \\
& ($\hbar/\sqrt{Q^2}$ is measure of resolution)\\
&\\
$\nu:=(Pq)/M\stackrel{\mathrm{lab}}{=}E-E^{\prime}$ & Energy transfer lepton $\rightarrow$ virtual photon\\
&\\
$y:=(Pq)/(Pk)\stackrel{\mathrm{lab}}{=}(E-E^{\prime})/E$ & Fractional energy of the virtual photon\\
&  \\
& \\
$x:=Q^2/(2Pq)=Q^2/(2M\nu)$ & Bjorken scaling variable \\
& (measure of inelasticity; in the infinite \\
&  momentum frame, fraction of nucleon\\
&  momentum carried by struck quark)\\
&\\\hline
\end{tabular}
\end{center}
\caption[Kinematic variables in inclusive lepton-nucleon DIS]{Kinematic variables in inclusive lepton-nucleon DIS, following the notations of Fig.~\ref{fig:elscat} and definitions of 4-momenta in Tab.~\ref{tab:elscat}. The proton mass is denoted as $M$. For fixed-target experiments, it is convenient to boost into the laboratory frame (``lab''). The $Q^2$ relation given for the lab frame neglects the lepton mass $m_\ell$ with respect to the lepton energies $E$ and $E^{\prime}$.
}
\label{tab:inclDISkine}
\end{table}

For a more extensive description with regards to inclusive DIS collider kinematics, the reader is referred to appendices A.2 and A.3 in the EIC Yellow Report \cite{EICYellowReport2021}. Note that the DIS $y$ variable $y=Q^2/(sx)$ is referred to as inelasticity in collider physics, with the center-of-mass (CM) energy $s=(p_1^2+p_2^2)$, with $p_1$ and $p_2$ the 4-momenta of the two involved particle species. Usually the square root of $s$ is given in units of energy and it reduces to $\sqrt{s}=\sqrt{4E_1E_2}$ in case of collider kinematics, with $E_2$ and $E_2$ the energies of the involved beams, and to $\sqrt{s}=\sqrt{2ME}$ in fixed-target kinematics, with $M$ the mass of the target and $E$ the beam energy.

\begin{figure}
\begin{center}
\includegraphics[width=0.8\textwidth]{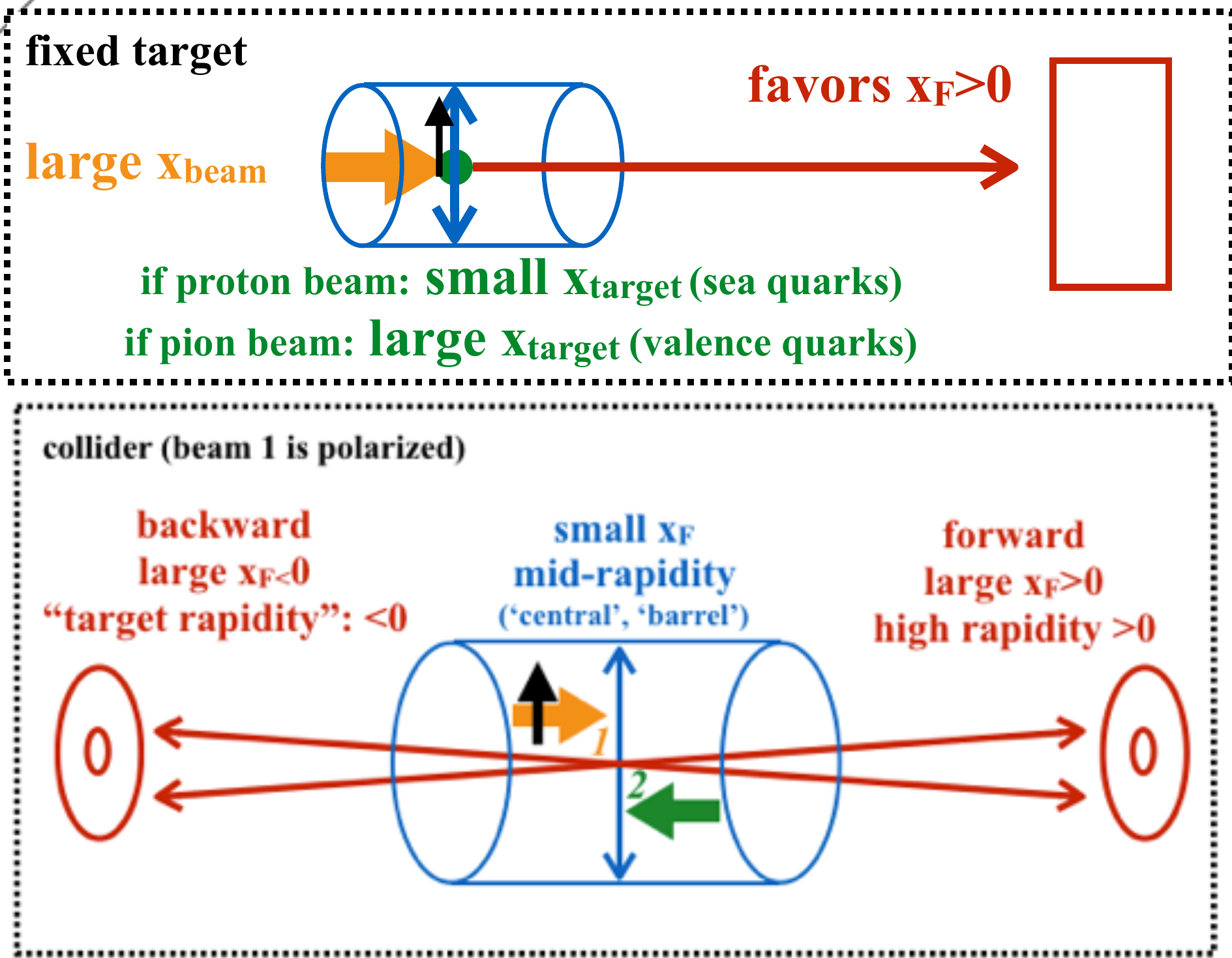}
\end{center}
\caption[Rapidity and Feynman-$x$]{Rapidity and Feynman-$x$ for fixed-target and collider topology.}
\label{fig:rapidity}
\end{figure}

It may be confusing that there is another variable $y$ in collider physics, which has nothing to do with the DIS-$y$. It is the rapidity: 
\begin{equation}
y=\frac{1}{2}\text{ln}\frac{1+\beta\cos(\theta)}{1-\beta\cos(\theta)}
\end{equation}
related to the polar angle $\theta$. For small masses $m\approx 0$, one defines the pseudo-rapidity 
\begin{equation}
\eta=-\ln\tan\frac{\theta}{2}.
\end{equation}
We speak of ``forwad'' large positive rapidity and ``backward'' for large negative rapidity. The Feynman-$x$ variable $x_\text{F}$ is
\begin{equation}
x_\text{F}  = x_1 - x_2,
\end{equation}
with $x_1$ the Bjorken-$x$ of parton 1 and $x_2$ that of parton 2. At a fixed-target experiment, $x_\text{F} = x_{\mathrm{beam}}-x_{\mathrm{target}}$. See also Fig.~\ref{fig:rapidity} for a geometry-motivated explanation of Feynman-$x$ and rapidity. 

\clearpage
\section{Elementary fermions}
\label{sec:elferm}
Elementary particles are those particles considered to have no substructure because no scattering experiment has to date revealed that they have one. Historically, atoms were literally introduced as elementary particles, ``atom'' referring to the Greek word for ``undividable'' (note that ``\emph{tomos}'' means ``slice'' and also appears in, e.g., ``tomography''). 
All particles are grouped into two categories: 
\begin{description}
\item {\bf Fermions} have a half-integer spin of $1/2\cdot n\hbar$ $(n = 1, 2, 3, \dots)$ and are subject to the Pauli exclusion principle: no two fermions may coincide in all quantum numbers (which enables, e.g., the structure of the nucleus' electron shell and its rich chemistry). Fermions obey the Fermi-Dirac statistics: 
\begin{equation}
\overline{n}_i=\frac{1}{\mathrm{e}^{(\epsilon_i-\mu)/k_\text{B}T}+1}.
\label{eq:fermistat}
\end{equation}
\item {\bf Bosons} have an integer spin of $n\cdot\hbar$ $(n = 0, 1, 2, \dots)$ and prefer to cluster in the \emph{same} quantum state (e.g., Lasers). Bosons obey the Bose-Einstein statistics:
\begin{equation}
\overline{n}_i=\frac{g_i}{\mathrm{e}^{(\epsilon_i-\mu)/k_\text{B}T}-1},
\label{eq:bosestat}
\end{equation}
with $g_i$ the degeneracy of state $i$.
\end{description}
Fermions and bosons can be elementary or composite. In this section we discuss elementary fermions. Together with the exchange bosons of the fundamental forces, which are discussed in Sec.~\ref{sec:fundforc}, they form the Standard Model of Particle Physics. Hadrons composed of quarks are treated in Sec.~\ref{sec:hadrons}. Hadrons can be fermions (baryons) or bosons (mesons). Every elementary particle has its anti-particle of the same mass but quantum numbers of the opposite sign, or ``anti'' in case of color charge (see Tab.~\ref{tab:conservation}). Composed anti-particles can be built from elementary anti-particles, for example an anti-hydrogen atom can be created from from an anti-proton and an anti-electron. There is a theorectical construct of a higher-level symmetry between particles, supersymmetry or SUSY, which assigns a bosonic partner to each fermion and a fermionic partner to each boson and as such unifies the description of matter and gauge-boson particles. Here we discuss only particles and not their superpartners. To date, no SUSY particle has been found in an experiment. 

As schematically shown in Tab.~\ref{tab:fermions}, there are two fundamental categories of elementary fermions: quarks, which carry the color charge of the strong interaction, and leptons, which do not carry color charge. There are six leptons and six quarks. Each category is found to consist of three generations of pairs with the following features: (1) The members of a generation are (much) heavier than the members of the previous generation of the same category. (2) Within a generation, the difference in electric charge is 1. (3) Within the same category, the generations are not qualitatively different. For example, a muon is really just a heavy electron. Many of the strange-quark phenomena (such as $\mathcal{CP}$ violation, see Sec.~\ref{sec:conqua}) appear again when studying bottom quarks, but at a heavier scale.
\begin{table}
\begin{center}
\begin{tabular}{c|ccc|ccc}
elementary & \small generation & \small generation & \small generation & electric & color  & \multirow{2}{*}{spin} \\
fermions & 1 & 2 & 3 & charge & charge &  \\\hline
\multirow{2}{*}{\textcolor{blue}{leptons}} &$\nu_\text{e}$&$\nu_\mu$&$\nu_\tau$&0&\multirow{2}{*}{-}&\multirow{2}{*}{1/2}\\
 &e&$\mu$&$\tau$&-1&&\\\hline
 \multirow{2}{*}{\textcolor{red}{quarks}} &u&c&t&+2/3&\multirow{2}{*}{r,g,b}&\multirow{2}{*}{1/2}\\
 &d&s&b&-1/3&&\\
 \end{tabular}
 \begin{tabular}{c|cc||c|cc}\hline\hline
\multirow{2}{*}{\textcolor{red}{quarks}}&current&constituent& \multirow{2}{*}{\textcolor{blue}{leptons}}&mass& life \\
&mass [MeV] & mass [MeV] &&[MeV]&time [s]\\\hline
d &5-15&$\approx$300 &$\nu_\text{e}$&small & \small can oscillate\\
u &2-8&$\approx$300&e&0.511&stable\\\hline
s &100-300&$\approx$450&$\nu_\mu$&small&\small can oscillate\\
c &\multicolumn{2}{c||}{$(1.0-1.6)\cdot10^3$}&$\mu$&105.7&$2.197\cdot10^{-6}$\\\hline
b &\multicolumn{2}{c||}{$(4.1-4.5)\cdot10^3$}&$\nu_\tau$&small&\small can oscillate\\
t &\multicolumn{2}{c||}{$(168-192)\cdot10^3$}&$\tau$&1776.8&$2.956\cdot10^{-13}$\\\hline
\end{tabular}
\caption[Elementary fermions.]{The twelve elementary fermions and their properties. There are three neutrino ($\nu$) flavors and three charged leptons (electron e, muon $\mu$, and tauon $\tau$). Quarks come in six flavors: up (u), down (d), charm (c), strange (s), top (t,) and bottom (b).
There is an analog table for the elementary anti-fermions (not included here). Anti-particles are usually written with a horizontal line over the letter, for example $\overline{\text{u}}$ for anti-u-quark. The anti-electron (e$^+$ or $\overline{\text{e}}$) is called positron.}
\label{tab:fermions}
\end{center}
\end{table}

Anticipating the strong nuclear force and hadrons discussed in Sec.~\ref{sec:hadrons}, the vast majority of hadron mass is dynamically generated when quarks come close to each other and form a bound state in the color field of the strong interaction. Only a tiny fraction of a hadron's mass is contributed by the bare quark masses, which are generated by the Higgs mechanism (see below). The latter mass is referred to as ``current mass'', while the mass contributed by each ``dressed quark'' including the mass generated by the interaction is called ``constituent mass''. 

All matter on Earth consists of u-quarks, d-quarks and electrons. Not only in our daily life, but much of the visible matter in the universe nowadays consists of only those three charged elementary fermions of the lightest generation 1, and the electron-neutrino to balance radioactive weak decays. With few exceptions, the members of the other two generations have existed only for a very short time after the Big Bang at very high temperatures and densities and have converted to members of generation 1 since then. In today's cold and expanding universe, the state of smallest possible mass is the stable one. Only hot spots like stars can produce particles from generation 2 (and to a much lesser extent of generation 3). In the late 1940's, such ``strange'' particles were discovered in cosmic rays. The only earthly hot spots are particle colliders in the Giga- or Tera-electronvolt regimes such as LEP, Tevatron, HERA, RHIC, and the LHC, where the ``universe is re-molten into the state it had at birth''.

The Nobel Prize in Physics in 2015 was awarded to Takaaki Kajita and Arthur B. McDonald ``for the discovery of neutrino oscillations, which shows that neutrinos have mass''. Neutrino oscillation refers to the feature that a neutrino changes its flavor ($\nu_\text{e}$, $\nu_\mu$, or $\nu_\tau$) depending on the distance it has traveled. This is possible because neutrinos have mixed flavor and mass eigenstates. 

The much-celebrated Higgs boson, whose discovery at CERN was announced by the CMS and ATLAS collaborations in 2012, is thought to provide the process by which the elementary particles attain their masses and is considered to bring closure to the Standard Model of Particle Physics (SM). The 2013 Nobel Prize in Physics was awarded to  to Francois Englert and Peter W. Higgs ``for the theoretical discovery of a mechanism that contributes to our understanding of the origin of mass of subatomic particles [...]''. Not much success has been made so far in experimentally identifying physics beyond the SM, but the searches continue.

\section{Fundamental forces}
\label{sec:fundforc}

\subsection{The four fundamental forces}
\label{sec:4forces}
In quantum field theory, a ``force'' or ``interaction'' is treated as the exchange of elementary virtual bosons with squared momentum transfer $Q^2$. These intermediate exchange vector bosons, or gauge particles, or ``force carriers'', couple to a charge (electric, weak, or color). The strength of each interaction is described by a coupling constant ``$\alpha$'', which generally depends on $Q^2$. The features of the four known fundamental forces are summarized in Tab.~\ref{tab:forces}. All electrically charged particles are subject to the electromagnetic force. Quarks and leptons are sensitive to the weak interaction. Only quarks and hadrons are sensitive to the strong nuclear force.

The gluon, the carrier of the strong nuclear force, was discovered at DESY in 1979 (TASSO collaboration at PETRA in e$^+$e$^-$ collisions and detection of 3-jet events). 
The W$^\pm$ and Z$^0$ bosons, the carriers of the weak nuclear force, were discovered at CERN in 1983 (UA1 and UA2 collaborations in proton-antiproton collisions at the SPS).

The gravitational force we will not discuss further here, as generations of physicists before us in this situation. There are major obstacles of transferring the concept of a quantum field theory to gravitation, which, in very short, is related to the incompatibility of ``wild quantum fluctuations'' at short distances and the ``desire'' of space-time (as in general relativity) to be curved smoothly.

\begin{table}
\begin{center}
\begin{tabular}{c|ccccccc}
      & \small exchange & \small boson &  & \small sensitive & {\small relative} & \small life& \\
force & \small boson(s) &  \small mass & charge &  \small particles &  {\small strength} & \small time &  \small reach \\
& \small (carrier) & \small [GeV] &&&& \small [s] & \small [m]\\ \hline
&&&&&&& \\
{\small electro-} & \multirow{2}{*}{\small photon $\gamma^\star$} &  \multirow{2}{*}{0} &  \multirow{2}{*}{{\small electric}} &  {\small electrically} &  \multirow{2}{*}{\small $10^{-2}$} &  \multirow{2}{*}{\small $10^{-16}$} &  \multirow{2}{*}{\small $\infty$}\\ 
{\small magnetic} &  &  & &  {\small charged} &  & & \\ 
&&&&&&& \\\hline
&&&&&&& \\
 \multirow{2}{*}{\small weak} &   W$^\pm$ &  \small 80.39  &  \multirow{2}{*}{weak} & {\small quarks \&} & \multirow{2}{*}{\small $10^{-5}$}& \multirow{2}{*}{\small $10^{-8}$}& \multirow{2}{*}{\small $10^{-18}$} \\ 
 &   Z$^0$ &  \small 91.19 && {\small leptons}   &&& \\ 
&&&&&&& \\\hline
&&&&&&& \\
 \multirow{2}{*}{strong} & \multirow{2}{*}{\small 8 gluons} &  \multirow{2}{*}{0} &  \multirow{2}{*}{\small color} &\small quarks \& &  \multirow{2}{*}{1} &  \multirow{2}{*}{\small $10^{-24}$} &  \multirow{2}{*}{\small $10^{-15}$}\\ 
&&&&\small hadrons&&& \\
&&&&&&& \\\hline
&&&&&&& \\
{\small gravitation}   & {\small graviton} & 0 & \small mass & all & \small $10^{-42}$ && \small $\infty$ \\ 
&&&&&&& \\\hline
\end{tabular}
\end{center}
\caption[Fundamental forces]{The four fundamental forces and their typical features. Table based in parts on Ref.~\cite{Close2004}.}
\label{tab:forces}
\end{table}

\subsection{QED vs.~QCD}
\label{sec:qedqcd}
Quantum electro dynamics (QED) is the U(1) gauge field theory of the electromagnetic interaction, which is mediated via virtual photons of zero mass and zero electrical charge. 
The QED coupling constant (often called electromagnetic fine structure constant) $\alpha_{\mathrm{em}}$ can be expressed as
\begin{equation}
\alpha_{\mathrm{em}}=\frac{e^2}{4\pi\epsilon_0\hbar c}\approx\frac{1}{137}.
\label{eq:alphaem}
\end{equation}
Equation~\ref{eq:alphaem} is an approximation that works well at low energies and in worldly applications. However in the exact calculation of the coupling constant at higher energies and smaller distances, radiative corrections to the basic Feynman diagrams of QED must be included, yielding a sum over all relevant QED higher-order diagrams. The coupling constant becomes $Q^2$-dependent and is therefore called ``running''.
To interpret this graphically, one can consider Eq.~\ref{eq:alphaem} as the limiting case for large distances from the electric charge. In a small volume around the electric charge, virtual electron-positron pairs make the vacuum a dielectric medium, thus screening the true charge and lowering its perceived magnitude at a distance \cite{PeskinSchroeder}. This phenomenon is referred to as vacuum polarization. Upon penetrating the screening volume, the true charge becomes apparent and $\alpha_{\mathrm{em}}$ rises with decreasing distance. Recalling that $\hbar/\sqrt{Q^2}$ is a measure of the spatial resolution, this means that $\alpha_{\mathrm{em}}$ rises with increasing $Q^2$. 

Quantum chromo dynamics (QCD), one the other hand, is the SU(3) gauge field theory of the strong nuclear interaction, which is mediated via gluons of zero mass and color/anti-color charge. The QCD ``running''  coupling ``constant'' reads:
\footnote{With $n_\text{f}$ the number of participating quarks and the scale $\Lambda$ the only free parameter of QCD. It has to be determined from experimental data, $\Lambda\approx 250$~MeV.}
\begin{equation}
\alpha_{\mathrm{S}}=\frac{12\pi}{(33-2n_\text{f})\cdot\mathrm{ln}(\frac{Q^2}{\Lambda^2})}.
\end{equation}
Because the QCD field exchange particles - the gluons - carry the charge that the very interaction is sensitive to, they also interact with each other, unlike the virtual photons in QED. Photons do not carry the (electric) charge of the (electromagnetic) interaction. In the language of the quantum field theory, this is reflected by the construction of QCD as a non-Abelian gauge theory - non-Abelian, i.e., the gauge fields do not commute with each other -, while QED is Abelian.

\begin{figure}
\begin{center}
\includegraphics[width=.9\textwidth]{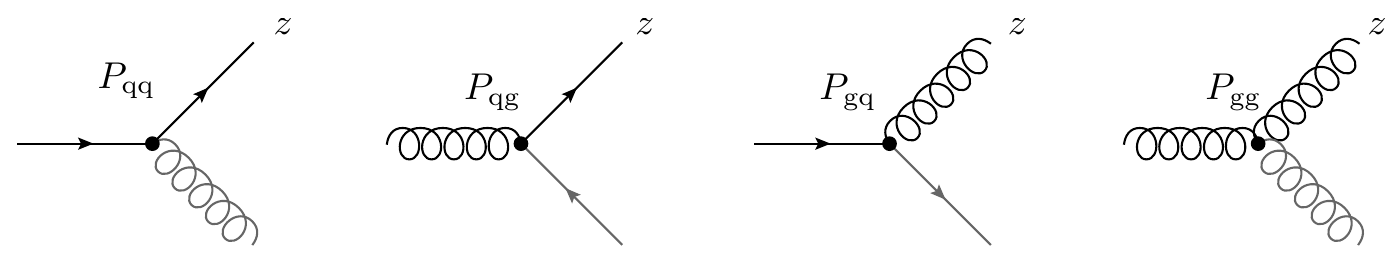}
\end{center}
\caption[Basic QCD interactions]{The four basic QCD interactions. Quarks are indicated by straight lines, gluons by curly lines.}
\label{fig:appqcd_processes}
\end{figure}

Of the four basic QCD interactions, only two are found to have corresponding counterparts in QED: gluon emission by a quark (bremsstrahlung in QED - a photon is radiated off an electron or a positron) and splitting of a gluon in a quark-antiquark pair (electron-positron pair production from a photon of sufficiently high energy). The other two QCD diagrams, gluon self-coupling via a 3- or 4-fold vertex, have no correspondence in QED because photons ``do not see'' (do not couple to) each other. The four QCD interactions are summarized in Fig.~\ref{fig:appqcd_processes}.

The self coupling of gluons has the peculiar effect that QCD can be described by two distinct dynamic regimes. As long as the quarks making up a proton are within the small radius of 10$^{-15}$\,m (1 fermi),
they exert relatively small forces onto each other, following a Coulomb-like QCD potential $\sim 1/r$ as in the case of QED. The situation changes drastically when one tries to remove a quark from the small radius around the proton, which starts happening at energies of around 1\,GeV (Eq.~\ref{eq:evm}). Then an additional linear component $\sim r$ in the QCD potential becomes relevant, which makes the potential energy in the color field grow linearly as the quarks are pulled apart. We can now define the two QCD regimes:
\begin{description}
\item Small distances $\leftrightarrow$ high energies $\leftrightarrow$ large $Q^2$, small coupling between quarks, pQCD applicable: {\bf asymptotic freedom}
\item Large distances $\leftrightarrow$ low energies $\leftrightarrow$ small $Q^2$, large coupling between quarks, pQCD \emph{not} applicable $\rightarrow$ lattice QCD: {\bf confinement}
\end{description}
Note that 1~fermi is about the distance light travels in $10^{-24}$~seconds, which is the typical reaction time we identified for the strong nuclear force in Tab.~\ref{tab:forces}.  
The Nobel Prize in Physics was in 2004 awarded ``for the discovery of asymptotic freedom in the theory of the strong interaction'' jointly to Gross, Politzer and Wilczek (Sec.~\ref{sec:nobel}).
In the regime of asymptotic freedom with small quark distances, perturbative QCD methods (pQCD) can be applied.
In order for the perturbation series to converge, $\alpha_\text{S}$ must be $\ll 1$, thus $Q^2\gg \Lambda^2\approx 0.06$\,GeV$^2$. In the non-perturbative regime, there has recently been made significant progress with lattice QCD, where quantum-mechanical calculations are performed on a discrete lattice of space-time points. 

\begin{figure}
\begin{center}
\includegraphics[width=0.95\textwidth]{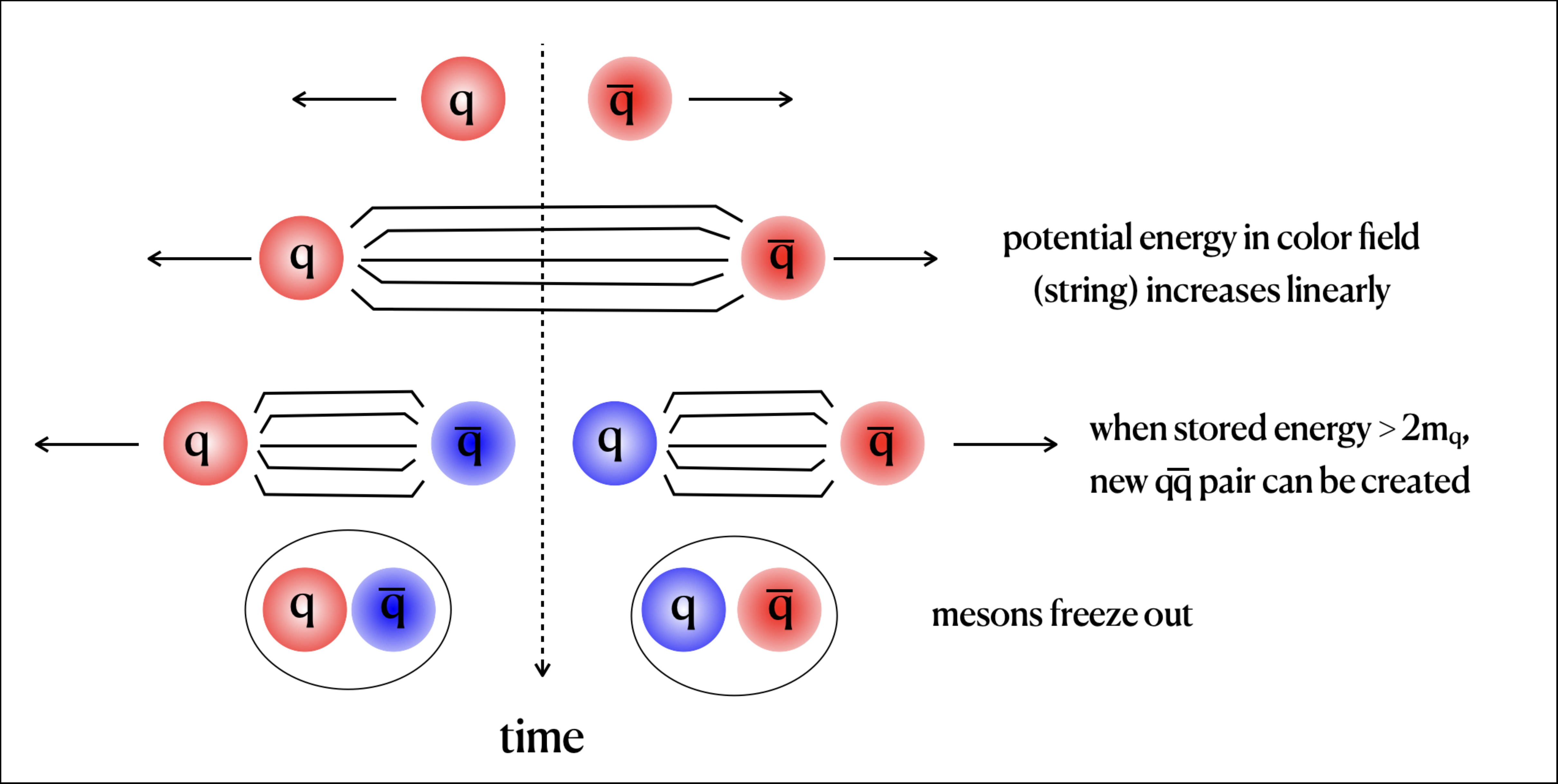}
\end{center}
\caption[String breaking in QCD hadronization]{String breaking in QCD as quarks are separated from each other.}
\label{fig:stringbreaking}
\end{figure}

What is happening when the quarks are pulled apart? Because of the attractive interactions (self coupling) between the exchanged gluons, the field lines of the color field between the quarks are squeezed into a tube or a string under tension, which stores potential energy as the quarks are pulled apart. In this context, the rubber band analogy is sometimes used 
- it doesn't get easier the more one pulls. 
As the quarks are separated further, the potential energy stored in the color field becomes sufficient to create quark-antiquark pairs and the string breaks into smaller strings, until there are only ``white'' (colorless) mesons and baryons left over \cite{Thomson2013}. This fragmentation process is illustrated in Fig.~\ref{fig:stringbreaking}.

No isolated color charge has been observed in particle detectors to date. The hadronization process results in jets (cone-shaped sprays) of hadrons, which are considered to be the macroscopic manifestation of perturbative QCD. 
\label{page:jets}
In the analysis of experimental high-energy physics data, a jet is a final-state construct introduced for the ease of interpreting the data and finding relevant signatures. Jet finding algorithms usually look for collimated sprays of particles in a cone with typical radii $R=\sqrt{\Delta\phi^2 + \Delta\eta^2}=0.4$ (with $\phi$ the azimuthal angle and $\eta$ the polar-angle related rapidity), applying purely geometrical and combinatorial considerations. Finding complete jets is difficult if the acceptance is limited and/or if the involved processes are too soft, which is why jet-analysis techniques are typically not applied at fixed-target nuclear experiments.

\section{Hadrons}
\label{sec:hadrons}
Particles composed of quarks are called hadrons. Quarks and gluons are generically referred to as partons. The quarks inside hadrons are held together by gluons, the force carriers of the strong nuclear interaction (Sec.~\ref{sec:4forces}), and they are subject to the peculiar effects of asymptotic freedom and confinement described in Sec.~\ref{sec:qedqcd}. 
Typical representatives of hadrons are baryons, three-quark compounds $|\text{qqq}\rangle$ (q~$\in\{\text{u,d,s,c,b,(t)}\} $) and thus fermions, and mesons, two-quark compounds $|\text{q}\bar{\text{q}}\rangle$ and thus bosons. Searches for exotic hadrons such as treta- or pentaquarks with four or five quarks are ongoing. 

The proton is the lightest baryon with a mass of 938\,MeV. The neutron is slightly heavier at around 940\,MeV. Protons and neutrons are collectively called nucleons. The proton is the only stable baryon,\footnote{No proton decay has been observed to date. The proton's lifetime as determined by experiments is currently beyond $10^{34}$~years.} unless it is very close to other protons at very high temperatures such as the center of our Sun. Then a proton can transmute into a neutron. The free neutron decays with half-life time of about 10~minutes into a proton, electron and anti-electron-neutrino unless it is bound inside a nucleus. All other baryons decay into lighter baryons. One can define a baryon as ``at the end of the decay chain is a proton'' and a meson as ``at the end of the decay chain there is no proton''. The stability of the proton is a consequence of the conservation of baryon number, see Sec.~\ref{sec:conqua}. 

There are no stable mesons. The lightest meson, the pion $|\pi\rangle=|\text{ud}\rangle$,\footnote{The three pions of different electrical charge are usually grouped together:\\ $|\pi^+\rangle=|\text{u}\bar{\text{d}}\rangle$, $|\pi^-\rangle=|\bar{\text{u}}\text{d}\rangle$ and $|\pi^0\rangle=1/\sqrt{2}(|\text{u}\bar{\text{u}}\rangle-|\text{d}\bar{\text{d}}\rangle)$} has a about mass of only 140~MeV, even though it contains two of the three quarks that make up the proton with almost seven times the mass of the pion. This mass difference is a manifestation of a large fraction of hadron mass being dynamically generated (see also Sec.~\ref{sec:qedqcd}). The kaon, $|\text{K}\rangle=|\text{us}\rangle$, has a mass of almost 500\,MeV - i.e., halfway between pion and proton and also not explainable by the bare quark masses only. 

Meson number does not have to be conserved and therefore mesons can decay into leptons. In general, anything that is not forbidden in particle physics will happen.  There may however be kinematically suppressed channels. Looking at a large number of decays of the same particle, the fractions for different decay channels are called \emph{branching ratios}. The charged pion decays in almost 100\% of the cases into a muon and (anti-)muon-neutrino, while the neutral pion decays to almost 100\% into 2 photons.

Only color-neutral, ``white'' hadrons are detected in our particle detectors: the 3 quarks in a baryon have different colors, the triplet of which combines to ``white'', and mesons are made of quark pairs of color plus anti-color, which effects to ``white''.

The name ``lepton'' comes from the Greek word for ``small'' or ``weak'' since leptons are elementary and do not interact via the strong force. The term ``hadron'' was coined after the Greek word for ``strong'' since all hadrons are subject to the strong nuclear force (see Sec.~\ref{sec:fundforc}). 
``Baryon'' contains the Greek word for ``heavy'' and ``meson'' that for ``middle'' - they are not as heavy as protons, but heavier than electrons.

\newpage
\subsection{The nucleon in a nutshell}
\label{sec:nucleon}
\begin{wrapfigure}{r}{0.5\textwidth}
\begin{center}
\includegraphics[width=0.5\textwidth]{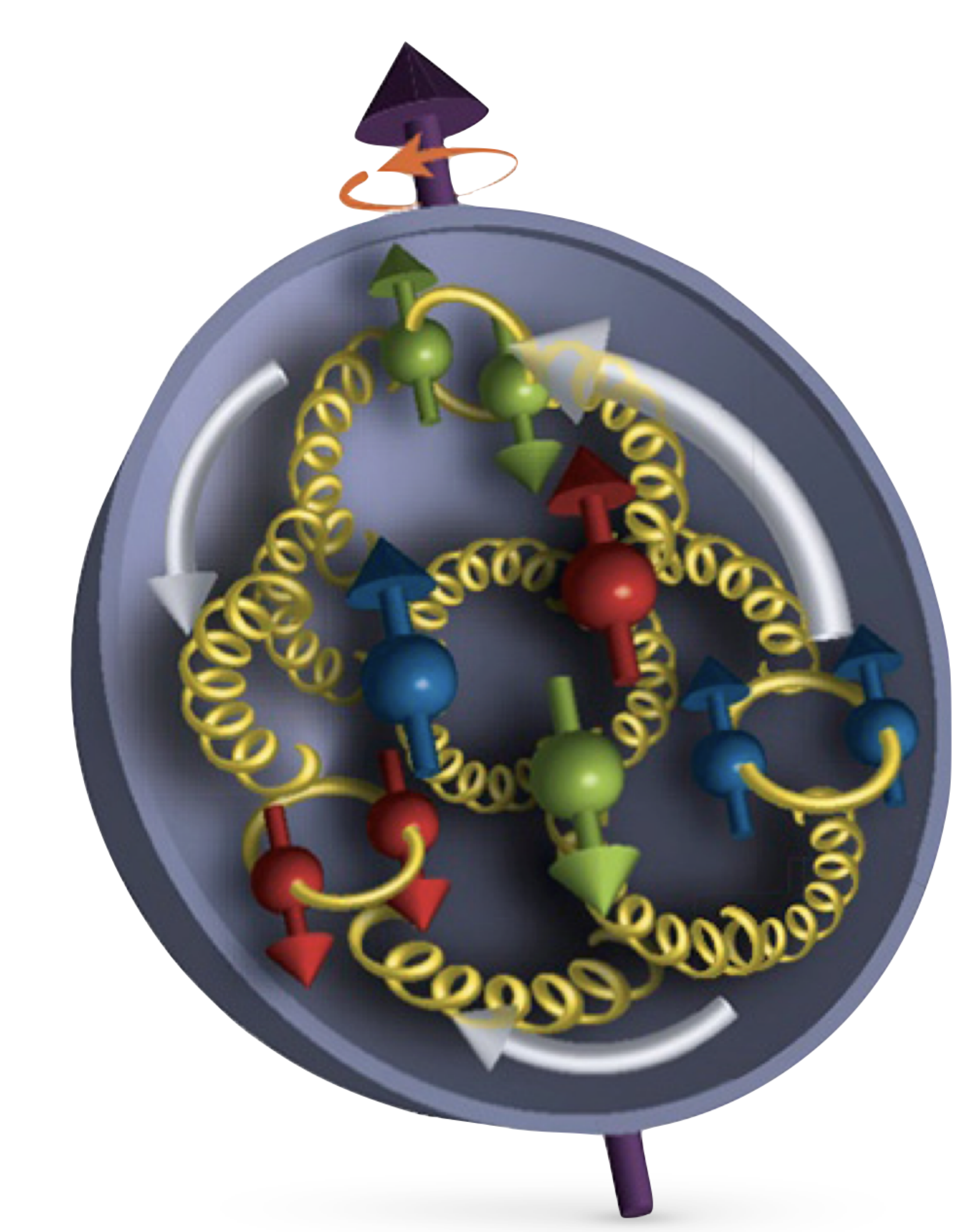}
\end{center}
\caption[Artistic rendering of the nucleon]{Artistic rendering of the nucleon with quarks and gluons.}
\label{fig:proton}
\end{wrapfigure}
The three valence quarks in the nucleon determine its quantum numbers (like electrical and color charge, baryon number, strong isospin and strangeness, and others, see Sec.~\ref{sec:conqua}). All other quarks in the nucleon - the sea quarks - occur as quark-antiquark pairs, the creation energy for which can be ``borrowed'' from the vacuum for a sufficiently short time (uncertainty principle). The sea quarks' effective quantum numbers average to zero and thus do not contribute to the nucleon's quantum numbers. Figure~\ref{fig:proton} represents a sketch of the nucleon's inside with its valence quarks with three different colors, quark-antiquark pairs with color-anticolor and the gluons as yellow wavy lines. Valence quarks, sea quarks and gluons possess spin and orbital angular momentum, which contribute to the nucleon's total angular momentum.

\subsection{Three quarks for Muster Mark}
\label{sec:quarks}

The title of this subsection stems from James Joyce’s \emph{Finnegans Wake}. This line inspired Murray Gell-Mann in 1964 to name the fundamental constituents of the nucleon ``quarks''. Let's take a quick walk through the exciting history of early particle physics. 

With the realization in the late 19th century that ``the atom'' is not an elementary particle and with the identification of the electron, the proton and the neutron, the old desire of mankind to explain matter in terms of just a handful ingredients flamed up again. This hope was short lived. Once more, nature turned out to be more complex. 

Electrons were first observed as cathode rays in the 1880's, but they were not yet identified as part of the atom. Based on the Geiger–Marsden experiments carried out in 1908 and following years (Rutherford gold-foil experiments), Ernest Rutherford established the modern atomic model in 1911: the mass of the atom is concentrated in its nucleus, which is positively charged. Electrons are negatively charged and orbit the nucleus. The atom is \emph{vastly} empty. Rutherford gave the proton its name because it is the nucleus of the lightest atom: ``proton'' comes from the Greek word for ``the first''. Coining the word ``neutron'', Rutherford also anticipated the existence of another proton-like, but electrically neutral particle since the masses of the investigated nuclei were larger than what would have been expected from their positive charge. Shortly after, James Chadwick experimentally identified the neutron in 1932. Together with photons (based on the Greek word for ``light''), electrons, protons and neutrons had become the well-known particles of daily life. It was also known that there must be a particle that is not detected but that carries away missing energy in the radioactive beta decay, and this particle was called neutrino.

Until 1947, the situation remained relatively simple. Cosmic ray experiments using cloud chambers had revealed the existence of a few ``exotic particles'': the positron (the first anti-particle), the muon (first 2nd-generation particle) and the pion (the first meson). The detection of the positron in 1933 (Carl Anderson at Caltech) was a triumph after Paul Dirac's prediction of anti-particles from 1928. The discovery of the neutral pion in 1947 (Cecil Powell at Bristol) matched well Hideki Yukawa's prediction from 1935 that mesons mediate the strong nuclear force. Something like such strong force should exist because otherwise the atomic nucleus would drift apart due to electrostatic repulsion between the positively charged protons. The muon had been discovered in cosmic rays already in 1937 (Carl Anderson and Seth Neddermeyer at Caltech) and was at first mistakenly identified as the Yukawa particle. Thus for every new discovery, there was a solid theoretical explanation available and the new exotic particles were not considered to be first glimpses of new physics, even though the muon did not really seem to fit in since it was not needed to explain the behavior of nuclei. 

For the discovery of the pion, the usage of photographic emulsion plates at high altitudes had been essential. It then also enabled the detection of ``strange particles'': the kaon (us) in 1947 and the Lambda baryon (uds) in 1950 with experimental signatures that did not fit to any of the previously discovered particles. These hadrons are usually generated via the strong interaction, but they decay via the weak interaction and they can create hadrons in their decay, even though the decay does not proceed via the strong interaction. This was considered to be strange. It was later realized that these particles contain a quark of the 2nd generation, which received the name ``strange quark''. By 1960, a whole ``zoo of particles'' was known. 

\begin{figure}
\begin{center}
\hspace{.11\textwidth}\includegraphics[width=0.79\textwidth]{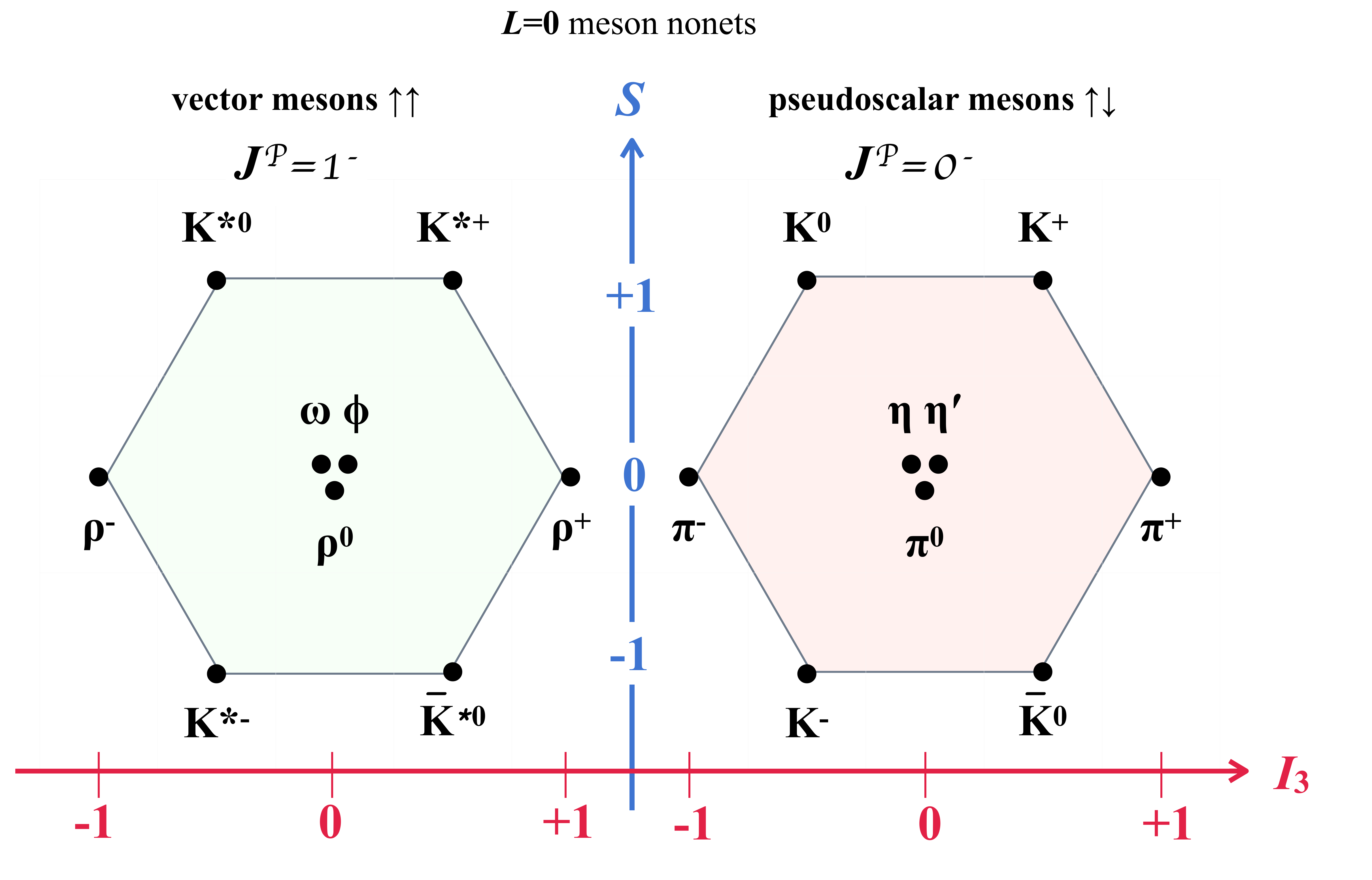}
\includegraphics[width=0.95\textwidth]{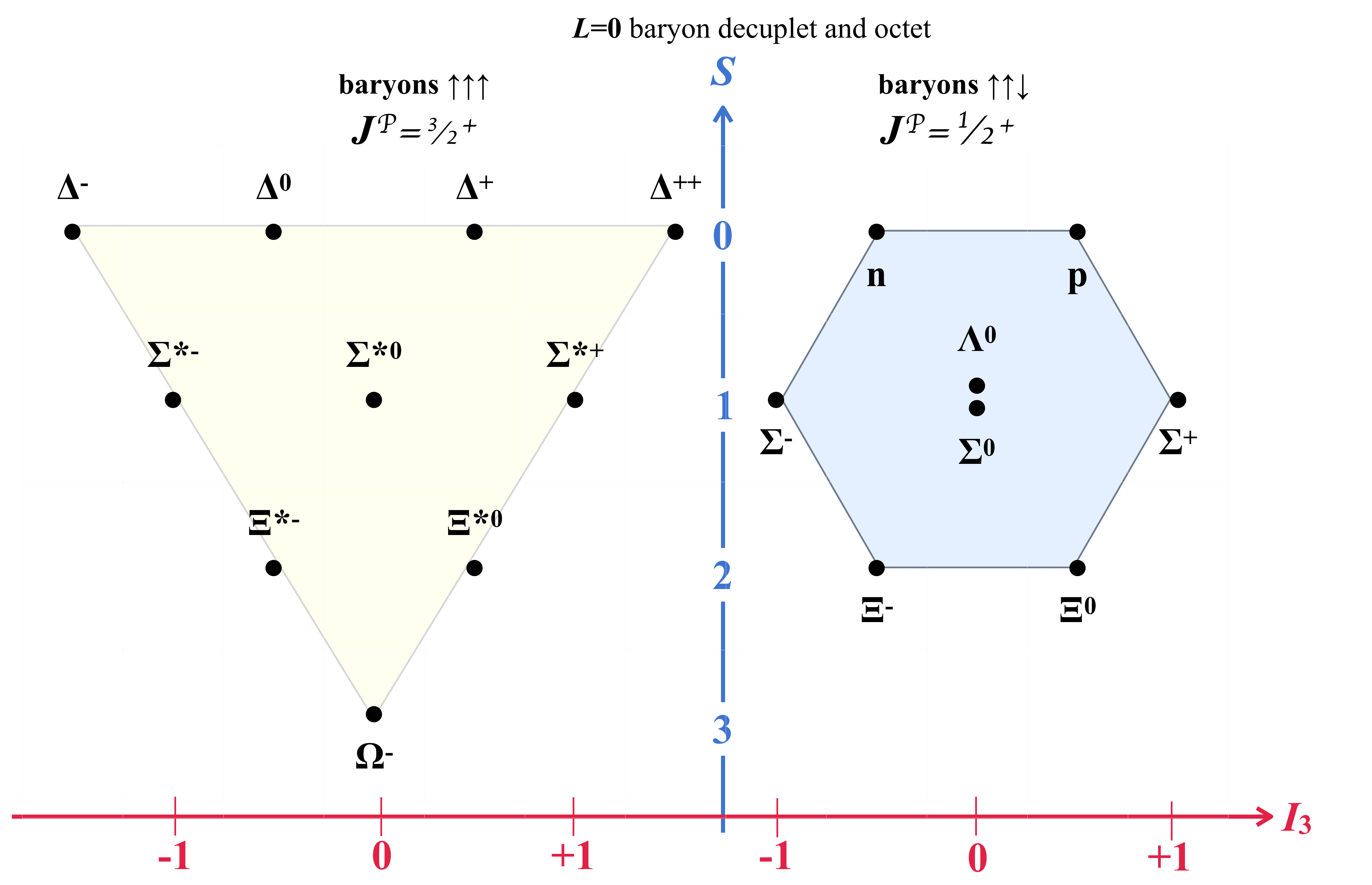}
\end{center}
\caption[Meson and baryon multiplets]{Meson and baryon multiplets for $L=0$ (no orbital excitation): meson nonets (top), baryon decuplet and baryon octet (bottom) ordered by the 3-component of the isospin $I_3$ (x-axis) and strangeness $S$ (y-axis). The vertical arrows above each table indicate the spin directions of the constituent quarks.}
\label{fig:multiplets}
\end{figure}

In 1961, Gell-Mann and Yuval Ne'eman realized that the hadrons could be ordered in groups, called multiplets, of 8, 9 or 10 members with zero average isospin 3-component $\langle I_3\rangle=0$, same hypercharge $Y$, and common spin-parity quantum number $J^{\mathcal{P}}$ (see Sec.~\ref{sec:conqua}), as shown in Fig.~\ref{fig:multiplets}. Electrical charge, strangeness and isospin are related through Eq.~\ref{eq:GMN}.
The three heavier quarks discovered after 1961 also found their place in the multiplet ordering scheme. The J$/\psi$ particle was discovered in the \emph{November revolution} of 1974 at Brookhaven National Lab in proton-beryllium scattering (Samuel C.~C.~Ting's MIT group) and SLAC in electron-positron collisions (Burton Richter) and was later realized to be a $\text{c}\overline{\text{c}}$ state, with the ``new'' charm quark forming a duo with the so-far unpartnered s-quark. Around the same time, the heaviest lepton was found - the $\tau$-lepton (Martin L. Perl and Y.-s. Tsai). The next heavier quark, the bottom quark, was found at Fermilab in 1977 in proton-copper/platinum scattering with the discovery of the $\Upsilon$ particle, a $\text{b}\overline{\text{b}}$ state (Leon Lederman). It took almost two decades for the heaviest quark, the top quark with a mass of a gold atom, to be found at Fermilab in 1995 in proton-antiproton collisions (Tevatron CDF and D0 collaborations). A complete description of meson and baryon multiplets can, e.g., be found in the Particle Data Group's (PDG) listings \cite{Zyla:2020zbs}.

This organization scheme is sometimes called the \emph{eightfold way} in allusion to the Noble Eightfold Path of Buddhism. The quark model was established in 1963 - hadrons are made of colored quarks; but hadrons are always color neutral. Nature's order scheme and symmetry was found to be represented once more by elements of group theory: the quantum numbers of hadrons turn out to be one-to-one related to members of the special unitary Lie group SU(3). The 1969 Nobel Prize in Physics was awarded to Gell-Mann ``for his contributions and discoveries concerning the classification of elementary particles and their interactions''. 

\section{Conservation laws}
\label{sec:conqua}

Conservation laws are based off symmetries we find in nature. The Emmy-Noether theorem teaches us that \emph{when there exists a symmetry transformation that does not change the Lagrange function, there is a corresponding conserved quantity.} Examples are: space and time are homogeneous, i.e., there is translational invariance, and it follows that momentum is conserved; space is isotropic, i.e., there is rotational invariance, and it follows angular momentum is conserved. Table \ref{tab:conservation} gives an overview over conserved quantities that are essential for particle physics. 

\begin{table}
\begin{center}
\begin{tabular}{cc|ccc|l}
symbol & name & {\small elmag?} & {\small weak?} & {\small strong?} & anti-particle remarks \\ \hline
\multirow{2}{*}{Q} & \small electrical &\multirow{2}{*}{$\times$} &\multirow{2}{*}{$\times$} &\multirow{2}{*}{$\times$} & \footnotesize charged anti-particles have \\
&\small charge&&&&\footnotesize opposite-sign charge\\\hline
\multirow{2}{*}{r,g,b} & \small color  & & & \multirow{2}{*}{$\times$} & \small anti-hadrons:  \\
&\small charge &&&&\small color$\rightarrow$anti-color  \\\hline
\multirow{2}{*}{$B$} & \small baryon & \multirow{2}{*}{$\times$} & \multirow{2}{*}{$\times$} & \multirow{2}{*}{$\times$} & \small $B=+1$ for baryons, \\
& \small number &&&&\small  $B=-1$ for anti-baryons\\\hline
& \small lepton &  &  &  & \small $\ell=(e,\mu,\tau)$  \\
$L_\ell$ &\small generation &$\times$&\Large $\text{\Lightning}$ &$\times$&\small $L_\ell=1$ for leptons \&\\
&\small number&&&& \small $L_\ell=-1$ for anti-leptons \\\hline
\multirow{2}{*}{$S$} & \multirow{2}{*}{\small strangeness} &\multirow{2}{*}{$\times$}&\multirow{2}{*}{\Large $\text{\Lightning}$} &\multirow{2}{*}{$\times$}& \small $s$-quark: $S=-1$,\\
&&&&& \small anti-$s$-quark: $S=+1$\\\hline
\multirow{2}{*}{$I$} & \small strong & & & \multirow{2}{*}{$\times$} & \footnotesize isospin n-tuplets contain  \\
&\small isospin &&&& \footnotesize their anti-particles\\ \hline
\multirow{2}{*}{$T$} & \small weak & & \multirow{2}{*}{$\times$} && \multirow{2}{*}{\small see remark for $I$}  \\
&\small isospin &&&&\\ \hline
 \multirow{2}{*}{$\mathcal{P}$} &  \multirow{2}{*}{\small parity} &  \multirow{2}{*}{$\times$}&  \multirow{2}{*}{\Large $\text{\Lightning}$}& \multirow{2}{*}{$\times$} & \small particle and anti-particle  \\
 &&&&& \small have opposite parity\\\hline
 \multirow{2}{*}{$\mathcal{C}$} & \small charge &  \multirow{2}{*}{$\times$}&  \multirow{2}{*}{\Large $\text{\Lightning}$}& \multirow{2}{*}{$\times$} &\footnotesize is the anti-particle operator   \\
 &\small conjugation&&&&\footnotesize also for neutral particles\\\hline
  \multirow{2}{*}{$\mathcal{CP}$} & \small charge conj. &  \multirow{2}{*}{$\times$}&  \multirow{2}{*}{\Large ($\text{\Lightning}$)}& \multirow{2}{*}{$\times$} &\footnotesize makes anti-particle    \\
 &\small \& parity&&&&\footnotesize \& mirrors\\\hline
   \multirow{2}{*}{$\mathcal{CPT}$} & \small $\mathcal{CP}$ \& time- &  \multirow{2}{*}{$\times$}&  \multirow{2}{*}{$\times$}& \multirow{2}{*}{$\times$} &\footnotesize makes anti-particle,    \\
 &\small reversal $\mathcal{T}$&&&&\footnotesize mirrors \& reverses time\\\hline
\end{tabular}
\end{center}
\caption[Conservation laws]{Conservation laws and symmetry violations in particle physics. ``$\times$'' indicates that a quantity is conserved for a specific interaction (see Tab.~\ref{tab:forces}) - electromagnetic (elmag), weak nuclear and strong nuclear. ``{\Large $\text{\Lightning}$}'' indicates that the quantity is not conserved and thus the underlying symmetry is violated, see text for more details. }
\label{tab:conservation}
\end{table}

\subsection{Particle types}
\label{sec:partypes}
The lepton generation number $L_\ell$ is not conserved when neutrinos oscillate between flavors (see Sec.~\ref{sec:elferm}). Charged leptons have not been observed to violate lepton-generation number conservation. In the weak interaction, strangeness changes by $\Delta S=1$ or $\Delta S=2$. Similarly, also the quantum numbers related to charm ($C$), bottom (bottomness $B^\prime$), and top (topness $T^\prime$) flavors are preserved under the strong and electromagnetic interactions, but not under weak interaction. The hypercharge $Y$,
\begin{equation}
Y=B+S+C+B^\prime+T^\prime,
\label{eq:hypercharge}
\end{equation}
and the weak hypercharge $Y_{\mathrm{W}}$,
\begin{equation}
Y_{\mathrm{W}}=2(Q-T_{3}),
\label{eq:weakhyper}
\end{equation}
are conserved in the strong, but not in the weak interaction. Another way of thinking about strangeness conservation is that spin flips can happen fast (e.g., the strong decay $\Delta^+\rightarrow \text{p}\pi^0$, effectively $\text{u}^\uparrow \text{u}^\uparrow \text{d}^\uparrow\rightarrow \text{u}^\uparrow \text{u}^\uparrow \text{d}^\downarrow +\pi^0$), while changing the quark generation takes a longer time (e.g., the weak decay $\Sigma^+\rightarrow \text{p}\pi^0$, $\text{u}^\uparrow \text{u}^\uparrow \text{s}^\downarrow\rightarrow \text{u}^\uparrow \text{u}^\uparrow \text{d}^\downarrow +\pi^0$, with strangeness non-conservation). 

\subsection{Angular momentum and isospin}
\label{sec:isospin}

Orbital angular momentum, spin, and isospin obey the same angular momentum algebra. With the generic angular momentum operator $\widehat{\vec{J}}$ and its 3-component $\widehat{J_3}$, one has the Lie algebra
\begin{equation}
[\widehat{J_i},\widehat{J_j}]=i\hbar\epsilon_{ijk}\widehat{J_k},
\label{eq:Liealg}
\end{equation}
from which the quantization of angular momentum in the microscopic world can be derived. Because $\widehat{\vec{J}}\;^2$ and $\widehat{J}_k$ commute, $[\widehat{\vec{J}}\;^2,\widehat{J}_k]=0$, they have common eigenvectors $|jm\rangle$ forming the eigenstates of angular momentum with eigenvalues $j$ and $m$:
\begin{eqnarray}
\widehat{\vec{J}}\;^2|jm\rangle=&j(j+1)\hbar^2&|jm\rangle, \label{eq:j}\\
\widehat{J}_3|jm\rangle=&m\hbar&|jm\rangle, \;\; m=-j...+j. \label{eq:m}
\end{eqnarray}
The algebra of angular momentum is described by the special unitary Lie group SU(2), which is generated by anti-Hermitian, trace-zero matrices, the Pauli spin matrices $\tau_1$, $\tau_2$ and $\tau_3$: 
\begin{equation}
\tau_1=\begin{pmatrix}0&1\\1&0\end{pmatrix}, \tau_2=\begin{pmatrix}0&-i\\i&0\end{pmatrix},  \tau_3=\begin{pmatrix}1&0\\0&-1\end{pmatrix}.
\label{eq:paulimat}
\end{equation}
The elements $U$ of SU(2) rotate one eigenstate into another: 
\begin{equation}
U=\mathrm{exp}(-i\theta n\tau)
\label{eq:Lierot}
\end{equation}
A prominent example of a quantity following algebra Eqs.~\ref{eq:Liealg}-\ref{eq:Lierot} is the particle's spin, sometimes called ``intrinsic angular momentum''. 

Isospin is a spin-like quantum number obeying angular momentum algebra with eigenvalues $I$ ($\equiv j$ from Eq.~\ref{eq:j}) and $I_3$ ($\equiv m$ from Eq.~\ref{eq:m}). Since $I_3$ can take values between $-I$ and $+I$, there are in total $2I+1$ possibilities. Historically, isospin was introduced to classify those hadrons that behave identical under the strong interaction into a multiplet with multiplicity $2I+1$, for example:
\begin{eqnarray}
I=\frac{1}{2}&:& (\text{p\,n}) \nonumber\\
I=1&:& (\pi^+ \pi^0 \pi^-) \nonumber\\
I=0&:& (\eta^0) \nonumber\\
I=\frac{3}{2}&:& (\Delta^{++}\Delta^+\Delta^0\Delta^-).
\end{eqnarray}
Initially called ``isotopic spin'', the name isospin can be explained as defining those particles that take the ``same location'' (isotopic in Greek), in analogy to the isotopes in the periodic table of elements. What is ``same number of protons'' there, is ``same isospin'' here. 
One ``obtains a neutron by rotating a proton in isospin space'' (the $\theta$ in Eq.~\ref{eq:Lierot}). Maybe best understood can isospin be in when considering that u-quarks have $I_3=+1/2$ and d-quarks $I_3=-1/2$, which is where they have their names from - ``isospin up'' and ``isospin down''. The quarks of heavier flavors (s, c, b, t) have $I_3=0$, while each carrying their own quantum number, for example strangeness $S$. Not only isospin, also  strangeness, the other ``flavornesses'', and parity are important for the classification scheme of hadrons into multiplets (see also Sec.~\ref{sec:quarks}). The Gell-Mann–Nishijima formula relates isospin, electrical charge $Q$ and strangeness: 
\begin{equation}
Q=I_3+\frac{B}{2}+\frac{S}{2}.
\label{eq:GMN}
\end{equation}
At applicable center-of-mass energies, the heavier flavors have to be added as spelled out in the hypercharge in Eq.~\ref{eq:hypercharge}. 

While $\vec{J}$ was above introduced as generic angular momentum, it usually denotes the total angular momentum of a particle $\vec{J}=\vec{s}+\vec{L}$, with $\vec{s}$ its spin\footnote{Here denoted with a small $s$ to avoid confusion with strangeness $S$} and $\vec{L}$ the orbital angular momentum. 
We have found an example of the coupling of two angular momenta. Couplings between spin and orbital angular momentum are called spin-orbit couplings. For example, the electrons' spin-orbit coupling creates the atom's fine structure in the shell of the hydrogen atom.

The spin-parity quantum number $J^{\mathcal{P}}$ used to classify particles is introduced in Sec.~\ref{sec:pct}.

\subsection{Parity, charge conjugation and time reversal}
\label{sec:pct}
We define the helicity $h$ of a particle, with $\vec{p}$ its 3-momentum vector and $\vec{s}$ its spin vector: 
\begin{equation}
h=\frac{\vec{s}\cdot\vec{p}}{|\vec{s}||\vec{p}|},
\label{eq:appheli}
\end{equation}
and will now take a closer look at three fundamental symmetry operations:
\begin{description}
\item {\bf Parity $\mathcal{P}$}. The operation that ``mirrors'' (changes the sign) of polar vectors (e.g., position and momentum vectors) and pseudo scalars (e.g., helicity, see Eq.~\ref{eq:appheli}), and leaves unchanged axial vectors\footnote{Axial vectors are vectors that point in the direction of a rotational axis, i.e., that have a sense of left and right - ``a right-hand helix stays a right-hand helix'' (e.g., angular momentum, magnetic field); the vector product of two same-kind vectors (polar~$\times$~polar or axial~$\times$~axial) is an axial vector, otherwise polar.} (e.g., angular momentum  $\vec{L}=\vec{r}\times\vec{p}$) and scalars.\footnote{e.g., numbers of things, densities, etc.} $\mathcal{P}=+1$ means even and $\mathcal{P}=-1$ odd parity. Every particle has its intrinsic parity: photons $\mathcal{P}=-1$; mesons $\mathcal{P}=(-1)^{L+1}$; baryons $\mathcal{P}=(-1)^{L}$. One bulk defines a particle's quantum numbers using the notation $J^{\mathcal{P}}$, with $J=s+L$ the total angular momentum from Sec.~\ref{sec:isospin}, or even $J^{\mathcal{PC}}$ (see next item for $\mathcal{C}$). For example, for the photon $J^{\mathcal{PC}}=1^{--}$ and for the $\pi^0$ meson $J^{\mathcal{PC}}=0^{-+}$. 
\item {\bf Charge conjugation $\mathcal{C}$}. The operation that changes the sign of all electrically charged particles. Effectively, it turns a
particle into its anti-particle: $\mathcal{C}|\text{particle}\rangle = |\text{anti-particle}\rangle$. All charge-like quantum numbers are changed (also lepton and baryon number). As $\mathcal{P}$, $\mathcal{C}$ is a discrete symmetry and multiplicative. 
\item {\bf Time reversal $\mathcal{T}$}. The operation that reverses the time arrow of the reaction. For example, the direction of momentum, orbital angular momentum, spin, and magnetic field vectors is reversed. The direction of, e.g., position, force, and electric field vectors is conserved under $\mathcal{T}$.
\end{description}
The weak interaction violates parity symmetry $\mathcal{P}$, often written as ``\cancel{$\mathcal{P}$}''. When we say that a process violates parity, we mean that the process looks differently in a mirror image of the universe.

In 1957, Chien-Shiung Wu (Madame Wu) studied the radioactive beta decay\footnote{On nucleon level: $\text{n}\rightarrow \text{pe}\bar{\nu}_\text{e}$; on quark level: $\text{d}\rightarrow \text{ue}\bar{\nu}_\text{e}$} of $^{60}$Co atoms to $^{60}$Ni. Her experimental finding was that the decay electrons are emitted preferably against the magnetic field direction. Together with total angular momentum conservation, one can conclude that electrons are always left-handed (negative helicity, see Eq.~\ref{eq:appheli}) and anti-neutrinos always right-handed (positive helicity).\footnote{Here, the mass of the electron is neglected and it is assumed to travel at $v=c$. Particles with mass$\neq 0$ can have the wrong helicity, like the right-handed muons from the pion decay (Fig.~\ref{fig:m2}). Because of the smallness of the electron mass, the pion decay into electrons is suppressed.} Any interaction that is parity conserving should equally couple to left- and right-handed particles. The weak interaction for charged currents (CC) is maximally parity violating. The W$^\pm$ couple only to left-handed fermions and right-handed anti-fermions. The Z$^0$ however (neutral weak current NC) couples also to right-handed fermions (partial parity violation), which is a manifestation of the close relation between the Z$^0$ and the photon in electroweak unification.

Also the charge symmetry $\mathcal{C}$ is broken in the weak interaction. What about the combined charge and parity symmetry, $\mathcal{CP}$, i.e., exchanging all particles with their anti-particles while also also looking at the mirrored situation? A left-handed fermion is turned by $\mathcal{CP}$ into a right-handed anti-fermion. There are no other combinations than left-handed fermion or right-handed anti-fermion, and $\mathcal{CP}$ is the symmetry operation between them. The $\mathcal{CP}$ symmetry is conserved in all interactions including the electromagnetic and strong interaction. An exception is the weak interaction, but only for weak neutral K-meson decays ($\text{K}^0$ / $\overline{\text{K}}^0$) and for the weak B-meson sector, where the combined symmetry is violated (\cancel{$\mathcal{CP}$}). If a $\mathcal{CP}$-violating reaction is looked at under time-reversed arrow, the symmetry is restored. The three-fold $\mathcal{CPT}$ symmetry is considered a fundamental property of physical laws.

\subsection{The weak force}
\label{sec:weak}
Examination of Tab.~\ref{tab:conservation} aids in understanding why the weak nuclear force is sometimes referred to as ``the force of change''. Not only can it change one form of particle into another - a nucleus of one chemical element into the nucleus of a different element, or a neutron into a proton, or a meson into a lepton, and others - but also does the weak force not preserve some of the quantities and symmetries that are conserved by all other forces (including gravity, which is not listed in the table). One other prominent feature of the weak force is its slowness (see Tab.~\ref{tab:forces}), which it owes its name to. If it weren't for the slow burning of our Sun's nuclear oven, a process that is driven by the weak process transmuting at high temperatures and high densities a proton into a neutron to form  deuteron, a heavy hydrogen nucleus, pp~$\rightarrow (\text{pn})\text{e}^+\nu_\text{e}$, you would not be reading this - the Sun would have run out of hydrogen ``fuel'' a long time ago.

\section{Nobel Prizes in Physics related to particle physics}
\label{sec:nobel}

\begin{figure}
\begin{center}
\includegraphics[width=0.98\textwidth]{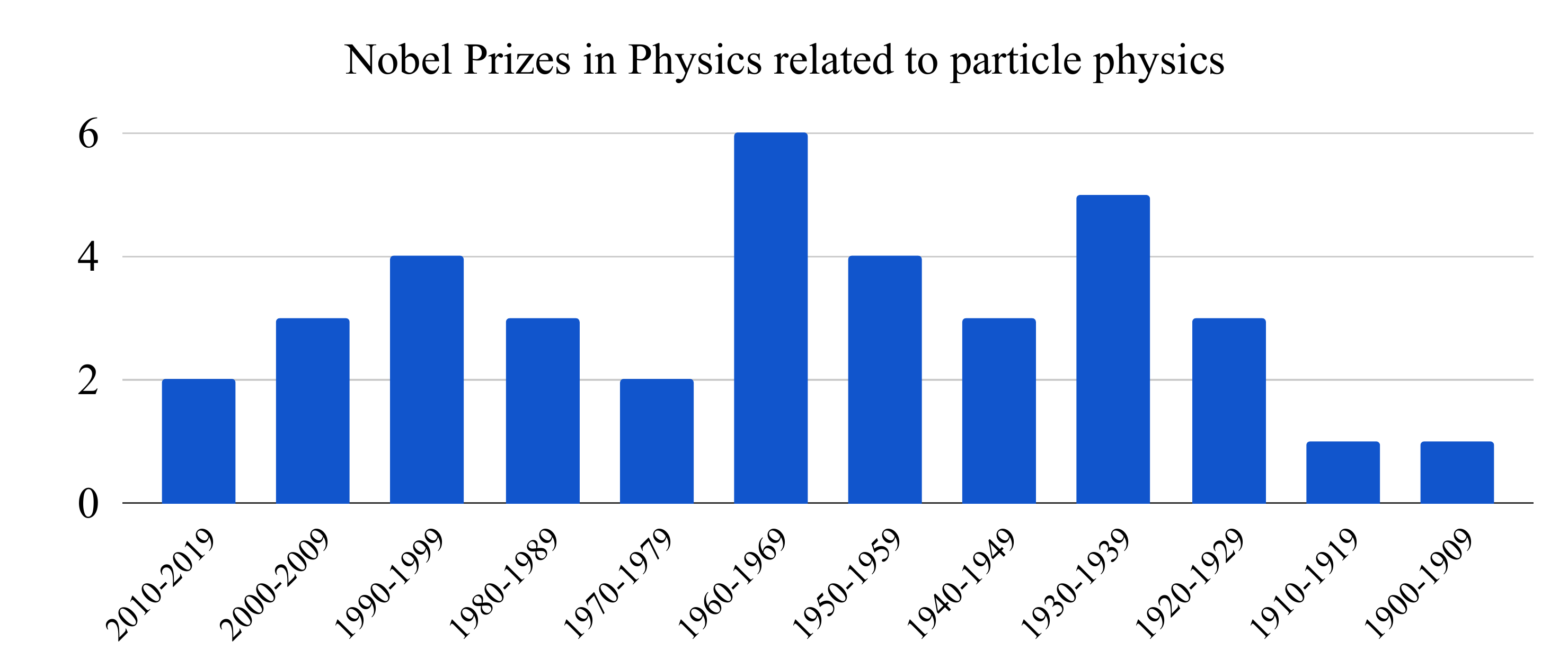}
\end{center}
\caption[Nobel Prizes in Physics related to particle physics]{Nobel Prizes in Physics related to particle physics, per decade.}
\label{fig:nobel}
\end{figure}

Figure~\ref{fig:nobel} shows the number of particle-physics-related Nobel Prizes in Physics per decade that were awarded for achievements in the theory sector, novel instrumentation, or major collaborative experimental efforts. For the early years, also general achievements are counted that were essential for particle physics. 

For more information on each prize, visit the Nobel Prize official web page.\footnote{At \href{https://www.nobelprize.org/prizes/physics/2015/summary}{https://www.nobelprize.org/prizes/physics/XXXX/summary}, where XXXX is the 4-digit year number} The laureates and their achievements are:

\begin{description}
\item {\bf 2015} to Takaaki Kajita and Arthur B. McDonald ``for the discovery of neutrino oscillations, which shows that neutrinos have mass''
\item  {\bf 2013} to Francois Englert and Peter W. Higgs ``for the theoretical discovery of a mechanism that contributes to our understanding of the origin of mass of subatomic particles, and which recently was confirmed through the discovery of the predicted fundamental particle, by the ATLAS and CMS experiments at CERN’s Large Hadron Collider"
\item {\bf 2008} to Yoichiro Nambu ``for the discovery of the mechanism of spontaneous broken symmetry in subatomic physics'' and

to Makoto Kobayashi and Toshihide Maskawa ``for the discovery of the origin of the broken symmetry which predicts the existence of at least three families of quarks in nature''
\item {\bf 2004} David J. Gross, H. David Politzer and Frank Wilczek  ``for the discovery of asymptotic freedom in the theory of the strong interaction''
\item {\bf 2002} to Raymond Davis Jr. and Masatoshi Koshiba ``for pioneering contributions to astrophysics, in particular for the detection of cosmic neutrinos'' and 

to Riccardo Giacconi ``for pioneering contributions to astrophysics, which have led to the discovery of cosmic X-ray sources''
\item {\bf 1999} to Gerardus `t Hooft and Martinus J.G. Veltman ``for elucidating the quantum structure of electroweak interactions in physics''

\item {\bf 1995} “for pioneering experimental contributions to lepton physics” 

to Martin L. Perl ``for the discovery of the tau lepton'' and 

to Frederick Reines ``for the detection of the neutrino''

\item {\bf 1992} to Georges Charpak ``for his invention and development of particle detectors, in particular the multiwire proportional chamber''
\item{\bf 1990} to Jerome I. Friedman, Henry W. Kendall and Richard E. Taylor ``for their pioneering investigations concerning deep inelastic scattering of electrons on protons and bound neutrons, which have been of essential importance for the development of the quark model in particle physics''
\item {\bf 1988} to Leon M. Lederman, Melvin Schwartz and Jack Steinberger ``for the neutrino beam method and the demonstration of the doublet structure of the leptons through the discovery of the muon neutrino''
\item {\bf 1984} to Carlo Rubbia and Simon van der Meer ``for their decisive contributions to the large project, which led to the discovery of the field particles W and Z, communicators of weak interaction''
\item {\bf 1980} to James Watson Cronin and Val Logsdon Fitch ``for the discovery of violations of fundamental symmetry principles in the decay of neutral K-mesons''
\item {\bf 1976} to Burton Richter and Samuel Chao Chung Ting ``for their pioneering work in the discovery of a heavy elementary particle of a new kind''
\item {\bf 1979} to Sheldon Lee Glashow, Abdus Salam and Steven Weinberg ``for their contributions to the theory of the unified weak and electromagnetic interaction between elementary particles, including, inter alia, the prediction of the weak neutral current''
\item {\bf 1969} to Murray Gell-Mann ``for his contributions and discoveries concerning the classification of elementary particles and their interactions''
\item {\bf 1968} to Luis Walter Alvarez ``for his decisive contributions to elementary particle physics, in particular the discovery of a large number of resonance states, made possible through his development of the technique of using hydrogen bubble chamber and data analysis''
\item {\bf 1965} to Sin-Itiro Tomonaga, Julian Schwinger and Richard P. Feynman ``for their fundamental work in quantum electrodynamics, with deep-ploughing consequences for the physics of elementary particles''
\item {\bf 1963} to Eugene Paul Wigner ``for his contributions to the theory of the atomic nucleus and the elementary particles, particularly through the discovery and application of fundamental symmetry principles'' and 

to Maria Goeppert Mayer and J. Hans D. Jensen ``for their discoveries concerning nuclear shell structure''
\item {\bf 1961} to Robert Hofstadter ``for his pioneering studies of electron scattering in atomic nuclei and for his thereby achieved discoveries concerning the structure of the nucleons'' and 

to  Rudolf Ludwig Moessbauer ``for his researches concerning the resonance absorption of gamma radiation and his discovery in this connection of the effect which bears his name''
\item {\bf 1960} to Donald Arthur Glaser ``for the invention of the bubble chamber''
\item {\bf 1959} to Emilio Gino Segre and Owen Chamberlain ``for their discovery of the antiproton''
\item {\bf 1958} to Pavel Alekseyevich Cherenkov, Il'ja Mikhailovich Frank and Igor Yevgenyevich Tamm ``for the discovery and the interpretation of the Cherenkov effect''
\item {\bf 1957} to Chen Ning Yang and Tsung-Dao (T.D.) Lee ``for their penetrating investigation of the so-called parity laws which has led to important discoveries regarding the elementary particles''
\item {\bf 1950} to Cecil Frank Powell ``for his development of the photographic method of studying nuclear processes and his discoveries regarding mesons made with this method''
\item {\bf 1949} to Hideki Yukawa ``for his prediction of the existence of mesons on the basis of theoretical work on nuclear forces''
\item {\bf 1948} to Patrick Maynard Stuart Blackett ``for his development of the Wilson cloud chamber method, and his discoveries therewith in the fields of nuclear physics and cosmic radiation''
\item {\bf 1945} to Wolfgang Pauli ``for the discovery of the Exclusion Principle, also called the Pauli Principle''
\item {\bf 1939} to Ernest Orlando Lawrence ``for the invention and development of the cyclotron and for results obtained with it, especially with regard to artificial radioactive elements''
\item {\bf 1936} to  Victor Franz Hess ``for his discovery of cosmic radiation'' and 

to Carl David Anderson ``for his discovery of the positron''
\item {\bf 1935} to James Chadwick ``for the discovery of the neutron''
\item {\bf 1933} to Erwin Schr\"odinger and Paul Adrien Maurice Dirac ``for the discovery of new productive forms of atomic theory''
\item {\bf 1932} to Werner Karl Heisenberg ``for the creation of quantum mechanics, the application of which has, inter alia, led to the discovery of the allotropic forms of hydrogen''
\item {\bf 1927} to Arthur Holly Compton ``for his discovery of the effect named after him'' and 

to Charles Thomson Rees Wilson ``for his method of making the paths of electrically charged particles visible by condensation of vapour''
\item {\bf 1922} to Niels Henrik David Bohr ``for his services in the investigation of the structure of atoms and of the radiation emanating from them''
\item {\bf 1921} to Albert Einstein ``for his services to Theoretical Physics, and especially for his discovery of the law of the photoelectric effect''
\item {\bf 1918} to Max Karl Ernst Ludwig Planck ``in recognition of the services he rendered to the advancement of Physics by his discovery of energy quanta''
\item {\bf 1903} to Antoine Henri Becquerel ``in recognition of the extraordinary services he has rendered by his discovery of spontaneous radioactivity'' and 

to Pierre Curie and Marie Curie, nee Sklodowska ``in recognition of the extraordinary services they have rendered by their joint researches on the radiation phenomena discovered by Professor Henri Becquerel''
\end{description}

\end{appendices}
\clearpage
\listoftables
\addcontentsline{toc}{section}{List of tables}
\listoffigures
\addcontentsline{toc}{section}{List of figures}

\end{document}